\newif\ifcomment
\commenttrue       

\newif\ifshowcomments
\showcommentstrue

\documentclass[12pt,a4paper]{article}
\usepackage[english]{babel}
\usepackage[ansinew]{inputenc}
\usepackage{amssymb,amsmath,bm,array,dsfont,graphicx,natbib}
\usepackage[colorlinks,citecolor=black,linkcolor=black,urlcolor=black]{hyperref}
\usepackage{authblk}
\usepackage{framed}
\usepackage{xcolor}
\usepackage[most]{tcolorbox}
\usepackage{bbm}
\usepackage{longtable}
\usepackage{rotating}
\usepackage{amsthm}
\usepackage{tcolorbox}
\usepackage[DIV=13]{typearea}
\usepackage[onehalfspacing]{setspace}
\usepackage{lscape}
\usepackage{layout}
\usepackage{booktabs}
\usepackage{subcaption}
\usepackage{nomencl}
\colorlet{shadecolor}{gray!25}
\usepackage{enumitem}

\newenvironment{remark}
{\par\medskip\noindent\textbf{Remark.}\ }
{\par\medskip}

\def\R{\mathbb R}
\newcommand{\norm}[1]{\left\lVert#1\right\rVert}

\newcommand{\bkappa}{\mathbf{\kappa}}

\newcommand{\EW}[2]{\bkappa_{#1}\left(#2\right)}
\newcommand{\EWg}[1]{\bkappa^{(#1)}}
\newcommand{\EWgt}[1]{\tilde{\bkappa}^{(#1)}}
\newcommand{\EWgh}[1]{\hat{\bkappa}^{(#1)}}

\newcommand{\EWgtD}[1]{\diag\left(\EWgt{#1}\right)}
\newcommand{\EWghD}[1]{\diag\left(\EWgh{#1}\right)}

\newcommand{\EWgti}[2]{\tilde{\kappa}_{#2}^{(#1)}}
\newcommand{\EWghi}[2]{\hat{\kappa}_{#2}^{(#1)}}

\newcommand{\newoperator}[3]{\newcommand*{#1}{\mathop{#2}#3}}
\newcommand{\renewoperator}[3]{\renewcommand*{#1}{\mathop{#2}#3}}
\renewoperator{\Re}{\mathrm{Re}}{\nolimits}
\renewoperator{\Im}{\mathrm{Im}}{\nolimits}
\makeatletter
\newcommand{\rd}{\@ifnextchar^{\DIfF}{\DIfF^{}}}
\def\DIfF^#1{%
   \mathop{\mathrm{\mathstrut d}}%
   \nolimits^{#1}\gobblespace}
\def\gobblespace{\futurelet\diffarg\opspace}
\def\opspace{%
   \let\DiffSpace\!%
   \ifx\diffarg(%
   \let\DiffSpace\relax
   \else
   \ifx\diffarg[%
   \let\DiffSpace\relax
   \else
   \ifx\diffarg\{%
   \let\DiffSpace\relax
   \fi\fi\fi\DiffSpace}

\newcommand{\E}{\operatorname{E}}

\newcommand{\Var}{\operatorname{Var}}

\newcommand{\Cov}{\operatorname{Cov}}

\newcommand{\diag}{\operatorname{diag}}

\newoperator{\ip}{\mathrm{int}}{\nolimits}

\newcommand{\tr}{\operatorname{tr}}

\newcommand{\EE}[1]{\E\left(#1\right)}

\newtheorem{theorem}{Theorem}[section]
\newtheorem{assumption}{Assumption}
\newcounter{subassumption}[assumption]
\renewcommand{\thesubassumption}{(\textit{\roman{subassumption}})}
\makeatletter
\renewcommand{\p@subassumption}{\theassumption}
\makeatother
\newcommand{\subasu}{
	\refstepcounter{subassumption}%
	\thesubassumption~\ignorespaces}

\newtheorem{lemma}{Lemma}[section]
\newtheorem{corollary}{Corollary}

\newcommand{\beq}{\begin{equation}}
\newcommand{\eeq}{\end{equation}}
\newcommand{\bal}{\begin{align*}}
\newcommand{\eal}{\end{align*}}

\newcommand{\bvec}{\begin{pmatrix}}
\newcommand{\evec}{\end{pmatrix}}
\newcommand{\bmat}{\begin{bmatrix}}
\newcommand{\emat}{\end{bmatrix}}

\newcommand{\bsmat}{\begin{smallmatrix}}
\newcommand{\esmat}{\end{smallmatrix}}

\makenomenclature
\title{Estimation and Model Selection in Factor Models with Long Memory}
\author[1]{Dominik Ammon\footnote{Corresponding author. E-mail: Dominik.Ammon@ur.de}}
\author[2]{Tobias Hartl}
\author[1]{Rolf Tschernig}

\affil[1]{University of Regensburg, 93053 Regensburg, Germany}
\affil[2]{Maastricht University, 6200 MD Maastricht, The Netherlands}
\date{\today}

\begin{document}
\maketitle

\thispagestyle{empty}
\setcounter{page}{0}
\begin{singlespace}
\paragraph{\bf Abstract.}
Panels in economics, finance and climate science are often driven by common components that differ in memory, such as common trends, cycles, or joint shocks. We study approximate factor models whose common factors are fractionally integrated with heterogeneous and unknown memory and develop the asymptotic theory of principal components for the panel in levels. Our theory builds on the eigenvalues of the sample covariance matrix being ordered by memory asymptotically: a nonstationary factor group with memory $d>1/2$ generates eigenvalues of order $NT^{2d-1}$, whereas stationary factors generate eigenvalues of order $N$. Principal components estimate the factors consistently up to rotation, while adjacent factor groups are separated at a rate given by their memory gap. The information criteria of Bai and Ng remain consistent for the total number of factors, with their penalties acting as thresholds on eigenvalues that are asymptotically ordered by factor memory. When the nonstationary factors lose their variance advantage, estimation in levels remains consistent, whereas after differencing the factors can become undetectable. A Monte Carlo study shows that factor-number selection in levels tolerates more idiosyncratic noise than selection after first differencing. For monthly temperatures at $308$ European weather stations, our methods select two factors. The first is a persistent common warming factor. The estimated common component implies that 2021-2025 was $2.15^\circ\mathrm{C}$ warmer than 1960-1990, with the strongest warming in Eastern Europe. After first differencing, the criterion selects one factor, and loses regional heterogeneity as compared to the model in levels.


\paragraph{\bf JEL-Classification.}
C32, C38, C51, C52
\end{singlespace}
\newpage

\section{Introduction}
\label{sec:introduction}
Applied time series analysis often assumes observables to be driven by components that differ in memory: While macroeconomic variables follow their potential growth path in the long-term, they are driven by the business cycle, season, and noise in the short-run \citep{Harvey1989, Mor2014}. Financial volatility is smooth at low- and rough at high-frequencies \citep{RayTsa2000,AsaMcA2015, GatJaiRo2018}. Climate is subject to warming trends, but also driven by persistent cycles, season, and noise. 
When a panel of $N$ series shares its trends, cycles, and noise, the decomposition implies a factor structure in which some common factors differ in memory. 

Large-panel factor theory does not cover this structure. Instead, for nonstationary panels, the common approach is to difference the data before the principal components are computed \citep{Bai2004Panic,Barigozzi2021, Ergemen2023}. For panels in which the nonstationarity is generated by the common factors, this is suboptimal: (i) integration orders are unobserved and thus must be estimated, so the appropriate degree of pre-differencing is itself subject to estimation error; (ii) when the factors have different integration orders, no single transformation returns a factor model in which all factors are $I(0)$; (iii) the signal of a smooth trend is weakened by taking differences, so that it may no longer be detected in finite samples; (iv) consistency of principal components and of factor-number criteria is not established for factors of heterogeneous memory.\footnote{Section~\ref{subsec:model_comparison} explains why the argument in \citet{Ergemen2023} does not provide it.}

We consider an approximate factor model 
\begin{align*}
	y_t &= \Lambda x_t + u_t = \Lambda^{(1)}x_t^{(1)} + \cdots + \Lambda^{(q)}x_t^{(q)} + u_t,
	\qquad t=1,\ldots,T,
\end{align*}
where each of the $s$ factors $x_{i,t}$ is fractionally integrated with its own memory parameter $d_i$ and its own short-run dynamics, factors of equal memory are collected in $q$ groups $x_t^{(j)}$ with memory $d^{(1)}>\cdots>d^{(q)}$, and the idiosyncratic component $u_t$ is stationary.\footnote{The latter rules out spurious factors, see \cite{OnaWan2021}.} The representation builds on the fractional components model of \citet{Hartl2023}, allows for common components that differ in memory, and nests the vast majority of stationary and nonstationary factor models in the literature, see Section~\ref{subsec:model_comparison} for a comparison. We develop the asymptotic estimation and factor selection theory for principal components analysis (PCA) without requiring prior knowledge about the integration orders or pre-differencing.

The paper makes four contributions.
First, the eigenvalues of the sample covariance matrix of the panel in levels are ordered by memory asymptotically: A nonstationary factor group with memory $d^{(j)}>1/2$ generates eigenvalues of order $NT^{2d^{(j)}-1}$, stationary factors generate eigenvalues of order $N$, and the idiosyncratic eigenvalues remain bounded both in $N$ and $T$. Diverging eigenvalues in $N$ allow us to separate the factors from the idiosyncratic component, differences in memory identify the factor groups from each other.

Second, we prove that principal components computed from the panel in levels estimate the factors consistently up to rotation.
The average squared estimation error is $O_p(N^{-1})$, which under $N/T\to0$ equals the rate of \citet{BaiNg2002} for stationary factors.
For nonstationary factors the error vanishes relative to the size of the factors, at the rate $T^{-(d^{(q^*)}-1/2)}$ for the full nonstationary block, where $d^{(q^*)}$ is the smallest nonstationary memory parameter.
The space spanned by the first $j$ nonstationary groups is separated from the next group at the rate $T^{-(d^{(j)}-d^{(j+1)})}$.
The separation of adjacent groups is therefore governed by the gap between their memory parameters, not by the level of persistence.
This parallels the fractionally cointegrated VAR of \citet{Johansen2008}, whose cointegration parameters are not identified when the cointegration gap is zero.

Third, we show that the penalized criteria of \citet{BaiNg2002}, in the form of the penalized value function and in logarithmic form, remain consistent for the total number of factors when they are applied to the panel in levels.
Because the eigenvalues are ordered by memory, the penalty acts as a threshold: a penalty of order $o(T^{2d^{(j)}-1})$ selects at least the factors of the first $j$ groups with probability approaching one, and the conditions of \citet{BaiNg2002} deliver the full number of factors.

Fourth, we study how differencing affects these results. Without a rescaling of the factors, differencing need not prevent consistent estimation, but it removes the time-series signal that yields the faster rates for nonstationary factors.
The consequences are larger when the nonstationary factors have no variance advantage.
We model this case by scaling the innovations of each nonstationary factor with $T^{1/2-d_i}$, so that the factors remain nonstationary while their sample second moments stay bounded; we call this the diminishing-signal regime.
In levels, principal components and the factor-number criteria remain consistent, because the factors are still pervasive in the cross-section.
After differencing, a nonstationary factor with memory $d_i$ generates an eigenvalue of order $NT^{1-2d_i}$.
If $N$ grows more slowly than $T^{2d_i-1}$, this eigenvalue vanishes, and a factor-number criterion applied to the differenced panel selects the wrong number of factors with probability approaching one.

A Monte Carlo study determines, for each panel size, the largest idiosyncratic standard deviation at which the criterion still selects the true number of factors in at least half of the replications. Within the family of data-generating mechanisms indexed by the idiosyncratic standard deviation, this allows us to identify the least favorable case in which the criterion still is informative.
In both regimes, estimation in levels tolerates more idiosyncratic noise than estimation after first differencing.
Under diminishing signals, the tolerable noise after differencing is close to zero and falls as $T$ grows, indicating that pre-differencing can cause model misspecification.

We apply the model to monthly temperatures at $308$ European weather stations from the GHCN-M database, covering January 1960 to December 2025.
The criterion selects two factors in levels.
The estimated common component implies that $2021$-$2025$ was on average $2.15^\circ\mathrm{C}$ warmer than $1960$-$1990$.
The first factor loads positively at $290$ of the $308$ stations, and its memory estimate from the low-frequency likelihood of \citet{MuellerWatson2008} is $1.15$, with a $95\%$ confidence interval of $[0.81,1.59]$.
The second factor contrasts northern with southern Europe, and its confidence interval for the memory parameter, $[0.31,1.12]$, includes stationary values. After first differencing, the criterion selects one factor. The cumulated common component still implies an average warming of $2.02^\circ\mathrm{C}$, but its regional distribution changes: it assigns less warming to southwestern and central Europe and more to northern Europe. 

The paper is organized as follows.
Section~\ref{sec:model} introduces the model, the assumptions, the two signal regimes and the estimators.
Section~\ref{sec:estimation_model_selection} derives the asymptotic theory for estimation and factor-number selection, first under standard signals and then under diminishing signals.
Section~\ref{sec:monte_carlo} reports the Monte Carlo study, Section~\ref{sec:climate_application} the application to European temperatures, and Section~\ref{sec:conclusion} concludes. Code and data to replicate all results are available at \url{https://github.com/D4monik/long-memory-factor-models}.
\section[A factor model with heterogeneous memory]{A factor model with heterogeneous memory}
\label{sec:model}

We use the following notation: For an $\left(N\times N\right)$ matrix $A$, $\norm{A}_2$ denotes the spectral norm and $\norm{A}_F = \left( \operatorname{tr}(A^T A) \right)^{1/2}$ denotes the Frobenius norm. For a symmetric matrix $A$, $\EW{1}{A} \geq \cdots \geq \EW{N}{A}$ denote its ordered eigenvalues. We write $V(A)_i$ for an eigenvector associated with $\EW{i}{A}$ and, for $1\leq a\leq b\leq N$, collect the eigenvectors corresponding to $\EW{a}{A},\ldots,\EW{b}{A}$ in $V(A)_{a:b}= \left( V(A)_a,\ldots,V(A)_b \right)$. For a random sequence $X_{N,T}$ and a positive deterministic sequence $a_{N,T}$, we write $X_{N,T}=O_p(a_{N,T})$ if $X_{N,T}/a_{N,T}$ is bounded in probability, $X_{N,T}=o_p(a_{N,T})$ if it vanishes in probability, and, for nonnegative random sequences, $X_{N,T}=O_e(a_{N,T})$ if it is bounded away from zero and infinity in probability.

\subsection[Fractional factors and memory groups]{Fractional factors and memory groups}
\label{subsec:model_dynamics}

We consider the factor model
\begin{align}
	y_t = \Lambda x_t + u_t,
	\qquad t = 1,\ldots,T,
	\label{eq:Factormodel_def}
\end{align}
where $y_t$ is an $(N\times 1)$-vector of observed variables, $x_t$ is an $(s \times 1)$-vector of latent factors, $\Lambda$ is the $\left(N \times s\right)$ loading matrix, and $u_t$ is the idiosyncratic component. The common component is denoted by $\chi_t = \Lambda x_t$. 

The factors are allowed to differ in persistence. We model the $i$th factor as a
fractionally integrated process with short-run dynamics and memory
parameter $d_i \in \R$:
\begin{align}
	x_{i,t} &= \mathcal B_{i,+}(L;d_i)\eta_{i,t},
	\qquad
	i=1,\ldots,s .
	\label{eq:factor_process}
\end{align}
Here $\eta_{i,t}$ is the innovation driving the $i$th factor, and
$\mathcal B_{i,+}(L;d_i)$ is a linear filter that combines short-run dynamics with
fractional integration of order $d_i$. Without the $+$-subscript, it exhibits the
polynomial expansion
\begin{align}
	\mathcal B_i(z;d_i) &= \psi_i(z)(1-z)^{-d_i} = \sum_{\ell=0}^{\infty} b_{i,\ell}(d_i)z^\ell,
	&
	b_{i,\ell}(d_i) &= \sum_{j=0}^{\ell} \psi_{i,j}\pi_{\ell-j}(-d_i),
	\label{eq:combined_filter}
\end{align}
where $\psi_i(z)=\sum_{j=0}^{\infty}\psi_{i,j}z^j$ is a short-run filter and
$\pi_\ell(d)$ are the coefficients of the fractional difference operator
\begin{align*}
	(1-L)^d &= \sum_{\ell=0}^{\infty} \pi_{\ell}(d)L^\ell,
	&
	\pi_{\ell}(d) &=
	\begin{cases}
		\frac{\ell-1-d}{\ell}\,\pi_{\ell-1}(d) & \ell=1,2,\ldots,\\[2pt]
		1 & \ell=0 .
	\end{cases}
\end{align*}
The motivation behind \eqref{eq:factor_process} and \eqref{eq:combined_filter} is that
$\psi_i(L)$ governs the transitory dynamics of the factor, whereas $d_i$ governs its
persistence: the higher $d_i$, the more persistent the effect of a past shock
$\eta_{i,t-\ell}$ on $x_{i,t}$. The $+$-subscript in \eqref{eq:factor_process} denotes
the truncation of the filter at $t\leq0$, so that
$
x_{i,t}
=
\sum_{\ell=0}^{t-1} b_{i,\ell}(d_i)\eta_{i,t-\ell} .
$
The truncated filter reflects the type-II definition of fractionally integrated
processes \citep{Marinucci1999} and is required to treat the asymptotically stationary
case alongside the nonstationary case. 

Equation \eqref{eq:factor_process} encompasses the factor specifications in the
literature. For $\psi_i(L)=1$ it reduces to the pure fractional specification, while if
$\psi_i(z)$ represents a stable and invertible finite-order ARMA transfer function, it yields a type-II ARFIMA factor. Applying the truncated fractional difference of order $d_i$
removes the fractional part and returns the short-run filter, since the coefficient
sequences $\pi(d_i)$ and $b_i(d_i)$ convolve to $\psi_i$
\begin{align*}
	\Delta_+^{d_i}x_{i,t}
	&=
	\sum_{m=0}^{t-1}\psi_{i,m}\eta_{i,t-m}
	=
	\psi_{i,+}(L)\eta_{i,t},
\end{align*}
which is an $I(0)$ process. Thus $d_i$ measures the fractional order of differencing
needed to remove the long-run dependence in $x_{i,t}$. Factors with $d_i<1/2$ are
asymptotically stationary, whereas factors with $d_i > 1/2$ are collected among the
nonstationary factors.

Without loss of generality, the factors are ordered by decreasing memory, so that $d_1\geq\cdots\geq d_s$. We collect factors with the same memory parameter into $q$ groups, and denote the size and memory of group $j$ by $s_j$ and $d^{(j)}$ respectively. It follows that $d^{(1)}>\cdots>d^{(q)}$ and $s=\sum_{j=1}^{q}s_j$. 
We denote the corresponding factor vector and loading block by $x_t^{(j)}$ and $\Lambda^{(j)}$, respectively, write $\Lambda^{(1:j)}=[\Lambda^{(1)},\ldots,\Lambda^{(j)}]$ for the first $j$ loading blocks collected together, and set $s^{(1:j)}=\sum_{k=1}^{j}s_k$, with $s^{(1:0)}=0$.
The common component can then be decomposed into a nonstationary part and a stationary part
\begin{align}
	y_t &= \chi_t^{\mathrm{ns}} + \chi_t^{\mathrm{s}} + u_t,
	\label{eq:factor_groups}
	\\
	\chi_t^{\mathrm{ns}}
	&= \sum_{j=1}^{q^*} \Lambda^{(j)}x_t^{(j)},
	\qquad
	\chi_t^{\mathrm{s}}
	=
	\sum_{j=q^*+1}^{q}
	\Lambda^{(j)}x_t^{(j)} .
	\nonumber
\end{align}
Here, $\chi_t^{\mathrm{ns}}$ collects the $q^*$ nonstationary factor groups, whereas $\chi_t^{\mathrm{s}}$ collects the stationary factor groups. The total number of nonstationary factors is $s^*=\sum_{j=1}^{q^*}s_j$. Empty sums are understood as zero, so that $q^*=0$ and $s^*=0$ if all factors are stationary, while $q^*=q$ and $s^*=s$ if all factor groups are nonstationary.

In what follows, parameters of the DGP are denoted with a zero subscript: $\Lambda_0$, $d_{0,i}$, $\Sigma_{u,0}$ and $s_0$ denote the true loading matrix, the true memory parameters, the true idiosyncratic covariance matrix and the true number of factors. Without the subscript, $\Lambda$, $d_i$ and $s$ refer to the model as stated here, and $s$ also serves as the candidate factor number in the model-selection criteria.

\subsection{Relation to existing models}
\label{subsec:model_comparison}
Our factor model framework builds on the four literature strands below.

\paragraph{Factor models with integer integration orders.}
	In approximate factor models, the common component generates eigenvalues of the covariance matrix that diverge with $N$, whereas the idiosyncratic eigenvalues remain bounded \citep{Chamberlain1983}. This separation underlies the estimation of factors by principal components \citep{Forni2001,Stock2002,BaiNg2002,Bai2003} and the criteria for their number \citep{BaiNg2002,Alessi2010,Onatski2010,Ahn2013}, which assume stationary factors. For $I(1)$ factors with stationary idiosyncratic errors, \citet{Bai2004} shows that the common eigenvalues are of order $NT$; \citet{Bai2004Panic}, \citet{PenPon2006}, \citet{Barigozzi2021} and \citet{Barigozzi2024} develop unit-root testing, dynamic factor analysis, impulse-response estimation and inference for nonstationary panels. Our model contains the stationary case for memory below $1/2$ and the $I(1)$ case for $d=1$. Nonstationary fractional memory generates eigenvalues of order $NT^{2d^{(j)}-1}$, so that, in addition to the separation in $N$, the time dimension orders the common components by memory and separates nested factor spaces asymptotically.

\paragraph{Fractional factor models.}
Closest to the present paper are \citet{LucVer2015}, \citet{Cheung2022}, and \citet{Ergemen2023}, who study long memory in factor models. Their factor estimates are based on a pre-differenced data set and build on the asymptotic theory established for stationary data. By contrast, we develop the asymptotic theory for factor estimation in levels, allowing also for nonstationary data. \citet{Ergemen2023} addresses the same problem as the present paper: a large panel whose common factors are fractionally integrated with possibly different memory parameters. His procedure estimates the loadings by principal components applied to the {first-differenced} panel,
following \citet{Barigozzi2021}, projects the panel in levels onto the estimated
loadings, and then estimates the memory of each factor by conditional
sum-of-squares. 

Three things distinguish our paper. First, as will be shown in Corollary~\ref{C:differencing_DS_non_detectability},
 pre-differencing can remove the eigenvalue separation between a nonstationary common factor and the idiosyncratic components, rendering the factor undetectable. In levels the same factor remains pervasive and the factor number is selected consistently.

Second, consistency of the factor estimators has not been established for factors of
heterogeneous memory. \citet{Ergemen2023} claims consistency carries over from the integer-integrated results of \citet{Barigozzi2021}, but his matching argument does not establish the conditions required by \citet{Barigozzi2021}. Their Assumption~1(b) requires the factor vector $x_t$
 to be $I(1)$ with a finite-degree ARIMA representation, and their
Assumption~1(d) requires $\Delta x_t$ to have a finite covariance matrix of full rank.
Individual factors may then be $I(0)$ or $I(1)$, but the differenced factor vector is
stationary. The matching step in \citet{Ergemen2023} verifies that the
{fractionally} differenced factors $\Delta^{d_{0, i}}x_{i,t}$ have short memory. The
principal components, however, are computed from the first differences of the observed
panel, $\Delta y_t$, whose factor component is $\Delta x_{i,t}\sim I(d_{0, i}-1)$, and
integer and fractional differencing coincide only if $d_{0, i}=1$. For $d_{0, i} \neq1$ the
differenced factor is not a finite-order ARMA process as required by
Assumption~1(b), and for $d_{0, i}\geq3/2$ it is even nonstationary. Consistency of the factor
estimators therefore does not follow from the result to which it is attributed, other
than in the case $d_{0, i}=1$, for which it is already proven by \citet{Barigozzi2021}.\footnote{The matching step is Appendix~A.1, p.~1494, of
\citet{Ergemen2023}; the admissible set for the memory parameters is described as
effectively unrestricted on p.~1487. In \citet{Barigozzi2021}, Assumption~1(b) and~1(d) are on
p.~458, the eigenvalue result on p.~459, and the exclusion of the $I(2)$ case in
Remark~3 on p.~459.} 
We solve this problem by deriving the asymptotic estimation theory for principal components when factors may be nonstationary and have heterogeneous memory.

Third, the number of factors is taken as known in this literature or estimated by
criteria developed for stationary panels, such as those of \citet{BaiNg2002}, \citet{Onatski2010} and
\citet{Ahn2013}. We contribute to model specification by showing that the penalized
Bai-Ng criterion remains consistent for $s_0$ when it is applied to the panel in
levels and some factor groups are fractionally nonstationary.
Theorem~\ref{T:Bai_Ng} shows that the penalty then acts as a threshold across the
eigenvalue hierarchy rather than as a single comparison between one common and one
idiosyncratic order: a penalty that is small relative to $T^{2d_0^{(q_0^*)}-1}$ selects
at least the nonstationary factor groups, whereas Bai-Ng-type penalty conditions
deliver the full factor dimension.

\paragraph{Fractional cointegration and the common-trends interpretation.}
	A vector of fractionally integrated series is cofractional if linear combinations of it
	have lower memory than the series themselves. \citet{Johansen2008} develops the
	representation theory for the fractionally cointegrated VAR (FCVAR), \citet{JohNie2012a}
	and \citet{Johansen2019} develop likelihood inference for stationary and nonstationary
	cointegrating relations, \citet{Hualde2011} and \citet{HuaNie2022} study likelihood and
	sum-of-squares estimation, \citet{CheHur2006} estimate cointegration subspaces
	semiparametrically, \citet{RobYaj2002} determine the cointegrating rank by grouping
	estimated memory parameters, and \citet{HuaRob2010} allow several cointegrating subspaces
	with distinct integration orders. The common component of \eqref{eq:Factormodel_def} inherits the cointegration structure
	of the fractional components model \citep[]{Hartl2023}: for $j=1,\ldots,q^*$,
	the projection $\Lambda_\perp^{(1:j)T}y_t$ onto the orthogonal complement of
	$\operatorname{span}(\Lambda^{(1:j)})$ removes the first $j$ factor groups and is
	integrated of order $d^{(j+1)}$, so that the nested complements are cointegration
	subspaces along which memory falls in steps $d^{(j)}-d^{(j+1)}$.
For two memory groups and as many components as variables, \citet[Section~2.1 and
Appendix~A]{Hartl2023} derive the error correction representation of the FCVAR, with
cointegration vectors spanning the orthogonal complement of the first-group loadings,
$d=d^{(1)}$ and cointegration gap $b=d^{(1)}-d^{(2)}$. In our panel, where $N$ exceeds the
number of factors, this gives a correspondence for the common component:
$\operatorname{span}(\Lambda^{(1)})$ takes the role of the common-trends space
$\operatorname{span}(\beta_\perp)$, $\Lambda_\perp^{(1:1)}$ that of the cointegration matrix $\beta$, and the memory
gap $d^{(1)}-d^{(2)}$ that of $b$. With more than two groups the correspondence ends: the
polynomial extension of \citet[Section~4]{Johansen2008} permits only the orders $d$,
$d-b$ and $d-2b$, whereas the increments $d^{(j)}-d^{(j+1)}$ of our hierarchy need not be
equal \citep[Section~2.1]{Hartl2023}.
	The FCVAR nevertheless indicates the limits of estimation. Its error-correction term
	enters through $L_b=1-\Delta^{b}$, which vanishes at $b=0$, so that $\alpha$ and $\beta$
	are not identified without a memory reduction. Likewise, adjacent nonstationary groups in
	our model are separated at the rate $T^{-(d^{(j)}-d^{(j+1)})}$, which is informative only
	if the memory gap is bounded away from zero.

\paragraph{Eigenanalysis and memory-based separation.}
 The idea of separating the common and idiosyncratic component by eigenanalysis underlies the work of  \citet{LamYao2012} for stationary and \citet{PenPon2006} for nonstationary data. \citet{ZhaRobYa2019} identify cointegration spaces by eigenanalysis with possibly different integration orders, including a fractional extension. Their setting is a noiseless rotation of the observed vector, without an idiosyncratic component, and their fractional results are
obtained for fixed dimension $N$. We use a related memory-based separation principle in an
approximate factor model, where the relevant objects are latent factor-loading spaces and
common components in a large panel with pervasive loadings and idiosyncratic errors.
This extends the sources of eigenvalue separation: differences in memory generate
time-series separation, while cross-sectional pervasiveness separates the stationary
common component from idiosyncratic variation. In contrast to \cite{ZhaRobYa2019}, this allows the cross-sectional dimension to grow with the time dimension.

\subsection[Assumptions]{Assumptions}
\label{subsec:assumptions}
We study factor estimation via PCA under the following set of assumptions. 
\begin{assumption}[Factors] \label{ass:factors}

	Let $\eta_t=(\eta_{1,t},\ldots,\eta_{s_0,t})^T$ collect the factor
	innovations, and let $\mathcal F_t=\sigma(\eta_r:r\leq t)$ denote their
	natural filtration.

	\subasu 
	For $i=1,\ldots,s_0$, let
	$\psi_i(z)=\sum_{j=0}^{\infty}\psi_{i,j}z^j$ with $\psi_{i,0}=1$. The $\psi_i(z)$ are deterministic and do not depend on $N$ or $T$.
	There exist constants $C_\psi<\infty$ and $c_\psi>0$ such that
	\[
	\max_{i\leq s_0}
	\sum_{j=0}^{\infty}(1+j)^2|\psi_{i,j}|
	\leq C_\psi,
	\qquad
	\min_{i\leq s_0}|\psi_i(1)|
	=
	\min_{i\leq s_0}
	\left|\sum_{j=0}^{\infty}\psi_{i,j}\right|
	\geq c_\psi.
	\]
	\label{ass:short_run_filters}

	\subasu The factor innovations are stationary and ergodic martingale
	difference sequences with respect to $\mathcal{F}_{t}$, i.e.
$
	\EE{\eta_{i,t}\mid\mathcal{F}_{t-1}}=0,$ and
	$
	0<\EE{\eta_{i,t}^2}<\infty,
$
	for all $i=1,\ldots,s_0$ and \mbox{t = 1,\ldots,T}.
	\label{ass:eta_mds}
	
	\subasu 	\label{ass:eta_future_covariances}
		The conditional covariance matrix of each future innovation vector is
		invariant to current information, i.e.
		$
		\EE{\eta_k\eta_k^T\mid\mathcal F_t}
		=
		\EE{\eta_k\eta_k^T}$, $k>t$.

	\subasu The factor innovations have finite and stationary moments up to order $4$ and have bounded fourth-order cumulant spectral densities. \label{ass:eta_fourth_moments}

	\subasu 
	Whenever $1/2<d_{0,i}\leq3/4$, there exists an even integer
		$m_i$ satisfying
		\[
		m_i>
		\frac{2}{2d_{0,i}-1},
		\]
		such that $\eta_{i,t}$ has finite and stationary moments up to
		order $m_i$ and possesses a bounded $m_i$th-order spectral density.
	\label{ass:eta_higher_moments}
\end{assumption}

Assumption~\ref{ass:short_run_filters} restricts the short-run dynamics to
	deterministic filters with sufficiently rapidly decaying coefficients. This includes standard ARMA specifications, whose impulse-response coefficients decay geometrically.
	The lower bound on the long-run multipliers $\psi_i(1)$ is the substantive
	restriction: it rules out cancellation of the fractional pole at frequency zero, so
	that $d_{0,i}$ remains the effective memory of the $i$th factor.

Assumptions~\ref{ass:eta_mds}--\ref{ass:eta_fourth_moments} are standard
	martingale-difference and moment conditions. For nonstationary factors with
	$d_{0,i}>3/4$, together with Assumption~\ref{ass:one_indep_all}, they are
	sufficient for the fractional invariance principle of \citet{Hosoya2005}, including positive definiteness of the innovation covariance matrix. Assumption~\ref{ass:eta_higher_moments} strengthens them when $1/2<d_{0,i}\leq3/4$. 
	 The required order $m_i$ diverges as $d_{0,i}\to1/2$.
	
	For stationary factor groups, $d_{0,i}<1/2$ together with
	Assumption~\ref{ass:short_run_filters} implies square summability of
	the combined coefficients $b_{i,\ell}(d_{0,i})$. The corresponding
	type-II process is therefore asymptotically equivalent in mean square
	to its stationary type-I counterpart, and stationarity and ergodicity
	yield convergence of sample second moments by the Birkhoff-Khinchin
	ergodic theorem \citep[Theorem 1]{CornfeldFominSinai1982}.

\begin{assumption}[Loadings] \label{ass:loadings}
	
	\subasu The factors are strong, i.e.\ $ \frac{\Lambda_0^T\Lambda_0}{N} \to \Sigma_{\Lambda,0} $ as $N\to\infty$, where $\Sigma_{\Lambda,0}$ is positive definite. \label{ass:strong_factors}
	
	\subasu The loading matrix $\Lambda_0$ has full column rank $s_0$ for all $N$. \label{ass:Lambda_full_rank}
\end{assumption}
Assumption \ref{ass:loadings} is standard in the approximate factor-model literature and ensures that the common factors can be separated from the idiosyncratic component in the cross-sectional dimension. Part \ref{ass:strong_factors} rules out weak factors. Since $\Lambda_0^T\Lambda_0/N$ converges to a positive definite matrix, the common component generates eigenvalues of cross-sectional order $N$. Part \ref{ass:Lambda_full_rank} excludes redundant factors and guarantees that the factor dimension $s_0$ is minimal.

\begin{assumption}[Idiosyncratic errors] \label{ass:idio_errors}
	
	\subasu The idiosyncratic errors are covariance stationary, 
$
	\EE{u_t}=0$,  $\EE{u_tu_t^T}=\Sigma_{u,0}, 
	$
	where $\Sigma_{u,0}$ does not depend on $t$. \label{ass:u_stationary}
	
	\subasu The fourth-order cumulants of the idiosyncratic errors are uniformly
	absolutely summable over lags: there exists a constant $C_4<\infty$,
	independent of $N$, such that
	\[
	\sup_{l,l'\leq N}\sup_t
	\sum_{h=-\infty}^{\infty}
	\left|
	\operatorname{cum}
	(u_{l,t},u_{l',t},u_{l,t-h},u_{l',t-h})
	\right|
	\leq C_4.
	\] \label{ass:u_fourth_cumulants}
	\subasu The autocovariances and cross-autocovariances of the idiosyncratic errors are uniformly absolutely summable: there exists a constant $C_2<\infty$, independent of $N$, such that
	\[
	\sup_{l,l'\leq N} \sum_{h=-\infty}^{\infty} \left| 
	\operatorname{Cov} \left( u_{l,t},u_{l',t-h} \right) \right| \leq C_2 .	\] 
	\label{ass:u_autocov_summability}
	
	\subasu The largest eigenvalue of $\Sigma_{u,0}$ is uniformly bounded. That is, there exists a constant $C_u<\infty$, independent of $N$, such that \[ \EW{1}{\Sigma_{u,0}} \leq C_u . \] \label{ass:Sigma_u}
	\subasu  Let
	\[
	\widetilde{\Sigma}_{u} = \frac{1}{T} \sum_{t=1}^T u_tu_t^T
	\]
	denote the sample covariance matrix of the true idiosyncratic component.
	$\widetilde{\Sigma}_{u}$ is uniformly
	well-conditioned with probability approaching one. There exist constants
	$0<\widetilde c_u<\widetilde C_u<\infty$, independent of $N$ and $T$, such
	that
	\[
	\Pr\left( \widetilde c_u \leq \EW{N}{\widetilde{\Sigma}_{u}} \leq \EW{1}{\widetilde{\Sigma}_{u}}
	\leq \widetilde C_u \right) \longrightarrow 1.
	\]
	\label{ass:Eigenvalue_Sigma_u_tilde}\vspace{-1\baselineskip}
\end{assumption}
 Assumptions~\ref{ass:u_fourth_cumulants} and \ref{ass:u_autocov_summability} allow for weak serial and cross-sectional dependence in the idiosyncratic component.
They are used in Lemma~\ref{L:Sigma_u_hat_rate} to obtain the covariance
approximation
\[
\frac{1}{N} \norm{ \widetilde{\Sigma}_{u} - \Sigma_{u,0} }_F = O_p\left(T^{-1/2}\right).
\]
The constants $C_4$ and $C_2$ must not depend on $N$ to ensure that the estimation error remains controlled as the cross-sectional dimension grows.
 Assumption~\ref{ass:Sigma_u} is the standard approximate-factor restriction that the idiosyncratic component does not generate pervasive eigenvalues, as in \citet{Chamberlain1983} and \citet{BaiNg2002}.

 Assumption~\ref{ass:Eigenvalue_Sigma_u_tilde} is the sample counterpart, and is two-sided. Its lower bound is stronger than what the factor-number selection argument of \citet{BaiNg2002} requires. 
In our proofs, we invert the empirical covariance matrix of the true $u_t$, which requires a lower bound on its eigenvalues.

\begin{remark}[Errors satisfying Assumption~\ref{ass:idio_errors}]
	\label{rem:idiosyncratic_example}
	Let $\varepsilon$ be an $\left(N\times T\right)$ matrix with i.i.d.\ entries of mean zero, unit variance,
	and finite fourth moment, and define $v=\varepsilon B_T$ and $u=A_Nv$, where
	$B_T^TB_T=\Gamma_T=\bigl(\gamma_u(t-s)\bigr)_{t,s\leq T}$. Assume that
	$\gamma_u(h)$ is absolutely summable, the corresponding spectral density $f_u$ is bounded
	away from zero and infinity, and the eigenvalues of $A_NA_N^T$ lie in $[a,\bar a]$ with
	$0<a\leq\bar a<\infty$. Then the rows of $v$ are independent copies of the same serially
	correlated process, while $A_N$ allows for contemporaneous correlation and heteroskedasticity
	across the panel.
	Since $\EE{u_tu_{t-h}^T}=\gamma_u(h)A_NA_N^T$, Assumptions~\ref{ass:u_stationary},
	\ref{ass:u_autocov_summability}, and \ref{ass:Sigma_u} hold. Finite fourth moments of
	the i.i.d.\ entries of $\varepsilon$ imply Assumption~\ref{ass:u_fourth_cumulants}.

Collecting the observations gives $u=A_N\varepsilon B_T$, so that
$\widetilde{\Sigma}_{u}
=A_N\bigl(T^{-1}\varepsilon B_TB_T^T\varepsilon^T\bigr)A_N^T$.
The eigenvalues of $B_TB_T^T$ are those of $\Gamma_T$ and therefore lie between the
infimum and the supremum of $2\pi f_u$, uniformly in $T$ \citep{GreSze1958}. Hence the
extreme eigenvalues of
$T^{-1}\varepsilon B_TB_T^T\varepsilon^T$ are bounded by the corresponding extreme
eigenvalues of $B_TB_T^T$ and $T^{-1}\varepsilon\varepsilon^T$. Since both extreme
eigenvalues of $T^{-1}\varepsilon\varepsilon^T$ converge to one when $N/T\to0$ by
\citet{BaiYin1993}, Part~\ref{ass:Eigenvalue_Sigma_u_tilde} follows.

Only the common autocovariance function is restrictive: it makes the cross-sectional
correlation pattern the same at every lag, so that
$\Var(\operatorname{vec}u)=\Gamma_T\otimes\bigl(A_NA_N^T\bigr)$, as in
\citet{Onatski2012}.
\end{remark}

\begin{remark}[Idiosyncratic memory]
	\label{rem:idiosyncratic_memory}
Assumption~\ref{ass:u_stationary} requires the errors to be covariance stationary. Analogously to \citet{Bai2004Panic} and \citet{Ergemen2023}, it can be relaxed to common idiosyncratic memory of order $d_{u, 0}$, $u_t^* \sim I(d_{u, 0})$, by pre-differencing the observable data: Let the observable panel be $y^*_t = \Lambda_0 x^*_t + u^*_t$. Then, applying $d_{u, 0}$-th differences to the left and right 
	and defining $y_t =\Delta_+^{d_{u, 0}} y^*_t $, $x_t =\Delta_+^{d_{u, 0}} x^*_t $, $u_t =\Delta_+^{d_{u, 0}} u^*_t $
	gives 
	\begin{align*}
		\Delta_+^{d_{u, 0}}y_t^* &= \Lambda_0 \Delta_+^{d_{u, 0}}x_t^* + \Delta_+^{d_{u, 0}}u_t^* , \\
		y_t &= \Lambda_0 x_t + u_t.
	\end{align*}
	Differencing has a price: to satisfy Assumption~\ref{ass:memory}, it requires $d_{u,0}<\min_i d_{0,i}+1/2$, and if it leaves
	every factor stationary, the separation between groups in the time dimension is lost; see
	Corollary~\ref{C:differencing_DS_non_detectability}. Ignoring the persistence is no remedy,
	since \citet{OnaWan2021} show that it generates spurious factors.
\end{remark}

\begin{assumption}[Independence] \label{ass:one_indep_all}
	The idiosyncratic errors $u_{t'}$ and the factor innovations $\eta_t$ are stochastically independent for all $t,t'=1,\ldots,T$. Moreover, $\eta_{i,t}$ is stochastically independent of $\eta_{i',t'}$ for all $i\neq i'$, $i,i'=1,\ldots,s_0$, and all $t,t'=1,\ldots,T$.
\end{assumption}
Assumption~\ref{ass:one_indep_all} is strong but standard in nonstationary factor models; \citet{Bai2004} likewise assumes mutual independence of the loadings, the innovations driving the common stochastic trends, and the idiosyncratic errors. The two parts of Assumption~\ref{ass:one_indep_all} do different work. Independence of $u_{t'}$ and $\eta_t$ removes the factor-idiosyncratic cross terms from the covariance decomposition of $y_t$; independence across $i$ makes $\Sigma_{\eta,0}$ diagonal and positive definite, as required by the joint fractional invariance principle. Both could be weakened to orthogonality or weak-dependence conditions, at the cost of less transparent eigenvalue arguments.

\begin{assumption}[Memory] \label{ass:memory}
	The first $s_0^*$ factors are nonstationary and satisfy
	\[
	d_{0,i}>1/2,
	\qquad
	i=1,\ldots,s_0^* .
	\]
	If stationary factors are present, their memory parameters satisfy
	\[
	-1/2 < d_{0,i} < 1/2,
	\quad
	i=s_0^*+1,\ldots,s_0 .
	\]
\end{assumption}
Assumption~\ref{ass:memory} ensures that nonstationary factors have diverging sample second moments, whereas stationary factors have finite second moments. The strict inequality $d_{0,i}>1/2$ excludes the logarithmic boundary case $d_{0,i}=1/2$. For stationary factors, $d_{0,i}<1/2$ together with Assumption~\ref{ass:short_run_filters} guarantees square summability of the combined coefficients $b_{i,\ell}(d_{0,i})$, while $d_{0,i}>-1/2$ keeps the process in the invertible fractional domain.
Within these bounds the most persistent stationary factor may exhibit long memory, $0<d_{0,s_0^*+1}<1/2$.

\subsection{Diminishing signals}
\label{subsec:signal_regimes}
Factor selection and estimation are relatively easy when the factor contribution dominates the sample covariance matrix of the observed panel and hard when its signal is weak. Section \ref{sec:monte_carlo} investigates how weak that signal may become before selection breaks down. This subsection lays the foundations: we introduce a modification to the factor equation \eqref{eq:factor_process}, which balances the (otherwise diverging) signal of the nonstationary factors and the noise component in the time domain. Loadings, memory parameters, and the idiosyncratic component are left unchanged.

The factors introduced in \eqref{eq:factor_process} are fractionally integrated processes with factor-specific short-run filters,
$x_{i,t}=\mathcal B_{i,+}(L;d_{0,i})\eta_{i,t}$, and memory $d_{0, i}$. For $\Var(\eta_{i, t}) = O_e(1)$,
Lemma~\ref{L:factor_second_moment_limits} and the ergodic theorem imply
\begin{align*}
	\frac{1}{T} \sum_{t=1}^T x_{i,t}^2 = \begin{cases}
		O_e\left(T^{2d_{0,i}-1}\right),
		& d_{0,i}>1/2, \\
		O_e(1),
		& d_{0,i}<1/2,
	\end{cases} 
\end{align*}
so that nonstationary factors carry a time series signal that grows with $T$,
on top of the cross-sectional signal that both groups share.

 Assumption \ref{ass:diminishing_signals} removes this time series advantage exactly: Rescaling the innovations of the nonstationary factors balances the signals of all factors and the noise component.
\begin{assumption}[Diminishing signals]
	\label{ass:diminishing_signals}
	For each nonstationary factor $i=1,\ldots,s_0^*$, the innovation is given by
	\begin{align*}
		\eta_{i,t,T} &= T^{-d_{0,i}+1/2}\eta_{i,t},
	\end{align*}
	where $\eta_{i,t}$ satisfies Assumption~\ref{ass:factors}. The corresponding
	factor is
	\begin{align*}
		x_{i,t,T} &= \mathcal B_{i,+}(L;d_{0,i})\eta_{i,t,T}.
	\end{align*}
	The stationary factors $i=s_0^*+1,\ldots,s_0$ remain unchanged.
\end{assumption}
Since $\mathcal B_{i,+}(L;d_{0,i})$ is linear,
\begin{align}
	x_{i,t,T} &= T^{-d_{0,i}+1/2}x_{i,t}, 	\qquad
	i=1,\ldots,s_0^* ,
	\label{eq:diminishing_factor_scaling}
\end{align}
so that $T^{-1}\sum_{t=1}^T x_{i,t,T}^2=O_e(1)$.

A nonstationary factor generates an eigenvalue of order $NT^{2d_{0,i}-1}$ in the standard regime and of order $N$ under Assumption~\ref{ass:diminishing_signals}, where it is separated from stationary variation by cross-sectional pervasiveness alone. Building on Assumption~\ref{ass:diminishing_signals}, Section~\ref{sec:monte_carlo} searches for the least favorable signal-to-noise ratio such that factor selection still yields credible results. This reflects the critique of  \citet{MUeller2008}, who shows that $I(0)$/$I(1)$ discrimination becomes infeasible once the nonstationary signal becomes sufficiently weak.

\subsection[Principal components and the factor-number criterion]{Principal components and the factor-number criterion}
\label{subsec:pca_factor_number}
We estimate loadings and factors by principal components computed from the sample covariance matrix
\begin{align}
	\widehat{\Sigma}_{y,T}
	&=
	\frac{1}{T}
	\sum_{t=1}^T
	y_t y_t^T.
	\label{eq:sample_covariance_y}
\end{align}
Let $\EW{1}{\widehat{\Sigma}_{y,T}}\geq\cdots\geq\EW{N}{\widehat{\Sigma}_{y,T}}$ denote its ordered eigenvalues, and let $V(\widehat{\Sigma}_{y,T})_{1:s}$ collect the corresponding eigenvectors. For a candidate number of factors $s$, the estimators are given by
\begin{align}
	\widehat{\Lambda}_s
	&= \sqrt{N}\,
	V(\widehat{\Sigma}_{y,T})_{1:s},
	&
	\frac{1}{N} \widehat{\Lambda}_s^T \widehat{\Lambda}_s &= I_s .
	\label{eq:pca_loading_normalization} 
\\ 
		\widehat{x}_{s,t} &= \bigl( \widehat{\Lambda}_s^T \widehat{\Lambda}_s \bigr)^{-1}
		\widehat{\Lambda}_s^T y_t = \frac{1}{N} \widehat{\Lambda}_s^T y_t ,
		&
		t &= 1,\ldots,T .
		\label{eq:pca_factor_estimator}
\end{align}
Therefore, the factor estimator corresponds to the least-squares coefficients from regressing $y_t$ on the loading estimator. By construction, the estimated factors are orthogonal, and their sample second moments equal the leading eigenvalues divided by $N$
\begin{align}
	\frac{1}{T} \sum_{t=1}^T \widehat{x}_{s,t} \widehat{x}_{s,t}^T
	&= \frac{1}{N} \operatorname{diag} \left(
	\EW{1}{\widehat{\Sigma}_{y,T}}, \ldots, \EW{s}{\widehat{\Sigma}_{y,T}}
	\right).
	\label{eq:pca_factor_second_moment_normalization}
	\end{align}
Since $\Lambda x_t = (\Lambda H^{-1})(H x_t)$ for any nonsingular matrix $H$, factors and loadings are identified only up to rotation. The normalizations \eqref{eq:pca_loading_normalization} and \eqref{eq:pca_factor_second_moment_normalization} determine the estimators but impose no restrictions on the true factors and loadings \citep{Barigozzi2026}. Our results therefore establish consistency of the factor estimator up to a random rotation. Otherwise our results concern objects that do not depend on the rotation: the factor-loading space, the associated projection and the common component.

 The fitted common component is the projection of $y_t$ onto the estimated loading space
 \begin{align*}
 	\widehat{\chi}_{s,t} &= \widehat{\Lambda}_s \widehat{x}_{s,t}
 	= \widehat P_s y_t,
 	&
 	\widehat P_s &= \widehat{\Lambda}_s \left( \widehat{\Lambda}_s^T \widehat{\Lambda}_s \right)^{-1}
 	\widehat{\Lambda}_s^T
 	= \frac{1}{N} \widehat{\Lambda}_s \widehat{\Lambda}_s^T .
 	\end{align*}
 Thus, PCA minimizes the average squared residual
 \begin{align}
 	\mathcal V(s) &= \frac{1}{NT} \sum_{t=1}^T
 	\left\| y_t-\widehat{\chi}_{s,t} \right\|^2_2
 	= \frac{1}{NT} \sum_{t=1}^T \left\| \left( I_N-\widehat P_s \right) y_t \right\|^2_2 ,
 	\label{eq:pca_value_function}
 \end{align}
which equals the average of the discarded eigenvalues
\begin{align}
	\mathcal V(s) &= \frac{1}{N} \sum_{j=s+1}^N \EW{j}{\widehat{\Sigma}_{y,T}}
	= \frac{1}{N} \left[ \operatorname{tr} \left(
	\widehat{\Sigma}_{y,T} \right) - \sum_{j=1}^s \EW{j}{\widehat{\Sigma}_{y,T}} \right].
	\label{eq:value_function_eigenvalue_representation}
\end{align}
Each additional component therefore reduces the value function by its eigenvalue divided by $N$
\begin{align}
	\mathcal V(s-1)-\mathcal V(s) &= \frac{1}{N} \EW{s}{\widehat{\Sigma}_{y,T}}.
	\label{eq:value_function_drop}
\end{align}
We select the number of factors by the penalized principal component criterion of \citet{BaiNg2002}. For a maximal candidate number $s_{\max}$, define
\begin{align}
	PC(s) &= \mathcal V(s) +	s\,g(N,T),
	\qquad
	s=0,1,\ldots,s_{\max},
	\label{eq:PC_criterion}
\end{align}
where $g(N,T)>0$ denotes the penalty term. The estimated number of factors is
\begin{align}
	\widehat{s} &= \arg\min_{0\leq s\leq s_{\max}} PC(s).
	\label{eq:factor_number_estimator}
\end{align}
As $N,T\to\infty$, the penalty satisfies
\begin{align}
	g(N,T) 	&\to 0,
	&
	\min\{N,T\}	g(N,T) &\to \infty .
	\label{eq:BN_penalty_conditions}
\end{align}
The first condition ensures that the penalty does not dominate the reduction in the value function obtained by including a true factor. The second condition ensures that the penalty dominates the spurious reduction obtained by adding components that only capture idiosyncratic variation.

Equations \eqref{eq:value_function_eigenvalue_representation} and \eqref{eq:value_function_drop} show that the criterion is driven by the empirical spectrum of the sample covariance matrix. If the population eigenvalues associated with the common and idiosyncratic components were known, the number of factors could be read from their separation. In applications, these eigenvalues are unobserved. The asymptotic results below show that the empirical spectrum contains enough information to estimate the number of factors consistently in the fractional setting, both when the common factors are stationary and when some factor groups are nonstationary.

\section[Asymptotic theory for factor estimation and selection]{Asymptotic theory for factor estimation and \mbox{selection}}
\label{sec:estimation_model_selection}

This section develops the asymptotic theory of principal component estimation and factor-number selection for the model of Section~\ref{sec:model}. The results are
organized around the eigenvalue orders of the sample covariance matrix $\widehat{\Sigma}_{y,T}$ in \eqref{eq:sample_covariance_y}. We call the model of Section~\ref{sec:model} without Assumption~\ref{ass:diminishing_signals} the \emph{standard-signal regime}. In that regime the eigenvalues form a staircase, collected in
Lemma~\ref{L:level_covariance_eigenvalue_orders}: The $s_j$ factors of a nonstationary group with memory $d_0^{(j)}$ generate eigenvalues of order $NT^{2d_0^{(j)}-1}$, the stationary common factors generate eigenvalues of order $N$, and the idiosyncratic eigenvalues remain of order one. Cross-sectional pervasiveness therefore allows us to separate factors from idiosyncratic noise, whereas differences in memory allow us to identify the different factor groups. 

Theorem~\ref{T:factor_convergence} shows that PCA in levels is consistent for the factor space up to rotation, at rates that follow the staircase: $O_p(N^{-1/2})$ for the full factor vector, and, for the nonstationary blocks, relative rates are governed by the memory gaps $d_0^{(j)}-d_0^{(j+1)}$, so that the factor group spaces are separated the more sharply the greater their memory gap.
Theorem~\ref{T:Bai_Ng} shows that the \citet{BaiNg2002} criterion remains consistent
when it is applied in levels, and that its penalty acts as a threshold placed in the
staircase: any penalty of order $o(T^{2d_0^{(q_0^*)}-1})$ already rules out missing a
nonstationary factor asymptotically, while recovering the full dimension $s_0$ requires the
\citet{BaiNg2002} penalty conditions. Pre-differencing the data is shown to worsen the convergence rates.

Section~\ref{subsec:DS_regime} considers the least favorable scenario in which the nonstationary factors are rescaled to neutralize the time-series eigenvalue growth that separates them in the standard-signal regime, thereby flattening the eigenvalue staircase. The attenuation relative to the standard-signal regime is stronger for factors with larger memory parameters. Under this \textit{diminishing-signal regime}, the nonstationary eigenvalues fall back to order $N$, and Theorem~\ref{T:factor_convergence_DS} and Corollary~\ref{C:Bai_Ng_DS} show that cross-sectional pervasiveness alone still suffices: PCA and the factor-number criterion remain consistent in levels, now at the stationary rate. Once data are differenced, separation from idiosyncratic variation in the \textit{diminishing-signal regime} is no longer guaranteed and depends on the relative growth rates of $N$ and $T$. If $N$ grows too slowly relative to $T$ for at least one nonstationary factor, a criterion applied to the differenced panel selects the wrong dimension with probability approaching one.

\subsection[The standard-signal regime]{The standard-signal regime}
\label{subsec:standard_regime}

Let

\[
z_t = x_t^{(1:q_0^*)} = \left( x_t^{(1)T},\ldots,x_t^{(q_0^*)T} \right)^T
\]
denote the full nonstationary factor vector, and let
\[
x_t^{(1:j)} = \left( x_t^{(1)T},\ldots,x_t^{(j)T} \right)^T,
\qquad j=1,\ldots,q_0^*-1,
\]
denote the vector collecting the first $j$ nonstationary factor groups. The corresponding empirical PCA scores are the estimator \eqref{eq:pca_factor_estimator} evaluated at the associated candidate dimensions
\[
\widehat{x}_t = \widehat{x}_{s_0,t} = N^{-1/2} V\left(\widehat{\Sigma}_{y,T}\right)_{1:s_0}^T y_t,
\qquad
\widehat{z}_t = \widehat{x}_{s_0^*,t} = N^{-1/2} V\left(\widehat{\Sigma}_{y,T}\right)_{1:s_0^*}^T y_t,
\]
and, for $j=1,\ldots,q_0^*-1$,
\[
\widehat{x}_t^{(1:j)} = \widehat{x}_{s^{(1:j)},t} = N^{-1/2} V\left(\widehat{\Sigma}_{y,T}\right)_{1:s^{(1:j)}}^T y_t .
\]

\begin{theorem}[Consistency of estimated factors]
	\label{T:factor_convergence}
	 Under Assumptions~\ref{ass:factors}--\ref{ass:memory} and  $N,T\to\infty$ with $N/T\to0$, there exist random nonsingular matrices $H_{0,T}$, $H_{z,T}$, and $H_{j,T}$ whose singular values are bounded away from zero and infinity in probability, such that the following statements hold.
	
	First, the full factor vector is estimated up to rotation and scale
	\begin{equation}
		\label{eq:factor_convergence_full}
		\widehat{x}_t = H_{0,T}x_t + e_{0,t},
		\qquad
		\norm{e_{0,t}}_F = O_p\left(N^{-1/2}\right),
	\end{equation}
	and
	\begin{equation}
		\label{eq:factor_convergence_full_average}
		\frac{1}{T} \sum_{t=1}^T
		\norm{
			\widehat{x}_t-H_{0,T}x_t
		}_F^2 =	O_p\left(N^{-1}\right).
	\end{equation}
	
	Second, the full nonstationary factor vector is estimated up to rotation and scale
	\begin{equation}
		\label{eq:factor_convergence_z} \widehat{z}_t
		= H_{z,T}z_t + e_{z,t},
		\qquad
		\norm{e_{z,t}}_F = O_p(1).
	\end{equation}
	Its average relative distortion satisfies
	\begin{equation}
		\label{eq:factor_convergence_z_average_relative}
		\left(
		\frac{\sum_{t=1}^T \norm{ \widehat{z}_t-H_{z,T}z_t }_F^2}
		{
			\sum_{t=1}^T\norm{ H_{z,T}z_t }_F^2}\right)^{1/2}
		= O_p\left(T^{-d_0^{(q_0^*)}+1/2}\right).
	\end{equation}
	
	Third, for each $j=1,\ldots,q_0^*-1$, the cumulative factor vector associated with the first $j$ nonstationary groups is estimated up to rotation and scale
	\begin{equation}
		\label{eq:factor_convergence_group}
		\widehat{x}_t^{(1:j)} = H_{j,T}x_t^{(1:j)} + e_{j,t},
		\qquad
		\norm{e_{j,t}}_F = O_p\left(t^{d_0^{(j+1)}-1/2} \right),
	\end{equation}
	and its average relative distortion satisfies
	\begin{equation}
		\label{eq:factor_convergence_group_average_relative}
		\left(
		\frac{\sum_{t=1}^T\norm{\widehat{x}_t^{(1:j)} -H_{j,T}x_t^{(1:j)}}_F^2}
		{
			\sum_{t=1}^T\norm{H_{j,T}x_t^{(1:j)}}_F^2
		}
		\right)^{1/2}
		=O_p\left(T^{-d_0^{(j)}+d_0^{(j+1)}}
		\right).
	\end{equation}
\end{theorem}

Theorem~\ref{T:factor_convergence} is proven in Appendix~\ref{sec:theorems}.

Equations~\eqref{eq:factor_convergence_full} and \eqref{eq:factor_convergence_full_average} yield pointwise and average squared errors of orders $O_p(N^{-1/2})$ and $O_p(N^{-1})$, respectively. Under $N/T\to0$, these coincide with the pointwise rate $\min\{\sqrt{N},T\}^{-1}$ of \citet{Bai2003} and the average squared-error rate $\min\{N,T\}^{-1}$ of \citet{BaiNg2002} for stationary factor models, and recover the corresponding rates of \citet{Bai2004} for $I(1)$ factors.

For nonstationary factors, \eqref{eq:factor_convergence_group} shows that the absolute estimation error need not vanish pointwise, since the contribution of the subsequent nonstationary factor groups may itself diverge. Consistency is therefore expressed relative to the size of the rotated factor block. Equation~\eqref{eq:factor_convergence_group_average_relative} shows that this relative error vanishes at a rate determined by the gap $d_0^{(j)}-d_0^{(j+1)}$ between adjacent memory groups. For the full nonstationary block, \eqref{eq:factor_convergence_z} and \eqref{eq:factor_convergence_z_average_relative} show that the relative error vanishes at a rate driven by the weakest nonstationary group. The groupwise results are stated for cumulative factor blocks because the cumulative factor-loading spaces are preserved, whereas an individual intermediate loading block need not span the corresponding original loading space; see the discussion preceding \eqref{eq:QR_cumulative_loading_spaces}. The dependence on memory gaps parallels \citet{ZhaRobYa2019}, whose convergence rates for integration-order subspaces are likewise governed by gaps between adjacent integration orders.

We next consider the selection of the number of factors. 

\begin{theorem}[Consistency of the Bai-Ng criterion under $N/T\to0$]
	\label{T:Bai_Ng}
	Suppose Assumptions~\ref{ass:factors}--\ref{ass:memory} hold and that $N,T\to\infty$ with $N/T\to0$. Let $s_{\max}\geq s_0$ be fixed, and let $\widehat{s}$ be defined
	by \eqref{eq:factor_number_estimator}.
	
If $s_0^*>0$, then, for every $j=1,\ldots,q_0^*$,
		\begin{align}
			g(N,T)
			=
			o\left(
			T^{2d_0^{(j)}-1}
			\right)
			\label{eq:standard_nonstationary_penalty_condition}
		\end{align}
		implies
		\[
		\Pr\left(
		\widehat{s}\geq s^{(1:j)}
		\right)
		\longrightarrow 1.
		\]
		If the penalty satisfies the Bai-Ng conditions \eqref{eq:BN_penalty_conditions}, then
	\begin{align}\label{eq:factor_number_result}
			\Pr\left(
		\widehat{s}=s_0
		\right)
		\longrightarrow
		1 .
	\end{align}
\end{theorem}
Theorem~\ref{T:Bai_Ng} is proven in Appendix~\ref{sec:theorems}.
Condition~\eqref{eq:standard_nonstationary_penalty_condition} gives the penalty a threshold interpretation: if the penalty is smaller than the value-function reduction generated by a nonstationary factor group, all factor groups above that threshold are selected. In particular, a penalty smaller than the contribution of the weakest nonstationary group ensures that all $s_0^*$ nonstationary factors are selected, without requiring the stronger conditions needed to recover the stationary factors.

The applied literature often prefers the logarithmic form of the Bai-Ng criteria
	\begin{align}
		IC(s) &= \log\mathcal V(s) + s\,g(N,T),
		\qquad
		s=0,1,\ldots,s_{\max},
		&
		\widehat{s}_{IC} &= \arg\min_{0\leq s\leq s_{\max}} IC(s),
		\label{eq:IC_criterion}
	\end{align}
	as it does not depend on the scale of the data. 
	\begin{corollary}[Consistency of the logarithmic criterion]
		\label{C:IC_criteria}
		Suppose Assumptions~\ref{ass:factors}--\ref{ass:memory} hold and that $N,T\to\infty$ with $N/T\to0$. Let $s_{\max}\geq s_0$ be fixed, let $\widehat{s}_{IC}$ be defined by \eqref{eq:IC_criterion}, and let the penalty satisfy the Bai-Ng conditions \eqref{eq:BN_penalty_conditions}. Then
		\[
		\Pr\left(\widehat{s}_{IC}=s_0\right) \longrightarrow 1 .
		\]
	\end{corollary}
	Corollary~\ref{C:IC_criteria} is proven in Appendix~\ref{sec:theorems}. As in \citet[Corollary~1]{BaiNg2002}, its consistency can be inferred from the intermediate results established in the proof of our Theorem~\ref{T:Bai_Ng}.
	The logarithm replaces the absolute reduction of the value function by its relative reduction $\mathcal V(s)/\mathcal V(s+1)$.
	The step from the PC to the IC criterion therefore requires $\mathcal V(s_0)$ to be bounded from above and $\mathcal V(s)$, $s\leq s_{\max}$, to be bounded away from zero in probability, see \citet[][proof of Corollary~1]{BaiNg2002}. Here, both follow from Assumption~\ref{ass:Eigenvalue_Sigma_u_tilde}.
	The groupwise threshold result in the first part of Theorem~\ref{T:Bai_Ng} is stated for the PC criterion. For the IC criterion the relevant quantities are ratios of value functions rather than absolute eigenvalue orders.
Together, Theorem~\ref{T:Bai_Ng} and Corollary~\ref{C:IC_criteria} show that, under the Bai-Ng penalty conditions in \eqref{eq:BN_penalty_conditions}, both the PC and the IC criterion consistently recover the full factor dimension, including stationary factors, also for fractional factor models.

Differencing, as suggested e.g.\ by \cite{Ergemen2023} for fractionally integrated factors, changes the source of identification. The stationary transformed panel then falls back to the approximate-factor setting of \citet{BaiNg2002}, where the average squared factor-estimation error is of order $\min\{N,T\}^{-1}$. The differenced idiosyncratic component needs no assumption beyond those already maintained: the coefficients of the fractional-differencing filter are absolutely summable \citep[Lemma~5.2]{Hassler2018},  so the largest eigenvalue of the differenced idiosyncratic sample covariance matrix remains $O_p(1)$; see Appendix~\ref{sec:diminishing_signals_appendix} for details.

Thus, differencing does not necessarily hinder consistency in the standard-signal regime. If the idiosyncratic component is stationary and the common factors are strong, PCA can still recover the common component from the differenced data. However, differencing removes the additional time-series signal contained in the original sample covariance matrix, thus lowering the convergence rate of the nonstationary factor estimators. 

\subsection{The diminishing-signal regime}
\label{subsec:DS_regime}

We now turn to the \textit{diminishing-signal} regime. Assumption~\ref{ass:diminishing_signals} rescales the innovations of the nonstationary factors so that their sample second moments remain of order one. The memory parameters are unchanged, and the factors remain nonstationary, but the time-series variance advantage present in the standard-signal regime is removed. The next result shows that PCA in levels still recovers the factor-loading space.

\begin{theorem}[PCA consistency under diminishing signals]
	\label{T:factor_convergence_DS}
	Suppose Assumptions~\ref{ass:factors}--\ref{ass:memory} and \ref{ass:diminishing_signals} hold, and let $N,T\to\infty$ with $N/T\to0$. Then there exists a random nonsingular matrix $H_{0,T}$ whose singular values are bounded away from zero and infinity in probability such that the PCA scores $ \widehat{x}_t$ from \eqref{eq:pca_factor_estimator} satisfy
	\begin{align}
		 \widehat{x}_t &= H_{0,T}x_{t,T} + e_{0,t},
		&
		\norm{e_{0,t}}_F &= O_p\left(N^{-1/2}\right),
		\label{eq:factor_convergence_DS}
	\end{align}
	and
	\begin{align}
		\frac{1}{T} \sum_{t=1}^T \norm{ \widehat{x}_{t,T}-H_{0,T}x_{t,T}}_F^2
		&= O_p\left(N^{-1}\right).
		\label{eq:factor_convergence_DS_average}
	\end{align}
\end{theorem}
Theorem~\ref{T:factor_convergence_DS} is proven in Appendix~\ref{sec:diminishing_signals_appendix}.
Under the diminishing-signal regime, PCA still estimates the common factors consistently up to rotation. However, the convergence rate now coincides with the stationary factor rate.
This is expected because the diminishing-signal scaling removes the divergence of the nonstationary factor variances. The nonstationary factors therefore no longer dominate stationary factors through increasing sample second moments. They remain pervasive in the cross-section, however, so PCA in levels still recovers the common factor space.

The same argument yields consistency of the Bai-Ng factor-number criterion.
\begin{corollary}[Bai-Ng consistency under diminishing signals]
	\label{C:Bai_Ng_DS}
	Suppose Assumptions~\ref{ass:factors}--\ref{ass:memory} and \ref{ass:diminishing_signals} hold, and let $N,T\to\infty$ with $N/T\to0$. Let $s_{\max}\geq s_0$ be fixed, and let $\widehat{s}$ be defined by \eqref{eq:factor_number_estimator}. If the penalty satisfies the Bai-Ng conditions \eqref{eq:BN_penalty_conditions}, then
	\[
	\Pr\left(\widehat{s}=s_0\right) \longrightarrow 1 .
	\]
\end{corollary}
Corollary~\ref{C:Bai_Ng_DS} is proven in Appendix~\ref{sec:diminishing_signals_appendix}.
The penalty conditions are again the Bai-Ng conditions. Thus, even under diminishing signals, the number of factors can be selected consistently from the panel in levels. The common component remains pervasive across the cross-section. However, diminishing signals come at the cost of removing the stronger rates of convergence present in the standard-signal regime.

The logarithmic criterion \eqref{eq:IC_criterion} remains consistent under diminishing signals.
\begin{corollary}[Consistency of the logarithmic criterion under diminishing signals]
	\label{C:IC_criteria_DS}
	Suppose Assumptions~\ref{ass:factors}--\ref{ass:memory} and \ref{ass:diminishing_signals} hold, and let $N,T\to\infty$ with $N/T\to0$. Let $s_{\max}\geq s_0$ be fixed, let $\widehat{s}_{IC}$ be defined by \eqref{eq:IC_criterion}, and let the penalty satisfy the Bai-Ng conditions \eqref{eq:BN_penalty_conditions}. Then
	\[
	\Pr\left(\widehat{s}_{IC}=s_0\right) \longrightarrow 1 .
	\]
\end{corollary}
Corollary~\ref{C:IC_criteria_DS} is proven in Appendix~\ref{sec:diminishing_signals_appendix}.

If, however, data are differenced in the diminishing-signal regime, nonstationary factors may become undetectable. To see the implication, consider fractional differencing by the largest memory parameter $d_0^{(1)}$. For a nonstationary factor $i=1,\ldots,s_0^*$,
\begin{align*}
	\Delta_+^{d_0^{(1)}}x_{i,t,T} &= \Delta_+^{d_0^{(1)}-d_{0,i}}\psi_{i,+}(L)\eta_{i,t,T}.
\end{align*}
The combined filter on the right-hand side does not depend on $T$, while
\begin{align*}
	\Var \left(\eta_{i,t,T}\right)
	&= T^{1-2d_{0,i}} \Var \left(\eta_{i,t}\right).
\end{align*}
Hence, the sample second moment of the differenced factor satisfies
	\begin{align}
		\frac{1}{T}\sum_{t=1}^T
		\left(\Delta_+^{d_0^{(1)}}x_{i,t,T}\right)^2
		&=
		O_e\left(T^{1-2d_{0,i}}\right).
		\label{eq:DS_differenced_factor_second_moment_text}
	\end{align}
By Assumption~\ref{ass:strong_factors}, the squared norm of the loading vector associated with factor $i$ is of order $N$. Combining this with \eqref{eq:DS_differenced_factor_second_moment_text}, the corresponding contribution of the common component to the covariance matrix has an eigenvalue of order $NT^{1-2d_{0,i}}$. If $NT^{1-2d_{0,i}}\to0$, this contribution vanishes. As a result, the factor no longer generates an eigenvalue separated from idiosyncratic variation, and a Bai-Ng type criterion applied to the differenced data cannot consistently detect all true factors.

\begin{corollary}[Loss of detectability after differencing]
	\label{C:differencing_DS_non_detectability} Suppose Assumptions~\ref{ass:factors}--\ref{ass:memory} and \ref{ass:diminishing_signals} hold, and let $N,T\to\infty$ with $N/T\to0$. Let $\widehat{s}$ denote the factor-number estimator \eqref{eq:factor_number_estimator} computed from the fractionally differenced panel $\Delta_+^{d_0^{(1)}}y_t$, with a penalty satisfying \eqref{eq:BN_penalty_conditions}. If
$
	{N}T^{1-2d_{0,i}} \to 0
$
	for some nonstationary factor $i\leq s_0^*$, then the eigenvalue generated by this factor in the fractionally differenced data
$
	\Delta_+^{d_0^{(1)}}y_t
$
is $o_p(1)$, whereas the largest eigenvalue of the differenced idiosyncratic sample covariance matrix remains $O_p(1)$. Consequently, the $s_0$th eigenvalue of the differenced sample covariance matrix is $O_p(1)$, and
	\[
	\Pr\left(\widehat{s}=s_0\right)\longrightarrow0 .
	\]
	A Bai-Ng type criterion applied to the differenced data therefore cannot
	consistently recover the factor dimension.
\end{corollary}

Corollary~\ref{C:differencing_DS_non_detectability} is proven in Appendix~\ref{sec:diminishing_signals_appendix}. In levels the factors stay detectable: they are pervasive in the cross-section, and the Bai-Ng criterion is consistent. Differencing removes the accumulation, so that only the scaled innovation variance is left, with an eigenvalue of order $N/T^{2d_{0,i}-1}$. Unless $N$ grows faster than $T^{2d_{0,i}-1}$, that eigenvalue vanishes and the factor is lost in the idiosyncratic variation. The analogous result holds for the log criteria. Differencing can hide a weak nonstationary common factor.\

\section{Monte Carlo evidence}
\label{sec:monte_carlo}
Building on consistency, we next address how useful our results are in finite samples. To foster interpretability, we ask how weak the common signal may become before model selection stops succeeding, and in doing so we deviate from the standard Monte Carlo design. In our case, a standard Monte Carlo study would fix the data-generating process and report how often the procedure selects the correct number of factors.  Thus, frequency is then influenced by the design as much as by the procedure, and comparing design variations becomes hard as soon as their frequencies approach zero or one. We invert the exercise: the probability of selecting the true factor number is held fixed at one half, and we solve for the largest idiosyncratic variance at which our factor selection criterion still attains the selected probability. Within the family of designs indexed by that variance, this determines the least favorable configuration the criterion can still handle, since any larger idiosyncratic variance implies a less than fifty percent chance to select the true factor number. For applied researchers, comparing the expected signal-to-noise ratio of their specific application to our simulation results directly answers whether our methods are capable of detecting the correct number of factors for their use case.

 In what follows, we consider a data-generating mechanism of the form
\begin{align*}
	y_t &= \Lambda_0 x_{t,T} + u_t,
	\qquad
	t=1,\ldots,T,
\end{align*}
where $s_0=3$, $u_t\sim \mathcal{N}(0,\sigma_{u,NT}^2 I_N)$, and the entries of $\Lambda_0$ are drawn independently from $U[0.5,1.5]$. The factors are generated as
\begin{align}
	x_{i,t,T} &= \Delta_+^{-d_{0,i}} \eta_{i,t,T},
	\qquad
	i=1,2,3,
	&
	\left(d_{0,1},d_{0,2},d_{0,3}\right) &= (1.25,0.75,0),
	\label{eq:MC_factor_DGP}
\end{align}
which corresponds to the special case $\psi_i(L)=1$ of the factor equation
\eqref{eq:factor_process}. Thus, the first two factors are nonstationary and the third factor is stationary, so that $q_0^*=2$ with $s_1=s_2=1$, $s_0^*=2$ and $s_0=3$. 

We consider two different specifications for \eqref{eq:MC_factor_DGP}: the \textit{standard-signal regime} assumes $\eta_{i,t,T}=\eta_{i,t}\sim\mathcal{N}(0,1)$ for all $i$.
In the \textit{diminishing-signal regime}, innovation variances are scaled according to Assumption~\ref{ass:diminishing_signals}. Specifically,
\begin{align}
	\eta_{i,t,T} &\sim \mathcal{N}(0,\sigma_{i,T}^2),
	&
	\sigma_{i,T}^2 &= \left( \max \left\{ T^{2d_{0,i}-1}, 1 \right\} \right)^{-1},
	\qquad i=1,2,3.
	\label{eq:MC_DS_scaling}
\end{align}

For factor-number selection, we apply the $IC_{p2}$ criterion of \citet{BaiNg2002} as given in equation~\eqref{eq:climate_ABC_criterion} for $c=1$ and allow for up to $s_{\max}=10$ factors. For each pair $(N,T)$ and each signal regime, we search for the largest idiosyncratic variance $\sigma_{u,NT}^2$ for which the criterion selects the true number of factors in at least $50\%$ of $1000$ replications.

To also study the impact of pre-differencing on model selection, we consider four designs: standard signals in levels, standard signals after taking first differences, diminishing signals in levels, and diminishing signals after taking first differences. The differenced specifications use $\Delta y_t=(1-L)y_t$ for factor estimation and selection. We use first differences because they are the most common empirical transformation for nonstationary panels. Under this transformation, the most persistent factor becomes $I(0.25)$, whereas the other two factors are over-differenced.\footnote{Differencing by the integration order $d = 1.25$ would further attenuate the signal of the nonstationary factor, requiring an even lower idiosyncratic variance for successful factor detection.} 

\begin{table}[htbp]
	\centering
	\scriptsize
	\setlength{\tabcolsep}{3pt}
	\renewcommand{\arraystretch}{1.08}
	\begin{minipage}[t]{0.48\textwidth}
		\centering
		\textit{Standard signals, levels}
		
		\vspace{2pt}
		
		\begin{tabular}{@{}lccccc@{}}
			\toprule
			& $N=50$ & $N=100$ & $N=200$ & $N=500$ & $N=1000$ \\[2pt]
			\midrule
			$T = 50$ & 0.938 & 1.096 & 1.202 & 1.285 & 1.311 \\ 
$T = 100$ & 1.131 & 1.242 & 1.432 & 1.603 & 1.671 \\ 
$T = 200$ & 1.250 & 1.452 & 1.636 & 1.957 & 2.115 \\ 
$T = 500$ & 1.362 & 1.654 & 1.986 & 2.388 & 2.760 \\ 
$T = 1000$ & 1.398 & 1.731 & 2.156 & 2.778 & 3.193 \\ 
			\bottomrule
		\end{tabular}
	\end{minipage}
	\hfill
	\begin{minipage}[t]{0.48\textwidth}
		\centering
		\textit{Standard signals, first differences}
		
		\vspace{2pt}
		
		\begin{tabular}{@{}lccccc@{}}
			\toprule
			& $N=50$ & $N=100$ & $N=200$ & $N=500$ & $N=1000$ \\[2pt]
			\midrule
			$T = 50$ & 0.601 & 0.706 & 0.784 & 0.839 & 0.856 \\ 
$T = 100$ & 0.701 & 0.772 & 0.906 & 1.027 & 1.069 \\ 
$T = 200$ & 0.775 & 0.908 & 1.025 & 1.232 & 1.330 \\ 
$T = 500$ & 0.827 & 1.022 & 1.236 & 1.489 & 1.723 \\ 
$T = 1000$ & 0.847 & 1.069 & 1.333 & 1.723 & 1.986 \\ 
			\bottomrule
		\end{tabular}
	\end{minipage}
	
	\vspace{0.8em}
	
	\begin{minipage}[t]{0.48\textwidth}
		\centering
		\textit{Diminishing signals, levels}
		
		\vspace{2pt}
		
		\begin{tabular}{@{}lccccc@{}}
			\toprule
			& $N=50$ & $N=100$ & $N=200$ & $N=500$ & $N=1000$ \\[2pt]
			\midrule
			$T = 50$ & 0.288 & 0.328 & 0.370 & 0.389 & 0.395 \\ 
$T = 100$ & 0.328 & 0.363 & 0.431 & 0.474 & 0.494 \\ 
$T = 200$ & 0.356 & 0.419 & 0.474 & 0.559 & 0.609 \\ 
$T = 500$ & 0.376 & 0.453 & 0.550 & 0.656 & 0.765 \\ 
$T = 1000$ & 0.395 & 0.494 & 0.625 & 0.791 & 0.906 \\ 
			\bottomrule
		\end{tabular}
	\end{minipage}
	\hfill
	\begin{minipage}[t]{0.48\textwidth}
		\centering
		\textit{Diminishing signals, first differences}
		
		\vspace{2pt}
		
		\begin{tabular}{@{}lccccc@{}}
			\toprule
			& $N=50$ & $N=100$ & $N=200$ & $N=500$ & $N=1000$ \\[2pt]
			\midrule
			$T = 50$ & 0.041 & 0.047 & 0.051 & 0.055 & 0.056 \\ 
$T = 100$ & 0.028 & 0.031 & 0.035 & 0.040 & 0.041 \\ 
$T = 200$ & 0.018 & 0.021 & 0.024 & 0.028 & 0.030 \\ 
$T = 500$ & 0.010 & 0.012 & 0.014 & 0.017 & 0.020 \\ 
$T = 1000$ & 0.006 & 0.008 & 0.009 & 0.012 & 0.014 \\
			\bottomrule
		\end{tabular}
	\end{minipage}
	\caption{
	Finite-sample detection thresholds. Each entry reports the largest calibrated idiosyncratic standard deviation $\sigma_{u,NT}$ for which the Bai-Ng $IC_{p2}$ criterion selects the true number of factors in at least $50\%$ of $1000$ replications. Results are reported by signal regime, transformation, and panel size.
}	
\label{tab:MC_g_values}
\end{table}
We observe two main patterns in Table~\ref{tab:MC_g_values}, where for readability we report standard deviations rather than variances.
First, under standard signals, the level specification tolerates substantially larger idiosyncratic variances than the differenced specification. For example, when $N=T=1000$, the calibrated value of $\sigma_{u,NT}$ is $3.193$ in levels and $1.986$ after first differencing. Thus, differencing does not eliminate the factors in this design, but makes them harder to detect. This aligns with the faster convergence rate of nonstationary factors as found in Theorem~\ref{T:factor_convergence}.

Second, under diminishing signals, the difference between levels and first differences becomes much larger. The level specification still improves with the panel dimensions, whereas the calibrated standard deviations in the differenced diminishing-signal design are close to zero and decrease as $T$ increases. This supports the finding in Corollary~\ref{C:differencing_DS_non_detectability}, that nonstationary factors with weak signals become undetectable once estimated on a data set in differences.

Overall, the simulations support the theoretical results in Section~\ref{sec:estimation_model_selection}: Differencing is relatively harmless when the nonstationary signal is strong, although it reduces detectability. When the nonstationary signal is weak, however, differencing can make factors undetectable for factor selection criteria.


\section{Measuring climate trends in Europe}
\label{sec:climate_application}

We apply our methods to investigate climate warming in Europe. Temperature records are a natural application for factor models with fractionally integrated factors: The monthly temperature at a weather station combines local weather, which changes from month to month, both regional and common seasonal and cyclical patterns that last for months or years, and common, permanent warming caused by rising greenhouse-gas concentrations. 

In a panel of station records, climate change is therefore the common and persistent component. Local weather, by contrast, is idiosyncratic and short-lived. This matches the structure of our model: a small number of pervasive, possibly nonstationary factors, and an idiosyncratic component with short memory.

Factor models are widely applied to extract common components from temperature panels.
	\citet{CheDolGonRam2023} apply the quantile factor models of \citet{CheDolGon2021} to the
	distributions of temperature across weather stations in both hemispheres, \citet{MumMar2023} estimate a dynamic factor
	model with stochastic volatility for temperature and precipitation across countries,
	\citet{RodRui2024} fit a multi-level dynamic factor model to temperatures in the Iberian
	Peninsula and find a considerable common trend with relevant regional components, and
	\citet{GosLopNg2025} document a common low-frequency temperature factor in U.S., European and
	international panels. In a related application, \citet{BenHilKoo2021} use a dynamic factor
	model of many macroeconomic predictors to forecast and nowcast U.S.\ CO$_2$ emissions.

We contribute to this literature by extracting and analyzing the common nonstationary component for a panel of weather stations across Europe. Subsection \ref{ss:app1} introduces the data, subsection \ref{subsec:climate_common_factor} discusses the number of factors, their dynamics, and the magnitude of realized warming across Europe, subsection \ref{subsec:climate_memory} details the memory diagnostics and subsection \ref{subsec:climate_differencing} compares the results to those of a model in first differences.

\subsection{Data and low-frequency transformation}\label{ss:app1}
For the analysis, we sample a panel of monthly mean temperature observations across European weather stations from the NOAA GHCN-M database \citep{Menne2018}, spanning January 1960 to December 2025. We retain stations observed at both sample endpoints, with at most $12$ missing observations, of which no more than two are consecutive. Missing observations satisfying these restrictions are filled by linear interpolation, resulting in a balanced panel of $N=308$ weather stations across Europe.

Variation in climate time series is predominantly driven by seasonality. Since our focus is on the long-run behavior of temperature, we filter out the short-run dynamics by applying the cosine-based projection of \citet{MuellerWatson2008, Mueller2017} to each station series. Denoting the temperature observed at station $l$ in month $t$ by $y_{l,t}$, the fitted low-frequency component used in the sequel is
\begin{align}
	y_{l,t}^{LF}
	&=
	\bar y_l
	+
	\sum_{j=1}^{J_L}
	Y_{l,j}\varphi_j\left(\frac{t-1/2}{T}\right),
	\label{eq:climate_low_frequency_component}
	\\
	\varphi_j(r)
	&=
	\sqrt{2}\cos(j\pi r),
	\qquad j=1,\ldots,J_L,
	\\
	Y_{l,j}
	&=
	\frac{1}{T}\sum_{\tau=1}^T
	\varphi_j\left(\frac{\tau-1/2}{T}\right)y_{l,\tau},
	\label{eq:climate_low_frequency_coefficients}
\end{align}
where $\bar y_l=T^{-1}\sum_{t=1}^T y_{l,t}$ is the station mean and $r\in[0,1]$.
The coefficients $Y_{l,j}$ in \eqref{eq:climate_low_frequency_coefficients} are the cosine transforms: weighted averages of the series, with weights given by $\varphi_j$ evaluated at the midpoints $(t-1/2)/T$ of the months.
The function $\varphi_j$ completes $j$ half-cycles over the sample, so it has a period of $2T/j$ months, and $Y_{l,j}$ measures the variation of $y_{l,t}$ at this period.
On the grid of midpoints the weights are orthonormal, $T^{-1}\sum_{t=1}^T\varphi_j\left(\frac{t-1/2}{T}\right)\varphi_k\left(\frac{t-1/2}{T}\right)=\mathbbm{1}(j=k)$ for $1\leq j,k<T$, and each of them sums to zero, so the transforms do not depend on the level of the series.
Together with the constant, the functions $\varphi_1,\ldots,\varphi_{T-1}$ form a basis of $\mathbb{R}^T$.
Hence $y_{l,t}^{LF}$ in \eqref{eq:climate_low_frequency_component} is the least-squares projection of $y_{l,t}$ on a constant and the first $J_L$ cosine functions, and the residual $y_{l,t}-y_{l,t}^{LF}$ is the sum of the cosine components with $j>J_L$, i.e.\ with periods shorter than $2T/J_L$ months.

\citet{MuellerWatson2008} choose cosine weights because the functions $\varphi_j$ are the eigenfunctions of the covariance kernel of a demeaned Wiener process.
As a consequence, the transforms of an $I(0)$ or an $I(1)$ series are uncorrelated, and those of a fractionally integrated series are nearly uncorrelated.
For a stationary series, the variance of a transform is approximately proportional to the spectral density at the frequency of $\varphi_j$, and a more persistent series has a larger variance in its first transforms relative to later ones.
Section~\ref{subsec:climate_memory} uses this property to estimate the memory of the factors.

\begin{figure}[htbp]
	\centering
	
	\begin{subfigure}[t]{0.49\textwidth}
		\centering
		\includegraphics[width=\linewidth]{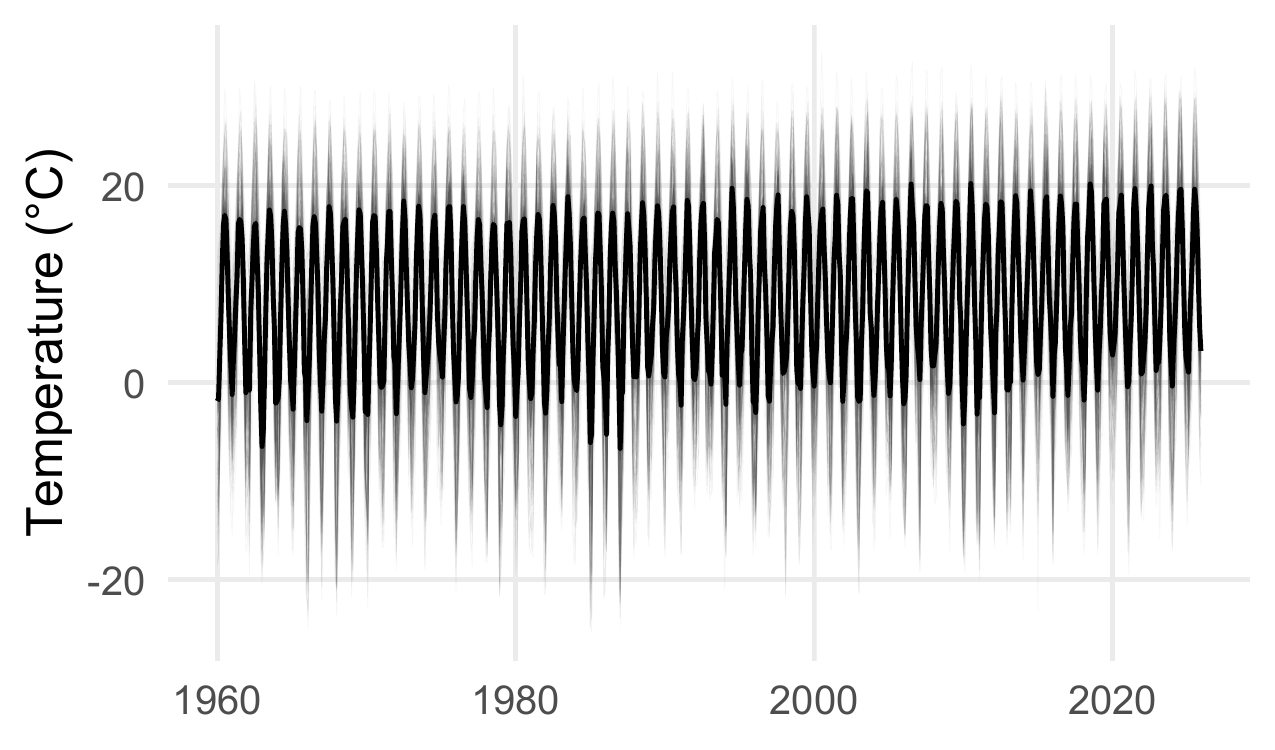}
		\caption{Monthly temperature levels}
		\label{fig:climate_temperature_levels}
	\end{subfigure}
	\hfill
	\begin{subfigure}[t]{0.49\textwidth}
		\centering
		\includegraphics[width=\linewidth]{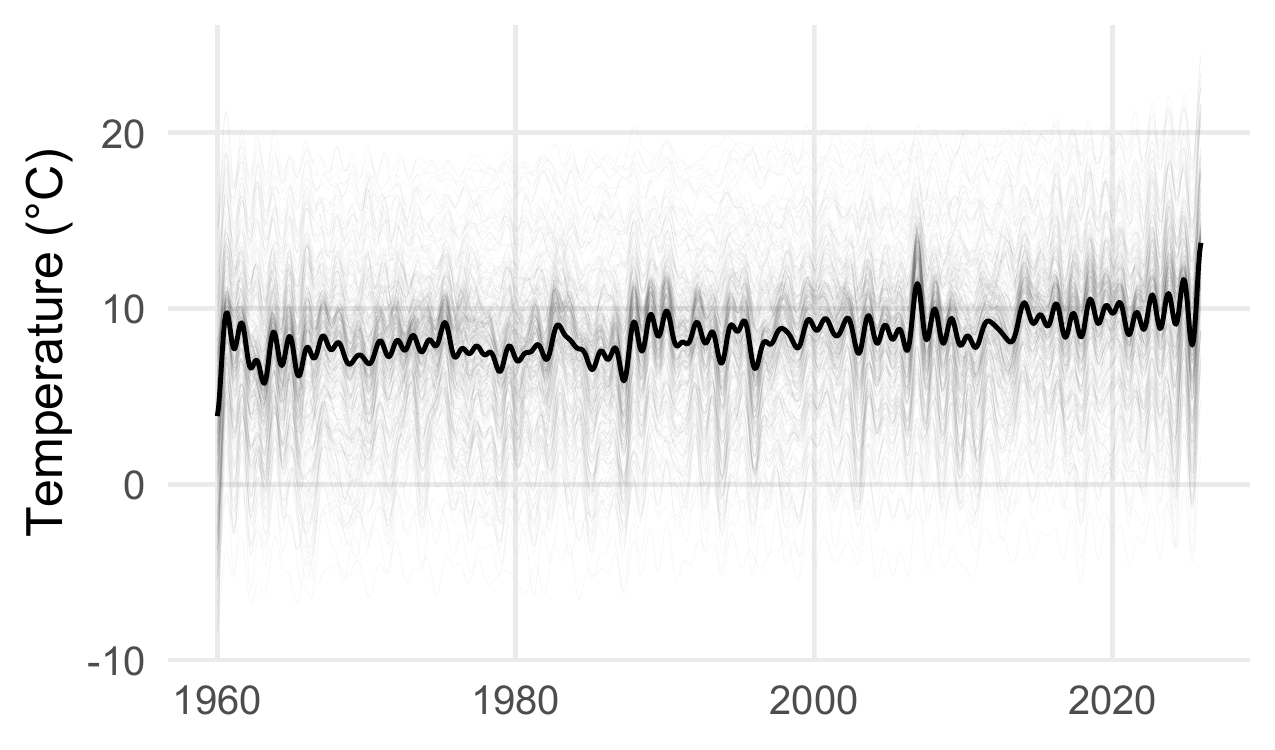}
		\caption{Low-frequency components}
		\label{fig:climate_lowpass}
	\end{subfigure}
	
	\caption{
		Observed temperatures and low-frequency components for the $308$ weather stations. Thin lines denote individual stations and the bold line denotes the cross-sectional mean.
	}
	\label{fig:climate_data}
\end{figure}

With $T=792$ months, $\varphi_j$ has a period of $1584/j$ months.
We set $J_L=125$, so that the shortest retained period is $1584/125\approx12.7 > 12.5$ months. Consequently, seasonal variation above the one-year frequency is projected out.
Figure~\ref{fig:climate_data} contrasts the monthly temperature series with their fitted low-frequency components. Panel (a) displays the predominant seasonal pattern, whereas trends and long-term cycles become visible from the low-frequency-transformed panel (b).

\subsection{Common factor structure of European temperatures}\label{subsec:climate_common_factor}
With a suitable dataset at hand, we next determine the number of factors using the Bai-Ng $IC_{p2}$ criterion, with the refinement suggested by \citet{Alessi2010},
on the low-frequency-transformed panel as described in subsection \ref{ss:app1}. For the value function $\mathcal{V}(s)$ from \eqref{eq:pca_value_function} and a penalty multiplier $c>0$, define
\begin{align}
	IC_c(s) &= \log \mathcal{V}(s) + c\,s \left(\frac{N+T}{NT}\right)
	\log\left(\min\{N,T\}\right),
	\qquad
	s=0,\ldots,s_{\max},
	\label{eq:climate_ABC_criterion}
\end{align}
and
\begin{align*}
	\widehat{s}(c) &= \arg\min_{0\leq s\leq s_{\max}} IC_c(s).
\end{align*}
We set $s_{\max}=20$. For $c=1$, \eqref{eq:climate_ABC_criterion} is the $IC_{p2}$ criterion of \citet{BaiNg2002}.
Theorem~\ref{T:Bai_Ng} and Corollary~\ref{C:IC_criteria} establish consistency of $\widehat{s}(c)$ for every fixed $c>0$. In finite samples, however, the selected number of factors depends on $c$. Following \citet{HallinLiska2007} and \citet{Alessi2010}, we calibrate $c$ by exploiting the stability of the number of common factors across subpanels of different sizes. A model selection criterion should therefore select the same number of factors in subpanels of different sizes.
Specifically, we apply \eqref{eq:climate_ABC_criterion} to $100$ randomly drawn subpanels for each size $n=\lceil N/2\rceil,\ldots,N-1$, and include the full panel once.
We select the factor number corresponding to the first interval of $c$ for which the subsampling standard deviation is zero and $\widehat{s}(c)<s_{\max}$.

\begin{figure}[htbp]
	\centering
    \begin{subfigure}[t]{0.48\textwidth}
		\centering
		    \vspace{0pt}
		\includegraphics[width=\linewidth]{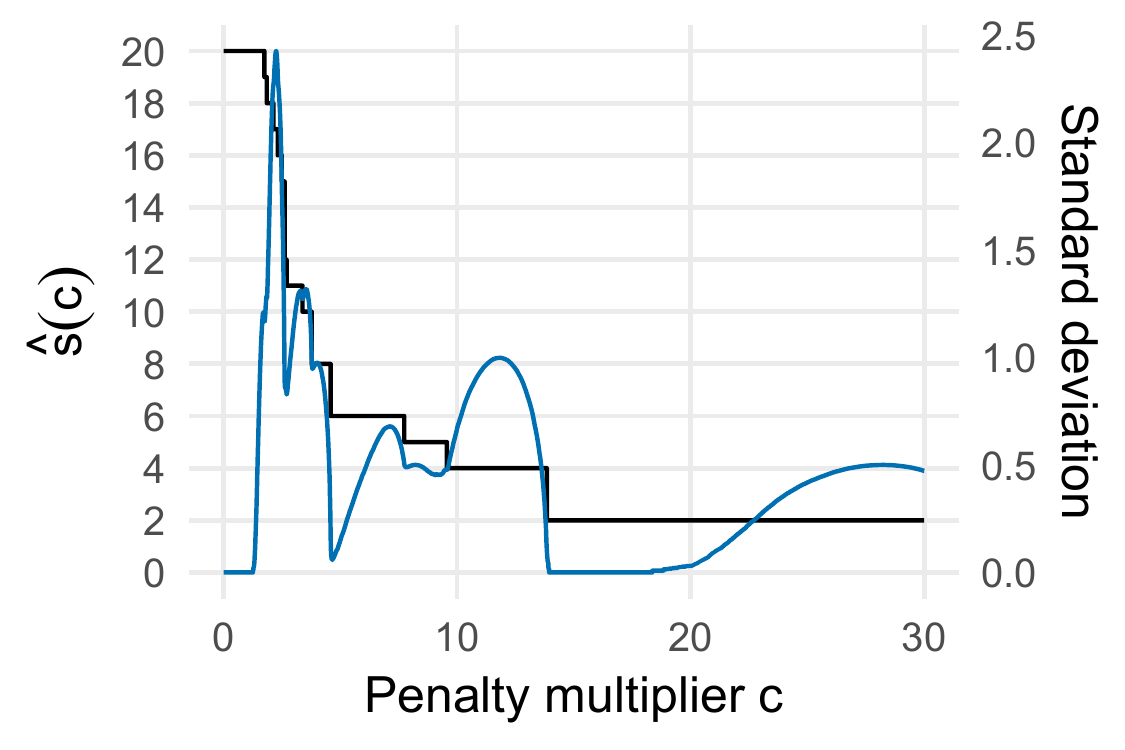}
		\caption{Selected number of factors}
		\label{fig:climate_abc_levels}
	\end{subfigure}
	\hfill
    \begin{subfigure}[t]{0.48\textwidth}
		\centering
		\vspace{0pt}
		\includegraphics[width=\linewidth]{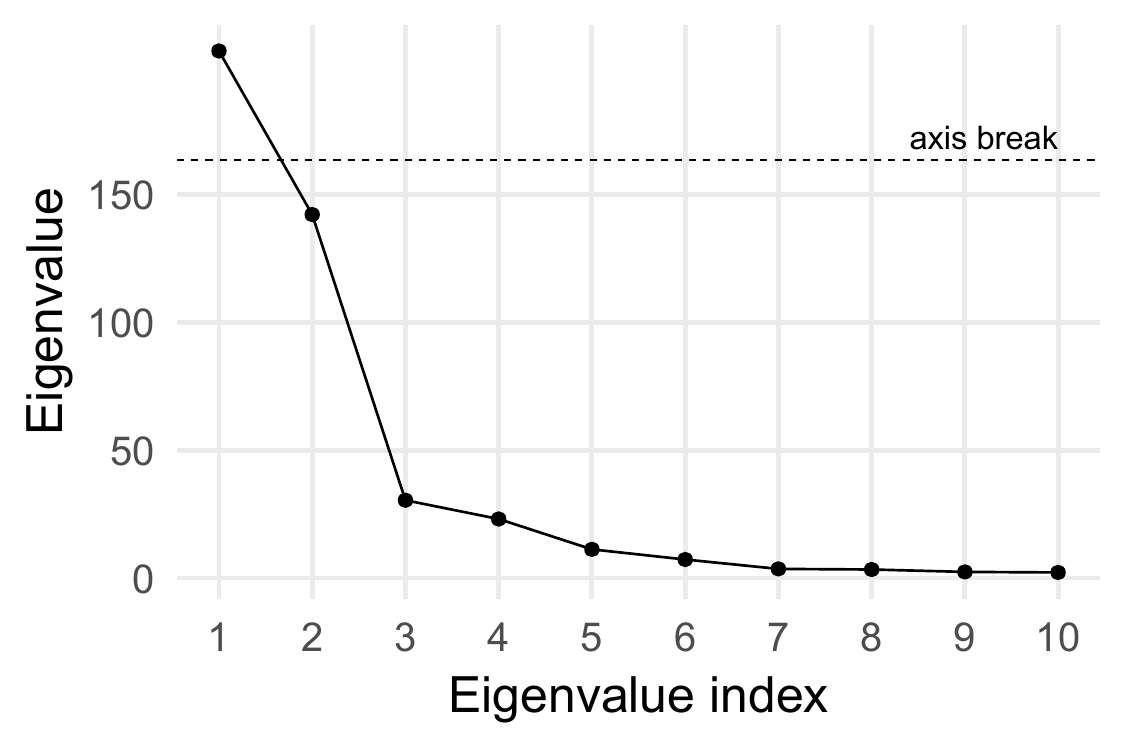}
		\caption{Leading PCA eigenvalues}
		\label{fig:climate_eigenvalues_levels}
	\end{subfigure}
	\caption{
		Factor-number selection for the low-frequency temperature panel. Panel~(a) shows the selected number of factors $\widehat{s}(c)$ (black line, left axis) and the subsampling standard deviation (blue line, right axis).
		Panel~(b) reports the leading eigenvalues of the sample second-moment matrix.
	}
	\label{fig:climate_factor_selection}
\end{figure}
Panel~(a) of Figure~\ref{fig:climate_factor_selection} shows that the criterion selects $s_{\max}$ factors for small values of $c$, while larger penalties progressively reduce $\widehat{s}(c)$. The first interval with zero subsampling standard deviation and $\widehat{s}(c)<s_{\max}$ is $c\in[13.9,18.3]$, where all subpanels select two factors. For larger $c$, the penalty starts to remove the second factor in the smaller subsamples, and the subsampling standard deviation increases again.
Panel~(b) supports this choice: the second eigenvalue is about $4.7$ times the third, whereas the eigenvalues from the third onwards decline gradually. Consequently, we set $\widehat s = 2.$

\begin{figure}[htbp]
	\centering
	\begin{subfigure}{0.48\textwidth}
		\centering
		\includegraphics[width=\linewidth]{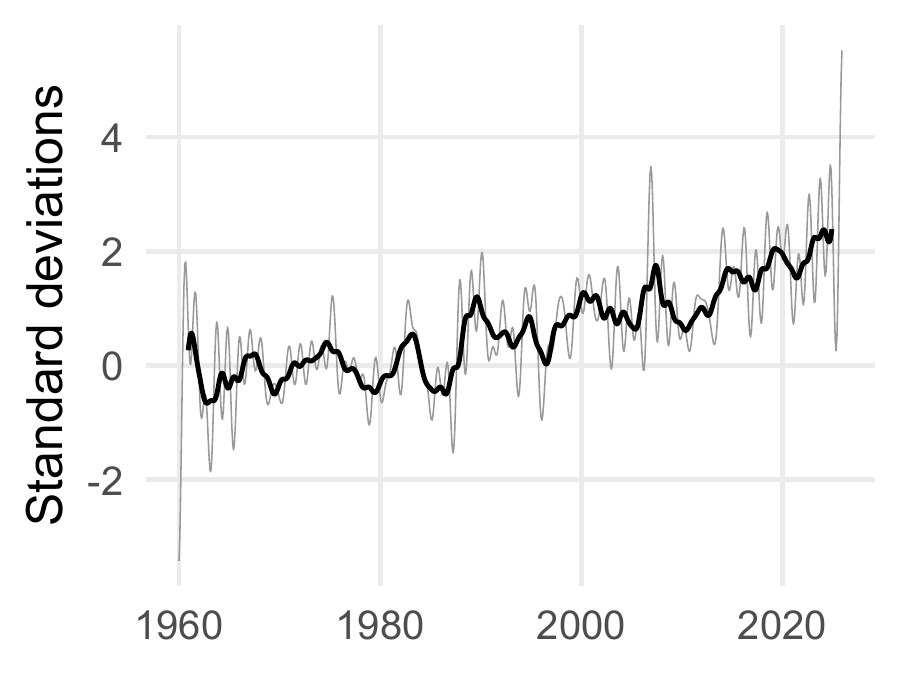}
		\caption{Factor 1}
		\label{fig:climate_factor_1}
	\end{subfigure}
	\hfill
	\begin{subfigure}{0.48\textwidth}
		\centering
		\includegraphics[width=\linewidth]{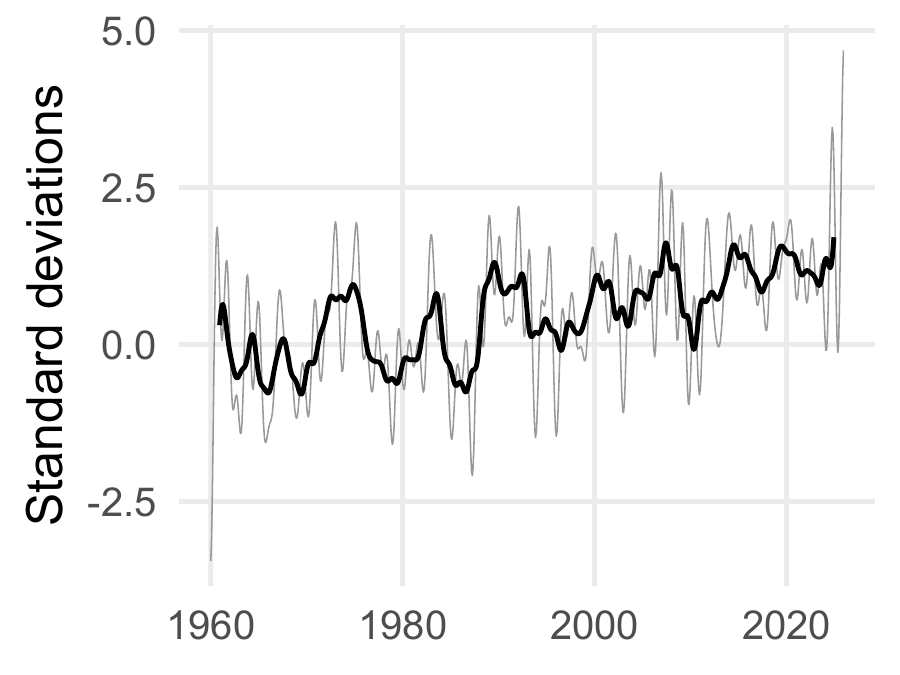}
		\caption{Factor 2}
		\label{fig:climate_factor_2}
	\end{subfigure}
	\caption{
		Estimated factors from PCA of the low-frequency temperature panel in levels.
		Each factor is centered by its 1960--1990 mean and scaled by its
		full-sample standard deviation. Gray lines report the monthly estimates
		and black lines two-sided $24$-month moving averages.
	}
	\label{fig:climate_factors}
\end{figure}

Figure~\ref{fig:climate_factors} plots the two estimated factors, with the sign of each chosen such that its linear trend is positive.
Both are close to their $1960$-$1990$ averages until the mid-1980s and increase steadily afterwards.
Relative to their $1960$-$1990$ means and measured in standard deviations over the full sample, the decade means of the first factor rise from $-0.17$ to $0.61$ in the 1990s and $2.07$ in $2020$-$2025$; those of the second factor rise from $-0.28$ to $0.47$ and $1.39$.

\begin{figure}[htbp]
	\centering
	\begin{subfigure}[t]{0.48\textwidth}
		\centering
		\includegraphics[width=\linewidth]{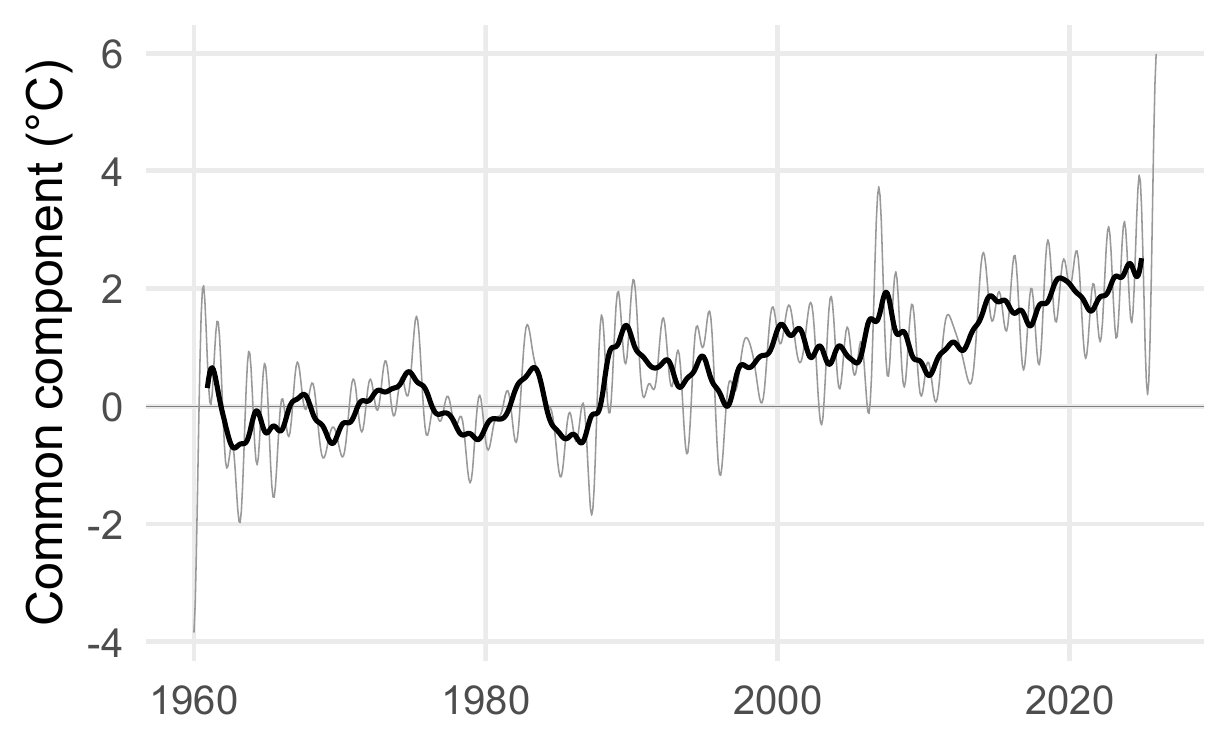}
		\caption{Sample-average common component}
		\label{fig:climate_common_signal}
	\end{subfigure}
	\hfill
	\begin{subfigure}[t]{0.48\textwidth}
		\centering
		\includegraphics[width=\linewidth]{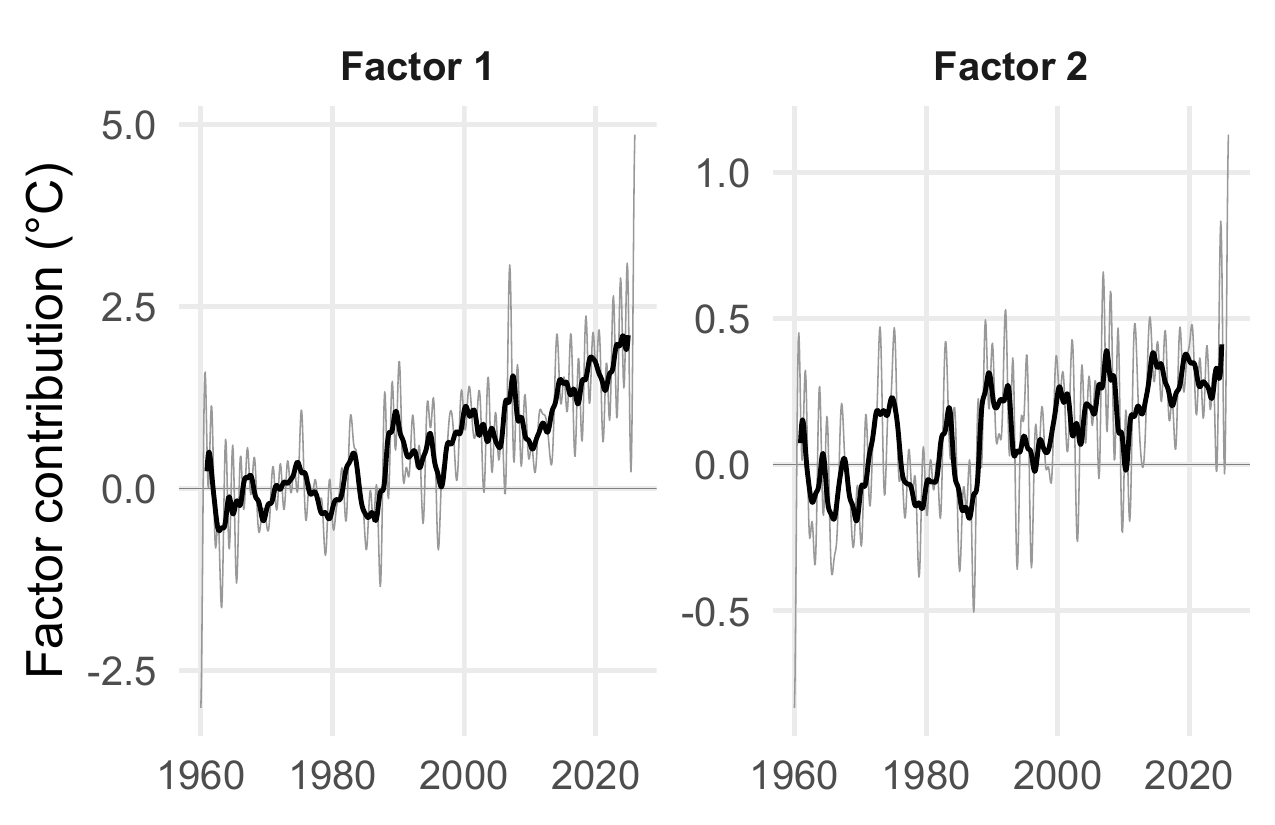}
		\caption{Factor contributions}
		\label{fig:climate_factor_contributions}
	\end{subfigure}
\caption{
	Estimated common component from PCA in levels. Panel~(a) shows the sample-average common component and Panel~(b) its factor-specific decomposition. For presentation, both are expressed relative to their $1960$-$1990$ means. In both panels, thin lines are monthly estimates and thick lines are two-sided $24$-month moving averages.
}
	\label{fig:climate_common}
\end{figure}

Figure~\ref{fig:climate_common}(a) shows the cross-sectional average of the estimated common component, $\bar\chi_t=N^{-1}\sum_{l=1}^N\widehat\chi_{l,t}$, relative to its $1960$-$1990$ mean.
Its average over $2021$-$2025$ exceeds the $1960$-$1990$ baseline by $2.15^\circ\mathrm{C}$.
The increase is concentrated in the second half of the sample: a linear trend fitted to $\bar\chi_t$ is $0.08^\circ\mathrm{C}$ per decade over $1960$-$1985$ and $0.49^\circ\mathrm{C}$ per decade over $1986$-$2025$.
The decadal averages of $\bar\chi_t$ lie $0.10^\circ\mathrm{C}$ above the baseline in the 1980s, $0.65^\circ\mathrm{C}$ in the 1990s, $1.13^\circ\mathrm{C}$ in the 2000s, and $1.44^\circ\mathrm{C}$ in the 2010s.
Panel~(b) splits the increase over 2021-2025 relative to the 1960-1990 baseline into the contributions of the two factors: the first factor accounts for $1.83^\circ\mathrm{C}$ ($85\%$) and the second for $0.32^\circ\mathrm{C}$ ($15\%$).
The shape of these series, flat for two decades and trending thereafter, is the kind of persistent behavior whose memory we examine in Section~\ref{subsec:climate_memory}.
\begin{figure}[htbp]
	\centering
	\begin{subfigure}{0.49\textwidth}
		\centering
		\includegraphics[width=\linewidth]{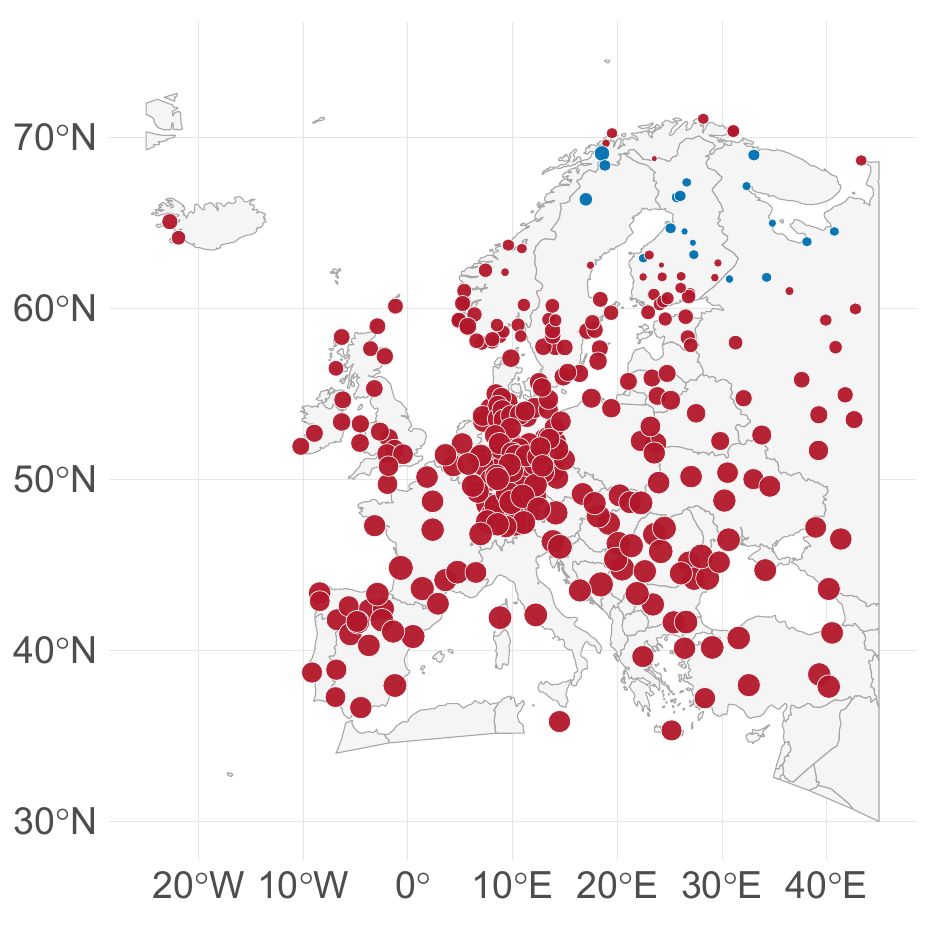}
		\caption{Factor 1}
		\label{fig:climate_factor1_loadings}
	\end{subfigure}
	\hfill
	\begin{subfigure}{0.49\textwidth}
		\centering
		\includegraphics[width=\linewidth]{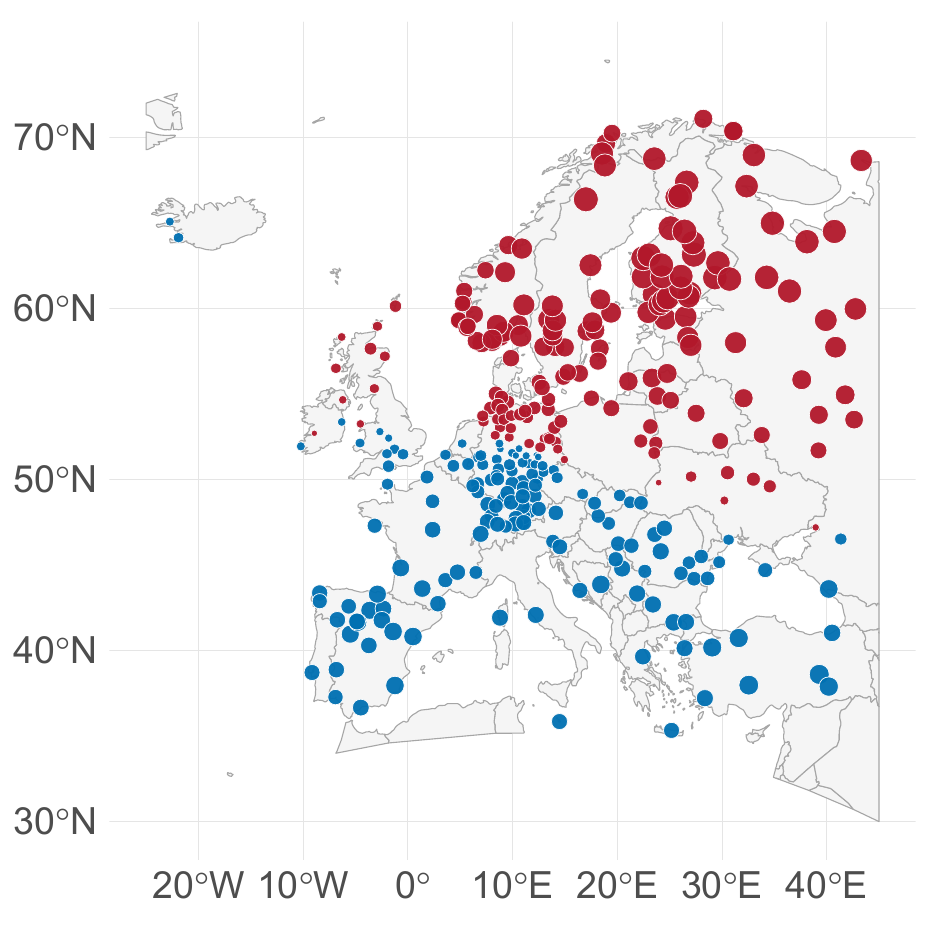}
		\caption{Factor 2}
		\label{fig:climate_factor2_loadings}
	\end{subfigure}
	\caption{
		Spatial loading patterns of the two estimated factors. Red points
		indicate positive loadings and blue points negative loadings; point size
		is proportional to the absolute loading.
	}
	\label{fig:climate_factor_loadings}
\end{figure}

Figure~\ref{fig:climate_factor_loadings} maps the loadings. The first factor loads positively at $290$ of the $308$ stations. Therefore, we call it the common warming factor. The $18$ stations with negative loadings all lie in northern Fennoscandia and northwestern Russia, the regions where the second factor exhibits the greatest loadings. The loadings of the second factor follow latitude: they are positive for the stations in Northern Europe, and negative for those in Southern Europe.

\begin{figure}[htbp]
	\centering
	\begin{subfigure}[t]{0.49\textwidth}
		\centering
		\includegraphics[width=\linewidth]{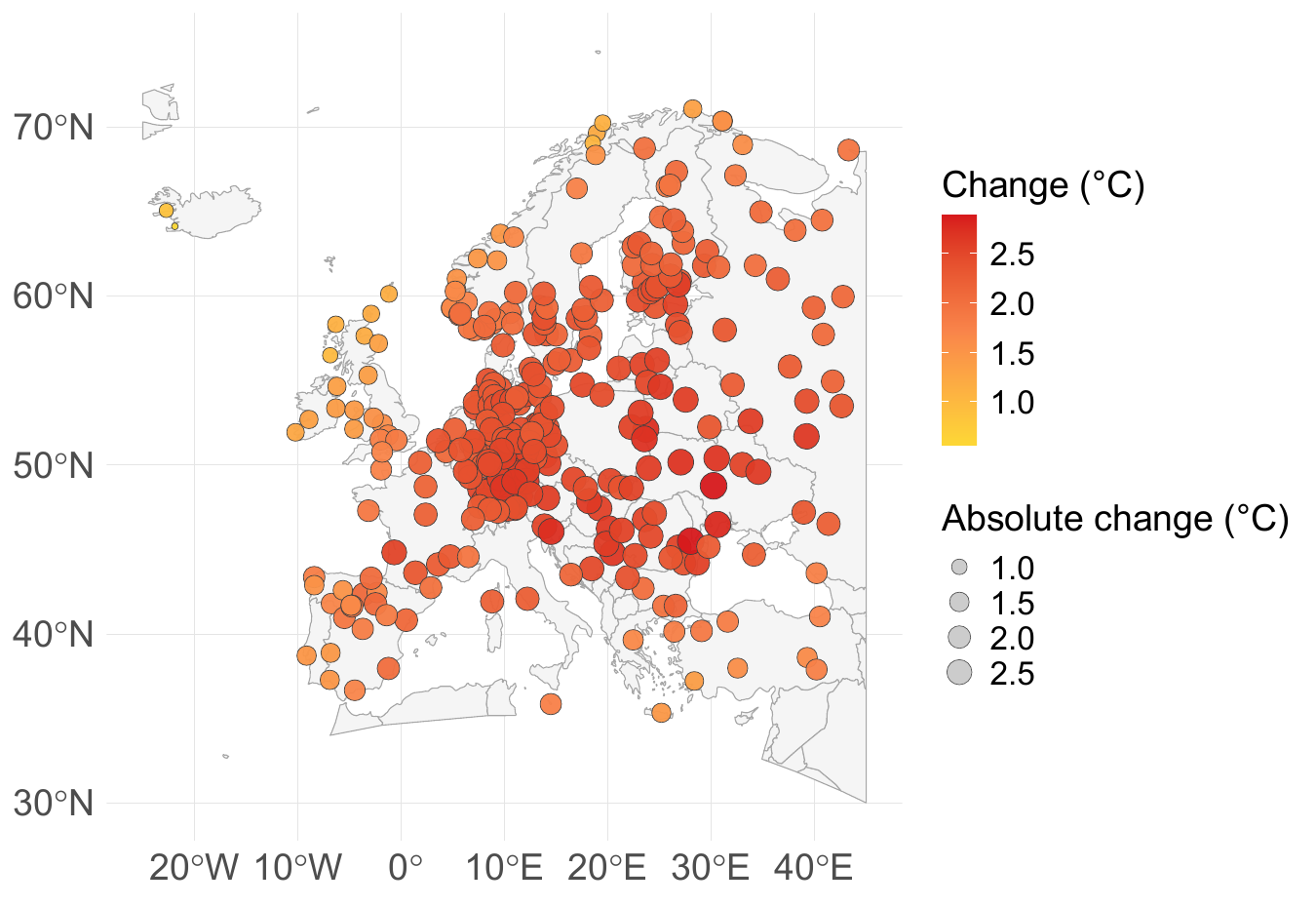}
		\caption{Station-specific common warming}
		\label{fig:climate_common_warming_map}
	\end{subfigure}
	\hfill
	\begin{subfigure}[t]{0.49\textwidth}
		\centering
		\includegraphics[width=\linewidth]{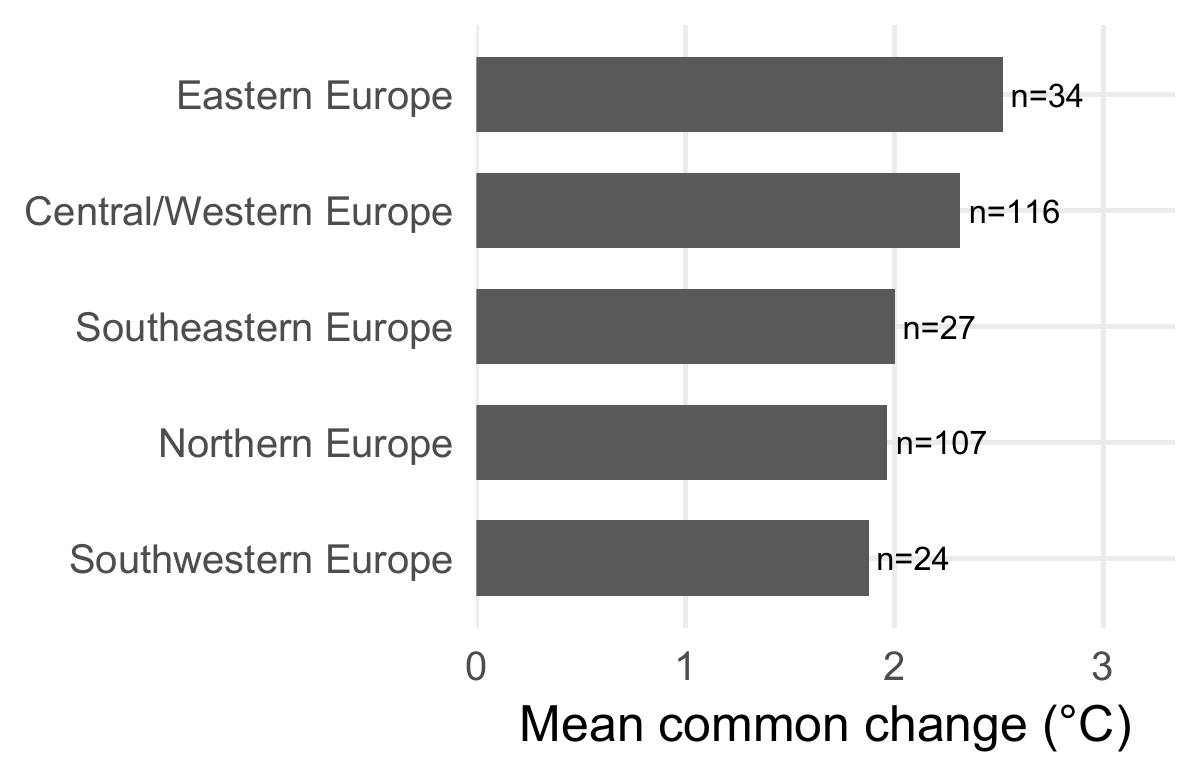}
		\caption{Regional averages}
		\label{fig:climate_region_warming}
	\end{subfigure}
	\caption{
		Spatial distribution of the estimated common warming. Panel~(a) reports station-specific common warming and Panel~(b) the corresponding regional averages. Warming is measured as the mean estimated common component in $2021$-$2025$ minus its mean in $1960$-$1990$.
	}
	\label{fig:climate_spatial_common}
\end{figure}

We measure the common warming at station $l$ as the mean of $\widehat\chi_{l,t}$ over $2021$-$2025$ minus its mean over $1960$-$1990$.
Figure~\ref{fig:climate_spatial_common} shows this quantity by station and averaged by region.\footnote{The regional classification is based on station coordinates. Northern Europe contains stations at or above $55^\circ$N. Stations below $45^\circ$N are assigned to Southwestern Europe west of $5^\circ$E and to Southeastern Europe otherwise. Among the remaining stations, those at or east of $20^\circ$E are assigned to Eastern Europe, with the remainder classified as Central/Western Europe.}
Common warming is positive at all $308$ stations.
It ranges from $0.56^\circ\mathrm{C}$ at Reykjav\'ik to $2.88^\circ\mathrm{C}$ at Gala\c{t}i, Romania, with a median of $2.27^\circ\mathrm{C}$. $80\%$ of the stations lie between $1.52^\circ\mathrm{C}$ and $2.58^\circ\mathrm{C}$.
The regional averages are largest in Eastern Europe ($2.52^\circ\mathrm{C}$) and Central/Western Europe ($2.31^\circ\mathrm{C}$) and smallest in Southeastern ($2.01^\circ\mathrm{C}$), Northern ($1.96^\circ\mathrm{C}$) and Southwestern Europe ($1.88^\circ\mathrm{C}$).

The two factors distribute the warming differently across regions.
The first factor contributes between $2.33^\circ\mathrm{C}$ and $2.77^\circ\mathrm{C}$ in all regions except Northern Europe, where it contributes $0.58^\circ\mathrm{C}$.
The second factor adds $1.39^\circ\mathrm{C}$ in Northern Europe and $0.20^\circ\mathrm{C}$ in Eastern Europe, and subtracts $0.15^\circ\mathrm{C}$ in Central/Western Europe and about $0.75^\circ\mathrm{C}$ in Southeastern and Southwestern Europe.
The first factor thus captures a warming common to continental and southern Europe, while the second shifts part of it from the south to the north, so that Northern Europe warms by a similar amount although it loads weakly on the first factor.

\subsection{Memory of the factors}
\label{subsec:climate_memory}
The factors are linear combinations of the low-frequency components of the station series.
Semiparametric estimators such as the local Whittle or the log-periodogram estimator use periodogram ordinates in a band of frequencies that also contains short-run variation, and such variation biases their estimates of the memory parameter downward \citep{Cheung2022}.
We therefore estimate the memory of the factors with the low-frequency approach of \citet{MuellerWatson2008}, which bases inference directly on the cosine transforms introduced in subsection~\ref{ss:app1}.

For the $i$th estimated factor $\widehat x_{i,t}$, define $\widehat X_{i,j}=T^{-1}\sum_{t=1}^T\varphi_j\left(\frac{t-1/2}{T}\right)\widehat x_{i,t}$, $j=1,\ldots,J$, as the coefficient of the projection on the $j$-th cosine wave, and collect these transforms in $\widehat X_i=(\widehat X_{i,1},\ldots,\widehat X_{i,J})^T$.

If the corresponding true factor satisfies $x_{i,t}=\Delta_+^{-d_{0,i}}e_t$ with $e_t$ weakly dependent, then at the true memory parameter $d=d_{0,i}$, $\widehat X_i$ is approximately $\mathcal N(0,\sigma^2\Omega(d_{0,i}))$. For likelihood evaluation, we use $\Omega(d)=T^{-2}\Phi^TA_dA_d^T\Phi$, where $\Phi$ is the $\left(T\times J\right)$ matrix with entries $\varphi_j((t-1/2)/T)$ and $A_d$ is the lower-triangular Toeplitz matrix of the coefficients of $(1-L)^{-d}$ \citep{MuellerWatson2008, Mueller2017}.
Because the $\varphi_j$ sum to zero, $\widehat X_i$ does not depend on the mean of the factor.
The scale $\sigma^2$ cannot be estimated consistently from a fixed number of transforms. Following \citet{MuellerWatson2008}, we therefore base inference on the scale-invariant vector $\widehat X_i/\|\widehat X_i\|$, whose density, as a function of $d$, gives the likelihood
\begin{align}
	L_i(d)
	&\propto
	\left|\Omega(d)\right|^{-1/2}
	\left(\widehat X_i^T\Omega(d)^{-1}\widehat X_i\right)^{-J/2}.
	\label{eq:climate_lowfreq_likelihood}
\end{align}
We maximize $L_i(d)$ over $d$ and invert the likelihood-ratio statistic for a $95\%$ confidence interval.
We compute $\Omega(d)$ for the type-II process of Section~\ref{sec:model}.

The choice of $J$ determines which periods enter the estimation: $\varphi_j$ has a period of $2T/j$ months.
Following \citet{MuellerWatson2008}, we set $J=13$, so that for our application the shortest period used is $2T/J\approx122$ months, or $10.2$ years. 
Variation at shorter periods therefore does not enter the estimates.

\begin{figure}[htbp]
	\centering
	\includegraphics[width=\linewidth]{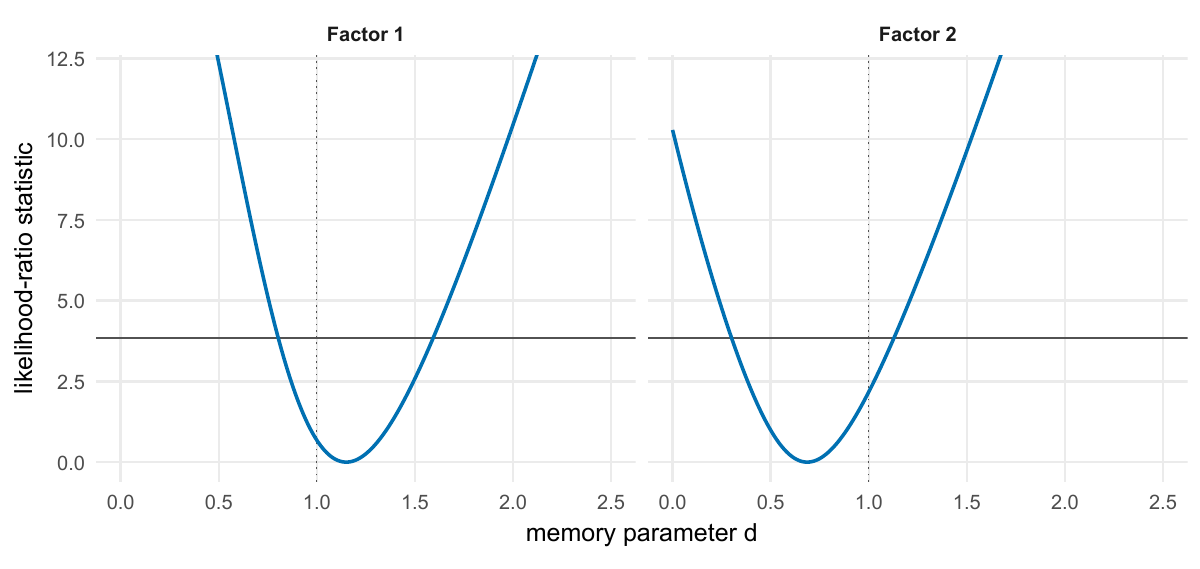}
	\caption{Likelihood-ratio statistics for the memory parameter $d$ of the two factors, based on the likelihood \eqref{eq:climate_lowfreq_likelihood} of the first $J=13$ cosine transforms. The horizontal line marks the $95\%$ critical value of the $\chi^2_1$ distribution, so that the $95\%$ confidence set is the range of $d$ for which the curve lies below it. The dotted vertical line marks $d=1$.}
	\label{fig:climate_memory_main}
\end{figure}

Figure~\ref{fig:climate_memory_main} shows the likelihood-ratio statistics. The likelihood treats the factors as if they were observed and thus ignores the estimation error generated by PCA. By Theorem~\ref{T:factor_convergence}, this error vanishes relative to the first factor, so that the confidence interval for the first factor is valid asymptotically.
For the first factor, the estimate is $\widehat d=1.15$ with $95\%$ confidence interval $[0.81,1.59]$.
For the second factor, the estimate is $\widehat d=0.69$ with interval $[0.31,1.12]$, which contains stationary values.

\subsection{Information loss in first differences}
\label{subsec:climate_differencing}
We finally apply the same analysis to the differenced low-frequency panel $\Delta y_{l,t}^{LF}=y_{l,t}^{LF}-y_{l,t-1}^{LF}$, $t=2,\ldots,T$.
Differencing is the standard way of handling nonstationary panels before principal components are computed \citep{Stock2002,Barigozzi2021,Ergemen2023}.

\begin{figure}[htbp]
	\centering
	\begin{subfigure}[t]{0.48\textwidth}
		\centering
		\includegraphics[width=\linewidth]{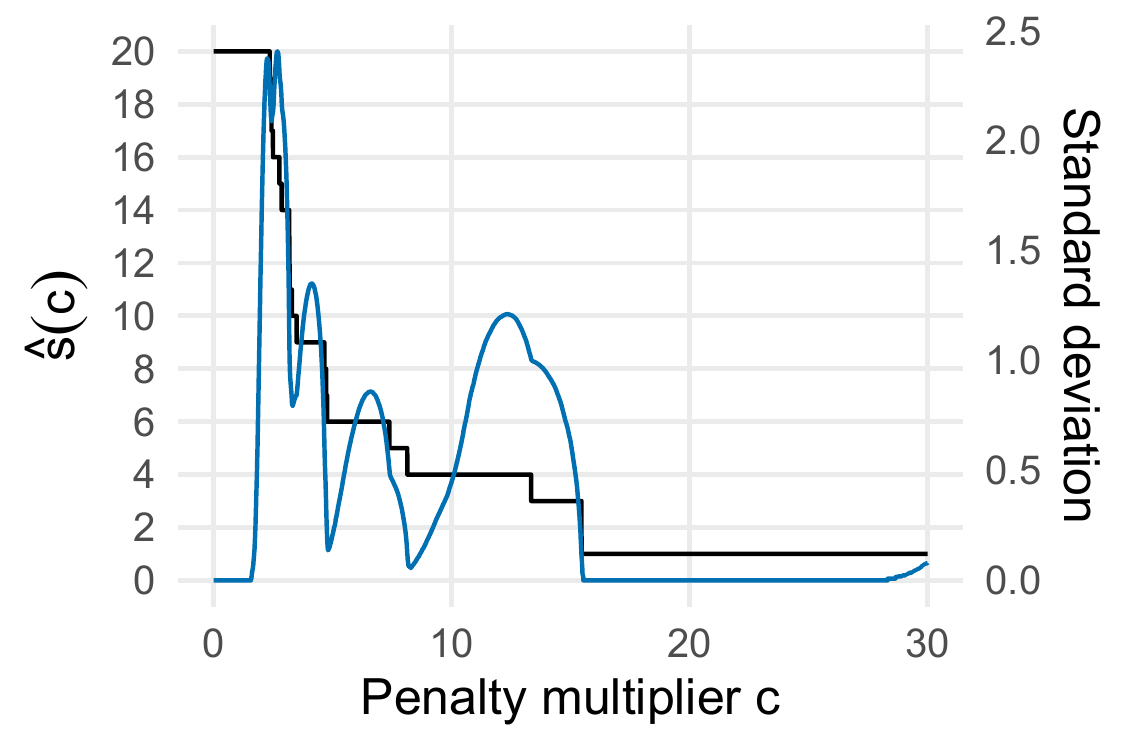}
		\caption{Selected number of factors}
		\label{fig:climate_diff_selection}
	\end{subfigure}
	\hfill
	\begin{subfigure}[t]{0.48\textwidth}
		\centering
		\includegraphics[width=\linewidth]{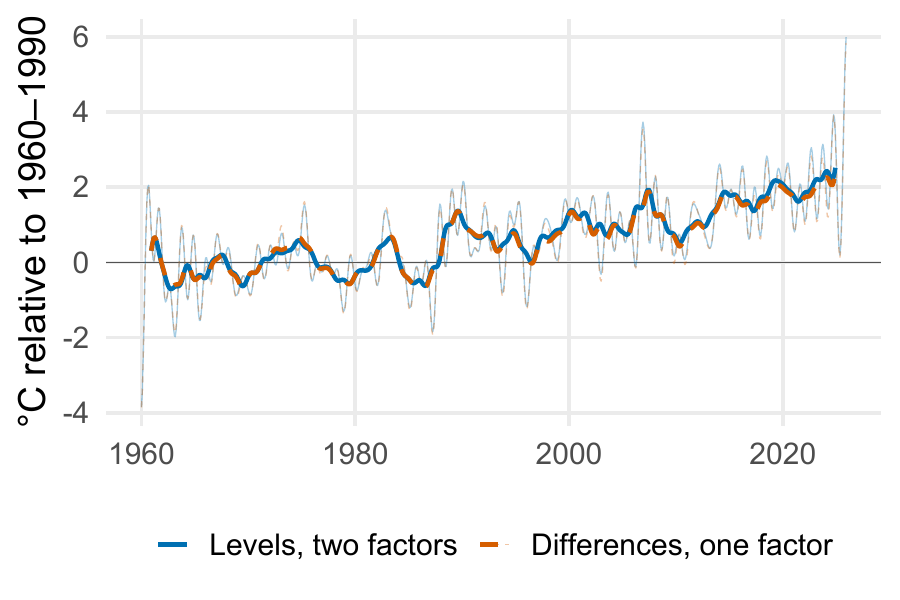}
		\caption{Levels and differences}
		\label{fig:climate_diff_common_compare}
	\end{subfigure}
	\caption{Effect of first differencing. Panel~(a) reports the selected number of factors $\widehat{s}(c)$ (black line, left axis) and the subsampling standard deviation
	(blue line, right axis) for the differenced low-frequency panel. Panel~(b) compares the sample-average common component from the two-factor model in levels with the cumulated common component from the one-factor model selected for the differenced panel, both relative to their $1960$-$1990$ means. Thin lines are monthly values, thick lines two-sided $24$-month moving averages.}
	\label{fig:climate_differencing}
\end{figure}

The selected number of factors drops from two to one, as indicated by the interval with zero subsampling standard deviation in Figure~\ref{fig:climate_differencing}(a). The eigenvalues stand in contrast to those of Figure~\ref{fig:climate_factor_selection}(b):
In levels, the second eigenvalue is about $4.7$ times the third; in differences, the second exceeds the third by only $20\%$, so that the second factor is no longer separated from idiosyncratic variation.
The first factor, whose memory estimate exceeds one, remains dominant after differencing, whereas the less persistent second factor loses the time-series signal that separates it in levels.
This is in line with the mechanism of Corollary~\ref{C:differencing_DS_non_detectability}: differencing reduces the accumulation of a persistent factor, and what remains of a weak factor does not stand out from the idiosyncratic component.
In what follows, we use the single factor selected for the differenced panel.

Differencing preserves the aggregate warming but not its regional composition.
The cumulated common component from the differenced panel implies an average warming of $2.02^\circ\mathrm{C}$ between $1960$-$1990$ and $2021$-$2025$, close to the $2.15^\circ\mathrm{C}$ from the level analysis and the $2.16^\circ\mathrm{C}$ observed in the low-frequency series, see Figure~\ref{fig:climate_differencing}(b).
Its distribution across Europe, however, differs substantially. As shown in Table~\ref{tab:climate_differencing_regions},
the differenced model assigns more warming to Northern Europe and much less to Central/Western and Southwestern Europe, and it reverses the relation between warming and latitude.
The level model reproduces the observed station warming much more closely: across stations, the correlation between observed and estimated common warming is $0.81$ for the level model and $0.43$ for the differenced model, with root mean squared deviations of $0.24^\circ\mathrm{C}$ and $0.58^\circ\mathrm{C}$.
The second factor, which the differenced panel no longer detects, is thus the component that carries the regional structure of European warming.

\begin{table}[htbp]
	\centering
	\small
	\begin{tabular}{@{}lccc@{}}
		\toprule
		& Observed & Levels, two factors & Differences, one factor \\
		\midrule
		Eastern Europe ($34$)          & $2.43$ & $2.52$ & $2.38$ \\
		Southwestern Europe ($24$)     & $2.22$ & $1.88$ & $1.09$ \\
		Central/Western Europe ($116$) & $2.16$ & $2.31$ & $1.81$ \\
		Southeastern Europe ($27$)     & $2.12$ & $2.01$ & $1.60$ \\
		Northern Europe ($107$)        & $2.08$ & $1.96$ & $2.45$ \\
		\midrule
		All stations ($308$)           & $2.16$ & $2.15$ & $2.02$ \\
		\bottomrule
	\end{tabular}
	\caption{Regional mean warming in $^\circ\mathrm{C}$, measured as the mean over $2021$-$2025$ minus the mean over $1960$-$1990$; number of stations in parentheses. Observed: low-frequency station series. Levels: estimated common component of the two-factor model in levels. Differences: cumulated common component of the one-factor model selected for the differenced panel.}
	\label{tab:climate_differencing_regions}
\end{table}

\begin{figure}[htbp]
	\centering
	\includegraphics[width=0.8\textwidth]{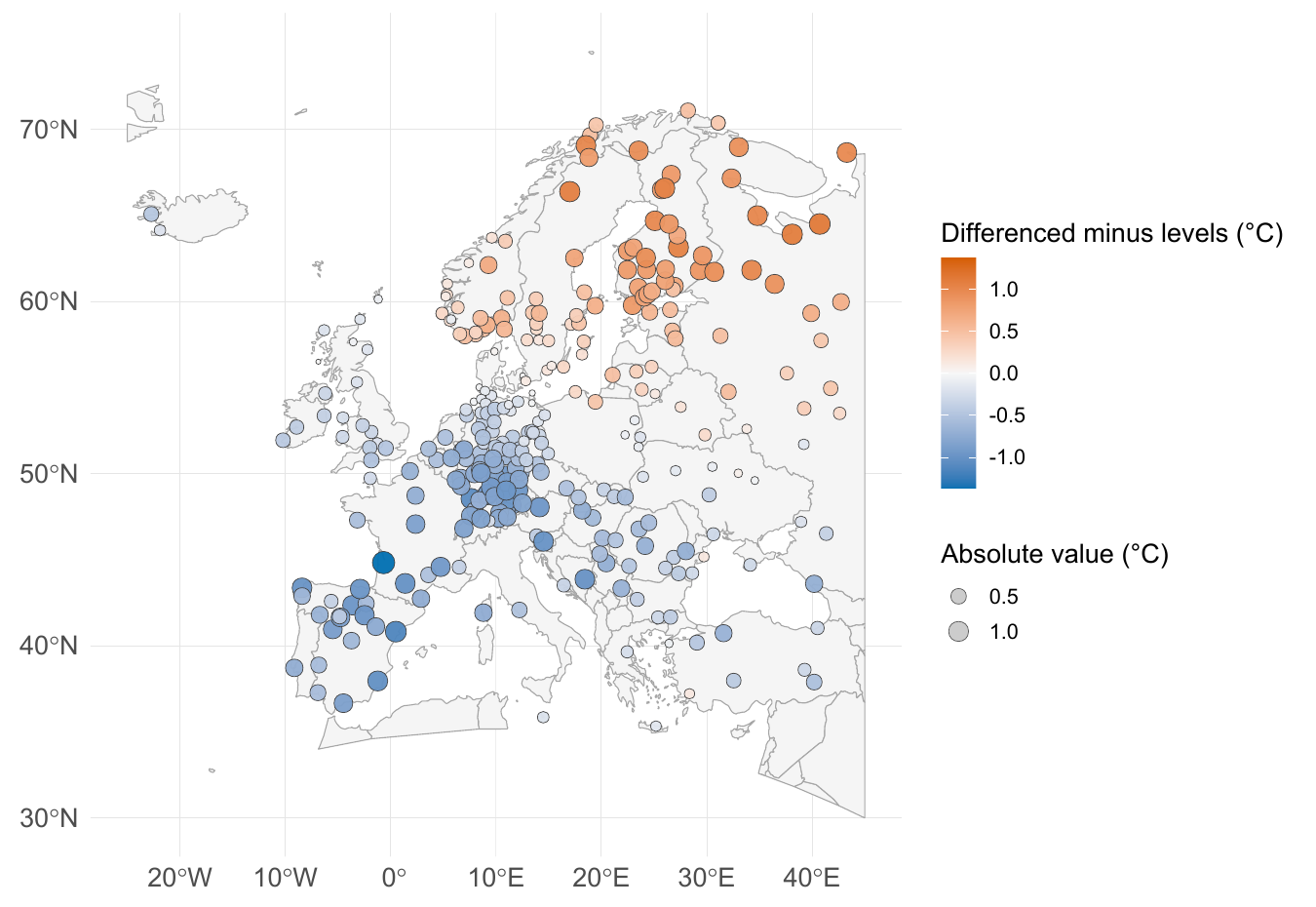}
	\caption{Station-specific common warming from the one-factor model selected for the differenced panel minus that from the two-factor model in levels. Warming is measured as the mean over $2021$-$2025$ minus the mean over $1960$-$1990$; point size is proportional to the absolute difference.}
	\label{fig:climate_differencing_map}
\end{figure}

Figure~\ref{fig:climate_differencing_map} maps the difference between the two estimates station by station.
Differencing reduces the estimated warming at $198$ of the $308$ stations, by up to $1.37^\circ\mathrm{C}$ in southwestern France, with the largest reductions in France, Spain and southwestern Germany, and raises it at the remaining $110$, by up to $1.12^\circ\mathrm{C}$ in Fennoscandia and northwestern Russia.

\section{Conclusion}
\label{sec:conclusion}

This paper develops the asymptotic theory of principal components for approximate factor models whose common factors are fractionally integrated with (possibly) heterogeneous memory. Factors of equal memory form groups, with unknown memory parameters that may take any value greater than $-1/2$, excluding $d=1/2$, while the idiosyncratic component is stationary. The theory applies to the panel in levels and requires neither knowledge of the integration orders nor a transformation, such as pre-differencing, of the data before estimation.

The results rest on the asymptotic ordering of the eigenvalues of the sample covariance matrix by memory. A nonstationary factor group with memory $d^{(j)}>1/2$ generates eigenvalues of order $NT^{2d^{(j)}-1}$, stationary factors generate eigenvalues of order $N$, and the idiosyncratic eigenvalues remain bounded. Divergence in $N$ separates the common from the idiosyncratic component, and differences in memory separate the factor groups from each other. Our three key results build upon this asymptotic ordering: First, principal components computed from the panel in levels estimate the factors consistently up to rotation, and the space of the first $j$ groups is separated from the next group at a rate determined by the memory gap $d^{(j)}-d^{(j+1)}$. Second, the penalized criteria of \citet{BaiNg2002}, in value-function and in logarithmic form, remain consistent for the number of factors, and their penalty acts as a threshold that selects the factor groups in the order of their memory. Third, in the diminishing-signal regime, where the nonstationary factors keep their memory but lose their variance advantage, estimation and factor-number selection in levels remain consistent because the factors are still pervasive in the cross-section. After differencing, a factor with memory $d_i$ generates an eigenvalue of order $NT^{1-2d_i}$. When $N$ grows more slowly than $T^{2d_i-1}$ this eigenvalue vanishes, and the criterion applied to the differenced panel does not select the true number of factors asymptotically.

The Monte Carlo study measures these effects as the largest idiosyncratic standard deviation at which the criterion still selects the true number of factors in half of the replications. Under standard signals, levels tolerate more noise than first differences at every panel size. Under diminishing signals, the tolerable noise in levels still increases with $N$ and $T$, whereas after differencing it is close to zero and falls as $T$ grows. 

In the application to monthly temperatures at $308$ European weather stations from 1960 to 2025, the criterion selects two factors in levels. The first factor loads positively at $290$ stations and has an estimated memory of $1.15$, with a $95\%$ confidence interval of $[0.81,1.59]$. It reflects common warming. The second factor contrasts northern with southern Europe, and the confidence interval of its memory includes stationary values. The estimated common component implies that 2021-2025 was $2.15^\circ\mathrm{C}$ warmer than 1960-1990, with the largest warming in Eastern Europe and the smallest in Southwestern Europe. After first differencing, the criterion selects one factor. The cumulated common component still implies an average warming of $2.02^\circ\mathrm{C}$, but it distributes the warming differently across regions. Across stations, its correlation with the observed warming is $0.43$, against $0.81$ for the level model. In this application differencing does not remove the aggregate trend, but it loses a factor and the regional pattern of the common component.

For applied work, the results imply that a panel whose nonstationarity comes from its common factors can be analyzed in levels. Pre-differencing requires a choice of the integration order, cannot return all factors to $I(0)$ when their memory differs, and reduces or removes the time-series signal of persistent factors. It should therefore be a modeling decision rather than a default step of the analysis.

Several questions remain open: Our theory assumes a stationary idiosyncratic component, so that panels with nonstationary idiosyncratic errors, as in the PANIC framework of \citet{Bai2004Panic}, need a different treatment. We establish consistency but no limiting distributions, so inference is left for further work. Finally, the threshold interpretation of the penalty suggests a procedure that selects not only the total number of factors but also the size of each memory group, which would give a data-driven estimate of how many common trends, cycles, and stationary components a panel contains.

\section*{Acknowledgments}
The research of this paper has partly been conducted while the second author was a visiting postdoc at Princeton University, and a research associate at the University of Regensburg and the Institute for Employment Research (IAB) Nuremberg.
The authors thank Matteo Barigozzi for valuable comments. Support through the project \textit{356439312} financed by the German Research Foundation (DFG) and the project \textit{VI.Veni.251E.025} financed by the Dutch Research Council (NWO) is gratefully acknowledged.

\section*{Data availability statement}
The climate data used in this study are publicly available from the NOAA National Centers for Environmental Information through the Global Historical Climatology Network--Monthly (GHCN-M), version 4 (\url{https://www.ncei.noaa.gov/pub/data/ghcn/v4/}). The GitHub repository \url{https://github.com/D4monik/long-memory-factor-models} provides the original data version used in this study, the processed climate data, and the code and results needed to reproduce the Monte Carlo study and the climate application.

\section*{Declaration of use of generative artificial intelligence}

During the preparation of this work, the authors used generative AI tools, including ChatGPT, Claude, and DeepL, to improve the language, clarity, and structure of the manuscript, to assist with code development and debugging, to explore research ideas and to critically examine mathematical arguments and proofs. 

The authors reviewed and edited all outputs as needed and take full responsibility for the theoretical results, computational analyses, and final content of the manuscript. No figures or tables were directly generated using generative AI.

\section*{Disclosure statement}
The authors report there are no competing interests to declare.
\appendix
\numberwithin{equation}{section}
\clearpage
\section{Notation and Definitions}
\label{sec:notation}

The model notation, factor-group structure, norms, stochastic orders,
and basic eigenvalue and eigenvector notation are introduced in
Section~\ref{sec:model}. We collect here the additional notation used
throughout the proofs. Whenever a factor group is absent, the corresponding blocks are omitted.

\begin{enumerate}
	
	\item \textbf{Empirical covariance matrices}
	
	The sample covariance matrix
	$\widehat{\Sigma}_{y,T}$ is defined in
	\eqref{eq:sample_covariance_y}. Empirical covariance and
	cross-covariance matrices involving latent DGP components are
	denoted by a tilde. For generic processes $a_t$ and $b_t$, write
	\[
	\widetilde{\Sigma}_{ab}
	=
	\frac{1}{T}
	\sum_{t=1}^T
	a_tb_t^T,
	\qquad
	\widetilde{\Sigma}_a
	=
	\widetilde{\Sigma}_{aa}.
	\]
	Thus, for example, $\widetilde{\Sigma}_z$,
	$\widetilde{\Sigma}_w$, $\widetilde{\Sigma}_u$, and
	$\widetilde{\Sigma}_{xu}$ denote the corresponding empirical
	covariance and cross-covariance matrices. To avoid unnecessarily
	long expressions, the sample-size subscript $T$ is suppressed for
	these latent covariance matrices. It is retained for
	$\widehat{\Sigma}_{y,T}$ and for the artificial covariance
	matrices associated with $y_t$ defined below. 
	
	\item \textbf{Scaling for asymptotics}
	
Motivated by the second-moment orders of Section~\ref{subsec:signal_regimes}, define
	\begin{align*}
		D_{T,0}
		&=
		\diag\left(
		D_{z,T,0},
		D_{w,0},
		D_{\perp,0}
		\right),
		\nonumber\\
		D_{z,T,0}
		&=
		\diag\left(
		T^{d_{0,1}-1/2},
		\ldots,
		T^{d_{0,s_0^*}-1/2}
		\right),
		\nonumber\\
		D_{w,0}
		&=
		I_{s_0-s_0^*},
		\qquad
		D_{\perp,0}
		=
		I_{N-s_0}.
	\end{align*}
	This scaling removes the diverging second-moment orders of the
	nonstationary coordinates while leaving the stationary and
	idiosyncratic coordinates unchanged.
	
	\item \textbf{QR decomposition and factor-loading spaces}
	
	For the loading matrix of the DGP, write the reduced QR decomposition as
	\[
	\Lambda_0
	=
	Q_0R_0,
	\qquad
	Q_0^TQ_0
	=
	I_{s_0},
	\]
	where $Q_0$ has orthonormal columns and $R_0$ is upper triangular.
	Let $Q_{\perp,0}$ denote an orthonormal basis for the orthogonal
	complement of $\operatorname{span}(Q_0)$.
	
	Since $\Lambda_0$ has full column rank,
	\[
	\operatorname{span}(\Lambda_0)
	=
	\operatorname{span}(Q_0).
	\]
	We refer to this $s_0$-dimensional subspace of $\mathbb R^N$ as
	the \emph{factor-loading space}. When there is no ambiguity, we
	use \emph{factor space} and \emph{loading space} as shorter terms
	for the same cross-sectional subspace.
	
	Premultiplication of the factor model by $Q_0^T$ gives
	\begin{align*}
		Q_0^Ty_t
		&=
		R_0x_t+Q_0^Tu_t.
	\end{align*}
	Moreover, Assumption~\ref{ass:strong_factors} implies
	\begin{align}
		\frac{1}{N}R_0^TR_0
		&=
		\frac{1}{N}\Lambda_0^T\Lambda_0
		\longrightarrow
		\Sigma_{\Lambda,0}.
		\label{eq:strong_factors_RR}
	\end{align}
	Thus, the QR decomposition replaces the generally nonorthonormal basis
		given by the columns of $\Lambda_0$ with the orthonormal basis given by
		the columns of $Q_0$, while $R_0$ contains the corresponding
		cross-sectional rescaling and change of coordinates.
	
	Partitioning the QR decomposition according to the factor groups
	in \eqref{eq:factor_groups} gives
	\begin{align*}
		\Lambda_0
		&=
		\left[
		\Lambda_0^{(1)},
		\ldots,
		\Lambda_0^{(q_0)}
		\right]
		\nonumber\\
		&=
		\left[
		Q_0^{(1)},
		\ldots,
		Q_0^{(q_0)}
		\right]
		\begin{pmatrix}
			R_0^{(1,1)}
			&
			R_0^{(1,2)}
			&
			\cdots
			&
			R_0^{(1,q_0)}
			\\
			0
			&
			R_0^{(2,2)}
			&
			\cdots
			&
			R_0^{(2,q_0)}
			\\
			\vdots
			&
			\vdots
			&
			\ddots
			&
			\vdots
			\\
			0
			&
			0
			&
			\cdots
			&
			R_0^{(q_0,q_0)}
		\end{pmatrix}.
	\end{align*}
	Since $R_0$ is block upper triangular,
		$\Lambda_0^{(j)} =
		\sum_{k=1}^j Q_0^{(k)}R_0^{(k,j)}, \
		j=1,\ldots,q_0.$
		Hence, $Q_0^{(j)}$ need not by itself span the same subspace as
		$\Lambda_0^{(j)}$. The cumulative factor-loading spaces, however, are
		preserved:
\begin{align}
	\operatorname{span}\left(
	\Lambda_0^{(1:j)}
	\right)
	=
	\operatorname{span}\left(
	Q_0^{(1:j)}
	\right),
	\qquad
	j=1,\ldots,q_0.
	\label{eq:QR_cumulative_loading_spaces}
\end{align} 
	This is why the groupwise consistency results below are stated for
	cumulative factor-loading spaces.
	
	\clearpage 
	
	\item \textbf{Nonstationary and stationary factor blocks}
	
	For proofs that only distinguish between nonstationary and
	stationary factor blocks of the DGP, write
	\[
	z_t
	=
	x_t^{(1:q_0^*)},
	\qquad
	w_t
	=
	x_t^{((q_0^*+1):q_0)}.
	\]
	and define
		\[
		Q_{z,0}
		=
		Q_0^{(1:q_0^*)},
		\qquad
		Q_{w,0}
		=
		Q_0^{((q_0^*+1):q_0)}.
		\]
	For the individual nonstationary groups, we also write
	\[
	z_t^{(j)}
	=
	x_t^{(j)},
	\qquad
	Q_{z,0}^{(j)}
	=
	Q_0^{(j)},
	\qquad
	j=1,\ldots,q_0^*.
	\]
	The corresponding QR representation is
	\begin{align}
		y_t
		&=
		\left[
		Q_{z,0},
		Q_{w,0}
		\right]
		\begin{pmatrix}
			R_{zz,0}
			&
			R_{zw,0}
			\\
			0
			&
			R_{ww,0}
		\end{pmatrix}
		\begin{pmatrix}
			z_t
			\\
			w_t
		\end{pmatrix}
		+
		u_t.
		\label{eq:z_w_QR}
	\end{align}
	
	The columns of $Q_{z,0}$ span the nonstationary factor-loading
	space. The columns of $Q_{w,0}$ span its orthogonal complement
	within the full factor-loading space. Hence,
	\[
	\operatorname{span}(Q_0)
	=
	\operatorname{span}(Q_{z,0})
	\oplus
	\operatorname{span}(Q_{w,0}).
	\]
	Because of the block $R_{zw,0}$,
	$\operatorname{span}(Q_{w,0})$ need not coincide with the span of
	the original stationary loading columns.
	
	For blockwise arguments in the sequel, define
	\[
	Q_{all,0}
	=
	\left[
	Q_{z,0},
	Q_{w,0},
	Q_{\perp,0}
	\right]
	=
	\left[
	Q_0,
	Q_{\perp,0}
	\right].
	\]
	We regard the orthogonal complement $Q_{\perp,0}$ as an additional block and set
	\[
	Q_{all,0}^{(j)}
	=
	Q_0^{(j)},
	\qquad
	j=1,\ldots,q_0,
	\qquad
	Q_{all,0}^{(q_0+1)}
	=
	Q_{\perp,0}.
	\]
	Since $Q_{all,0}$ is orthogonal,
	\begin{align}
		Q_{all,0}^Ty_t
		&=
		\begin{pmatrix}
			R_{zz,0}z_t+R_{zw,0}w_t
			\\
			R_{ww,0}w_t
			\\
			0
		\end{pmatrix}
		+
		\begin{pmatrix}
			Q_{z,0}^Tu_t
			\\
			Q_{w,0}^Tu_t
			\\
			Q_{\perp,0}^Tu_t
		\end{pmatrix}.
		\label{eq:Q_all_y}
	\end{align}
	Thus, only the first block contains nonstationary factor
	variation, although it may also contain stationary variation
	through $R_{zw,0}w_t$. The second block contains only stationary
	common variation, while the third contains no common-factor
	variation.
	
	\item \textbf{Matrix square roots}
	
	For a symmetric positive semidefinite $(n\times n)$ matrix $A$,
	write its spectral decomposition as
	\[
	A
	=
	V(A)
	\diag\left(
	\EW{1}{A},
	\ldots,
	\EW{n}{A}
	\right)
	V(A)^T.
	\]
	Its positive semidefinite square root is obtained by taking the square
	root of each eigenvalue:
	\begin{align*}
		A^{1/2}
		&=
		V(A)
		\diag\left(
		\sqrt{\EW{1}{A}},
		\ldots,
		\sqrt{\EW{n}{A}}
		\right)
		V(A)^T.
	\end{align*}
	Thus,
	\[
	A^{1/2}A^{1/2}
	=
	A.
	\]
	If $A$ is positive definite, its inverse square root is defined
	analogously as
	\begin{align*}
		A^{-1/2}
		&=
		V(A)
		\diag\left(
		\frac{1}{\sqrt{\EW{1}{A}}},
		\ldots,
		\frac{1}{\sqrt{\EW{n}{A}}}
		\right)
		V(A)^T,
	\end{align*}
	so that
	\[
	A^{-1/2}AA^{-1/2}
	=
	I_n.
	\]
	Whenever an inverse or inverse square root of a random covariance
		matrix is used below, the proofs establish that its smallest
		eigenvalue is bounded away from zero in probability, possibly after
		the appropriate scaling. Thus, the matrix is positive definite with
		probability approaching one.

\item \textbf{Canonical angles}

Let $A$ and $\widetilde A$ be symmetric matrices, and let
$V(A)^{(j)}$ and $V(\widetilde A)^{(j)}$ contain the eigenvectors
spanning their respective $j$th eigenspaces, each of dimension $s_j$.
Following \citet{Davis1970} and \citet{Dopico2000}, define the
canonical angles
\[
0
\leq
\theta_1
\leq
\cdots
\leq
\theta_{s_j}
\leq
\frac{\pi}{2}
\]
by
\begin{align}
	\cos\left(\theta_i\right)
	&=
	\sqrt{
		\EW{i}{
			V(A)^{(j)T}
			V(\widetilde A)^{(j)}
			V(\widetilde A)^{(j)T}
			V(A)^{(j)}
		}
	},
	\qquad
	i=1,\ldots,s_j.
	\label{eq:canonical_angle_cosine}
\end{align}
The corresponding canonical-angle matrix is
\begin{align*}
	\Theta\left(
	V(A)^{(j)},
	V(\widetilde A)^{(j)}
	\right)
	&=
	\diag\left(
	\theta_1,\ldots,\theta_{s_j}
	\right).
\end{align*}
Thus, the eigenvalues in
\eqref{eq:canonical_angle_cosine} are the squared cosines of the
canonical angles.

Let $V(\widetilde A)^{(j),\perp}$ contain all eigenvectors of
$\widetilde A$ except those in $V(\widetilde A)^{(j)}$. Since
\[
V(\widetilde A)^{(j)}
V(\widetilde A)^{(j)T}
+
V(\widetilde A)^{(j),\perp}
V(\widetilde A)^{(j),\perp T}
=
I,
\]
the eigenvalues of
\[
V(A)^{(j)T}
V(\widetilde A)^{(j),\perp}
V(\widetilde A)^{(j),\perp T}
V(A)^{(j)}
\]
are the squared sines of the same canonical angles. In particular,
\begin{align*}
	\norm{
		\sin\Theta\left(
		V(A)^{(j)},
		V(\widetilde A)^{(j)}
		\right)
	}_F
	&=
	\norm{
		V(A)^{(j)T}
		V(\widetilde A)^{(j),\perp}
	}_F =
	\norm{
	V(\widetilde A)^{(j),\perp T} V(A)^{(j)}
	}_F.
\end{align*}
The distance equals zero if the two eigenspaces coincide and equals
$\sqrt{s_j}$ if they are orthogonal.

The same notation is used when one of the subspaces is represented
directly by a matrix with orthonormal columns, such as $Q_{z,0}$ or
$Q_0$. In this case, the corresponding matrix simply replaces the
eigenvector matrix in the definitions above. The same convention
applies to cumulative blocks, such as $V(A)^{(1:j)}$ and
$Q_{z,0}^{(1:j)}$, with dimension $s^{(1:j)}$.

	\item \textbf{Artificial covariance matrices}
	
	To isolate the nonstationary factor-loading space from the
	remaining coordinates, define
	\begin{align}
		\widetilde{\Sigma}_{\widetilde{Qy},T}
		&=
		\nonumber\\[-0.5em]
		&\quad
		\begin{pmatrix}
			R_{zz,0}\widetilde{\Sigma}_zR_{zz,0}^T
			&
			0
			&
			0
			\\
			0
			&
			R_{ww,0}\widetilde{\Sigma}_wR_{ww,0}^T
			+
			Q_{w,0}^T\widetilde{\Sigma}_uQ_{w,0}
			&
			Q_{w,0}^T\widetilde{\Sigma}_uQ_{\perp,0}
			\\
			0
			&
			Q_{\perp,0}^T\widetilde{\Sigma}_uQ_{w,0}
			&
			Q_{\perp,0}^T\widetilde{\Sigma}_uQ_{\perp,0}
		\end{pmatrix}.
		\label{eq:Sigma_widetilde_Qy_def}
	\end{align}
	The corresponding matrix in the original coordinates is
	\begin{align}
		\widetilde{\Sigma}_{\tilde y,T}
		&=
		Q_{all,0}
		\widetilde{\Sigma}_{\widetilde{Qy},T}
		Q_{all,0}^T.
		\label{eq:Sigma_tilde_y_def}
	\end{align}
	The matrix in \eqref{eq:Sigma_widetilde_Qy_def} is block diagonal
	between the nonstationary coordinates and the joint
	stationary-idiosyncratic coordinates.
	
	For the separation of the full factor-loading space from its
	orthogonal complement, define
	\begin{align}
		\widetilde{\Sigma}_{\widetilde{Qy},T}^{\mathrm{bd}}
		&=
		\nonumber\\[-0.5em]
		&\quad
		\begin{pmatrix}
			R_{zz,0}\widetilde{\Sigma}_zR_{zz,0}^T
			&
			0
			&
			0
			\\
			0
			&
			R_{ww,0}\widetilde{\Sigma}_wR_{ww,0}^T
			+
			Q_{w,0}^T\widetilde{\Sigma}_uQ_{w,0}
			&
			0
			\\
			0
			&
			0
			&
			Q_{\perp,0}^T\widetilde{\Sigma}_uQ_{\perp,0}
		\end{pmatrix}.
		\label{eq:Sigma_widetilde_Qy_bd_def}
	\end{align}
	Its counterpart in the original coordinates is
	\begin{align}
		\widetilde{\Sigma}_{\tilde y,T}^{\mathrm{bd}}
		&=
		Q_{all,0}
		\widetilde{\Sigma}_{\widetilde{Qy},T}^{\mathrm{bd}}
		Q_{all,0}^T.
		\label{eq:Sigma_tilde_y_bd_def}
	\end{align}
The two matrices differ only through the off-diagonal empirical
idiosyncratic covariance between the stationary factor-loading space
and its orthogonal complement. In particular,
\begin{align}
	\widetilde{\Sigma}_{\widetilde{Qy},T}
	-
	\widetilde{\Sigma}_{\widetilde{Qy},T}^{\mathrm{bd}}
	&=
	\begin{pmatrix}
		0
		&
		0
		&
		0
		\\
		0
		&
		0
		&
		Q_{w,0}^T\widetilde{\Sigma}_uQ_{\perp,0}
		\\
		0
		&
		Q_{\perp,0}^T\widetilde{\Sigma}_uQ_{w,0}
		&
		0
	\end{pmatrix}.
	\label{eq:Sigma_tilde_Qy_bd_difference}
\end{align}
	
	Finally, to separate the individual nonstationary factor groups,
	define
	\begin{align}
		\widetilde{\widetilde{\Sigma}}_
		{\widetilde{\widetilde{Qy}},T}
		&=
		\nonumber\\[-0.5em]
		&\quad
		\begin{pmatrix}
			\displaystyle
			\diag_{j=1,\ldots,q_0^*}
			\left(
			R_0^{(j,j)}
			\widetilde{\Sigma}_{z^{(j)}}
			\left(R_0^{(j,j)}\right)^T
			\right)
			&
			0
			&
			0
			\\
			0
			&
			R_{ww,0}\widetilde{\Sigma}_wR_{ww,0}^T
			+
			Q_{w,0}^T\widetilde{\Sigma}_uQ_{w,0}
			&
			Q_{w,0}^T\widetilde{\Sigma}_uQ_{\perp,0}
			\\
			0
			&
			Q_{\perp,0}^T\widetilde{\Sigma}_uQ_{w,0}
			&
			Q_{\perp,0}^T\widetilde{\Sigma}_uQ_{\perp,0}
		\end{pmatrix}.
		\label{eq:Sigma_tildetilde_tildetilde_Qy}
	\end{align}
	Here, the upper-left entry denotes the block-diagonal matrix over
	$j=1,\ldots,q_0^*$. The corresponding matrix in the original
	coordinates is
	\begin{align}
		\widetilde{\widetilde{\Sigma}}_{\tilde{\tilde y},T}
		&=
		Q_{all,0}
		\widetilde{\widetilde{\Sigma}}_
		{\widetilde{\widetilde{Qy}},T}
		Q_{all,0}^T.
		\label{eq:Sigma_tildetilde_tildetilde_y}
	\end{align}
	The first two artificial covariance matrices are used in
		Theorem~\ref{T:Factor_space} to separate, respectively, the
		nonstationary factor-loading space and the full factor-loading space
		from their complements. The third is used in
		Theorem~\ref{T:Factor_space_2} to separate the individual
		nonstationary factor-loading groups.
	
\end{enumerate}

\clearpage
\clearpage
\section{Proofs and auxiliary results for the standard-signal regime}
\label{sec:theorems}

\subsection{Factor-loading-space consistency}
\label{sec:Factor-loading-space}
\begin{theorem}[Consistency of the factor-loading spaces]
	\label{T:Factor_space}
	Suppose Assumptions~\ref{ass:factors}-\ref{ass:memory} hold and
	that $N,T \to \infty$ with $N/T\to0$. If $s_0^*>0$, then
	\begin{align}
		\norm{
			\sin\Theta\left(
			V\left(
			\widehat{\Sigma}_{y,T}
			\right)^{(1:q_0^*)},
			Q_{z,0}
			\right)
		}_F
		&=
		o_p\left(
		T^{-d_0^{(q_0^*)}+1/2}
		\right).
		\label{eq:Theorem_Qz_app}
	\end{align}
	If $s_0^*<s_0$, then
	\begin{align}
		\norm{
			\sin\Theta\left(
			V\left(
			\widehat{\Sigma}_{y,T}
			\right)^{(1:q_0)},
			Q_0
			\right)
		}_F
		&=
		o_p\left(
		N^{-1/2}
		\right).
		\label{eq:Theorem_Qq_app}
	\end{align}
\end{theorem}

\begin{proof}
Suppose first that $s_0^*>0$. By
\eqref{eq:Sigma_widetilde_Qy_def} and
\eqref{eq:Sigma_tilde_y_def},
$\widetilde{\Sigma}_{\tilde y,T}$ is block diagonal in the basis
$Q_{all,0}$ between the nonstationary factor-loading space
$\operatorname{span}(Q_{z,0})$ and its orthogonal complement.
Moreover, Lemma~\ref{L:level_covariance_eigenvalue_orders} shows that
the eigenvalues associated with the nonstationary block dominate those
of the remaining blocks. Hence, with probability approaching one,
\[
\operatorname{span}\left(
V\left(
\widetilde{\Sigma}_{\tilde y,T}
\right)^{(1:q_0^*)}
\right)
=
\operatorname{span}(Q_{z,0}).
\]
Therefore,
	\begin{align*}
		&\norm{
			\sin\Theta\left(
			V\left(
			\widehat{\Sigma}_{y,T}
			\right)^{(1:q_0^*)},
			Q_{z,0}
			\right)
		}_F =
		\norm{
			\sin\Theta\left(
			V\left(
			\widehat{\Sigma}_{y,T}
			\right)^{(1:q_0^*)},
			V\left(
			\widetilde{\Sigma}_{\tilde y,T}
			\right)^{(1:q_0^*)}
			\right)
		}_F.
	\end{align*}
With $\rho_{z,T}$ defined in \eqref{eq:relative_gap_z_definition}, the relative eigenspace perturbation bound of \citet[Theorem~2.4]{Dopico2000} gives
	\begin{align}
		&\norm{
			\sin\Theta\left(
			V\left(
			\widehat{\Sigma}_{y,T}
			\right)^{(1:q_0^*)},
			Q_{z,0}
			\right)
		}_F\leq
		\frac{1}{\rho_{z,T}}
		\norm{
			\widetilde{\Sigma}_{\tilde y,T}^{-1/2}
			\left(
			\widehat{\Sigma}_{y,T}
			-
			\widetilde{\Sigma}_{\tilde y,T}
			\right)
			\widehat{\Sigma}_{y,T}^{-1/2}
		}_F.
		\label{eq:Dopico_Theorem_Qz}
	\end{align}
	By \eqref{eq:eigenvalue_order_gap_z_uniform},
	\begin{align}
		\rho_{z,T}^{-1}
		&=
		O_p\left(
		T^{-d_0^{(q_0^*)}+1/2}
		\right),
		\label{eq:rho_z_rate}
	\end{align}
	while Lemma~\ref{L:Sigma_y_hat_tilde_diff_N} gives
	\begin{align}
		\norm{
			\widetilde{\Sigma}_{\tilde y,T}^{-1/2}
			\left(
			\widehat{\Sigma}_{y,T}
			-
			\widetilde{\Sigma}_{\tilde y,T}
			\right)
			\widehat{\Sigma}_{y,T}^{-1/2}
		}_F
		&=
		o_p(1).
		\label{eq:weighted_perturbation_tilde_y}
	\end{align}
	Inserting \eqref{eq:rho_z_rate} and
	\eqref{eq:weighted_perturbation_tilde_y} into
	\eqref{eq:Dopico_Theorem_Qz} proves
	\eqref{eq:Theorem_Qz_app}.
	
Suppose now that $s_0^*<s_0$. By
\eqref{eq:Sigma_widetilde_Qy_bd_def} and
\eqref{eq:Sigma_tilde_y_bd_def},
$\widetilde{\Sigma}_{\tilde y,T}^{\mathrm{bd}}$ is block diagonal in
the basis $Q_{all,0}$ between the factor-loading space
$\operatorname{span}(Q_0)$ and its orthogonal complement.
Lemma~\ref{L:level_covariance_eigenvalue_orders} shows that the
eigenvalues associated with the factor block dominate those of the
idiosyncratic block. Hence, with probability approaching one,
\[
\operatorname{span}\left(
V\left(
\widetilde{\Sigma}_{\tilde y,T}^{\mathrm{bd}}
\right)^{(1:q_0)}
\right)
=
\operatorname{span}(Q_0).
\]
Hence,
	\begin{align*}
		\norm{
			\sin\Theta\left(
			V\left(
			\widehat{\Sigma}_{y,T}
			\right)^{(1:q_0)},
			Q_0
			\right)
		}_F
		&=
		\norm{
			\sin\Theta\left(
			V\left(
			\widehat{\Sigma}_{y,T}
			\right)^{(1:q_0)},
			V\left(
			\widetilde{\Sigma}_{\tilde y,T}^{\mathrm{bd}}
			\right)^{(1:q_0)}
			\right)
		}_F.
	\end{align*}
With $\rho_{0,N}$ defined in \eqref{eq:relative_gap_factor_idiosyncratic_definition}, another application of \citet[Theorem~2.4]{Dopico2000} yields
	\begin{align}
		\norm{
			\sin\Theta\left(
			V\left(
			\widehat{\Sigma}_{y,T}
			\right)^{(1:q_0)},
			Q_0
			\right)
		}_F
		&\leq
		\frac{1}{\rho_{0,N}}
		\norm{
			\left(
			\widetilde{\Sigma}_{\tilde y,T}^{\mathrm{bd}}
			\right)^{-1/2}
			\left(
			\widehat{\Sigma}_{y,T}
			-
			\widetilde{\Sigma}_{\tilde y,T}^{\mathrm{bd}}
			\right)
			\widehat{\Sigma}_{y,T}^{-1/2}
		}_F.
		\label{eq:Dopico_Theorem_Q0}
	\end{align}
	By \eqref{eq:eigenvalue_order_gap_factor_idiosyncratic},
	\begin{align}
		\rho_{0,N}^{-1}
		&=
		O_p\left(
		N^{-1/2}
		\right),
		\label{eq:rho_0_rate}
	\end{align}
	and Lemma~\ref{L:Sigma_y_hat_tilde_bd_diff_N} gives
	\begin{align}
		\norm{
			\left(
			\widetilde{\Sigma}_{\tilde y,T}^{\mathrm{bd}}
			\right)^{-1/2}
			\left(
			\widehat{\Sigma}_{y,T}
			-
			\widetilde{\Sigma}_{\tilde y,T}^{\mathrm{bd}}
			\right)
			\widehat{\Sigma}_{y,T}^{-1/2}
		}_F
		&=
		o_p(1).
		\label{eq:weighted_perturbation_tilde_y_bd}
	\end{align}
	Inserting \eqref{eq:rho_0_rate} and
	\eqref{eq:weighted_perturbation_tilde_y_bd} into
	\eqref{eq:Dopico_Theorem_Q0} proves
	\eqref{eq:Theorem_Qq_app}.
\end{proof}

We now refine the result within the nonstationary factor space.
Theorem~\ref{T:Factor_space} establishes consistency of the full
nonstationary factor space, but does not distinguish the individual
nonstationary factor groups.

By \eqref{eq:Sigma_tildetilde_tildetilde_Qy} and
\eqref{eq:Sigma_tildetilde_tildetilde_y},
\[
\widetilde{\widetilde{\Sigma}}_{\tilde{\tilde y},T}
=
Q_{all,0}
\widetilde{\widetilde{\Sigma}}_{
	\widetilde{\widetilde{Qy}},T
}
Q_{all,0}^T.
\]
The covariance matrix
$\widetilde{\widetilde{\Sigma}}_{
	\widetilde{\widetilde{Qy}},T
}$
is block diagonal with respect to the individual nonstationary factor
groups and the remaining stationary-idiosyncratic block. Since
$Q_{all,0}$ is orthogonal and the groups are ordered by decreasing
memory, it follows that, for every $j=1,\ldots,q_0^*$,
$
V\left(
\widetilde{\widetilde{\Sigma}}_{\tilde{\tilde y},T}
\right)^{(j)}
$
spans the same subspace as $Q_{z,0}^{(j)}$. Consequently,
$
V\left(
\widetilde{\widetilde{\Sigma}}_{\tilde{\tilde y},T}
\right)^{(1:j)}
$
spans the same subspace as $Q_{z,0}^{(1:j)}$.

The next theorem combines this block structure with the relative
eigenvalue-gap rate in
\eqref{eq:eigenvalue_order_gap_adjacent_groups} and the relative
perturbation bound in
Lemma~\ref{L:Sigma_y_hat_tildetilde_diff_N}.

\begin{theorem}[Separation of nonstationary factor groups]
	\label{T:Factor_space_2}
	Suppose Assumptions~\ref{ass:factors}-\ref{ass:memory} hold and
	that $N,T \to \infty$ with $N/T\to0$. Then, for every
	$j=1,\ldots,q_0^*-1$,
	\begin{align}
		\norm{
			\sin\Theta\left(
			V\left(
			\widehat{\Sigma}_{y,T}
			\right)^{(1:j)},
			Q_{z,0}^{(1:j)}
			\right)
		}_F
		&=
		O_p\left(
		T^{-d_0^{(j)}+d_0^{(j+1)}}
		\right).
		\label{eq:Theorem_Qj_app}
	\end{align}
\end{theorem}

\begin{proof}
	Fix $j\in\{1,\ldots,q_0^*-1\}$. By \eqref{eq:Sigma_tildetilde_tildetilde_y},
	$\widetilde{\widetilde{\Sigma}}_{\tilde{\tilde y},T}$ is block diagonal in the
	basis $Q_{\mathrm{all},0}$ with respect to the individual
	nonstationary factor groups and the remaining
	stationary-idiosyncratic block. Since the groups are ordered by
	decreasing memory,
$
	V\left(
	\widetilde{\widetilde{\Sigma}}_{\tilde{\tilde y},T}
	\right)^{(1:j)}
$
	spans the same subspace as $Q_{z,0}^{(1:j)}$. Hence,
	\begin{align*}
		\norm{
			\sin\Theta\left(
			V\left(
			\widehat{\Sigma}_{y,T}
			\right)^{(1:j)},
			Q_{z,0}^{(1:j)}
			\right)
		}_F
		&=
		\norm{
			\sin\Theta\left(
			V\left(
			\widehat{\Sigma}_{y,T}
			\right)^{(1:j)},
			V\left(
			\widetilde{\widetilde{\Sigma}}_{
				\tilde{\tilde y},T
			}
			\right)^{(1:j)}
			\right)
		}_F.
	\end{align*}
	
	With $\rho_{j,T}$ defined in
	\eqref{eq:relative_gap_adjacent_groups_definition}, the relative
	eigenspace perturbation bound of
	\citet[Theorem~2.4]{Dopico2000} gives
	\begin{align}
		\norm{
			\sin\Theta\left(
			V\left(
			\widehat{\Sigma}_{y,T}
			\right)^{(1:j)},
			Q_{z,0}^{(1:j)}
			\right)
		}_F
		&\leq
		\frac{1}{\rho_{j,T}}
		\norm{
			\widetilde{\widetilde{\Sigma}}_{
				\tilde{\tilde y},T
			}^{-1/2}
			\left(
			\widehat{\Sigma}_{y,T}
			-
			\widetilde{\widetilde{\Sigma}}_{
				\tilde{\tilde y},T
			}
			\right)
			\widehat{\Sigma}_{y,T}^{-1/2}
		}_F.
		\label{eq:Dopico_Theorem_Qj}
	\end{align}
	
	By \eqref{eq:eigenvalue_order_gap_adjacent_groups},
	\begin{align*}
		\rho_{j,T}^{-1}
		&=
		O_p\left(
		T^{-d_0^{(j)}+d_0^{(j+1)}}
		\right).
	\end{align*}
	Moreover, Lemma~\ref{L:Sigma_y_hat_tildetilde_diff_N} gives
	\begin{align*}
		\norm{
			\widetilde{\widetilde{\Sigma}}_{
				\tilde{\tilde y},T
			}^{-1/2}
			\left(
			\widehat{\Sigma}_{y,T}
			-
			\widetilde{\widetilde{\Sigma}}_{
				\tilde{\tilde y},T
			}
			\right)
			\widehat{\Sigma}_{y,T}^{-1/2}
		}_F
		&=
		O_p(1).
	\end{align*}
	Inserting these two bounds into
	\eqref{eq:Dopico_Theorem_Qj} proves
	\eqref{eq:Theorem_Qj_app}.
\end{proof}

\begin{corollary}[Cross-block projections of estimated eigenspaces]
	\label{cor:factorspaces}
	Suppose Assumptions~\ref{ass:factors}-\ref{ass:memory} hold and that $N,T \to \infty$ with $N/T\to0$. Then the following cross-block projection
	bounds hold.
	
	For distinct nonstationary groups
	$j,j'=1,\ldots,q_0^*$ with $j\neq j'$,
	\begin{align}
		\norm{
			\left(
			Q_{z,0}^{(j')}
			\right)^T
			V\left(
			\widehat{\Sigma}_{y,T}
			\right)^{(j)}
		}_F
		&=
		O_p\left(
		T^{-\left|
			d_0^{(j)}-d_0^{(j')}
			\right|}
		\right).
		\label{eq:Dopico_z_j}
	\end{align}
	
	If $s_0^*<s_0$, then, for each $j=1,\ldots,q_0^*$, the projections
	between the $j$-th nonstationary group and the stationary factor space
	satisfy
\begin{align}
	\norm{
		\left(
		Q_{all,0}^{((q_0^*+1):q_0)}
		\right)^T
		V\left(
		\widehat{\Sigma}_{y,T}
		\right)^{(j)}
	}_F
	&=
	o_p\left(
	T^{-\left(
		d_0^{(j)}-1/2
		\right)}
	\right),
	\label{eq:Dopico_z_w}
	\\
	\norm{
		\left(
		V\left(
		\widehat{\Sigma}_{y,T}
		\right)^{((q_0^*+1):q_0)}
		\right)^T
		Q_{z,0}^{(j)}
	}_F
	&=
	O_p\left(
	T^{-\left(
		d_0^{(j)}-1/2
		\right)}
	\right).
	\label{eq:Dopico_w_z}
\end{align}
	
	For each $j=1,\ldots,q_0^*$, the projections between the $j$-th
	nonstationary group and the idiosyncratic space satisfy
\begin{align}
	\norm{
		\left(
		Q_{all,0}^{(q_0+1)}
		\right)^T
		V\left(
		\widehat{\Sigma}_{y,T}
		\right)^{(j)}
	}_F
	&=
	o_p\left(
	N^{-1/2}
	T^{-\left(
		d_0^{(j)}-1/2
		\right)}
	\right),
	\label{eq:Dopico_z_u}
	\\
	\norm{
		\left(
		V\left(
		\widehat{\Sigma}_{y,T}
		\right)^{(q_0+1)}
		\right)^T
		Q_{z,0}^{(j)}
	}_F
	&=
	O_p\left(
	N^{-1/2}
	T^{-\left(
		d_0^{(j)}-1/2
		\right)}
	\right).
	\label{eq:Dopico_qj}
\end{align}
	
	Finally, if $s_0^*<s_0$, then the projections between the estimated
	stationary factor space and the true idiosyncratic space satisfy
	\begin{align}
		\norm{
			\left(
			Q_{all,0}^{((q_0^*+1):q_0)}
			\right)^T
			V\left(
			\widehat{\Sigma}_{y,T}
			\right)^{(q_0+1)}
		}_F
		&=
		o_p\left(
		N^{-1/2}
		\right),
		\label{eq:Dopico_qq_star}
		\\
		\norm{
			\left(
			V\left(
			\widehat{\Sigma}_{y,T}
			\right)^{((q_0^*+1):q_0)}
			\right)^T
			Q_{all,0}^{(q_0+1)}
		}_F
		&=
		o_p\left(
		N^{-1/2}
		\right).
		\label{eq:Dopico_q_star_q}
	\end{align}
\end{corollary}

\begin{proof}
	We repeatedly apply Lemma~\ref{L:Dopico_Extension}, using
	$\widetilde{\widetilde{\Sigma}}_{\tilde{\tilde y},T}$ for comparisons in which
	the reference eigenspace is a nonstationary factor group and
	$\widetilde{\Sigma}^{\mathrm{bd}}_{\tilde y,T}$ for comparisons in which
	the reference eigenspace is the stationary factor or
	idiosyncratic space. By
	Lemma~\ref{L:Sigma_y_hat_tildetilde_diff_N},
	\begin{align}
		\norm{
			\widetilde{\widetilde{\Sigma}}_{\tilde{\tilde y},T}^{-1/2}
			\left(
			\widehat{\Sigma}_{y,T}
			-
			\widetilde{\widetilde{\Sigma}}_{\tilde{\tilde y},T}
			\right)
			\widehat{\Sigma}_{y,T}^{-1/2}
		}_F
		&=
		O_p(1),
		\label{eq:cor_tildetilde_perturbation}
	\end{align}
	while Lemma~\ref{L:Sigma_y_hat_tilde_bd_diff_N} gives
	\begin{align}
		\norm{
			\left(
			\widetilde{\Sigma}_{\tilde y,T}^{\mathrm{bd}}
			\right)^{-1/2}
			\left(
			\widehat{\Sigma}_{y,T}
			-
			\widetilde{\Sigma}_{\tilde y,T}^{\mathrm{bd}}
			\right)
			\widehat{\Sigma}_{y,T}^{-1/2}
		}_F
		&=
		o_p(1).
		\label{eq:cor_tilde_bd_perturbation}
	\end{align}
	We first compare two distinct nonstationary groups $j$ and $j'$.
	The eigenspace $
	V\left(
	\widetilde{\widetilde{\Sigma}}_{\tilde{\tilde y},T}
	\right)^{(j')}
$
	spans the same subspace as $Q_{z,0}^{(j')}$. Define
	\begin{align*}
		\rho_{j',j,T}
		&=
		\min_{\substack{
				\tilde\kappa
				\in
				\EWgt{j'}\left(
				\widetilde{\widetilde{\Sigma}}_{\tilde{\tilde y},T}
				\right)
				\\
				\hat\kappa
				\in
				\EWgh{j}\left(
				\widehat{\Sigma}_{y,T}
				\right)
		}}
		\frac{
			\left|
			\tilde\kappa-\hat\kappa
			\right|
		}{
			\sqrt{
				\tilde\kappa\hat\kappa
			}
		}.
	\end{align*}
By
\eqref{eq:level_covariance_nonstationary_eigenvalue_orders},
the eigenvalues in groups $j$ and $j'$ are of orders
$NT^{2d_0^{(j)}-1}$ and $NT^{2d_0^{(j')}-1}$, respectively.
Since $d_0^{(j)}\neq d_0^{(j')}$, their ratio diverges. Hence,
	\begin{align*}
		\rho_{j',j,T}^{-1}
		&=
		O_p\left(
		T^{-\left|
			d_0^{(j)}-d_0^{(j')}
			\right|}
		\right).
	\end{align*}
	Lemma~\ref{L:Dopico_Extension} and
	\eqref{eq:cor_tildetilde_perturbation} therefore yield
	\eqref{eq:Dopico_z_j}.
	
	Suppose next that $s_0^*<s_0$ and fix
	$j\in\{1,\ldots,q_0^*\}$. To prove
	\eqref{eq:Dopico_z_w}, we use
	$\widetilde{\Sigma}_{\tilde y,T}^{\mathrm{bd}}$, whose stationary
	reference eigenspace spans the same subspace as
	$Q_{all,0}^{((q_0^*+1):q_0)}$. Define
	\begin{align*}
		\rho_{w,j,T}
		&=
		\min_{\substack{
				\tilde\kappa
				\in
				\EWgt{((q_0^*+1):q_0)}\left(
				\widetilde{\Sigma}_{\tilde y,T}^{\mathrm{bd}}
				\right)
				\\
				\hat\kappa
				\in
				\EWgh{j}\left(
				\widehat{\Sigma}_{y,T}
				\right)
		}}
		\frac{
			\left|
			\tilde\kappa-\hat\kappa
			\right|
		}{
			\sqrt{
				\tilde\kappa\hat\kappa
			}
		}.
	\end{align*}
By \eqref{eq:level_covariance_nonstationary_eigenvalue_orders} and \eqref{eq:level_covariance_stationary_eigenvalue_orders}, the group-$j$ and stationary eigenvalues are of orders $NT^{2d_0^{(j)}-1}$ and $N$, respectively. Hence,
	\begin{align*}
		\rho_{w,j,T}^{-1}
		&=
		O_p\left(
		T^{-d_0^{(j)}+1/2}
		\right).
	\end{align*}
	Lemma~\ref{L:Dopico_Extension} and
	\eqref{eq:cor_tilde_bd_perturbation} imply the stronger bound
	\[
	\norm{
		\left(
		Q_{all,0}^{((q_0^*+1):q_0)}
		\right)^T
		V\left(
		\widehat{\Sigma}_{y,T}
		\right)^{(j)}
	}_F
	=
	o_p\left(
	T^{-d_0^{(j)}+1/2}
	\right),
	\]
	which proves \eqref{eq:Dopico_z_w}.
	
	To prove \eqref{eq:Dopico_w_z}, we instead use
	$\widetilde{\widetilde{\Sigma}}_{\tilde{\tilde y},T}$, because
	its group-$j$ reference eigenspace spans the same subspace as
	$Q_{z,0}^{(j)}$. Define
	\begin{align*}
		\rho_{j,w,T}
		&=
		\min_{\substack{
				\tilde\kappa
				\in
				\EWgt{j}\left(
				\widetilde{\widetilde{\Sigma}}_{\tilde{\tilde y},T}
				\right)
				\\
				\hat\kappa
				\in
				\EWgh{((q_0^*+1):q_0)}\left(
				\widehat{\Sigma}_{y,T}
				\right)
		}}
		\frac{
			\left|
			\tilde\kappa-\hat\kappa
			\right|
		}{
			\sqrt{
				\tilde\kappa\hat\kappa
			}
		}.
	\end{align*}
The same eigenvalue orders imply
	\begin{align}
		\rho_{j,w,T}^{-1}
		&=
		O_p\left(
		T^{-d_0^{(j)}+1/2}
		\right).
		\label{eq:cor_gap_zw}
	\end{align}
	Combining Lemma~\ref{L:Dopico_Extension},
	\eqref{eq:cor_tildetilde_perturbation}, and
	\eqref{eq:cor_gap_zw} proves \eqref{eq:Dopico_w_z}.
	
	We next compare group $j$ with the idiosyncratic space. For
	\eqref{eq:Dopico_z_u}, we use
	$\widetilde{\Sigma}_{\tilde y,T}^{\mathrm{bd}}$, whose
	idiosyncratic reference eigenspace spans the same subspace as
	$Q_{all,0}^{(q_0+1)}$. Define
	\begin{align*}
		\rho_{u,j,T}
		&=
		\min_{\substack{
				\tilde\kappa
				\in
				\EWgt{q_0+1}\left(
				\widetilde{\Sigma}_{\tilde y,T}^{\mathrm{bd}}
				\right)
				\\
				\hat\kappa
				\in
				\EWgh{j}\left(
				\widehat{\Sigma}_{y,T}
				\right)
		}}
		\frac{
			\left|
			\tilde\kappa-\hat\kappa
			\right|
		}{
			\sqrt{
				\tilde\kappa\hat\kappa
			}
		}.
	\end{align*}
By \eqref{eq:level_covariance_nonstationary_eigenvalue_orders} and \eqref{eq:level_covariance_idiosyncratic_eigenvalue_orders}, the group-$j$ and idiosyncratic eigenvalues are of orders $NT^{2d_0^{(j)}-1}$ and one, respectively. Therefore,
	\begin{align*}
		\rho_{u,j,T}^{-1}
		&=
		O_p\left(
		N^{-1/2}
		T^{-d_0^{(j)}+1/2}
		\right).
	\end{align*}
	Lemma~\ref{L:Dopico_Extension} and
	\eqref{eq:cor_tilde_bd_perturbation} give the stronger bound
	\[
	\norm{
		\left(
		Q_{all,0}^{(q_0+1)}
		\right)^T
		V\left(
		\widehat{\Sigma}_{y,T}
		\right)^{(j)}
	}_F
	=
	o_p\left(
	N^{-1/2}
	T^{-d_0^{(j)}+1/2}
	\right),
	\]
	and hence \eqref{eq:Dopico_z_u}.
	
	For \eqref{eq:Dopico_qj}, use the group-$j$ reference eigenspace
	of
	$\widetilde{\widetilde{\Sigma}}_{\tilde{\tilde y},T}$ and define
	\begin{align*}
		\rho_{j,u,T}
		&=
		\min_{\substack{
				\tilde\kappa
				\in
				\EWgt{j}\left(
				\widetilde{\widetilde{\Sigma}}_{\tilde{\tilde y},T}
				\right)
				\\
				\hat\kappa
				\in
				\EWgh{q_0+1}\left(
				\widehat{\Sigma}_{y,T}
				\right)
		}}
		\frac{
			\left|
			\tilde\kappa-\hat\kappa
			\right|
		}{
			\sqrt{
				\tilde\kappa\hat\kappa
			}
		}.
	\end{align*}
The same eigenvalue orders imply
	\begin{align*}
		\rho_{j,u,T}^{-1}
		&=
		O_p\left(
		N^{-1/2}
		T^{-d_0^{(j)}+1/2}
		\right).
	\end{align*}
	Lemma~\ref{L:Dopico_Extension} and
	\eqref{eq:cor_tildetilde_perturbation} prove
	\eqref{eq:Dopico_qj}.
	
	Finally, suppose $s_0^*<s_0$. Both stationary-idiosyncratic
	comparisons use
	$\widetilde{\Sigma}_{\tilde y,T}^{\mathrm{bd}}$. Define
	\begin{align*}
		\rho_{w,u,N}
		&=
		\min_{\substack{
				\tilde\kappa
				\in
				\EWgt{((q_0^*+1):q_0)}\left(
				\widetilde{\Sigma}_{\tilde y,T}^{\mathrm{bd}}
				\right)
				\\
				\hat\kappa
				\in
				\EWgh{q_0+1}\left(
				\widehat{\Sigma}_{y,T}
				\right)
		}}
		\frac{
			\left|
			\tilde\kappa-\hat\kappa
			\right|
		}{
			\sqrt{
				\tilde\kappa\hat\kappa
			}
		},
		\\
		\rho_{u,w,N}
		&=
		\min_{\substack{
				\tilde\kappa
				\in
				\EWgt{q_0+1}\left(
				\widetilde{\Sigma}_{\tilde y,T}^{\mathrm{bd}}
				\right)
				\\
				\hat\kappa
				\in
				\EWgh{((q_0^*+1):q_0)}\left(
				\widehat{\Sigma}_{y,T}
				\right)
		}}
		\frac{
			\left|
			\tilde\kappa-\hat\kappa
			\right|
		}{
			\sqrt{
				\tilde\kappa\hat\kappa
			}
		}.
	\end{align*}
By \eqref{eq:level_covariance_stationary_eigenvalue_orders} and \eqref{eq:level_covariance_idiosyncratic_eigenvalue_orders}, the stationary and idiosyncratic eigenvalues are of orders $N$ and one, respectively. Hence,
	\begin{align*}
		\rho_{w,u,N}^{-1}
		&=
		O_p\left(
		N^{-1/2}
		\right),
		&
		\rho_{u,w,N}^{-1}
		&=
		O_p\left(
		N^{-1/2}
		\right).
	\end{align*}
	Applying Lemma~\ref{L:Dopico_Extension} and
	\eqref{eq:cor_tilde_bd_perturbation} in the two directions proves
	\eqref{eq:Dopico_qq_star} and
	\eqref{eq:Dopico_q_star_q}.
\end{proof}
\subsection{Proofs of the main results}
\label{sec:Proofs-main-results}
The factors are identified only up to nonsingular transformations. For any
nonsingular matrix $H$,
\[
\Lambda_0x_t
=
\left(
\Lambda_0H^{-1}
\right)
Hx_t.
\]
Principal components can therefore recover only rotated and scaled versions of
the factors, with the inverse transformation absorbed into the loading matrix.
In Theorem~\ref{T:factor_convergence}, this indeterminacy is represented by the
random nonsingular matrices $H_{0,T}$, $H_{z,T}$, and $H_{j,T}$.

The estimated factor scores are normalized as
\[
N^{-1/2}
V\left(
\widehat{\Sigma}_{y,T}
\right)^T
y_t.
\]
This normalization is natural under Assumption~\ref{ass:strong_factors}, because
the projection of the loading matrix onto the corresponding empirical
eigenvectors is of order $N^{1/2}$.

The theorem gives both pointwise and average results. The pointwise bounds
describe the estimation error at a given time $t$, whereas the average bounds
control the complete estimated factor path over $t=1,\ldots,T$. The theorem
states the groupwise results for cumulative nonstationary spaces. Individual
groups can subsequently be obtained from the differences between successive
cumulative spaces.

\subsubsection{Proof of Theorem~\ref{T:factor_convergence}}
\begin{proof}[Proof of Theorem~\ref{T:factor_convergence}]
	\noindent\textit{The full factor vector.}
	Define
	\[
	H_{0,T}
	=
	N^{-1/2}
	V\left(
	\widehat{\Sigma}_{y,T}
	\right)^{(1:q_0)T}
	\Lambda_0.
	\]
	Since $y_t=\Lambda_0x_t+u_t$,
	\[
	\widehat{x}_t
	=
	H_{0,T}x_t
	+
	e_{0,t},
	\qquad
	e_{0,t}
	=
	N^{-1/2}
	V\left(
	\widehat{\Sigma}_{y,T}
	\right)^{(1:q_0)T}
	u_t.
	\]
	
Collecting \eqref{eq:Dopico_z_u} over the nonstationary groups and
adding \eqref{eq:Dopico_q_star_q} when stationary factors are
present yields
	\begin{align}
		\norm{
			V\left(
			\widehat{\Sigma}_{y,T}
			\right)^{(1:q_0)T}
			Q_{\perp,0}
		}_F
		&=
		o_p\left(
		N^{-1/2}
		\right).
		\label{eq:full_factor_idiosyncratic_projection}
	\end{align}
	
	Using
	\[
	I_N
	=
	Q_0Q_0^T
	+
	Q_{\perp,0}Q_{\perp,0}^T,
	\]
	we obtain
	\[
	\begin{aligned}
		e_{0,t}
		=
		N^{-1/2}
		V\left(
		\widehat{\Sigma}_{y,T}
		\right)^{(1:q_0)T}
		Q_0Q_0^Tu_t
		&+
		N^{-1/2}
		V\left(
		\widehat{\Sigma}_{y,T}
		\right)^{(1:q_0)T}
		Q_{\perp,0}Q_{\perp,0}^Tu_t.
	\end{aligned}
	\]
	By Assumption~\ref{ass:Sigma_u},
	\begin{align}
		\EE{
			\norm{Q_0^Tu_t}_F^2
		}
		&=
		\tr\left(
		Q_0^T\Sigma_{u,0}Q_0
		\right)
		\leq
		s_0
		\EW{1}{\Sigma_{u,0}}
		=
		O(1),
		\nonumber\\
		\EE{
			\norm{Q_{\perp,0}^Tu_t}_F^2
		}
		&=
		\tr\left(
		Q_{\perp,0}^T
		\Sigma_{u,0}
		Q_{\perp,0}
		\right)
		\leq
		(N-s_0)
		\EW{1}{\Sigma_{u,0}}
		=
		O(N). \nonumber
	\end{align}
Therefore,
\begin{align}
	\norm{Q_0^Tu_t}_F
=
O_p(1),
\qquad
\norm{Q_{\perp,0}^Tu_t}_F
=
O_p\left(
N^{1/2}
\right).
		\label{eq:factor_convergence_projected_u_pointwise}
\end{align}
	Using
	\[
	\norm{
		V\left(
		\widehat{\Sigma}_{y,T}
		\right)^{(1:q_0)T}
		Q_0
	}_F
	\leq
	\sqrt{s_0}
	\]
together with
\eqref{eq:full_factor_idiosyncratic_projection} and \eqref{eq:factor_convergence_projected_u_pointwise} show that the two components of $e_{0,t}$ are respectively of orders
$O_p(N^{-1/2})$ and $o_p(N^{-1/2})$. Therefore,
\[
\norm{e_{0,t}}_F
=
O_p\left(
N^{-1/2}
\right).
\]
	This proves \eqref{eq:factor_convergence_full}.
	
	For the average statement,
	\begin{align*}
		\frac{1}{T}
		\sum_{t=1}^T
		\norm{e_{0,t}}_F^2
		&=
		\frac{1}{N}
		\tr\left[
		V\left(
		\widehat{\Sigma}_{y,T}
		\right)^{(1:q_0)T}
		\widetilde{\Sigma}_{u}
		V\left(
		\widehat{\Sigma}_{y,T}
		\right)^{(1:q_0)}
		\right]
		\nonumber\\
		&\leq
		\frac{s_0}{N}
		\EW{1}{
			\widetilde{\Sigma}_{u}
		}
		=
		O_p\left(
		N^{-1}
		\right)
	\end{align*}
	by Assumption~\ref{ass:Eigenvalue_Sigma_u_tilde}. This proves
	\eqref{eq:factor_convergence_full_average}.
	
	It remains to verify the well-conditioning of $H_{0,T}$. Writing
	$\Lambda_0=Q_0R_0$ gives
	\[
	H_{0,T}
	=
	V\left(
	\widehat{\Sigma}_{y,T}
	\right)^{(1:q_0)T}
	Q_0
	\left(
	N^{-1/2}R_0
	\right).
	\]
	If $s_0^*<s_0$, \eqref{eq:Theorem_Qq_app} implies that the
	canonical angles between
	$V(\widehat{\Sigma}_{y,T})^{(1:q_0)}$ and $Q_0$ converge to zero.
	If $s_0^*=s_0$, the same conclusion follows from
	\eqref{eq:Theorem_Qz_app}, because then $Q_0=Q_{z,0}$. Thus, the
	singular values of
$
	V\left(
	\widehat{\Sigma}_{y,T}
	\right)^{(1:q_0)T}
	Q_0
$
	converge to one in probability.
	
	Moreover,
	\begin{align}
		\frac{1}{N}R_0^TR_0
		&=
		\frac{1}{N}\Lambda_0^T\Lambda_0
		\longrightarrow
		\Sigma_{\Lambda,0}.
		\label{eq:factor_convergence_loading_limit}
	\end{align}
	Since $\Sigma_{\Lambda,0}$ is positive definite by
	Assumption~\ref{ass:strong_factors}, the singular values of
	$N^{-1/2}R_0$ are bounded away from zero and infinity. It follows
	that the singular values of $H_{0,T}$ are bounded away from zero
	and infinity in probability.
	
For the factor-score bounds below, we also record the orders of the
blocks in the QR decomposition
\begin{align}
	\norm{R_{zz,0}}_F
	&=
	O\left(
	N^{1/2}
	\right),
	&
	\norm{R_{zw,0}}_F
	&=
	O\left(
	N^{1/2}
	\right),
	&
	\norm{R_{ww,0}}_F
	&=
	O\left(
	N^{1/2}
	\right),
	\label{eq:factor_convergence_R_orders}
\end{align}
where terms corresponding to absent blocks are omitted.
	
	\medskip
	\noindent\textit{The full nonstationary factor vector.}
	Suppose $s_0^*>0$ and define
	\[
	H_{z,T}
	=
	N^{-1/2}
	V\left(
	\widehat{\Sigma}_{y,T}
	\right)^{(1:q_0^*)T}
	Q_{z,0}R_{zz,0}.
	\]
	The block representation in \eqref{eq:z_w_QR} gives
	\[
	y_t
	=
	Q_{z,0}R_{zz,0}z_t
	+
	Q_{z,0}R_{zw,0}w_t
	+
	Q_{w,0}R_{ww,0}w_t
	+
	u_t,
	\]
	where the terms involving $w_t$ are omitted when $s_0^*=s_0$.
	Consequently,
	\begin{align}
		\widehat z_t-H_{z,T}z_t
		&=
		N^{-1/2}
		V\left(
		\widehat{\Sigma}_{y,T}
		\right)^{(1:q_0^*)T}
		Q_{z,0}R_{zw,0}w_t
		\nonumber\\
		&\quad+
		N^{-1/2}
		V\left(
		\widehat{\Sigma}_{y,T}
		\right)^{(1:q_0^*)T}
		Q_{w,0}R_{ww,0}w_t
		\nonumber\\
		&\quad+
		N^{-1/2}
		V\left(
		\widehat{\Sigma}_{y,T}
		\right)^{(1:q_0^*)T}
		u_t
		\nonumber\\
		&=
		e_{z,t}.
		\label{eq:nonstationary_factor_error_decomposition}
	\end{align}
	
	By \eqref{eq:stationary_type_II_covariance_bound} with $s=t$ and
	\eqref{eq:stationary_filter_square_summability},
$
	\norm{w_t}_F
	=
	O_p(1).
$
	Since
	\[
	\norm{
		V\left(
		\widehat{\Sigma}_{y,T}
		\right)^{(1:q_0^*)T}
		Q_{z,0}
	}_F
	\leq
	\sqrt{s_0^*},
	\]
	\eqref{eq:factor_convergence_R_orders} gives
	\begin{align}
		&N^{-1/2}
		\norm{
			V\left(
			\widehat{\Sigma}_{y,T}
			\right)^{(1:q_0^*)T}
			Q_{z,0}R_{zw,0}w_t
		}_F
		=
		O_p(1).
		\label{eq:nonstationary_factor_inspace_stationary_error}
	\end{align}
	
	By \eqref{eq:Dopico_z_w},
	\begin{align}
		\norm{
			V\left(
			\widehat{\Sigma}_{y,T}
			\right)^{(1:q_0^*)T}
			Q_{w,0}
		}_F
		&=
		o_p\left(
		T^{-d_0^{(q_0^*)}+1/2}
		\right).
		\label{eq:nonstationary_stationary_projection}
	\end{align}
	Together with \eqref{eq:factor_convergence_R_orders}, this implies
	\begin{align}
		N^{-1/2}
		\norm{
			V\left(
			\widehat{\Sigma}_{y,T}
			\right)^{(1:q_0^*)T}
			Q_{w,0}R_{ww,0}w_t
		}_F
		&=
		o_p\left(
		T^{-d_0^{(q_0^*)}+1/2}
		\right).
		\label{eq:nonstationary_factor_outside_stationary_error}
	\end{align}
	
	Finally, Assumption~\ref{ass:Sigma_u} gives
	\[
	\EE{\norm{u_t}_F^2}
	=
	\tr(\Sigma_{u,0})
	=
	O(N),
	\]
	and hence
	\begin{align}
		N^{-1/2}
		\norm{
			V\left(
			\widehat{\Sigma}_{y,T}
			\right)^{(1:q_0^*)T}
			u_t
		}_F
		&\leq
		N^{-1/2}
		\norm{u_t}_F
		=
		O_p(1).
		\label{eq:nonstationary_factor_pointwise_idiosyncratic_error}
	\end{align}
	Combining
	\eqref{eq:nonstationary_factor_inspace_stationary_error},
	\eqref{eq:nonstationary_factor_outside_stationary_error}, and
	\eqref{eq:nonstationary_factor_pointwise_idiosyncratic_error}
	proves
	\[
	\norm{e_{z,t}}_F
	=
	O_p(1),
	\]
	and hence \eqref{eq:factor_convergence_z}.
	
	For the average error,
	\eqref{eq:stationary_factor_covariance_limit} gives
	\begin{align}
		\sum_{t=1}^T
		\norm{w_t}_F^2
		&=
		O_p(T)
		\label{eq:stationary_factor_sum_of_squares}
	\end{align}
	when stationary factors are present. Therefore,
	\begin{align*}
		&\sum_{t=1}^T
		\norm{
			N^{-1/2}
			V\left(
			\widehat{\Sigma}_{y,T}
			\right)^{(1:q_0^*)T}
			Q_{z,0}R_{zw,0}w_t
		}_F^2
		=
		O_p(T).
	\end{align*}
	Similarly,
	\eqref{eq:nonstationary_stationary_projection},
	\eqref{eq:factor_convergence_R_orders}, and
	\eqref{eq:stationary_factor_sum_of_squares} imply
	\begin{align*}
		&\sum_{t=1}^T
		\norm{
			N^{-1/2}
			V\left(
			\widehat{\Sigma}_{y,T}
			\right)^{(1:q_0^*)T}
			Q_{w,0}R_{ww,0}w_t
		}_F^2 =
		o_p\left(
		T^{2-2d_0^{(q_0^*)}}
		\right).
	\end{align*}
	Moreover, Assumption~\ref{ass:Eigenvalue_Sigma_u_tilde} gives
	\begin{align*}
		\sum_{t=1}^T
		\norm{
			N^{-1/2}
			V\left(
			\widehat{\Sigma}_{y,T}
			\right)^{(1:q_0^*)T}
			u_t
		}_F^2
		&=
		\frac{T}{N}
		\tr\left[
		V\left(
		\widehat{\Sigma}_{y,T}
		\right)^{(1:q_0^*)T}
		\widetilde{\Sigma}_{u}
		V\left(
		\widehat{\Sigma}_{y,T}
		\right)^{(1:q_0^*)}
		\right]
		\nonumber\\
		&=
		O_p\left(
		\frac{T}{N}
		\right).
	\end{align*}
	Using the error decomposition
	\eqref{eq:nonstationary_factor_error_decomposition} and
	$\norm{a+b+c}_F^2
	\leq
	3\norm{a}_F^2+
	3\norm{b}_F^2+
	3\norm{c}_F^2$, we obtain
	\begin{align}
		\sum_{t=1}^T
		\norm{e_{z,t}}_F^2
		&=
		O_p(T)
		+
		o_p\left(
		T^{2-2d_0^{(q_0^*)}}
		\right)
		+
		O_p\left(
		\frac{T}{N}
		\right)=
		O_p(T).
		\label{eq:nonstationary_factor_average_error}
	\end{align}
	
	Next, we look at the well-conditioning of $H_{z,T}$.
	\[
	H_{z,T}
	=
	V\left(
	\widehat{\Sigma}_{y,T}
	\right)^{(1:q_0^*)T}
	Q_{z,0}
	\left(
	N^{-1/2}R_{zz,0}
	\right).
	\]
	By \eqref{eq:Theorem_Qz_app}, the singular values of
$
	V\left(
	\widehat{\Sigma}_{y,T}
	\right)^{(1:q_0^*)T}
	Q_{z,0}
$
	converge to one in probability. Furthermore,
	\eqref{eq:factor_convergence_loading_limit} implies that
	\[
	\frac{1}{N}R_{zz,0}^TR_{zz,0}
	\]
	converges to the corresponding positive-definite principal
	submatrix of $\Sigma_{\Lambda,0}$. Hence, the singular values of
	$H_{z,T}$ are bounded away from zero and infinity in probability.
	
	We continue with $\sum_{t = 1}^{T}\norm{H_{z,T}z_t}_F^2.$
	By \eqref{eq:nonstationary_factor_covariance_limit} and the
	almost-sure positive definiteness of $S_z$,
	\begin{align*}
		\sum_{t=1}^T
		\norm{
			z_t^{(q_0^*)}
		}_F^2
		&=
		O_e\left(
		T^{2d_0^{(q_0^*)}}
		\right).
	\end{align*}
	Consequently,
	\[
	\begin{aligned}
		\sum_{t=1}^T
		\norm{
			H_{z,T}z_t
		}_F^2
		&\geq
		\EW{s_0^*}{
			H_{z,T}^TH_{z,T}
		}
		\sum_{t=1}^T
		\norm{
			z_t^{(q_0^*)}
		}_F^2.
	\end{aligned}
	\]
	Since the smallest eigenvalue of $H_{z,T}^TH_{z,T}$ is bounded
	away from zero in probability,
	\begin{align}
		\left(
		\sum_{t=1}^T
		\norm{
			H_{z,T}z_t
		}_F^2
		\right)^{-1}
		&=
		O_p\left(
		T^{-2d_0^{(q_0^*)}}
		\right).
		\label{eq:nonstationary_factor_denominator_inverse}
	\end{align}
	Combining
	\eqref{eq:nonstationary_factor_average_error} and
	\eqref{eq:nonstationary_factor_denominator_inverse} yields
	\[
	\left(
	\frac{
		\displaystyle
		\sum_{t=1}^T
		\norm{
			\widehat z_t-H_{z,T}z_t
		}_F^2
	}{
		\displaystyle
		\sum_{t=1}^T
		\norm{
			H_{z,T}z_t
		}_F^2
	}
	\right)^{1/2}
	=
	O_p\left(
	T^{-d_0^{(q_0^*)}+1/2}
	\right).
	\]
	This proves
	\eqref{eq:factor_convergence_z_average_relative}.
	
	\medskip
	\noindent\textit{Cumulative nonstationary factor groups.}
	Fix $j\in\{1,\ldots,q_0^*-1\}$ and define
	\[
	H_{j,T}
	=
	N^{-1/2}
	V\left(
	\widehat{\Sigma}_{y,T}
	\right)^{(1:j)T}
	Q_{z,0}^{(1:j)}
	R_{zz,0}^{(1:j,1:j)}.
	\]
	Since
$
	\Lambda_0^{(1:j)}
	=
	Q_{z,0}^{(1:j)}
	R_{zz,0}^{(1:j,1:j)},
$
	the decomposition
	\[
	y_t
	=
	\Lambda_0^{(1:j)}x_t^{(1:j)}
	+
	\Lambda_0^{((j+1):q_0)}
	x_t^{((j+1):q_0)}
	+
	u_t
	\]
	implies
	\begin{align*}
		\widehat{x}_t^{(1:j)}
		-
		H_{j,T}x_t^{(1:j)}
		&=
		N^{-1/2}
		V\left(
		\widehat{\Sigma}_{y,T}
		\right)^{(1:j)T}
		\Lambda_0^{((j+1):q_0)}
		x_t^{((j+1):q_0)}
		\nonumber\\
		&\quad+
		N^{-1/2}
		V\left(
		\widehat{\Sigma}_{y,T}
		\right)^{(1:j)T}
		u_t
		\nonumber\\
		&=
		e_{j,t}.
	\end{align*}
	
Applying \eqref{eq:nonstationary_factor_FCLT} with sample size $t$
and evaluating the limiting process at $u=1$ gives
	\begin{align*}
		\norm{x_t^{(k)}}_F
		&=
		O_p\left(
		t^{d_0^{(k)}-1/2}
		\right),
		\qquad
		k=1,\ldots,q_0^*.
	\end{align*}
	The stationary factors satisfy
	$\norm{w_t}_F=O_p(1)$ by
	\eqref{eq:stationary_type_II_covariance_bound} and
	\eqref{eq:stationary_filter_square_summability}.
	Furthermore, \eqref{eq:factor_convergence_loading_limit} implies
	\[
	\norm{\Lambda_0^{(k)}}_F
	=
	O\left(
	N^{1/2}
	\right),
	\qquad
	k=1,\ldots,q_0.
	\]
	Using the ordering of the memory parameters, we obtain
	\begin{align*}
		N^{-1/2}
		\norm{
			V\left(
			\widehat{\Sigma}_{y,T}
			\right)^{(1:j)T}
			\Lambda_0^{((j+1):q_0)}
			x_t^{((j+1):q_0)}
		}_F
		=
		O_p\left(
		t^{d_0^{(j+1)}-1/2}
		\right).
	\end{align*}
	Moreover,
	\[
	N^{-1/2}
	\norm{
		V\left(
		\widehat{\Sigma}_{y,T}
		\right)^{(1:j)T}
		u_t
	}_F
	\leq
	N^{-1/2}
	\norm{u_t}_F
	=
	O_p(1).
	\]
Since $t^{d_0^{(j+1)}-1/2}\geq1$, the idiosyncratic component is of no larger order. Hence,
	\[
	\norm{e_{j,t}}_F
	=
	O_p\left(
	t^{d_0^{(j+1)}-1/2}
	\right),
	\]
	which proves \eqref{eq:factor_convergence_group}.
	
	For the average error,
	\eqref{eq:nonstationary_factor_covariance_limit} implies
	\begin{align}
		\sum_{t=1}^T
		\norm{x_t^{(k)}}_F^2
		&=
		O_e\left(
		T^{2d_0^{(k)}}
		\right),
		\qquad
		k=1,\ldots,q_0^*,
		\label{eq:nonstationary_group_second_moment_orders}
	\end{align}
	while \eqref{eq:stationary_factor_covariance_limit} gives
	\[
	\sum_{t=1}^T
	\norm{w_t}_F^2
	=
	O_p(T).
	\]
	Using the loading orders implied by
	\eqref{eq:factor_convergence_loading_limit}, the omitted common
	component therefore satisfies
	\begin{align*}
		\sum_{t=1}^T
		\norm{
			N^{-1/2}
			V\left(
			\widehat{\Sigma}_{y,T}
			\right)^{(1:j)T}
			\Lambda_0^{((j+1):q_0)}
			x_t^{((j+1):q_0)}
		}_F^2=
		O_p\left(
		T^{2d_0^{(j+1)}}
		\right).
	\end{align*}
	In addition, Assumption~\ref{ass:Eigenvalue_Sigma_u_tilde} gives
	\begin{align*}
		\sum_{t=1}^T
		\norm{
			N^{-1/2}
			V\left(
			\widehat{\Sigma}_{y,T}
			\right)^{(1:j)T}
			u_t
		}_F^2
		&=
		\frac{T}{N}
		\tr\left[
		V\left(
		\widehat{\Sigma}_{y,T}
		\right)^{(1:j)T}
		\widetilde{\Sigma}_{u}
		V\left(
		\widehat{\Sigma}_{y,T}
		\right)^{(1:j)}
		\right]
		\nonumber\\
		&=
		O_p\left(
		\frac{T}{N}
		\right).
	\end{align*}
	Because $d_0^{(j+1)}>1/2$, the omitted common component dominates
	the stationary and idiosyncratic contributions. Hence,
	\begin{align}
		\sum_{t=1}^T
		\norm{
			\widehat{x}_t^{(1:j)}
			-
			H_{j,T}x_t^{(1:j)}
		}_F^2
		&=
		O_p\left(
		T^{2d_0^{(j+1)}}
		\right).
		\label{eq:cumulative_factor_average_error}
	\end{align}
	
	Finally,
	\[
	H_{j,T}
	=
	V\left(
	\widehat{\Sigma}_{y,T}
	\right)^{(1:j)T}
	Q_{z,0}^{(1:j)}
	\left(
	N^{-1/2}
	R_{zz,0}^{(1:j,1:j)}
	\right).
	\]
	By Theorem~\ref{T:Factor_space_2}, the singular values of
$
	V\left(
	\widehat{\Sigma}_{y,T}
	\right)^{(1:j)T}
	Q_{z,0}^{(1:j)}
$
	converge to one in probability. Furthermore,
	\eqref{eq:factor_convergence_loading_limit} implies that
	\[
	\frac{1}{N}
	\left(
	R_{zz,0}^{(1:j,1:j)}
	\right)^T
	R_{zz,0}^{(1:j,1:j)}
	\]
	converges to the corresponding positive-definite principal
	submatrix of $\Sigma_{\Lambda,0}$. Hence, the singular values of
	$H_{j,T}$ are bounded away from zero and infinity in probability.
	
	By \eqref{eq:nonstationary_group_second_moment_orders}, we have
	\[
	\begin{aligned}
		\sum_{t=1}^T
		\norm{
			H_{j,T}x_t^{(1:j)}
		}_F^2
		&\geq
		\EW{s_1+\cdots+s_j}{
			H_{j,T}^TH_{j,T}
		}
		\sum_{t=1}^T
		\norm{x_t^{(j)}}_F^2.
	\end{aligned}
	\]
	Since the smallest eigenvalue of $H_{j,T}^TH_{j,T}$ is bounded
	away from zero in probability,
	\begin{align}
		\left(
		\sum_{t=1}^T
		\norm{
			H_{j,T}x_t^{(1:j)}
		}_F^2
		\right)^{-1}
		&=
		O_p\left(
		T^{-2d_0^{(j)}}
		\right).
		\label{eq:cumulative_factor_denominator_inverse}
	\end{align}
	Combining
	\eqref{eq:cumulative_factor_average_error} and
	\eqref{eq:cumulative_factor_denominator_inverse} gives
	\[
	\left(
	\frac{
		\displaystyle
		\sum_{t=1}^T
		\norm{
			\widehat{x}_t^{(1:j)}
			-
			H_{j,T}x_t^{(1:j)}
		}_F^2
	}{
		\displaystyle
		\sum_{t=1}^T
		\norm{
			H_{j,T}x_t^{(1:j)}
		}_F^2
	}
	\right)^{1/2}
	=
	O_p\left(
	T^{-d_0^{(j)}+d_0^{(j+1)}}
	\right).
	\]
	This proves
	\eqref{eq:factor_convergence_group_average_relative}.
\end{proof}
At the true factor dimension, the PCA estimator of the common
component is
\[
\widehat{\chi}_{s_0,t}
=
\widehat{\Lambda}_{s_0}
\widehat{x}_{s_0,t}.
\]
By \eqref{eq:pca_loading_normalization} and
\eqref{eq:pca_factor_estimator},
\begin{align}
	\widehat{\chi}_{s_0,t}
	&=
	V\left(
	\widehat{\Sigma}_{y,T}
	\right)^{(1:q_0)}
	V\left(
	\widehat{\Sigma}_{y,T}
	\right)^{(1:q_0)T}
	y_t
	=
	\widehat P_{s_0}y_t,
	\label{eq:true_dimension_common_component_projection}
\end{align}
where $\widehat P_{s_0}$ is the orthogonal projector onto the
estimated $s_0$-dimensional factor space. The following corollary
establishes average consistency of this fitted common component.

\begin{corollary}[Average common-component consistency]
	\label{cor:common_component_consistency}
	Suppose Assumptions~\ref{ass:factors}-\ref{ass:memory} hold and
	that $N,T \to \infty$ with $N/T\to0$. Then
	\begin{align}
		\frac{1}{NT}
		\sum_{t=1}^T
		\norm{
			\chi_t-\widehat{\chi}_{s_0,t}
		}_F^2
		&=
		O_p\left(
		N^{-1}
		\right).
		\label{eq:Bai_Ng_common_component_rate}
	\end{align}
\end{corollary}

\begin{proof}
	By \eqref{eq:true_dimension_common_component_projection} and
	$y_t=\chi_t+u_t$,
	\[
	\chi_t-\widehat{\chi}_{s_0,t}
	=
	\left(
	I_N-\widehat P_{s_0}
	\right)\chi_t
	-
	\widehat P_{s_0}u_t.
	\]
	We bound the two terms separately.
	
	Write $\Lambda_0=Q_0R_0$. The block-upper-triangular structure of
	$R_0$ gives
	\[
	\chi_t
	=
	\sum_{j=1}^{q_0^*}
	Q_{z,0}^{(j)}
	\left(
	R_0x_t
	\right)^{(j)}
	+
	Q_{all,0}^{((q_0^*+1):q_0)}
	\left(
	R_0x_t
	\right)^{((q_0^*+1):q_0)},
	\]
	where the last term is omitted when $s_0^*=s_0$.
	
	Assumption~\ref{ass:strong_factors} implies that every fixed block
	of $R_0$ has Frobenius norm of order $N^{1/2}$. Together with
	\eqref{eq:nonstationary_factor_covariance_limit} and
	\eqref{eq:stationary_factor_covariance_limit}, this gives
	\begin{align}
		\frac{1}{T}
		\sum_{t=1}^T
		\norm{
			\left(
			R_0x_t
			\right)^{(j)}
		}_F^2
		&=
		O_p\left(
		NT^{2d_0^{(j)}-1}
		\right),
		\qquad
		j=1,\ldots,q_0^*,
\nonumber
		\\
		\frac{1}{T}
		\sum_{t=1}^T
		\norm{
			\left(
			R_0x_t
			\right)^{((q_0^*+1):q_0)}
		}_F^2
		&=
		O_p(N),
		\label{eq:common_component_rotated_stationary_order}
	\end{align}
	where \eqref{eq:common_component_rotated_stationary_order} applies
	only when stationary factors are present.
	
	Moreover,
	\[
	I_N-\widehat P_{s_0}
	=
	V\left(
	\widehat{\Sigma}_{y,T}
	\right)^{(q_0+1)}
	V\left(
	\widehat{\Sigma}_{y,T}
	\right)^{(q_0+1)T}.
	\]
	Hence, by \eqref{eq:Dopico_qj},
	\begin{align*}
		\norm{
			\left(
			I_N-\widehat P_{s_0}
			\right)
			Q_{z,0}^{(j)}
		}_F
		&=
		O_p\left(
		N^{-1/2}
		T^{-d_0^{(j)}+1/2}
		\right),
		\qquad
		j=1,\ldots,q_0^*.
	\end{align*}
	Combining the projection and second-moment bounds therefore gives
	\begin{align}
		\frac{1}{NT}
		\sum_{t=1}^T
		\norm{
			\left(
			I_N-\widehat P_{s_0}
			\right)
			Q_{z,0}^{(j)}
			\left(
			R_0x_t
			\right)^{(j)}
		}_F^2
		&=
		O_p\left(
		N^{-1}
		\right).
		\label{eq:common_component_nonstationary_error}
	\end{align}
	
	When stationary factors are present,
	\eqref{eq:Dopico_qq_star} gives
	\[
	\norm{
		\left(
		I_N-\widehat P_{s_0}
		\right)
		Q_{all,0}^{((q_0^*+1):q_0)}
	}_F
	=
	o_p\left(
	N^{-1/2}
	\right).
	\]
	Together with
	\eqref{eq:common_component_rotated_stationary_order}, this implies
	\begin{align}
		\frac{1}{NT}
		\sum_{t=1}^T
		\norm{
			\left(
			I_N-\widehat P_{s_0}
			\right)
			Q_{all,0}^{((q_0^*+1):q_0)}
			\left(
			R_0x_t
			\right)^{((q_0^*+1):q_0)}
		}_F^2
		&=
		o_p\left(
		N^{-1}
		\right).
		\label{eq:common_component_stationary_error}
	\end{align}
	Since the number of factor blocks is fixed,
	\eqref{eq:common_component_nonstationary_error} and
	\eqref{eq:common_component_stationary_error} yield
	\begin{align}
		\frac{1}{NT}
		\sum_{t=1}^T
		\norm{
			\left(
			I_N-\widehat P_{s_0}
			\right)
			\chi_t
		}_F^2
		&=
		O_p\left(
		N^{-1}
		\right).
		\label{eq:common_component_projection_error}
	\end{align}
	
	For the projected idiosyncratic component,
	Assumption~\ref{ass:Eigenvalue_Sigma_u_tilde} gives
	\begin{align}
		\frac{1}{NT}
		\sum_{t=1}^T
		\norm{
			\widehat P_{s_0}u_t
		}_F^2
		=
		\frac{1}{N}
		\tr\left(
		\widehat P_{s_0}
		\widetilde{\Sigma}_{u}
		\right)
		&\leq
		\frac{s_0}{N}
		\EW{1}{
			\widetilde{\Sigma}_{u}
		}
		=
		O_p\left(
		N^{-1}
		\right).
		\label{eq:common_component_idiosyncratic_error}
	\end{align}
	Combining
	\eqref{eq:common_component_projection_error} and
	\eqref{eq:common_component_idiosyncratic_error} proves
	\eqref{eq:Bai_Ng_common_component_rate}.
\end{proof}

\subsubsection{Proof of Theorem~\ref{T:Bai_Ng}}
\begin{proof}[Proof of Theorem~\ref{T:Bai_Ng}]
	The proof follows the two statements of the theorem. We first establish the groupwise no-underestimation result. We then impose the additional conditions in \eqref{eq:BN_penalty_conditions} and establish consistency
	for the full factor dimension.
	
	\paragraph{No groupwise underestimation}
	Fix $j=1,\ldots,q_0^*$. Suppose $s_0^*>0$ and let $s<s^{(1:j)}$. By the definition of the PCA
	value function,
	\begin{align}
		PC(s)-PC(s+1)
		&=
		\mathcal V(s)-\mathcal V(s+1)-g(N,T)
		\nonumber\\
		&=
		\frac{1}{N}
		\EW{s+1}{
			\widehat{\Sigma}_{y,T}
		}
		-
		g(N,T).
		\label{eq:Bai_Ng_adjacent_criterion_difference}
	\end{align}
	Since the empirical eigenvalues are ordered decreasingly,
	\[
	\frac{1}{N}
	\EW{s+1}{
		\widehat{\Sigma}_{y,T}
	}
	\geq
	\frac{1}{N}
	\EW{s^{(1:j)}}{
		\widehat{\Sigma}_{y,T}
	}.
	\]
	By
\eqref{eq:level_covariance_nonstationary_eigenvalue_orders},
\[
\frac{1}{N}
\EW{s^{(1:j)}}{
	\widehat{\Sigma}_{y,T}
}
=
O_e\left(
T^{2d_0^{(j)}-1}
\right).
\]
Hence, if
\[
g(N,T)
=
o\left(
T^{2d_0^{(j)}-1}
\right),
\]
	then
	\[
	PC(s)-PC(s+1)>0
	\]
	with probability approaching one. Since $s^{(1:j)}$ is fixed, this holds simultaneously for all $s=0,\ldots,s^{(1:j)}-1$. Hence,
	\[
		\Pr\left(
		\widehat{s}\geq s^{(1:j)}
		\right)
		\longrightarrow
		1,
		\]
	which proves the first statement of the theorem.
	
	We now impose the additional conditions in \eqref{eq:BN_penalty_conditions}. Since $N/T\to0$, we have
	$\min\{N,T\}=N$ for all sufficiently large $N,T$, so that
	\[
	\min\{N,T\}g(N,T)\to\infty
	\]
	is equivalent to
	\[
	Ng(N,T)\to\infty.
	\]
	Moreover, $g(N,T)\to0$ implies
	\[
	g(N,T)
	=
	o\left(
	T^{2d_0^{(q_0^*)}-1}
	\right)
	\]
	whenever $s_0^*>0$. Hence, the preceding no-underestimation result
	for the nonstationary factors remains valid.
	
	\paragraph{No underestimation of the remaining stationary factors.\\}
	Suppose first that $0<s_0^*<s_0$ and let
$
	s_0^*
	\leq
	s
	<
	s_0.
$
	By \eqref{eq:Bai_Ng_adjacent_criterion_difference} and the ordering of
	the empirical eigenvalues,
	\[
	PC(s)-PC(s+1)
	\geq
	\frac{1}{N}
	\EW{s_0}{
		\widehat{\Sigma}_{y,T}
	}
	-
	g(N,T).
	\]
	By \eqref{eq:level_covariance_stationary_eigenvalue_orders},
	\[
	\frac{1}{N}
	\EW{s_0}{
		\widehat{\Sigma}_{y,T}
	}
	=
	O_e(1).
	\]
	Since $g(N,T)\to0$, it follows that
	\[
	PC(s)-PC(s+1)>0
	\]
	with probability approaching one, simultaneously for all
	$s=s_0^*,\ldots,s_0-1$.
	 Together with the preceding result for the nonstationary factors,
		this rules out underestimation when $0<s_0^*<s_0$.
	
	If $s_0^*=0$, all factors are stationary and the same argument
		applies directly to every $s=0,\ldots,s_0-1$. If $s_0^*=s_0$, the
		preceding nonstationary-factor argument applies to all factors. Hence,
		under \eqref{eq:BN_penalty_conditions}, underestimation of the
		full factor dimension is ruled out in all cases:
		\begin{align}
			\label{eq:Bai_Ng_no_underestimation}
			\Pr\left(
			\widehat{s}<s_0
			\right)
			\longrightarrow
			0 .
	\end{align}
	
	\paragraph{No overestimation.}
	For a candidate factor dimension $s$, define
	\[
	\widehat P_s
	=
	V\left(
	\widehat{\Sigma}_{y,T}
	\right)_{1:s}
	V\left(
	\widehat{\Sigma}_{y,T}
	\right)_{1:s}^T
	\]
	and
	\[
	\widehat{\chi}_{s,t}
	=
	\widehat P_s y_t.
	\]
	Then
	\[
	\mathcal V(s)
	=
	\frac{1}{NT}
	\sum_{t=1}^T
	\norm{
		y_t-\widehat{\chi}_{s,t}
	}_F^2.
	\]
	
	Fix $s>s_0$. Since the principal-component spaces are nested,
	\[
	\widehat P_s\widehat P_{s_0}
	=
	\widehat P_{s_0}\widehat P_s
	=
	\widehat P_{s_0}.
	\]
	Consequently,
	$\widehat P_s-\widehat P_{s_0}$ is the orthogonal projector onto
	the additional $s-s_0$ empirical principal components. Moreover,
	\[
	\widehat{\chi}_{s,t}
	-
	\widehat{\chi}_{s_0,t}
	=
	\left(
	\widehat P_s-\widehat P_{s_0}
	\right)y_t
	\]
	is orthogonal to
	\[
	y_t-\widehat{\chi}_{s,t}
	=
	\left(
	I_N-\widehat P_s
	\right)y_t.
	\]
	The Pythagorean identity therefore gives
	\begin{align}
		\mathcal V(s_0)- \mathcal V(s)
		&=
		\frac{1}{NT}
		\sum_{t=1}^T
		\norm{
			\widehat{\chi}_{s,t}
			-
			\widehat{\chi}_{s_0,t}
		}_F^2.
		\label{eq:overestimation_value_difference}
	\end{align}
	
Since
\[
\widehat{\chi}_{s_0,t}
=
\widehat P_{s_0}y_t
\]
lies in the smaller empirical principal-component space, it is
annihilated by the projector onto the additional components. Indeed,
nestedness implies
\[
\begin{aligned}
	\left(
	\widehat P_s-\widehat P_{s_0}
	\right)
	\widehat{\chi}_{s_0,t}
	&=
	\left(
	\widehat P_s-\widehat P_{s_0}
	\right)
	\widehat P_{s_0}y_t
	\\
	&=
	\left(
	\widehat P_s\widehat P_{s_0}
	-
	\widehat P_{s_0}^2
	\right)y_t
	=
	0.
\end{aligned}
\]
Using $y_t=\chi_t+u_t$ and adding and subtracting
$\widehat{\chi}_{s_0,t}$ in the common component, we therefore obtain
\[
\begin{aligned}
	\widehat{\chi}_{s,t}
	-
	\widehat{\chi}_{s_0,t}
	&=
	\left(
	\widehat P_s-\widehat P_{s_0}
	\right)y_t
	\\
	&=
	\left(
	\widehat P_s-\widehat P_{s_0}
	\right)
	\left(
	\chi_t-\widehat{\chi}_{s_0,t}
	\right)
	+
	\left(
	\widehat P_s-\widehat P_{s_0}
	\right)
	\widehat{\chi}_{s_0,t}
	+
	\left(
	\widehat P_s-\widehat P_{s_0}
	\right)
	u_t
	\\
	&=
	\left(
	\widehat P_s-\widehat P_{s_0}
	\right)
	\left(
	\chi_t-\widehat{\chi}_{s_0,t}
	\right)
	+
	\left(
	\widehat P_s-\widehat P_{s_0}
	\right)
	u_t.
\end{aligned}
\]
Because $\widehat P_s-\widehat P_{s_0}$ is an orthogonal projector,
it is a contraction. Hence,
\[
\begin{aligned}
	\norm{
		\widehat{\chi}_{s,t}
		-
		\widehat{\chi}_{s_0,t}
	}_F^2
	&\leq
	2
	\norm{
		\left(
		\widehat P_s-\widehat P_{s_0}
		\right)
		\left(
		\chi_t-\widehat{\chi}_{s_0,t}
		\right)
	}_F^2
	+
	2
	\norm{
		\left(
		\widehat P_s-\widehat P_{s_0}
		\right)
		u_t
	}_F^2
		\\
	&\leq
	2
	\norm{
		\chi_t-\widehat{\chi}_{s_0,t}
	}_F^2
	+
	2
	\norm{
		\left(
		\widehat P_s-\widehat P_{s_0}
		\right)
		u_t
	}_F^2.
\end{aligned}
\]
	Hence, by \eqref{eq:overestimation_value_difference},
	\begin{align}
		\mathcal V(s_0) - \mathcal V(s)
		\leq
		\frac{2}{NT}
		\sum_{t=1}^T
		\norm{
			\chi_t-\widehat{\chi}_{s_0,t}
		}_F^2
		&+
		\frac{2}{NT}
		\sum_{t=1}^T
		\norm{
			\left(
			\widehat P_s-\widehat P_{s_0}
			\right)
			u_t
		}_F^2.
		\label{eq:Bai_Ng_overestimation_decomposition}
	\end{align}
	
By Corollary~\ref{cor:common_component_consistency},
\eqref{eq:Bai_Ng_common_component_rate} gives the required order for
the first term in \eqref{eq:Bai_Ng_overestimation_decomposition}.

	Furthermore,
	$\widehat P_s-\widehat P_{s_0}$ is an orthogonal projector of
	rank $s-s_0$, so
	\begin{align}
		\frac{1}{NT}
		\sum_{t=1}^T
		\norm{
			\left(
			\widehat P_s-\widehat P_{s_0}
			\right)
			u_t
		}_F^2
		&=
		\frac{1}{N}
		\tr\left[
		\left(
		\widehat P_s-\widehat P_{s_0}
		\right)
		\widetilde{\Sigma}_{u}
		\right]
		\nonumber\\
		&\leq
		\frac{s-s_0}{N}
		\EW{1}{
			\widetilde{\Sigma}_{u}
		}
		=
		O_p\left(
		N^{-1}
		\right)
		\label{eq:Bai_Ng_projected_idiosyncratic_rate}
	\end{align}
	by Assumption~\ref{ass:Eigenvalue_Sigma_u_tilde}.
	
	Combining
	\eqref{eq:Bai_Ng_overestimation_decomposition}-%
	\eqref{eq:Bai_Ng_projected_idiosyncratic_rate} gives
	\[
	0
	\leq
	\mathcal V(s_0)-\mathcal V(s)
	=
	O_p\left(
	N^{-1}
	\right).
	\]
	Since $s_{\max}$ is fixed,
	\begin{align}
		\max_{s_0<s\leq s_{\max}}
		\left[
		\mathcal V(s_0)-\mathcal V(s)
		\right]
		&=
		O_p\left(
		N^{-1}
		\right).
		\label{eq:Bai_Ng_uniform_overestimation_gain}
	\end{align}
	Therefore, uniformly over $s=s_0+1,\ldots,s_{\max}$,
	\[
	\begin{aligned}
		PC(s)-PC(s_0)
		&=
		\left(
		s-s_0
		\right)
		g(N,T)
		-
		\left[
	\mathcal V(s_0)-\mathcal V(s)
		\right]
		\\
		&=
		\left(
		s-s_0
		\right)
		g(N,T)
		-
		O_p\left(
		N^{-1}
		\right).
	\end{aligned}
	\]
	By \eqref{eq:BN_penalty_conditions}, the penalty dominates the
	reduction in the value function. Hence,
	\[
	\min_{s_0<s\leq s_{\max}}
	\left[
	PC(s)-PC(s_0)
	\right]
	>
	0
	\]
	with probability approaching one, and therefore
	\begin{align}
		\Pr\left(
		\widehat{s}>s_0
		\right)
		&\longrightarrow
		0.
		\label{eq:Bai_Ng_no_overestimation}
	\end{align}
	
	Combining
	\eqref{eq:Bai_Ng_no_overestimation} and
	\eqref{eq:Bai_Ng_no_underestimation} proves
	\[
	\Pr\left(
	\widehat{s}=s_0
	\right)
	\longrightarrow
	1.
	\]
\end{proof}

\subsubsection{Proof of Corollary~\ref{C:IC_criteria}}
\begin{proof}[Proof of Corollary~\ref{C:IC_criteria}]
We adapt the proof of \citet[Corollary~1]{BaiNg2002} and use two results from the proof of Theorem~\ref{T:Bai_Ng}.

First, if $s_0 \geq 1$,
	\[
	\ell_{N,T} = N^{-1}\kappa_{s_0}\!\left(\widehat{\Sigma}_{y,T}\right)
	\]
	is bounded away from zero in probability: by \eqref{eq:level_covariance_stationary_eigenvalue_orders} if $s_0^*<s_0$, and by \eqref{eq:level_covariance_nonstationary_eigenvalue_orders}, together with
	$T^{2d_0^{(q_0^*)}-1}\geq 1$, if $s_0^*=s_0$.
	Second,
	\[
	\max_{s_0<s\leq s_{\max}} \left[V(s_0)-V(s)\right] = O_p(N^{-1})
	\]
	by \eqref{eq:Bai_Ng_uniform_overestimation_gain}.
	Since $N/T\to0$, \eqref{eq:BN_penalty_conditions} implies $g(N,T)\to0$ and $Ng(N,T)\to\infty$.

To control the logarithmic criterion, we require an upper bound for $V(s_0)$ and a positive lower bound for $V(s)$. By Assumption~\ref{ass:Eigenvalue_Sigma_u_tilde}, 
\[\widetilde c_u\leq\EW{N}{\widetilde\Sigma_u}\leq\EW{1}{\widetilde\Sigma_u}\leq\widetilde C_u,
\]
with probability approaching one.

First, by \eqref{eq:pca_value_function} and \eqref{eq:value_function_eigenvalue_representation}, $\mathcal V(s_0)$ minimizes $N^{-1}\tr[(I_N-P)\widehat{\Sigma}_{y,T}]$ over orthogonal projection matrices $P$ of rank $s_0$. Choose the projection matrix $P_{\Lambda_0}=\Lambda_0\left(\Lambda_0^T\Lambda_0\right)^{-1}\Lambda_0^T$ onto the column space of $\Lambda_0$, which has rank $s_0$ by Assumption~\ref{ass:Lambda_full_rank} ($P_{\Lambda_0}=0$ if $s_0=0$). Since $\chi_t=\Lambda_0x_t$, $(I_N-P_{\Lambda_0})y_t=(I_N-P_{\Lambda_0})u_t$, and therefore
\[
\mathcal V(s_0)
\leq
\frac{1}{N}\tr\left[(I_N-P_{\Lambda_0})\widetilde\Sigma_u\right]
\leq
\frac{1}{N}\tr\left(\widetilde\Sigma_u\right)
\leq
\EW{1}{\widetilde\Sigma_u}
\leq
\widetilde C_u,
\]
where the second inequality holds because $\tr(P_{\Lambda_0}\widetilde\Sigma_uP_{\Lambda_0})\geq0$.
Second, $\widehat{\Sigma}_{y,T}-\widetilde\Sigma_u=T^{-1}\sum_{t=1}^T(\chi_t\chi_t^T+\chi_tu_t^T+u_t\chi_t^T)$ is symmetric, and its columns lie in the span of the columns of $\Lambda_0$ and of $T^{-1}\sum_{t=1}^Tu_tx_t^T$, so its rank is at most $2s_0$. Weyl's inequality $\EW{i+j-1}{A+B}\leq\EW{i}{A}+\EW{j}{B}$, applied with $A=\widehat{\Sigma}_{y,T}$, $B=\widetilde\Sigma_u-\widehat{\Sigma}_{y,T}$ and $j=2s_0+1$, gives $\EW{i+2s_0}{\widetilde\Sigma_u}\leq\EW{i}{\widehat{\Sigma}_{y,T}}$ for $i\leq N-2s_0$, because $B$ has at most $2s_0$ positive eigenvalues. As the eigenvalues of $\widehat{\Sigma}_{y,T}$ are nonnegative, \eqref{eq:value_function_eigenvalue_representation} yields, for $s\leq s_{\max}$ and $N\geq2(s_{\max}+2s_0)$,
\[
\mathcal V(s)
\geq
\frac{1}{N}\sum_{i=s+1}^{N-2s_0}\EW{i+2s_0}{\widetilde\Sigma_u}
\geq
\frac{N-s-2s_0}{N}\EW{N}{\widetilde\Sigma_u}
\geq
\frac{\widetilde c_u}{2}.
\]

\paragraph{No underestimation.} Let $s_0\geq1$ and $s<s_0$, and put $a_s=\mathcal V(s)-\mathcal V(s+1)=N^{-1}\EW{s+1}{\widehat{\Sigma}_{y,T}}\geq\ell_{N,T}$. By \eqref{eq:value_function_eigenvalue_representation}, $\mathcal V(s+1)=N^{-1}\sum_{i=s+2}^{s_0}\EW{i}{\widehat{\Sigma}_{y,T}}+\mathcal V(s_0)\leq(s_0-1)a_s+\mathcal V(s_0)$. Since $x\mapsto x/((s_0-1)x+\widetilde C_u)$ is increasing on $(0,\infty)$
\[
IC(s)-IC(s+1)
=
\log\left(1+\frac{a_s}{\mathcal V(s+1)}\right)-g(N,T)
\geq
\log\left(1+\frac{\ell_{N,T}}{(s_0-1)\ell_{N,T}+\widetilde C_u}\right)-g(N,T)
\]
for all $s<s_0$ with probability $\to 1$. By (i), the logarithm on the right is bounded away from zero in probability, and $g(N,T)\to0$. Hence, with probability approaching one, $IC(0)>IC(1)>\cdots>IC(s_0)$, and $\Pr(\widehat{s}_{IC}<s_0)\to0$.

\paragraph{No overestimation.} Let $s_0<s\leq s_{\max}$. Then $\mathcal V(s)\leq\mathcal V(s_0)$, and $\log(1+x)\leq x$ for $x\geq0$ and $g(N,T)>0$ give
\begin{align*}
IC(s)-IC(s_0)
&=
(s-s_0)g(N,T)-\log\left(1+\frac{\mathcal V(s_0)-\mathcal V(s)}{\mathcal V(s)}\right)
\\
&\geq
g(N,T)-\frac{2}{\widetilde c_u}\max_{s_0<s'\leq s_{\max}}\left[\mathcal V(s_0)-\mathcal V(s')\right],
\end{align*}
with probability $\to 1$. 
By (ii), the maximum on the right is $O_p(N^{-1})$. Since $Ng(N,T)\to\infty$, the right-hand side is positive with probability approaching one, uniformly in $s$, and $\Pr(\widehat{s}_{IC}>s_0)\to0$.
Together, $\Pr(\widehat{s}_{IC}=s_0)\to1$.
\end{proof}

\subsection{Auxiliary lemmas}
The following auxiliary lemmas collect the probabilistic and matrix-analytic results used in the proofs. They are included for readers who wish to follow the technical details in Appendices~\ref{sec:Factor-loading-space} and \ref{sec:Proofs-main-results}. Readers interested only in the main results may skip this section.

\subsubsection{Factor filters and second-moment limits}
\label{subsec:factor_second_moment_limits}
The perturbation arguments in Section~\ref{sec:perturbation} require the asymptotic behavior of the empirical second moments of the factor blocks. For the nonstationary factors, Lemma~\ref{L:combined_filter_coefficients} verifies the power-law expansion required for the fractional invariance principle of \citet{Hosoya2005}. For the stationary factors, Lemma~\ref{L:stationary_filter_coefficients} establishes the coefficient-decay and truncation bounds needed to compare the observed type-II factors with stationary type-I counterparts. The resulting weak-convergence and second-moment limits are collected in Lemma~\ref{L:factor_second_moment_limits}.

\begin{lemma}[Power-law expansion of the nonstationary filter coefficients]
	\label{L:combined_filter_coefficients}
	Suppose Assumptions~\ref{ass:short_run_filters} and \ref{ass:memory} hold. Then there exists a constant $C_b<\infty$ such that, for all $i=1,\ldots,s_0^*$ and $\ell\geq1$,
	\begin{align}
		\left|
		b_{i,\ell}(d_{0,i}) - \frac{\psi_i(1)}{\Gamma(d_{0,i})} \ell^{d_{0,i}-1}
		\right|
		&\leq
		C_b\ell^{d_{0,i}-2}.
		\label{eq:combined_filter_power_law}
	\end{align}
\end{lemma}

\begin{proof}
	Fix $i\in\{1,\ldots,s_0^*\}$ and write $d=d_{0,i}>1/2$. We first compare the combined coefficient $b_{i,\ell}(d)$ with $\psi_i(1)\pi_\ell(-d)$ and then use the standard expansion of the fractional coefficients.
	
	The fractional coefficients satisfy
	\begin{align}
		\pi_\ell(-d) &= \frac{\Gamma(\ell+d)}{\Gamma(d)\Gamma(\ell+1)}
		=
		\frac{\ell^{d-1}}{\Gamma(d)} + O(\ell^{d-2}),
		\label{eq:fractional_coefficient_expansion}\\
		|\pi_\ell(-d)| &= O\left((1+\ell)^{d-1}\right).
		\label{eq:fractional_coefficient_bound}
	\end{align}
	Moreover, the recursion for the fractional coefficients gives
	\[
	\pi_r(-d)-\pi_{r-1}(-d) = \frac{d-1}{r}\pi_{r-1}(-d).
	\]
	It follows from \eqref{eq:fractional_coefficient_bound} that
	\[
	\left|\pi_r(-d)-\pi_{r-1}(-d)\right| = O(r^{d-2}).
	\]
	Consequently, for $0\leq j\leq\ell/2$, telescoping yields
	\begin{align}
		\left|\pi_{\ell-j}(-d)-\pi_\ell(-d)\right|
		&\leq
		\sum_{r=\ell-j+1}^{\ell}\left|\pi_r(-d)-\pi_{r-1}(-d)\right|
		\nonumber\\
		&= O\left(\sum_{r=\ell-j+1}^{\ell}r^{d-2}\right)
		= O(j\ell^{d-2}),
		\label{eq:fractional_coefficient_shift}
	\end{align}
	because $\ell/2\leq r\leq\ell$ throughout the sum.
	
	By Assumption~\ref{ass:short_run_filters}, the series defining $\psi_i(1)$ converges absolutely. Hence,
	\begin{align}
		b_{i,\ell}(d)-\psi_i(1)\pi_\ell(-d)
		={}&
		\sum_{j=0}^{\ell} \psi_{i,j}
		\left[\pi_{\ell-j}(-d)-\pi_\ell(-d)\right]
		-	\pi_\ell(-d) \sum_{j=\ell+1}^{\infty}\psi_{i,j}.
		\label{eq:combined_filter_decomposition}
	\end{align}
	Let $m_\ell	=\left\lfloor\frac{\ell}{2}\right\rfloor$ and split the first sum in \eqref{eq:combined_filter_decomposition} at $m_\ell$.
	
	For $j\leq m_\ell$, the shift bound in \eqref{eq:fractional_coefficient_shift} gives
	\begin{align}
		&\sum_{j=0}^{m_\ell} |\psi_{i,j}|
		\left|\pi_{\ell-j}(-d)-\pi_\ell(-d)
		\right|
		\leq
		C\ell^{d-2} \sum_{j=0}^{m_\ell}j|\psi_{i,j}|
		= O(\ell^{d-2}),
		\label{eq:combined_filter_small_j}
	\end{align}
	where the last equality follows from Assumption~\ref{ass:short_run_filters}.
	
	For $j>m_\ell$, the index $\ell-j$ may be close to zero, so the shift bound is no longer useful. We instead use the decay of the short-run coefficients. Assumption~\ref{ass:short_run_filters} implies
	\[
	|\psi_{i,j}| \leq C(1+j)^{-2}.
	\]
	Since $j>m_\ell$ implies $(1+j)^{-2}=O(\ell^{-2})$, it follows from \eqref{eq:fractional_coefficient_bound} that
	\begin{align}
		\sum_{j=m_\ell+1}^{\ell}
		|\psi_{i,j}| |\pi_{\ell-j}(-d)|
		&= O\left(\ell^{-2} \sum_{j=m_\ell+1}^{\ell} |\pi_{\ell-j}(-d)| \right)
		\nonumber\\
		&=
		O\left(\ell^{-2} \sum_{r=0}^{\lfloor\ell/2\rfloor}
		(1+r)^{d-1}\right)
		= O(\ell^{d-2}),
				\label{eq:combined_filter_large_j}
	\end{align}
	because
	\[
	\sum_{r=0}^{\lfloor\ell/2\rfloor} (1+r)^{d-1} = O(\ell^d).
	\]
	
	The weighted summability condition also gives
	\begin{align*}
		\sum_{j>m_\ell}|\psi_{i,j}|
		&\leq
		\frac{1}{(1+m_\ell)^2} \sum_{j=0}^{\infty} (1+j)^2|\psi_{i,j}|
		= O(\ell^{-2}).
	\end{align*}
	Therefore, by \eqref{eq:fractional_coefficient_bound},
	\begin{align}
		|\pi_\ell(-d)| \sum_{j>m_\ell}|\psi_{i,j}|
		&= O(\ell^{d-3}).
		\label{eq:combined_filter_fractional_tail}
	\end{align}
	This controls both the terms involving $\pi_\ell(-d)$ for $j>m_\ell$ and the omitted tail $j>\ell$ in \eqref{eq:combined_filter_decomposition}.
	
	Combining \eqref{eq:combined_filter_small_j}, \eqref{eq:combined_filter_large_j}, and \eqref{eq:combined_filter_fractional_tail} gives
	\begin{align}
		b_{i,\ell}(d) &= \psi_i(1)\pi_\ell(-d) + O(\ell^{d-2}).
		\label{eq:combined_filter_fractional_approximation}
	\end{align}
	Substituting \eqref{eq:fractional_coefficient_expansion} into \eqref{eq:combined_filter_fractional_approximation} yields

\begin{align} \label{eq:combined_filter_power_law_asymptotic}
	b_{i,\ell}(d) = \frac{\psi_i(1)}{\Gamma(d)} \ell^{d-1} + O(\ell^{d-2}). 
\end{align}
	
	The number of nonstationary factors and their memory parameters are fixed, while Assumption~\ref{ass:short_run_filters} holds uniformly over $i$. 
Hence, the asymptotic bound in \eqref{eq:combined_filter_power_law_asymptotic} holds uniformly over $i=1,\ldots,s_0^*$. Thus, there exist constants $C_b<\infty$ and $\ell_0<\infty$ such that
	\[
	\left|
	b_{i,\ell}(d_{0,i})
	-\frac{\psi_i(1)}{\Gamma(d_{0,i})}\ell^{d_{0,i}-1}
	\right|
	\leq C_b\ell^{d_{0,i}-2}
	\]
	for all $i=1,\ldots,s_0^*$ and all $\ell\geq\ell_0$. Increasing $C_b$ if necessary to cover the finitely many remaining values $\ell=1,\ldots,\ell_0-1$ proves \eqref{eq:combined_filter_power_law} for every $\ell\geq1$.
\end{proof}

\begin{lemma}[Decay and truncation of the stationary filter coefficients]
	\label{L:stationary_filter_coefficients}
	Suppose Assumptions~\ref{ass:short_run_filters} and \ref{ass:memory} hold. Then there exists a constant $C_b<\infty$ such that, for all $i=s_0^*+1,\ldots,s_0$ and $\ell\geq0$,
	\begin{align}
		\left| b_{i,\ell}(d_{0,i})\right|
		&\leq
		C_b(1+\ell)^{d_{0,i}-1}.
		\label{eq:stationary_filter_coefficient_bound}
	\end{align}
	In particular,
	\begin{align}
		\sum_{\ell=0}^{\infty} b_{i,\ell}(d_{0,i})^2
		&< \infty.
		\label{eq:stationary_filter_square_summability}
	\end{align}
	Moreover,
	\begin{align}
		\frac{1}{T} \sum_{t=1}^{T}
		\sum_{\ell=t}^{\infty} b_{i,\ell}(d_{0,i})^2
		&=
		O\left(T^{2\max\{d_{0,i},0\}-1}\right) = o(1).
		\label{eq:stationary_filter_average_tail}
	\end{align}
\end{lemma}

\begin{proof}
	Fix $i\in\{s_0^*+1,\ldots,s_0\}$ and write $d=d_{0,i}\in(-1/2,1/2)$. We first establish \eqref{eq:stationary_filter_coefficient_bound}.
	
	Suppose $d\neq0$. The standard gamma-ratio bound gives
	\begin{align}
		|\pi_\ell(-d)|
		&\leq
		C(1+\ell)^{d-1},
		\qquad \ell\geq0;
		\label{eq:stationary_fractional_coefficient_bound}
	\end{align}
	see, for example, \citet[equation~(5.25)]{Hassler2018}. By the definition of the combined coefficients,
	\[
	b_{i,\ell}(d) =
	\sum_{j=0}^{\ell} \psi_{i,j}\pi_{\ell-j}(-d).
	\]
	Let $m_\ell=\lfloor\ell/2\rfloor$ and write
	\begin{align*}
		|b_{i,\ell}(d)|
		&\leq
		\sum_{j=0}^{m_\ell} |\psi_{i,j}| |\pi_{\ell-j}(-d)|
		+
		\sum_{j=m_\ell+1}^{\ell} |\psi_{i,j}| |\pi_{\ell-j}(-d)|.
	\end{align*}
	
	For $j\leq m_\ell$, $1+\ell-j \geq \frac{1+\ell}{2}$. 
	Since $d-1<0$,
	\[
	(1+\ell-j)^{d-1} 	\leq 2^{1-d}(1+\ell)^{d-1}.
	\]
	Hence, by \eqref{eq:stationary_fractional_coefficient_bound} and absolute summability of the short-run coefficients,
	\begin{align}
		\sum_{j=0}^{m_\ell} |\psi_{i,j}| |\pi_{\ell-j}(-d)|
		&\leq
		C(1+\ell)^{d-1} \sum_{j=0}^{\infty}|\psi_{i,j}|
		=
		O\left( (1+\ell)^{d-1} \right).
		\label{eq:stationary_coefficient_small_j}
	\end{align}
	
	For $j>m_\ell$, Assumption~\ref{ass:short_run_filters} implies $|\psi_{i,j}|\leq C(1+\ell)^{-2}$. Therefore,
	\begin{align}
		\sum_{j=m_\ell+1}^{\ell} |\psi_{i,j}| |\pi_{\ell-j}(-d)|
		&\leq
		\frac{C}{(1+\ell)^2} \sum_{r=0}^{\lfloor\ell/2\rfloor} (1+r)^{d-1}.
		\label{eq:stationary_coefficient_large_j}
	\end{align}
	If $0<d<1/2$, the sum on the right is $O((1+\ell)^d)$,  so \eqref{eq:stationary_coefficient_large_j} is $O((1+\ell)^{d-2})$, and therefore also $O((1+\ell)^{d-1})$.

	If $-1/2<d<0$, the sum is $O(1)$, so \eqref{eq:stationary_coefficient_large_j} is $O((1+\ell)^{-2})$, and therefore $O((1+\ell)^{d-1})$
	because $-2<d-1$. Together with \eqref{eq:stationary_coefficient_small_j}, this proves \eqref{eq:stationary_filter_coefficient_bound} when $d\neq0$.
	
	If $d=0$, then
	\[
	\pi_0(0)=1, \qquad
	\pi_\ell(0)=0, \quad \ell\geq1,
	\]
	so that
	\[
	b_{i,\ell}(0)=\psi_{i,\ell}.
	\]
	Assumption~\ref{ass:short_run_filters} gives $|\psi_{i,\ell}|\leq C(1+\ell)^{-2}$, which also implies \eqref{eq:stationary_filter_coefficient_bound}.
	
	Since $d<1/2$, we have $2d-2<-1$. Thus,
	\begin{align*}
		\sum_{\ell=0}^{\infty} b_{i,\ell}(d)^2
		&\leq
		C \sum_{\ell=0}^{\infty} (1+\ell)^{2d-2} < \infty,
	\end{align*}
	which proves \eqref{eq:stationary_filter_square_summability}.
	
	For the average truncation error, reversing the order of summation gives
	\begin{align*}
		\frac{1}{T}
		\sum_{t=1}^{T} \sum_{\ell=t}^{\infty} b_{i,\ell}(d)^2
		&= 
		\frac{1}{T} \sum_{\ell=1}^{T} \ell b_{i,\ell}(d)^2
		+ \sum_{\ell=T+1}^{\infty} b_{i,\ell}(d)^2.
	\end{align*}
	
	If $0<d<1/2$, then
	\begin{align*}
		\frac{1}{T} \sum_{\ell=1}^{T} \ell b_{i,\ell}(d)^2
		&\leq
		\frac{C}{T} \sum_{\ell=1}^{T} \ell^{2d-1} = O(T^{2d-1}),
		\label{eq:stationary_average_tail_positive_first}\\
		\sum_{\ell=T+1}^{\infty} b_{i,\ell}(d)^2
		&\leq C \sum_{\ell=T+1}^{\infty} \ell^{2d-2}
		= O(T^{2d-1}).
	\end{align*}
	
	If $-1/2<d<0$, then
	\begin{align*}
		\frac{1}{T} \sum_{\ell=1}^{T} \ell b_{i,\ell}(d)^2
		&\leq
		\frac{C}{T} \sum_{\ell=1}^{T} \ell^{2d-1}
		= O(T^{-1}), \\
		\sum_{\ell=T+1}^{\infty} b_{i,\ell}(d)^2
		&= O(T^{2d-1}) = o(T^{-1}).
	\end{align*}
	
	Finally, if $d=0$, then $b_{i,\ell}(0)=\psi_{i,\ell}$ and
	\[
	b_{i,\ell}(0)^2 = O\left((1+\ell)^{-4}\right).
	\]
	Consequently,
	\begin{align*}
		\frac{1}{T} \sum_{\ell=1}^{T} \ell b_{i,\ell}(0)^2
		&\leq
		\frac{C}{T} \sum_{\ell=1}^{\infty} \ell^{-3}
		= O(T^{-1}),
		\label{eq:stationary_average_tail_zero_first}\\
		\sum_{\ell=T+1}^{\infty} b_{i,\ell}(0)^2
		&\leq
		C \sum_{\ell=T+1}^{\infty} \ell^{-4}
		= O(T^{-3}).
	\end{align*}
	
	The three cases prove \eqref{eq:stationary_filter_average_tail}. Uniformity over the fixed set of stationary factors follows from Assumption~\ref{ass:short_run_filters}.
\end{proof}

We next define the limiting objects. Whenever $s_0^*>0$, let
\[
d_0^*
=
\left(
d_{0,1},\ldots,d_{0,s_0^*}
\right)^T
\]
and
\[
\Sigma_{\eta,0}^*
=
\Var\left(
\left(
\eta_{1,t},\ldots,\eta_{s_0^*,t}
\right)^T
\right).
\]
By Assumptions~\ref{ass:eta_mds} and \ref{ass:one_indep_all}, $\Sigma_{\eta,0}^*$ is diagonal and positive definite.

For $a\in\mathbb R$, let $\lfloor a\rfloor$ denote the greatest
integer less than or equal to $a$, and let
$B(\cdot,\Sigma_{\eta,0}^*)$ denote an $s_0^*$-dimensional Brownian
motion with covariance matrix $\Sigma_{\eta,0}^*$, that is,
\[
\mathbb{E}\left[
B(u,\Sigma_{\eta,0}^*)
B(v,\Sigma_{\eta,0}^*)^T
\right]
=
\min\{u,v\}\Sigma_{\eta,0}^*.
\]

For $0\leq r\leq u\leq1$, define
\[
W(u,r,d_0^*)
=
\diag\left(
(u-r)^{d_{0,1}-1},
\ldots,
(u-r)^{d_{0,s_0^*}-1}
\right)
\]
and the type-II fractional Brownian motion
\begin{align*}
	F_B(u,d_0^*,\Sigma_{\eta,0}^*)
	&=
	\int_0^u
	W(u,r,d_0^*)
	\,dB(r,\Sigma_{\eta,0}^*).
\end{align*}
Finally, let
\[
G(d_0^*)
=
\diag\left(
\Gamma(d_{0,1}),
\ldots,
\Gamma(d_{0,s_0^*})
\right)
\]
and
\[
\Psi_0^*
=
\diag\left(
\psi_1(1),
\ldots,
\psi_{s_0^*}(1)
\right).
\]

Whenever $s_0^*<s_0$, Lemma~\ref{L:stationary_filter_coefficients}
implies that the following infinite-past representation converges in
mean square. Define the stationary type-I counterpart of factor $i$ by
\begin{align}
	\bar x_{i,t}
	&=
	\sum_{\ell=0}^{\infty}
	b_{i,\ell}(d_{0,i})\eta_{i,t-\ell},
	\qquad
	i=s_0^*+1,\ldots,s_0,
	\label{eq:stationary_type_I_factor}
\end{align}
and collect these factors in
\[
\bar w_t
=
\left(
\bar x_{s_0^*+1,t},
\ldots,
\bar x_{s_0,t}
\right)^T.
\]
In particular,
\[
\mathbb{E}\norm{\bar w_t}_F^2<\infty \qquad\text{for all }t.
\]

\begin{lemma}[Weak convergence and covariance limits of the factor blocks]
	\label{L:factor_second_moment_limits}
	Suppose Assumptions~\ref{ass:factors}, \ref{ass:one_indep_all}, and \ref{ass:memory} hold.
	If $s_0^*>0$, then
	\begin{align}
		D_{z,T,0}^{-1} z_{\lfloor T\cdot\rfloor}
		&\Rightarrow
		\Psi_0^* G(d_0^*)^{-1}F_B(\cdot,d_0^*,\Sigma_{\eta,0}^*)
		\label{eq:nonstationary_factor_FCLT}
	\end{align}
	in $D([0,1],\mathbb R^{s_0^*})$ equipped with the Skorokhod $J_1$ topology. Consequently,
\begin{align}
	D_{z,T,0}^{-1}\widetilde\Sigma_zD_{z,T,0}^{-1}
	&\Rightarrow
	S_z,
	\label{eq:nonstationary_factor_covariance_limit}
\end{align}
where
\begin{align*}
	S_z
	&=
	\int_0^1
	\Psi_0^*G(d_0^*)^{-1}
	F_B(u,d_0^*,\Sigma_{\eta,0}^*)
	\nonumber\\
	&\qquad\qquad\times
	F_B(u,d_0^*,\Sigma_{\eta,0}^*)^T
	G(d_0^*)^{-1}(\Psi_0^*)^T
	\,du.
\end{align*}
	
	If $s_0^*<s_0$, then
	\begin{align}
		\widetilde\Sigma_w
		= \frac{1}{T} \sum_{t=1}^{T}w_tw_t^T &\overset{p}{\longrightarrow}
		S_w = \EE{\bar w_t\bar w_t^T}.
		\label{eq:stationary_factor_covariance_limit}
	\end{align}
	The random matrix $S_z$ is positive definite almost surely, while $S_w$ is deterministic, finite, and positive definite.
\end{lemma}

\begin{proof}
	We first consider the nonstationary factor block. By Assumption~\ref{ass:memory}, $d_{0,i}>1/2$ for $i=1,\ldots,s_0^*$. Moreover, Lemma~\ref{L:combined_filter_coefficients} gives
	\begin{align}
		b_{i,\ell}(d_{0,i})
		&=
		\frac{\psi_i(1)}{\Gamma(d_{0,i})} \ell^{d_{0,i}-1} + O(\ell^{d_{0,i}-2}).
		\label{eq:Hosoya_filter_condition}
	\end{align}
This is the coefficient condition in \citet[equation~(2)]{Hosoya2005}, apart from Hosoya's normalization that the leading coefficient is positive. If $\psi_i(1)<0$, the theorem can be applied to the sign-adjusted process $\widetilde x_{i,t}=-x_{i,t}$, whose leading coefficient is $|\psi_i(1)|/\Gamma(d_{0,i})>0$. The weak limit for the original factor is then obtained by reversing the sign of the limiting process.
	
	We next verify Hosoya's assumptions on the nonstationary factor innovations. By Assumption~\ref{ass:eta_mds}, for every $k>t$,
	\begin{align*}
		\EE{\eta_k\mid\mathcal F_t} &= 0,
		\qquad
		\Var\left(
		\EE{\eta_k\mid\mathcal F_t}
		\right) = 0.
	\end{align*}
	Hence Hosoya's Assumption A(i) holds with zero approximation error.
	
	For equal future dates, Assumption~\ref{ass:eta_future_covariances} gives
	\begin{align*}
		\EE{
			\eta_k\eta_k^T
			\mid\mathcal F_t
		}
		&=
		\EE{\eta_k\eta_k^T},
		\qquad k>t.
	\end{align*}
	If $k>m>t$, then $\eta_m$ is $\mathcal F_{k-1}$-measurable and the martingale-difference property implies
	\[
	\EE{\eta_k\eta_m^T \mid\mathcal F_t } = 0 = \EE{\eta_k\eta_m^T}.
	\]
	Thus Hosoya's Assumption A(ii) also holds with zero approximation error. Assumption A(iii) follows from Assumption~\ref{ass:eta_fourth_moments} and \citet[Lemma~1]{Hosoya2005}. For Assumption A(iv), the fourth-order conditions in Assumption~\ref{ass:eta_fourth_moments} and \citet[Theorem~3]{Hosoya2005} are sufficient when $d_{0,i}>3/4$. When $1/2<d_{0,i}\leq3/4$, Assumption~\ref{ass:eta_higher_moments} and \citet[Theorem~3]{Hosoya2005} ensure Assumption A(iv).
	
	Theorem~2 of \citet{Hosoya2005}, together with \eqref{eq:Hosoya_filter_condition}, therefore yields \eqref{eq:nonstationary_factor_FCLT}.
	
	To derive the covariance limit, define
	\[
	\Phi(f) = \int_0^1 f(u)f(u)^T\,du.
	\]
	Convergence in the $J_1$ topology to a continuous function implies uniform convergence \citep[Section~12, p.~124]{Billingsley1999}. If $f_n$ converges uniformly to $f$, then
	\begin{align*}
		\norm{\Phi(f_n)-\Phi(f)}_F
		&\leq
		\left(
		\norm{f_n}_{\infty}+ \norm{f}_{\infty}
		\right)
		\norm{f_n-f}_{\infty}
		\longrightarrow
		0.
	\end{align*}
	Thus $\Phi$ is continuous at every continuous function under the $J_1$ topology. The limiting process in \eqref{eq:nonstationary_factor_FCLT} has continuous sample paths almost surely. The continuous mapping theorem \citep[Theorem~2.7, pp.~21-22]{Billingsley1999} therefore gives
	\begin{align}
		&\int_0^1
		D_{z,T,0}^{-1} z_{\lfloor Tu\rfloor} z_{\lfloor Tu\rfloor}^T D_{z,T,0}^{-1}
		\,du
		\Rightarrow
		S_z.
		\label{eq:nonstationary_covariance_integral_limit}
	\end{align}
	
	The integral averages the step process over the intervals $[t/T,(t+1)/T)$ and therefore contains the observations $z_0,\ldots,z_{T-1}$:
	\begin{align*}
		\int_0^1
		D_{z,T,0}^{-1} z_{\lfloor Tu\rfloor} z_{\lfloor Tu\rfloor}^T D_{z,T,0}^{-1}
		\,du
		= \frac{1}{T}
		\sum_{t=0}^{T-1} D_{z,T,0}^{-1} z_tz_t^T D_{z,T,0}^{-1}.
	\end{align*}
	By contrast, $\widetilde\Sigma_z$ contains $z_1,\ldots,z_T$. Under the type-II convention, $z_0=0$, so the two expressions differ only by the final observation:
	\begin{align}
		&D_{z,T,0}^{-1} \widetilde\Sigma_z D_{z,T,0}^{-1}
		-
		\int_0^1 D_{z,T,0}^{-1} z_{\lfloor Tu\rfloor} z_{\lfloor Tu\rfloor}^T D_{z,T,0}^{-1}
		\,du =
		\frac{1}{T} D_{z,T,0}^{-1} z_Tz_T^T D_{z,T,0}^{-1}.
		\label{eq:nonstationary_covariance_boundary_term}
	\end{align}
	By \eqref{eq:nonstationary_factor_FCLT},
	\[
	D_{z,T,0}^{-1}z_T = O_p(1).
	\]
	The Frobenius norm of the right-hand side of \eqref{eq:nonstationary_covariance_boundary_term} is therefore $O_p(T^{-1})=o_p(1)$. Combining this with \eqref{eq:nonstationary_covariance_integral_limit} proves \eqref{eq:nonstationary_factor_covariance_limit}.
	
Let
\[
Z(u)
=
\Psi_0^*G(d_0^*)^{-1}
F_B(u,d_0^*,\Sigma_{\eta,0}^*).
\]
For any fixed points $
0<u_1<\cdots<u_{s_0^*}\leq1,$
the finite-dimensional distribution
\[
\bigl(
Z(u_1)^T,\ldots,Z(u_{s_0^*})^T
\bigr)^T
\]
is nondegenerate Gaussian. This follows from the nondegeneracy of the
finite-dimensional distributions of the type-II fractional Brownian
motion, the positive definiteness of $\Sigma_{\eta,0}^*$, and the
nonsingularity of $\Psi_0^*G(d_0^*)^{-1}$. Indeed, the $s_0^*$ coordinates
of $Z$ are driven by independent Brownian motions, and within coordinate $i$ the
kernels $r\mapsto(u_k-r)^{d_{0,i}-1}\mathbf 1\{r<u_k\}$, $k=1,\ldots,s_0^*$, are
linearly independent in $L^2[0,1]$ for distinct $u_k$, so the covariance matrix of
the stacked vector is nonsingular. Hence the random matrix
\[
\bigl(
Z(u_1),\ldots,Z(u_{s_0^*})
\bigr)
\]
has a density on $\mathbb R^{(s_0^*)^2}$ and is therefore nonsingular
almost surely, since the set of singular matrices has Lebesgue measure
zero.

Now suppose that $S_z$ were singular. Since
\[
S_z
=
\int_0^1 Z(u)Z(u)^T\,du,
\]
there would exist $a\neq0$ such that
\[
0
=
a^TS_za
=
\int_0^1
\bigl(a^TZ(u)\bigr)^2\,du.
\]
Because $Z$ has continuous sample paths, this implies
\[
a^TZ(u)=0
\qquad
\text{for every }u\in[0,1].
\]
In particular,
\[
a^T
\bigl(
Z(u_1),\ldots,Z(u_{s_0^*})
\bigr)
=
0,
\]
contradicting the almost-sure nonsingularity of this matrix. Therefore,
$S_z$ is positive definite almost surely.
	
	We finally consider the stationary factor block. By \eqref{eq:stationary_filter_square_summability}, the type-I process $\bar w_t$ in \eqref{eq:stationary_type_I_factor} is stationary and has finite second moments. As a measurable time-invariant function of the stationary and ergodic innovation process, it is also ergodic.
	
	The difference between the type-II and type-I versions of factor $i$ is
	\begin{align*}
		x_{i,t}-\bar x_{i,t} &=
		- \sum_{\ell=t}^{\infty} b_{i,\ell}(d_{0,i})\eta_{i,t-\ell}.
	\end{align*}
	The martingale-difference property implies temporal uncorrelatedness of the innovations. Therefore,
	\begin{align}
		\EE{|x_{i,t}-\bar x_{i,t}|^2}
		&=
		\EE{\eta_{i,t}^2} \sum_{\ell=t}^{\infty} b_{i,\ell}(d_{0,i})^2.
		\label{eq:type_II_type_I_mean_square}
	\end{align}
	Summing \eqref{eq:type_II_type_I_mean_square} over the finitely many stationary factors and applying \eqref{eq:stationary_filter_average_tail} gives
	\begin{align*}
		\frac{1}{T} \sum_{t=1}^{T} \EE{ \norm{w_t-\bar w_t}^2}
		&= o(1).
	\end{align*}
	Markov's inequality then yields
	\begin{align*}
		\frac{1}{T} \sum_{t=1}^{T} \norm{w_t-\bar w_t}^2
		&= o_p(1).
	\end{align*}
	
	The Birkhoff-Khinchin ergodic theorem gives
	\begin{align}
		\frac{1}{T} \sum_{t=1}^{T} \bar w_t\bar w_t^T
		&\overset{\mathrm{a.s.}}{\longrightarrow}
		\EE{\bar w_t\bar w_t^T} = S_w.
		\label{eq:stationary_type_I_covariance_limit}
	\end{align}
	In particular,
	\[
	\frac{1}{T} \sum_{t=1}^{T} \norm{\bar w_t}^2
	= O_p(1).
	\]
	Moreover,
	\begin{align*}
		\frac{1}{T} \sum_{t=1}^{T} \norm{w_t}^2
		&\leq
		\frac{2}{T} \sum_{t=1}^{T}
		\norm{w_t-\bar w_t}^2 + \frac{2}{T} \sum_{t=1}^{T} \norm{\bar w_t}^2
		= O_p(1).
	\end{align*}
	
	Using
	\[
	w_tw_t^T-\bar w_t\bar w_t^T
	= (w_t-\bar w_t)w_t^T + \bar w_t(w_t-\bar w_t)^T
	\]
	and Cauchy-Schwarz, we obtain
	\begin{align}
		&\left\|
		\frac{1}{T} \sum_{t=1}^{T}
		\left( w_tw_t^T-\bar w_t\bar w_t^T \right)
		\right\|_F
		\nonumber\\
		&\qquad\leq
		\left(
		\frac{1}{T} \sum_{t=1}^{T} \norm{w_t-\bar w_t}^2
		\right)^{1/2}
		\left[ \left(
		\frac{1}{T}\sum_{t=1}^{T}\norm{w_t}^2
		\right)^{1/2}
		+
		\left(
		\frac{1}{T}\sum_{t=1}^{T}\norm{\bar w_t}^2
		\right)^{1/2}
		\right]
= o_p(1).
		\label{eq:stationary_type_I_type_II_covariance_difference}
	\end{align}
	Combining
	\eqref{eq:stationary_type_I_covariance_limit} and \eqref{eq:stationary_type_I_type_II_covariance_difference} proves \eqref{eq:stationary_factor_covariance_limit}.
	
Finally, Assumption~\ref{ass:one_indep_all} implies that the stationary factor coordinates are mutually independent, so $S_w$ is diagonal. Moreover, for each stationary factor, temporal uncorrelatedness of the innovations gives
\[
\EE{\bar x_{i,t}^2}
=
\EE{\eta_{i,t}^2}
\sum_{\ell=0}^{\infty}
b_{i,\ell}(d_{0,i})^2
\geq
\EE{\eta_{i,t}^2}
b_{i,0}(d_{0,i})^2
>
0,
\]
because $b_{i,0}(d_{0,i})=\psi_{i,0}=1$. Hence, $S_w$ is finite and positive definite.
\end{proof}
\subsubsection{Empirical covariance and cross-covariance bounds}
This subsection derives bounds for empirical covariances and
cross-covariances involving $z_t$, $w_t$, and $u_t$. We first record
the Frobenius-norm approximation for the empirical idiosyncratic
covariance implied by Assumptions~\ref{ass:u_stationary}-\ref{ass:u_autocov_summability}. The remaining lemmas
provide the blockwise covariance rates used in the relative
perturbation arguments in the next subsection.

\begin{lemma}[Empirical idiosyncratic covariance approximation]
	\label{L:Sigma_u_hat_rate}
	Under Assumptions \ref{ass:u_stationary}, 
	\ref{ass:u_fourth_cumulants}, and \ref{ass:u_autocov_summability} it holds that
	\[
	\left\| \widetilde{\Sigma}_u-\Sigma_{u,0} \right\|_F
	=
	O_p\left(\frac{N}{\sqrt{T}}\right).
	\]
\end{lemma}

\begin{proof}
	Let
	\[
	\Delta_{ll'}=\frac{1}{T}\sum_{t=1}^{T}u_{l,t}u_{l',t}
	-\EE{u_{l,t}u_{l',t}}
	\]
	be the $(l,l')$-th entry of $\widetilde{\Sigma}_u-\Sigma_{u,0}$. By Assumption \ref{ass:u_stationary}, $\EE{\Delta_{ll'}}=0$, and therefore
	\[
	\EE{\Delta_{ll'}^2}
	=
	\frac{1}{T^2}\sum_{t=1}^{T}\sum_{s=1}^{T} \operatorname{Cov}(u_{l,t}u_{l',t},u_{l,s}u_{l',s}).
	\]
	For $h=t-s$, define
	\[
	\Gamma_{ll',t}(h)
	=
	\operatorname{Cov}(u_{l,t}u_{l',t},u_{l,t-h}u_{l',t-h}).
	\]
	Using the fourth-order cumulant decomposition and $\EE{u_t}=0$ gives
	\[
	\Gamma_{ll',t}(h)
	=
	\operatorname{cum}(u_{l,t},u_{l',t},u_{l,t-h},u_{l',t-h})
	+
	\gamma_{ll}(h)\gamma_{l'l'}(h)
	+
	\gamma_{ll'}(h)\gamma_{l'l}(h),
	\]
	where
	\[
	\gamma_{ll'}(h)=\operatorname{Cov}(u_{l,t},u_{l',t-h}).
	\]
	We next show that $\Gamma_{ll',t}(h)$ is absolutely summable, uniformly in $t,l,l'$
	and in $N$. By Assumption \ref{ass:u_fourth_cumulants},
	\[
	\sup_{l,l'\leq N} \sup_t
	\sum_{h=-\infty}^{\infty}
	\left|
	\operatorname{cum}(u_{l,t},u_{l',t},u_{l,t-h},u_{l',t-h})
	\right|
	\leq C_4 .
	\]
	Moreover, Assumption \ref{ass:u_autocov_summability} implies
	\[
	\sup_{l,l'\leq N}
	\sum_{h=-\infty}^{\infty}
	|\gamma_{ll'}(h)|\leq C_2,
	\qquad
	\sup_{l,l'\leq N}\sup_{h\in\mathbb{Z}}
	|\gamma_{ll'}(h)|\leq C_2 ,
	\]
	where the second bound follows immediately from the first.
	Hence the product terms are uniformly absolutely summable as
	\[
	\sum_{h=-\infty}^{\infty}
	|\gamma_{ll}(h)\gamma_{l'l'}(h)|
	\leq
	\sup_{h\in\mathbb{Z}}|\gamma_{l'l'}(h)|
	\sum_{h=-\infty}^{\infty}|\gamma_{ll}(h)|
	\leq C_2^2 .
	\]
	The same argument applies to
	$\gamma_{ll'}(h)\gamma_{l'l}(h)$. Thus, setting $C_\Gamma=C_4+2C_2^2$, which does not
	depend on $N$,
	\[
  \sup_{l,l'\leq N}\sup_t
	\sum_{h=-\infty}^{\infty}
	|\Gamma_{ll',t}(h)|
	\leq C_\Gamma .
	\]
Consequently, for fixed $t$, the change of variables $h=t-s$ maps
$s=1,\ldots,T$ to \mbox{$h=t-T,\ldots,t-1$.} Hence,
\[
\begin{aligned}
	\EE{\Delta_{ll'}^2}
	&=
	\frac{1}{T^2}
	\sum_{t=1}^{T}
	\sum_{h=t-T}^{t-1}
	\Gamma_{ll',t}(h) 
	\leq
	\frac{1}{T^2}
	\sum_{t=1}^{T}
	\sum_{h=t-T}^{t-1}
	\left|\Gamma_{ll',t}(h)\right| \\
	&\leq
	\frac{1}{T^2}
	\sum_{t=1}^{T}
	\sum_{h=-\infty}^{\infty}
	\left|\Gamma_{ll',t}(h)\right| 
	\leq
	\frac{1}{T^2}
	\sum_{t=1}^{T} C_\Gamma \\
	&=
	\frac{C_\Gamma}{T}
	=
	O(T^{-1}),
\end{aligned}
\]
uniformly in $N$ and $l,l'\leq N$.
	
	Since
	\[
	\left\|
	\widetilde{\Sigma}_u-\Sigma_{u,0}
	\right\|_F^2
	=
	\sum_{l=1}^{N}\sum_{l'=1}^{N}\Delta_{ll'}^2,
	\]
	we obtain
	\[
	\EE{
		\left\|
		\widetilde{\Sigma}_u-\Sigma_{u,0}
		\right\|_F^2
	}
	=
	\sum_{l=1}^{N}\sum_{l'=1}^{N}
	\EE{\Delta_{ll'}^2}
	\leq \frac{N^2C_\Gamma}{T}
	=
	O\left(\frac{N^2}{T}\right).
	\]
	By Markov's inequality, this second-moment bound implies
	\[
	\left\|\widetilde{\Sigma}_u-\Sigma_{u,0}\right\|_F^2
	= O_p\left(\frac{N^2}{T}\right),
	\]
	and hence
	\[
	\left\|\widetilde{\Sigma}_u-\Sigma_{u,0}\right\|_F
	= O_p\left(\frac{N}{\sqrt{T}}\right).
	\]
\end{proof}

\begin{lemma}[Empirical covariances of $z_t$ and $u_t$]
	\label{L:sum_z_u}
	Suppose Assumptions~\ref{ass:factors},
	\ref{ass:u_stationary},
	\ref{ass:u_autocov_summability},
	\ref{ass:one_indep_all}, and \ref{ass:memory} hold.
	Then, uniformly over
	$i=1,\ldots,s_0^*$ and $l=1,\ldots,N$,
	\begin{align}
		\left|
		\frac{1}{T^{d_{0,i}+1/2}}
		\sum_{t=1}^{T}z_{i,t}u_{l,t}
		\right|
		&=
		O_p(T^{-1/2}),
		\label{eq:sum_z_u_scaled}\\
		\left|
		\frac{1}{T}
		\sum_{t=1}^{T}z_{i,t}u_{l,t}
		\right|
		&=
		O_p(T^{d_{0,i}-1}).
		\label{eq:sum_z_u}
	\end{align}
\end{lemma}
\begin{proof}
	Fix $i\in\{1,\ldots,s_0^*\}$ and
	$l\in\{1,\ldots,N\}$. By the type-II representation and
	\eqref{eq:combined_filter_power_law},
	\[
	|b_{i,r}(d_{0,i})|
	\leq
	C(1+r)^{d_{0,i}-1},
	\qquad r\geq0.
	\]
	Since the martingale-difference innovations are uncorrelated over
	time,
	\begin{align}
		\EE{z_{i,t}^2}
		&=
		\EE{\eta_{i,t}^2}
		\sum_{r=0}^{t-1}
		b_{i,r}(d_{0,i})^2
\leq
		C
		\sum_{r=0}^{t-1}
		(1+r)^{2d_{0,i}-2}
		=
		O\left(t^{2d_{0,i}-1}\right).
		\label{eq:nonstationary_factor_second_moment_bound}
	\end{align}
	
	By Assumptions~\ref{ass:u_stationary} and
\ref{ass:one_indep_all},
\begin{align}
	\EE{
		\left(
		\sum_{t=1}^{T}z_{i,t}u_{l,t}
		\right)^2
	}
	&=
	\sum_{t=1}^{T}
	\sum_{s=1}^{T}
	\EE{z_{i,t}z_{i,s}}
	\gamma_{ll}(t-s),
	\label{eq:sum_z_u_second_moment}
\end{align}
where
\[
\gamma_{ll}(h)
=
\Cov(u_{l,t},u_{l,t-h}).
\]
By Cauchy-Schwarz and
\eqref{eq:nonstationary_factor_second_moment_bound},
\begin{align}
	\left|
	\EE{z_{i,t}z_{i,s}}
	\right|
	&\leq
	\left(
	\EE{z_{i,t}^2}
	\EE{z_{i,s}^2}
	\right)^{1/2}\leq
	C(ts)^{d_{0,i}-1/2}
	\leq
	CT^{2d_{0,i}-1},
	\qquad t,s\leq T.
	\label{eq:nonstationary_factor_cross_moment_bound}
\end{align}
Substituting this bound into \eqref{eq:sum_z_u_second_moment} and using Assumption~\ref{ass:u_autocov_summability} gives
\begin{align}
	\EE{
		\left(
		\sum_{t=1}^{T}z_{i,t}u_{l,t}
		\right)^2
	}
	&\leq
	CT^{2d_{0,i}-1}
	\sum_{t=1}^{T}
	\sum_{s=1}^{T}
	|\gamma_{ll}(t-s)|
	\nonumber\\
	&\leq
	CT^{2d_{0,i}-1}
	\sum_{t=1}^{T}
	\sum_{h=-\infty}^{\infty}
	|\gamma_{ll}(h)|
	\nonumber\\
	&\leq
	CT^{2d_{0,i}}
	\sum_{h=-\infty}^{\infty}
	|\gamma_{ll}(h)|
	=
	O\left(T^{2d_{0,i}}\right).
	\label{eq:sum_z_u_variance_bound}
\end{align}

Since
\[
\EE{
	\sum_{t=1}^{T}z_{i,t}u_{l,t}
}
=
0,
\]
Chebyshev's inequality and
\eqref{eq:sum_z_u_variance_bound} imply
\begin{align}
	\sum_{t=1}^{T}z_{i,t}u_{l,t}
	&=
	O_p\left(T^{d_{0,i}}\right).
	\label{eq:sum_z_u_unscaled_rate}
\end{align}
	Dividing \eqref{eq:sum_z_u_unscaled_rate} by
	$T^{d_{0,i}+1/2}$ and by $T$, respectively, proves the two stated
	bounds.
	
	The constants are uniform over $l$ by
	Assumption~\ref{ass:u_autocov_summability} and over the fixed set
	$i=1,\ldots,s_0^*$ by the uniform conditions in
	Assumption~\ref{ass:factors}.
\end{proof}

\begin{lemma}[Covariance and partial-sum bounds for stationary factors]
	\label{L:stationary_factor_dependence}
	Suppose Assumptions~\ref{ass:factors} and
	\ref{ass:memory} hold. Then, for
	$i=s_0^*+1,\ldots,s_0$,
	\begin{align}
		\sum_{t=1}^{T}
		\sum_{s=1}^{T}
		\left|
		\EE{x_{i,t}x_{i,s}}
		\right|
		&=
		O\left(
		T^{2\max\{d_{0,i},0\}+1}
		\right).
		\label{eq:stationary_type_II_double_covariance_sum}
	\end{align}
	Consequently,
	\begin{align}
		\Var\left(
		\sum_{t=1}^{T}x_{i,t}
		\right)
		&=
		O\left(
		T^{2\max\{d_{0,i},0\}+1}
		\right),
		\label{eq:stationary_type_II_partial_sum_variance}\\
		\sum_{t=1}^{T}x_{i,t}
		&=
		O_p\left(
		T^{\max\{d_{0,i},0\}+1/2}
		\right).
		\label{eq:stationary_type_II_partial_sum_rate}
	\end{align}
\end{lemma}
\begin{proof}
	Fix $i\in\{s_0^*+1,\ldots,s_0\}$. For $t\geq s$, the type-II
	representation and temporal uncorrelatedness of the innovations give
	\begin{align*}
		\EE{x_{i,t}x_{i,s}}
		&=
		\EE{\eta_{i,t}^2}
		\sum_{\ell=0}^{s-1}
		b_{i,\ell}(d_{0,i})
		b_{i,\ell+t-s}(d_{0,i}).
	\end{align*}
	Therefore,
	\begin{align}
		\left|
		\EE{x_{i,t}x_{i,s}}
		\right|
		&\leq
		\EE{\eta_{i,t}^2}
		\sum_{\ell=0}^{\infty}
		\left|
		b_{i,\ell}(d_{0,i})
		b_{i,\ell+t-s}(d_{0,i})
		\right|.
		\label{eq:stationary_type_II_covariance_bound}
	\end{align}
	
	Suppose first that $d_{0,i}\leq0$. If $d_{0,i}<0$, then
	\eqref{eq:stationary_filter_coefficient_bound} implies
	\[
	\sum_{\ell=0}^{\infty}
	\left|
	b_{i,\ell}(d_{0,i})
	\right|
	<
	\infty,
	\]
	because $d_{0,i}-1<-1$. If $d_{0,i}=0$, then
	$b_{i,\ell}(0)=\psi_{i,\ell}$, and the same conclusion follows from
	Assumption~\ref{ass:short_run_filters}.
	
	Using symmetry in $t$ and $s$ and collecting the pairs
	$(t,s)$ according to the lag $h=|t-s|$, we obtain
	\begin{align}
		\sum_{t=1}^{T}
		\sum_{s=1}^{T}
		\left|
		\EE{x_{i,t}x_{i,s}}
		\right|
		&\leq
		\EE{\eta_{i,t}^2}
		\left[
		T
		\sum_{\ell=0}^{\infty}
		\left|
		b_{i,\ell}(d_{0,i})
		\right|^2
		\right.
		\nonumber\\
		&\hspace{25mm}\left.
		+
		2
		\sum_{h=1}^{T-1}
		(T-h)
		\sum_{\ell=0}^{\infty}
		\left|
		b_{i,\ell}(d_{0,i})
		b_{i,\ell+h}(d_{0,i})
		\right|
		\right]
		\nonumber\\
		&\leq
		T\EE{\eta_{i,t}^2}
		\left[
		\sum_{\ell=0}^{\infty}
		\left|
		b_{i,\ell}(d_{0,i})
		\right|^2
		\right.
		\nonumber\\
		&\hspace{25mm}\left.
		+
		2
		\sum_{h=1}^{\infty}
		\sum_{\ell=0}^{\infty}
		\left|
		b_{i,\ell}(d_{0,i})
		b_{i,\ell+h}(d_{0,i})
		\right|
		\right]
		\nonumber\\
		&=
		T\EE{\eta_{i,t}^2}
		\left(
		\sum_{\ell=0}^{\infty}
		\left|
		b_{i,\ell}(d_{0,i})
		\right|
		\right)^2
		=
		O(T).
		\label{eq:stationary_type_II_double_covariance_nonpositive}
	\end{align}
	
	Now suppose that $0<d_{0,i}<1/2$. By
	\eqref{eq:stationary_filter_coefficient_bound}, for $h\geq0$,
	\begin{align}
		\sum_{\ell=0}^{\infty}
		\left|
		b_{i,\ell}(d_{0,i})
		b_{i,\ell+h}(d_{0,i})
		\right|
		&\leq
		C
		\sum_{\ell=0}^{\infty}
		(1+\ell)^{d_{0,i}-1}
		(1+\ell+h)^{d_{0,i}-1}
		\nonumber\\
		&\leq
		C(1+h)^{d_{0,i}-1}
		\sum_{\ell=0}^{h}
		(1+\ell)^{d_{0,i}-1} +
		C
		\sum_{\ell=h+1}^{\infty}
		(1+\ell)^{2d_{0,i}-2}
		\nonumber\\
		&=
		O\left(
		(1+h)^{2d_{0,i}-1}
		\right).
		\label{eq:stationary_filter_coefficient_convolution}
	\end{align}
	Here, the first term has the stated order because
	\[
	\sum_{\ell=0}^{h}
	(1+\ell)^{d_{0,i}-1}
	=
	O\left((1+h)^{d_{0,i}}\right).
	\]
	For the second term, since $2d_{0,i}-2<-1$, the integral
	comparison gives
	\[
	\sum_{\ell=h+1}^{\infty}
	(1+\ell)^{2d_{0,i}-2}
	\leq
	\int_h^\infty
	(1+x)^{2d_{0,i}-2}\,dx
	=
	\frac{
		(1+h)^{2d_{0,i}-1}
	}{
		1-2d_{0,i}
	}.
	\]
	Thus, both terms are of order
	$O\bigl((1+h)^{2d_{0,i}-1}\bigr)$.
	
Using the same decomposition into diagonal and off-diagonal lags
as in the derivation of
\eqref{eq:stationary_type_II_double_covariance_nonpositive} gives
\begin{align}
	\sum_{t=1}^{T}
	\sum_{s=1}^{T}
	\left|
	\EE{x_{i,t}x_{i,s}}
	\right|
	&\leq
	\EE{\eta_{i,t}^2}
	\left[
	T
	\sum_{\ell=0}^{\infty}
	b_{i,\ell}(d_{0,i})^2
	\right.
	\nonumber\\
	&\hspace{25mm}\left.
	+
	2
	\sum_{h=1}^{T-1}
	(T-h)
	\sum_{\ell=0}^{\infty}
	\left|
	b_{i,\ell}(d_{0,i})
	b_{i,\ell+h}(d_{0,i})
	\right|
	\right]
	\nonumber\\
	&\leq
	CT
	\sum_{h=0}^{T-1}
	(1+h)^{2d_{0,i}-1} =
	O\left(
	T^{2d_{0,i}+1}
	\right).
	\label{eq:stationary_type_II_double_covariance_positive}
\end{align}
Here, the second inequality follows from
\eqref{eq:stationary_filter_coefficient_convolution} and
$T-h\leq T$.

Combining
\eqref{eq:stationary_type_II_double_covariance_nonpositive}
and
\eqref{eq:stationary_type_II_double_covariance_positive}
proves
\eqref{eq:stationary_type_II_double_covariance_sum}.

Since $x_{i,t}$ has mean zero,
\begin{align*}
	\Var\left(
	\sum_{t=1}^{T}x_{i,t}
	\right)
	&=
	\sum_{t=1}^{T}
	\sum_{s=1}^{T}
	\EE{x_{i,t}x_{i,s}}
\leq
	\sum_{t=1}^{T}
	\sum_{s=1}^{T}
	\left|
	\EE{x_{i,t}x_{i,s}}
	\right|.
\end{align*}
Thus,
\eqref{eq:stationary_type_II_partial_sum_variance}
follows from
\eqref{eq:stationary_type_II_double_covariance_sum},
and Chebyshev's inequality yields
\eqref{eq:stationary_type_II_partial_sum_rate}.
\end{proof}


\begin{lemma}[Empirical covariances of $z_t$ and $w_t$]
	\label{L:zw}
	Suppose Assumptions~\ref{ass:factors},
	\ref{ass:one_indep_all}, and \ref{ass:memory} hold.
	Then, for
	$i=1,\ldots,s_0^*$ and
	$i'=s_0^*+1,\ldots,s_0$,
	\begin{align}
		\frac{1}{T^{d_{0,i}+1/2}}
		\sum_{t=1}^{T}
		z_{i,t}w_{i',t}
		&=
		O_p\left(
		T^{\max\{d_{0,i'},0\}-1/2}
		\right),
		\label{eq:zw_scaled}\\
		\frac{1}{T}
		\sum_{t=1}^{T}
		z_{i,t}w_{i',t}
		&=
		O_p\left(
		T^{d_{0,i}+\max\{d_{0,i'},0\}-1}
		\right).
		\label{eq:zw}
	\end{align}
\end{lemma}

\begin{proof}
	Fix $i\in\{1,\ldots,s_0^*\}$ and
	$i'\in\{s_0^*+1,\ldots,s_0\}$. By
	Assumption~\ref{ass:one_indep_all},
	\eqref{eq:nonstationary_factor_cross_moment_bound}, and
	\eqref{eq:stationary_type_II_double_covariance_sum},
	\begin{align*}
		\EE{
			\left|
			\sum_{t=1}^{T}
			z_{i,t}w_{i',t}
			\right|^2
		}
		&=
		\sum_{t=1}^{T}
		\sum_{s=1}^{T}
		\EE{z_{i,t}z_{i,s}}
		\EE{w_{i',t}w_{i',s}}
\leq
		CT^{2d_{0,i}-1}
		\sum_{t=1}^{T}
		\sum_{s=1}^{T}
		\left|
		\EE{w_{i',t}w_{i',s}}
		\right|
		\nonumber\\
		&=
		O\left(
		T^{2d_{0,i}
			+2\max\{d_{0,i'},0\}}
		\right).
	\end{align*}
	Markov's inequality therefore yields
	\[
	\sum_{t=1}^{T}z_{i,t}w_{i',t}
	=
	O_p\left(
	T^{d_{0,i}+\max\{d_{0,i'},0\}}
	\right).
	\]
	Dividing by $T^{d_{0,i}+1/2}$ and by $T$, respectively, proves
	\eqref{eq:zw_scaled} and \eqref{eq:zw}.
\end{proof}

\begin{lemma}[Empirical covariances of $w_t$ and $u_t$]
	\label{L:wu}
	Suppose Assumptions~\ref{ass:factors},
	\ref{ass:u_stationary},
	\ref{ass:u_autocov_summability},
	\ref{ass:one_indep_all}, and \ref{ass:memory} hold.
	Then, uniformly over $i=s_0^*+1,\ldots,s_0$ and $l=1,\ldots,N$,
	\begin{align}
		\frac{1}{T}
		\sum_{t=1}^{T}
		w_{i,t}u_{l,t}
		&=
		O_p\left(T^{-1/2}\right).
		\label{eq:wu}
	\end{align}
\end{lemma}

\begin{proof}
	Fix $i\in\{s_0^*+1,\ldots,s_0\}$ and
	$l\in\{1,\ldots,N\}$. By
	\eqref{eq:stationary_filter_square_summability},
	\[
	\sup_{t\geq1}\EE{w_{i,t}^2}
	\leq
	\EE{\eta_{i,t}^2}
	\sum_{r=0}^{\infty}
	b_{i,r}(d_{0,i})^2
	<
	\infty.
	\]
	Let
	\[
	\gamma_{ll}(h)
	=
	\Cov(u_{l,t},u_{l,t-h}).
	\]
	By Assumptions~\ref{ass:u_stationary} and
	\ref{ass:one_indep_all}, the summands have mean zero and
	\begin{align*}
		\EE{
			\left|
			\frac{1}{T}
			\sum_{t=1}^{T}
			w_{i,t}u_{l,t}
			\right|^2
		}
		&=
		\frac{1}{T^2}
		\sum_{t=1}^{T}
		\sum_{s=1}^{T}
		\EE{w_{i,t}w_{i,s}}
		\gamma_{ll}(t-s)
		\\
		&\leq
		\frac{C}{T^2}
		\sum_{t=1}^{T}
		\sum_{s=1}^{T}
		|\gamma_{ll}(t-s)|
\leq
		\frac{C}{T}
		\sum_{h=-\infty}^{\infty}
		|\gamma_{ll}(h)|
		=
		O(T^{-1}),
	\end{align*}
	where Cauchy-Schwarz was used in the first inequality and
	Assumption~\ref{ass:u_autocov_summability} in the last step.
	Chebyshev's inequality proves \eqref{eq:wu}.
\end{proof}

\subsubsection{Relative covariance perturbation bounds}
\label{sec:perturbation}
The preceding covariance bounds provide the rates needed to compare
the empirical covariance matrix with the artificial covariance
matrices defined in Appendix~\ref{sec:notation}. We now establish the
corresponding relative perturbation bounds.

\begin{lemma}[Relative covariance perturbation]
	\label{L:Sigma_y_hat_tilde_diff_N}
	Suppose Assumptions~\ref{ass:factors}-\ref{ass:memory} hold and
	that $N,T \to \infty$ with $N/T\to0$. Then
	\begin{align}
		\norm{
			\widetilde{\Sigma}_{\tilde y,T}^{-1/2}
			\Delta\Sigma_{y,T}
			\widehat{\Sigma}_{y,T}^{-1/2}
		}_F
		&=
		o_p(1),
		\label{eq:relative_covariance_perturbation}
	\end{align}
	where
	\[
	\Delta\Sigma_{y,T}
	=
	\widehat{\Sigma}_{y,T}
	-
	\widetilde{\Sigma}_{\tilde y,T}.
	\]
\end{lemma}

\begin{proof}
	Without loss of generality, suppose that both the nonstationary and
	stationary factor blocks are nonempty. If either block is absent,
	the argument below applies after deleting the corresponding rows,
	columns, and terms.
	
	The proof proceeds by normalizing the observed and artificial
	covariance matrices so that their factor blocks have nondegenerate
	orders. We first derive blockwise perturbation bounds and establish the
	required invertibility of the normalized covariance matrices. These
	results yield a relative perturbation bound. Orthogonality of
	$Q_{all,0}$ and cyclicity of the trace then reduce the original claim
	to this normalized bound.
	
	Let
	\[
	\hat A
	=
	N^{-1}D_{T,0}^{-1}
	Q_{all,0}^T
	\widehat{\Sigma}_{y,T}
	Q_{all,0}
	D_{T,0}^{-1},
	\]
	and
	\[
	\tilde A
	=
	N^{-1}D_{T,0}^{-1}
	Q_{all,0}^T
	\widetilde{\Sigma}_{\tilde y,T}
	Q_{all,0}
	D_{T,0}^{-1}
	=
	N^{-1}D_{T,0}^{-1}
	\widetilde{\Sigma}_{\widetilde{Qy},T}
	D_{T,0}^{-1}.
	\]
	Set
	\[
	\Delta A
	=
	\hat A-\tilde A.
	\]
	By the definitions of $\hat A$ and $\tilde A$,
	\[
	\widehat{\Sigma}_{y,T}
	=
	NQ_{all,0}D_{T,0}
	\hat A
	D_{T,0}Q_{all,0}^T,
	\]
	\[
	\widetilde{\Sigma}_{\tilde y,T}
	=
	NQ_{all,0}D_{T,0}
	\tilde A
	D_{T,0}Q_{all,0}^T,
	\]
	and therefore
	\begin{align}
		\Delta\Sigma_{y,T}
		&=
		NQ_{all,0}D_{T,0}
		\Delta A
		D_{T,0}Q_{all,0}^T.
		\label{eq:Delta_Sigma_y_Delta_A}
	\end{align}
	
	Using \eqref{eq:Q_all_y},
	\[
	Q_{all,0}^Ty_t
	=
	\begin{pmatrix}
		R_{zz,0}z_t+R_{zw,0}w_t+Q_{z,0}^Tu_t\\
		R_{ww,0}w_t+Q_{w,0}^Tu_t\\
		Q_{\perp,0}^Tu_t
	\end{pmatrix},
	\]
	we obtain
	\[
	\hat A
	=
	\begin{pmatrix}
		\hat A_{zz}
		&
		\hat A_{zw}^T
		&
		\hat A_{z\perp}^T\\
		\hat A_{zw}
		&
		\hat A_{ww}
		&
		\hat A_{w\perp}^T\\
		\hat A_{z\perp}
		&
		\hat A_{w\perp}
		&
		\hat A_{\perp\perp}
	\end{pmatrix},
	\]
	where
	\[
	\begin{aligned}
		\hat A_{zz}
		&=
		\frac{1}{NT}
		\sum_{t=1}^{T}
		D_{z,T,0}^{-1}
		\left(
		R_{zz,0}z_t
		+
		R_{zw,0}w_t
		+
		Q_{z,0}^Tu_t
		\right)\\
		&\qquad\qquad\times
		\left(
		R_{zz,0}z_t
		+
		R_{zw,0}w_t
		+
		Q_{z,0}^Tu_t
		\right)^T
		D_{z,T,0}^{-1},\\
		\hat A_{zw}
		&=
		\frac{1}{NT}
		\sum_{t=1}^{T}
		\left(
		R_{ww,0}w_t
		+
		Q_{w,0}^Tu_t
		\right)\\
		&\qquad\qquad\times
		\left(
		R_{zz,0}z_t
		+
		R_{zw,0}w_t
		+
		Q_{z,0}^Tu_t
		\right)^T
		D_{z,T,0}^{-1},\\
		\hat A_{z\perp}
		&=
		\frac{1}{NT}
		\sum_{t=1}^{T}
		Q_{\perp,0}^Tu_t
		\left(
		R_{zz,0}z_t
		+
		R_{zw,0}w_t
		+
		Q_{z,0}^Tu_t
		\right)^T
		D_{z,T,0}^{-1},\\
		\hat A_{ww}
		&=
		\frac{1}{NT}
		\sum_{t=1}^{T}
		\left(
		R_{ww,0}w_t
		+
		Q_{w,0}^Tu_t
		\right)
		\left(
		R_{ww,0}w_t
		+
		Q_{w,0}^Tu_t
		\right)^T,\\
		\hat A_{w\perp}
		&=
		\frac{1}{NT}
		\sum_{t=1}^{T}
		Q_{\perp,0}^Tu_t
		\left(
		R_{ww,0}w_t
		+
		Q_{w,0}^Tu_t
		\right)^T,\\
		\hat A_{\perp\perp}
		&=
		\frac{1}{NT}
		\sum_{t=1}^{T}
		Q_{\perp,0}^T
		u_tu_t^T
		Q_{\perp,0}.
	\end{aligned}
	\]
	
	The artificial matrix is
	\[
	\tilde A
	=
	\begin{pmatrix}
		\tilde A_{zz}
		&
		0
		&
		0\\
		0
		&
		\tilde A_{ww}
		&
		\tilde A_{w\perp}^T\\
		0
		&
		\tilde A_{w\perp}
		&
		\tilde A_{\perp\perp}
	\end{pmatrix},
	\]
	where
	\[
	\begin{aligned}
		\tilde A_{zz}
		&=
		\frac{1}{N}
		D_{z,T,0}^{-1}
		R_{zz,0}
		\widetilde{\Sigma}_z
		R_{zz,0}^T
		D_{z,T,0}^{-1},\\
		\tilde A_{ww}
		&=
		\frac{1}{N}
		R_{ww,0}
		\widetilde{\Sigma}_w
		R_{ww,0}^T
		+
		\frac{1}{N}
		Q_{w,0}^T
		\widetilde{\Sigma}_u
		Q_{w,0},\\
		\tilde A_{w\perp}
		&=
		\frac{1}{N}
		Q_{\perp,0}^T
		\widetilde{\Sigma}_u
		Q_{w,0},\\
		\tilde A_{\perp\perp}
		&=
		\frac{1}{N}
		Q_{\perp,0}^T
		\widetilde{\Sigma}_u
		Q_{\perp,0}.
	\end{aligned}
	\]
	Thus,
	\[
	\Delta A
	=
	\begin{pmatrix}
		\Delta A_{zz}
		&
		\Delta A_{zw}^T
		&
		\Delta A_{z\perp}^T\\
		\Delta A_{zw}
		&
		\Delta A_{ww}
		&
		\Delta A_{w\perp}^T\\
		\Delta A_{z\perp}
		&
		\Delta A_{w\perp}
		&
		0
	\end{pmatrix}.
	\]
	The $\perp\perp$-block is zero because the artificial covariance
	matrix retains the empirical idiosyncratic covariance in this
	block.
	
	We first collect a few bounds that are used repeatedly. By
	Assumption~\ref{ass:strong_factors},
	\begin{align}
		\norm{R_{zz,0}}_F
		&=
		O(\sqrt N),
		&
		\norm{R_{zw,0}}_F
		&=
		O(\sqrt N),
		&
		\norm{R_{ww,0}}_F
		&=
		O(\sqrt N).
		\label{eq:loading_block_bounds}
	\end{align}
	Because the factors are ordered by decreasing memory and both
$R_{zz,0}$ and $R_{zz,0}^{-1}$ are upper triangular, the diagonal
rescaling $D_{z,T,0}^{-1}(\,\cdot\,)D_{z,T,0}$ does not increase
the magnitude of any nonzero entry of either matrix. Indeed, a
nonzero $(i,j)$th entry can occur only if $i\leq j$, in which case
$d_{0,i}\geq d_{0,j}$ and $T^{d_{0,j}-d_{0,i}}\leq1$. Since the factor dimension is fixed, these entrywise bounds,
Assumption~\ref{ass:strong_factors}, and the block-upper-triangular
QR structure give
		\begin{align}
			\norm{
				N^{-1/2}
				D_{z,T,0}^{-1}
				R_{zz,0}
				D_{z,T,0}
			}_2
			+
			\norm{
				\left(
				N^{-1/2}
				D_{z,T,0}^{-1}
				R_{zz,0}
				D_{z,T,0}
				\right)^{-1}
			}_2
			&=
			O(1),
			\label{eq:scaled_Rzz_bounds}\\
			\norm{
				N^{-1/2}
				D_{z,T,0}^{-1}
				R_{zw,0}
			}_F
			&=
			O\left(
			T^{-d_{0,s_0^*}+1/2}
			\right).
			\label{eq:scaled_Rzw_bound} \\
					\norm{
				N^{-1/2}R_{ww,0}
			}_2
			+
			\norm{
				\left(
				N^{-1/2}R_{ww,0}
				\right)^{-1}
			}_2
			&=
			O(1).
			\label{eq:scaled_Rww_bounds}
		\end{align}
	
	Assumption~\ref{ass:Eigenvalue_Sigma_u_tilde} further implies
	\begin{align}
		\norm{\widetilde{\Sigma}_u}_2
		&=
		O_p(1).
		\label{eq:empirical_idiosyncratic_spectral_bound}
	\end{align}
	Because $Q_{z,0}$ and $Q_{w,0}$ have fixed numbers of orthonormal
columns, \eqref{eq:empirical_idiosyncratic_spectral_bound} also
controls the Frobenius norms of projected idiosyncratic blocks
having $Q_{z,0}$ or $Q_{w,0}$ on at least one side. For example,
\[
\norm{
	Q_{\perp,0}^T
	\widetilde{\Sigma}_u
	Q_{z,0}
}_F
\leq
\sqrt{s_0^*}\,
\norm{\widetilde{\Sigma}_u}_2
=
O_p(1),
\]
and the analogous bound holds with $Q_{w,0}$ in place of
$Q_{z,0}$.

Applying Lemma~\ref{L:zw} entrywise and summing over the fixed
number of entries gives
	\begin{align}
		\norm{
			D_{z,T,0}^{-1}
			\frac{1}{T}
			\sum_{t=1}^{T}
			z_tw_t^T
		}_F
		&=
		O_p\left(
		T^{\max\{d_{0,s_0^*+1},0\}-1/2}
		\right).
		\label{eq:zw_matrix_rate}
	\end{align}
	
	For the $zu$ terms, the second-moment bound in the proof of
	Lemma~\ref{L:sum_z_u} yields
	\[
	\EE{
		\left|
		\sum_{t=1}^{T}
		z_{i,t}u_{l,t}
		\right|^2
	}
	=
	O\left(T^{2d_{0,i}}\right)
	\]
	uniformly over $i=1,\ldots,s_0^*$ and $l=1,\ldots,N$.
	Consequently,
	\begin{align*}
		&\EE{
			\norm{
				D_{z,T,0}^{-1}
				\frac{1}{T}
				\sum_{t=1}^{T}
				z_tu_t^T
			}_F^2
		}
=
		\sum_{i=1}^{s_0^*}
		\sum_{l=1}^{N}
		T^{-2d_{0,i}+1}
		\frac{1}{T^2}
		\EE{
			\left|
			\sum_{t=1}^{T}
			z_{i,t}u_{l,t}
			\right|^2
		}
=
		O\left(\frac{N}{T}\right).
	\end{align*}
	Hence, by Markov's inequality,
	\begin{align}
		\norm{
			D_{z,T,0}^{-1}
			\frac{1}{T}
			\sum_{t=1}^{T}
			z_tu_t^T
		}_F
		&=
		O_p\left(
		\sqrt{\frac{N}{T}}
		\right).
		\label{eq:zu_matrix_rate}
	\end{align}
	
	Similarly, the second-moment bound in the proof of
	Lemma~\ref{L:wu} gives
	\[
	\EE{
		\left|
		\frac{1}{T}
		\sum_{t=1}^{T}
		w_{i,t}u_{l,t}
		\right|^2
	}
	=
	O(T^{-1})
	\]
	uniformly over $i=s_0^*+1,\ldots,s_0$ and $l=1,\ldots,N$.
	Therefore,
	\begin{align}
		\EE{
			\norm{
				\frac{1}{T}
				\sum_{t=1}^{T}
				w_tu_t^T
			}_F^2
		}
		&=
		O\left(\frac{N}{T}\right),
		\nonumber\\
		\norm{
			\frac{1}{T}
			\sum_{t=1}^{T}
			w_tu_t^T
		}_F
		&=
		O_p\left(
		\sqrt{\frac{N}{T}}
		\right).
		\label{eq:wu_matrix_rate}
	\end{align}
	
	We now bound the blocks of $\Delta A$. The order below each
	underbrace refers to the Frobenius norm of the corresponding term.
	
	For the $zz$-block, $\tilde A_{zz}$ contains the leading $zz$
	term. Hence,
	\[
	\begin{aligned}
		\Delta A_{zz}
		&=
		\underbrace{
			\frac{1}{NT}
			\sum_{t=1}^{T}
			D_{z,T,0}^{-1}
			R_{zw,0}w_tw_t^TR_{zw,0}^T
			D_{z,T,0}^{-1}
		}_{
			O_p\left(
			T^{-2d_{0,s_0^*}+1}
			\right)
			\text{ by
				Lemma~\ref{L:factor_second_moment_limits}
				and \eqref{eq:scaled_Rzw_bound}}
			}
		\\
		&\quad+
		\underbrace{
			\frac{1}{N}
			D_{z,T,0}^{-1}
			Q_{z,0}^T
			\widetilde{\Sigma}_u
			Q_{z,0}
			D_{z,T,0}^{-1}
		}_{
			O_p\left(
			N^{-1}T^{-2d_{0,s_0^*}+1}
			\right)
\text{ by
	\eqref{eq:empirical_idiosyncratic_spectral_bound}}
		}
		\\
		&\quad+
		\underbrace{
			\frac{1}{NT}
			\sum_{t=1}^{T}
			D_{z,T,0}^{-1}
			R_{zz,0}z_tw_t^TR_{zw,0}^T
			D_{z,T,0}^{-1}
		}_{
			O_p\left(
			T^{\max\{d_{0,s_0^*+1},0\}
				-d_{0,s_0^*}}
			\right)
		\text{ by
				\eqref{eq:scaled_Rzz_bounds}, 
					\eqref{eq:scaled_Rzw_bound}, and \eqref{eq:zw_matrix_rate}}
		}
		\\
		&\quad+
		\underbrace{
			\frac{1}{NT}
			\sum_{t=1}^{T}
			D_{z,T,0}^{-1}
			R_{zw,0}w_tz_t^TR_{zz,0}^T
			D_{z,T,0}^{-1}
		}_{
			O_p\left(
			T^{\max\{d_{0,s_0^*+1},0\}
				-d_{0,s_0^*}}
			\right)
\text{ by
	\eqref{eq:scaled_Rzz_bounds},
	\eqref{eq:scaled_Rzw_bound}, and \eqref{eq:zw_matrix_rate}}
		}
		\\
		&\quad+
		\underbrace{
			\frac{1}{NT}
			\sum_{t=1}^{T}
			D_{z,T,0}^{-1}
			R_{zz,0}z_tu_t^TQ_{z,0}
			D_{z,T,0}^{-1}
		}_{
			O_p\left(
			T^{-d_{0,s_0^*}}
			\right)
\text{ by
	\eqref{eq:scaled_Rzz_bounds}
	and \eqref{eq:zu_matrix_rate}}
		}
		\\
		&\quad+
		\underbrace{
			\frac{1}{NT}
			\sum_{t=1}^{T}
			D_{z,T,0}^{-1}
			Q_{z,0}^Tu_tz_t^TR_{zz,0}^T
			D_{z,T,0}^{-1}
		}_{
			O_p\left(
			T^{-d_{0,s_0^*}}
			\right)
\text{ by
	\eqref{eq:scaled_Rzz_bounds}
	and \eqref{eq:zu_matrix_rate}}
		}
		\\
		&\quad+
		\underbrace{
			\frac{1}{NT}
			\sum_{t=1}^{T}
			D_{z,T,0}^{-1}
			R_{zw,0}w_tu_t^TQ_{z,0}
			D_{z,T,0}^{-1}
		}_{
			O_p\left(
			T^{1/2-2d_{0,s_0^*}}
			\right)
\text{ by
	\eqref{eq:scaled_Rzw_bound}
	and \eqref{eq:wu_matrix_rate}}
}
		\\
		&\quad+
		\underbrace{
			\frac{1}{NT}
			\sum_{t=1}^{T}
			D_{z,T,0}^{-1}
			Q_{z,0}^Tu_tw_t^TR_{zw,0}^T
			D_{z,T,0}^{-1}
		}_{
			O_p\left(
			T^{1/2-2d_{0,s_0^*}}
			\right)
\text{ by
	\eqref{eq:scaled_Rzw_bound}
	and \eqref{eq:wu_matrix_rate}}
		}.
	\end{aligned}
	\]
	Therefore,
	\begin{align}
		\norm{\Delta A_{zz}}_F
		&\leq
		O_p\left(
		T^{-2d_{0,s_0^*}+1}
		+
		N^{-1}T^{-2d_{0,s_0^*}+1}
		+
		T^{\max\{d_{0,s_0^*+1},0\}
			-d_{0,s_0^*}}
		\right.
		\nonumber\\
		&\hspace{33mm}\left.
		+
		T^{-d_{0,s_0^*}}
		+
		T^{1/2-2d_{0,s_0^*}}
		\right)
		=
		o_p(1),
		\label{eq:Delta_A_zz_rate}
	\end{align}
	because $d_{0,s_0^*}>1/2$ and
	$\max\{d_{0,s_0^*+1},0\}<1/2$.
	
	For the $zw$-block, $\tilde A_{zw}=0$, so that
	$\Delta A_{zw}=\hat A_{zw}$. Thus,
	\[
	\begin{aligned}
		\Delta A_{zw}
		&=
		\underbrace{
			\frac{1}{NT}
			\sum_{t=1}^{T}
			R_{ww,0}w_tz_t^TR_{zz,0}^T
			D_{z,T,0}^{-1}
		}_{
			O_p\left(
			T^{\max\{d_{0,s_0^*+1},0\}-1/2}
			\right)
\text{ by
	\eqref{eq:scaled_Rww_bounds},
	\eqref{eq:scaled_Rzz_bounds}, and
	\eqref{eq:zw_matrix_rate}}
		}
		\\
		&\quad+
		\underbrace{
			\frac{1}{NT}
			\sum_{t=1}^{T}
			Q_{w,0}^Tu_tz_t^TR_{zz,0}^T
			D_{z,T,0}^{-1}
		}_{
			O_p(T^{-1/2})
\text{ by
	\eqref{eq:scaled_Rzz_bounds}
	and \eqref{eq:zu_matrix_rate}}
		}
		\\
		&\quad+
		\underbrace{
			\frac{1}{NT}
			\sum_{t=1}^{T}
			R_{ww,0}w_tw_t^TR_{zw,0}^T
			D_{z,T,0}^{-1}
		}_{
			O_p\left(
			T^{-d_{0,s_0^*}+1/2}
			\right)
\text{ by
	Lemma~\ref{L:factor_second_moment_limits},
	\eqref{eq:scaled_Rww_bounds}, and
	\eqref{eq:scaled_Rzw_bound}}
		}
		\\
		&\quad+
		\underbrace{
			\frac{1}{NT}
			\sum_{t=1}^{T}
			Q_{w,0}^Tu_tw_t^TR_{zw,0}^T
			D_{z,T,0}^{-1}
		}_{
			O_p\left(
			T^{-d_{0,s_0^*}}
			\right)
\text{ by
	\eqref{eq:scaled_Rzw_bound}
	and \eqref{eq:wu_matrix_rate}}
		}
		\\
		&\quad+
		\underbrace{
			\frac{1}{NT}
			\sum_{t=1}^{T}
			R_{ww,0}w_tu_t^TQ_{z,0}
			D_{z,T,0}^{-1}
		}_{
			O_p\left(
			T^{-d_{0,s_0^*}}
			\right)
\text{ by
	\eqref{eq:scaled_Rww_bounds}
	and \eqref{eq:wu_matrix_rate}}
}
\\
		&\quad+
		\underbrace{
			\frac{1}{N}
			Q_{w,0}^T
			\widetilde{\Sigma}_u
			Q_{z,0}
			D_{z,T,0}^{-1}
		}_{
			O_p\left(
			N^{-1}T^{-d_{0,s_0^*}+1/2}
			\right)
\text{ by
	\eqref{eq:empirical_idiosyncratic_spectral_bound}}
}.
	\end{aligned}
	\]
	Hence,
	\begin{align}
		\norm{\Delta A_{zw}}_F
		&\leq
		O_p\left(
		T^{\max\{d_{0,s_0^*+1},0\}-1/2}
		+
		T^{-1/2}
		+
		T^{-d_{0,s_0^*}+1/2}
		\right.
		\nonumber\\
		&\hspace{34mm}\left.
		+
		T^{-d_{0,s_0^*}}
		+
		N^{-1}T^{-d_{0,s_0^*}+1/2}
		\right)
		=
		o_p(1).
		\label{eq:Delta_A_zw_rate}
	\end{align}
	
	For the $ww$-block, the artificial matrix already contains the
	pure $ww$ and $uu$ terms. Therefore,
	\[
	\begin{aligned}
		\Delta A_{ww}
		&=
		\underbrace{
			\frac{1}{N}
			R_{ww,0}
			\left(
			\frac{1}{T}
			\sum_{t=1}^{T}
			w_tu_t^T
			\right)
			Q_{w,0}
		}_{
			O_p(T^{-1/2})
\text{ by
	\eqref{eq:scaled_Rww_bounds}
	and \eqref{eq:wu_matrix_rate}}
		}
		\\
		&\quad+
		\underbrace{
			\frac{1}{N}
			Q_{w,0}^T
			\left(
			\frac{1}{T}
			\sum_{t=1}^{T}
			u_tw_t^T
			\right)
			R_{ww,0}^T
		}_{
			O_p(T^{-1/2})
\text{ by
	\eqref{eq:scaled_Rww_bounds}
	and \eqref{eq:wu_matrix_rate}}
		}.
	\end{aligned}
	\]
	Thus,
	\begin{align}
		\norm{\Delta A_{ww}}_F
		&=
		O_p(T^{-1/2})
		=
		o_p(1).
		\label{eq:Delta_A_ww_rate}
	\end{align}
	
	For the $z\perp$-block, $\tilde A_{z\perp}=0$, and
	\[
	\begin{aligned}
		\Delta A_{z\perp}
		&=
		\underbrace{
			\frac{1}{NT}
			\sum_{t=1}^{T}
			Q_{\perp,0}^Tu_tz_t^TR_{zz,0}^T
			D_{z,T,0}^{-1}
		}_{
			O_p(T^{-1/2})
\text{ by
	\eqref{eq:scaled_Rzz_bounds}
	and \eqref{eq:zu_matrix_rate}}
		}
		\\
		&\quad+
		\underbrace{
			\frac{1}{NT}
			\sum_{t=1}^{T}
			Q_{\perp,0}^Tu_tw_t^TR_{zw,0}^T
			D_{z,T,0}^{-1}
		}_{
			O_p\left(
			T^{-d_{0,s_0^*}}
			\right)
\text{ by
	\eqref{eq:scaled_Rzw_bound}
	and \eqref{eq:wu_matrix_rate}}
}
		\\
		&\quad+
		\underbrace{
			\frac{1}{N}
			Q_{\perp,0}^T
			\widetilde{\Sigma}_u
			Q_{z,0}
			D_{z,T,0}^{-1}
		}_{
			O_p\left(
			N^{-1}T^{-d_{0,s_0^*}+1/2}
			\right)
\text{ by
	\eqref{eq:empirical_idiosyncratic_spectral_bound}}
		}.
	\end{aligned}
	\]
	Consequently,
	\begin{align}
		\norm{\Delta A_{z\perp}}_F
		&\leq
		O_p\left(
		T^{-1/2}
		+
		T^{-d_{0,s_0^*}}
		+
		N^{-1}T^{-d_{0,s_0^*}+1/2}
		\right).
		\label{eq:Delta_A_zperp_rate}
	\end{align}
	
	For the $w\perp$-block, the artificial matrix already contains the
	$uu$ term. Hence,
	\begin{align*}
		\Delta A_{w\perp}
		&=
		\underbrace{
			\frac{1}{NT}
			\sum_{t=1}^{T}
			Q_{\perp,0}^Tu_tw_t^TR_{ww,0}^T
		}_{
			O_p(T^{-1/2})
\text{ by
	\eqref{eq:scaled_Rww_bounds}
	and \eqref{eq:wu_matrix_rate}}
		},
	\end{align*}
	and therefore
	\begin{align}
		\norm{\Delta A_{w\perp}}_F
		&=
		O_p(T^{-1/2}).
		\label{eq:Delta_A_wperp_rate}
	\end{align}
	Finally,
	\[
	\Delta A_{\perp\perp}
	=
	0.
	\]
	
We now establish the conditioning of the factor blocks of the
reference matrix. First, write
\[
\tilde A_{zz}
=
\left(
N^{-1/2}
D_{z,T,0}^{-1}
R_{zz,0}
D_{z,T,0}
\right)
\left(
D_{z,T,0}^{-1}
\widetilde{\Sigma}_z
D_{z,T,0}^{-1}
\right)
\left(
N^{-1/2}
D_{z,T,0}^{-1}
R_{zz,0}
D_{z,T,0}
\right)^T.
\]
By \eqref{eq:scaled_Rzz_bounds}, the singular values of
\[
N^{-1/2}
D_{z,T,0}^{-1}
R_{zz,0}
D_{z,T,0}
\]
are bounded away from zero and infinity. Moreover,
Lemma~\ref{L:factor_second_moment_limits} gives
\[
D_{z,T,0}^{-1}
\widetilde{\Sigma}_z
D_{z,T,0}^{-1}
\Rightarrow
S_z,
\]
where $S_z$ is positive definite almost surely. Since the dimension
of the nonstationary factor block is fixed, its smallest and largest
eigenvalues are therefore bounded away from zero and infinity in
probability. Consequently,
\begin{align}
	\kappa_{s_0^*}(\tilde A_{zz})
	&=
	O_e(1),
	&
	\kappa_1(\tilde A_{zz})
	&=
	O_e(1).
	\label{eq:tilde_Azz_conditioning}
\end{align}

For the stationary factor block,
\[
\tilde A_{ww}
=
\left(
N^{-1/2}R_{ww,0}
\right)
\widetilde{\Sigma}_w
\left(
N^{-1/2}R_{ww,0}
\right)^T
+
\frac{1}{N}
Q_{w,0}^T
\widetilde{\Sigma}_u
Q_{w,0}.
\]
By \eqref{eq:scaled_Rww_bounds} and
Lemma~\ref{L:factor_second_moment_limits}, the eigenvalues of the
first term are bounded away from zero and infinity in probability.
The second term is positive semidefinite and, by
\eqref{eq:empirical_idiosyncratic_spectral_bound}, satisfies
\[
\norm{
	\frac{1}{N}
	Q_{w,0}^T
	\widetilde{\Sigma}_u
	Q_{w,0}
}_2
\leq
\frac{1}{N}
\norm{\widetilde{\Sigma}_u}_2
=
O_p(N^{-1}).
\]
It follows that
\begin{align}
	\kappa_{s_0-s_0^*}(\tilde A_{ww})
	&=
	O_e(1),
	&
	\kappa_1(\tilde A_{ww})
	&=
	O_e(1).
	\label{eq:tilde_Aww_conditioning}
\end{align}
	
	For the $\perp\perp$-block,
	\[
	\tilde A_{\perp\perp}
	=
	\frac{1}{N}
	Q_{\perp,0}^T
	\widetilde{\Sigma}_u
	Q_{\perp,0}.
	\]
	Since $Q_{\perp,0}$ has orthonormal columns, the Poincar\'e
	separation theorem gives
	\begin{align*}
		\kappa_N(\widetilde{\Sigma}_u)
		&\leq
		\kappa_{N-s_0}\left(
		Q_{\perp,0}^T
		\widetilde{\Sigma}_u
		Q_{\perp,0}
		\right) \leq
		\kappa_1\left(
		Q_{\perp,0}^T
		\widetilde{\Sigma}_u
		Q_{\perp,0}
		\right)
		\leq
		\kappa_1(\widetilde{\Sigma}_u).
	\end{align*}
	Hence, by
	Assumption~\ref{ass:Eigenvalue_Sigma_u_tilde}, with probability
	approaching one,
	\begin{align*}
		\widetilde c_u
		&\leq
		\kappa_{N-s_0}\left(
		Q_{\perp,0}^T
		\widetilde{\Sigma}_u
		Q_{\perp,0}
		\right) \leq
		\kappa_1\left(
		Q_{\perp,0}^T
		\widetilde{\Sigma}_u
		Q_{\perp,0}
		\right)
		\leq
		\widetilde C_u.
	\end{align*}
	Equivalently,
	\begin{align}
		\frac{\widetilde c_u}{N}
		&\leq
		\kappa_{N-s_0}
		\left(
		\tilde A_{\perp\perp}
		\right)
		\leq
		\kappa_1
		\left(
		\tilde A_{\perp\perp}
		\right)
		\leq
		\frac{\widetilde C_u}{N}
		\label{eq:tilde_Sigma_u_EW}
	\end{align}
	with probability approaching one.
	
	To account for the eigenvalues of order $N^{-1}$ in the
	idiosyncratic block, define
	\begin{align}
		W_N
		&=
		\begin{pmatrix}
			I_{s_0^*} & 0 & 0\\
			0 & I_{s_0-s_0^*} & 0\\
			0 & 0 & \sqrt N I_{N-s_0}
		\end{pmatrix}.
		\label{eq:W_N_def}
	\end{align}
	Then
	\[
	W_N\tilde A W_N
	=
	\begin{pmatrix}
		\tilde A_{zz}
		&
		0
		&
		0\\
		0
		&
		\tilde A_{ww}
		&
		\sqrt N\,\tilde A_{w\perp}^T\\
		0
		&
		\sqrt N\,\tilde A_{w\perp}
		&
		N\tilde A_{\perp\perp}
	\end{pmatrix},
	\]
	and
	\[
	W_N\Delta A W_N
	=
	\begin{pmatrix}
		\Delta A_{zz}
		&
		\Delta A_{zw}^T
		&
		\sqrt N\,\Delta A_{z\perp}^T\\
		\Delta A_{zw}
		&
		\Delta A_{ww}
		&
		\sqrt N\,\Delta A_{w\perp}^T\\
		\sqrt N\,\Delta A_{z\perp}
		&
		\sqrt N\,\Delta A_{w\perp}
		&
		0
	\end{pmatrix}.
	\]

	The scaling by $W_N$ brings the $\perp\perp$-block of $\tilde{A}$, whose
eigenvalues are of order $N^{-1}$, to order one. We now show
directly that the scaled reference matrix is well-conditioned.

Consider the lower-right block of $W_N\tilde A W_N$. It admits
the factorization
\begin{align}
	\begin{pmatrix}
		\tilde A_{ww}
		&
		\sqrt N\,\tilde A_{w\perp}^T\\
		\sqrt N\,\tilde A_{w\perp}
		&
		N\tilde A_{\perp\perp}
	\end{pmatrix}
&=
	\begin{pmatrix}
		N^{-1/2}R_{ww,0}
		&
		N^{-1/2}Q_{w,0}^T\\
		0
		&
		Q_{\perp,0}^T
	\end{pmatrix}
	\begin{pmatrix}
		\widetilde{\Sigma}_w & 0\\
		0 & \widetilde{\Sigma}_u
	\end{pmatrix}
	\begin{pmatrix}
		N^{-1/2}R_{ww,0}
		&
		N^{-1/2}Q_{w,0}^T\\
		0
		&
		Q_{\perp,0}^T
	\end{pmatrix}^{T}.
	\label{eq:scaled_lower_block_factorization}
\end{align}
Because the columns of $Q_{w,0}$ and $Q_{\perp,0}$ are mutually
orthonormal,
\begin{align*}
	\begin{pmatrix}
		N^{-1/2}R_{ww,0}
		&
		N^{-1/2}Q_{w,0}^T\\
		0
		&
		Q_{\perp,0}^T
	\end{pmatrix}
	\begin{pmatrix}
		N^{-1/2}R_{ww,0}
		&
		N^{-1/2}Q_{w,0}^T\\
		0
		&
		Q_{\perp,0}^T
	\end{pmatrix}^{T} =
	\begin{pmatrix}
		N^{-1}R_{ww,0}R_{ww,0}^T
		+
		N^{-1}I_{s_0-s_0^*}
		&
		0\\
		0
		&
		I_{N-s_0}
	\end{pmatrix}.
\end{align*}
It follows from \eqref{eq:scaled_Rww_bounds} that there exist
constants $0<c<C<\infty$ such that, for every conformable vector
$x$,
\[
c\norm{x}^2
\leq
\norm{
	\begin{pmatrix}
		N^{-1/2}R_{ww,0}
		&
		N^{-1/2}Q_{w,0}^T\\
		0
		&
		Q_{\perp,0}^T
	\end{pmatrix}^{T}
	x
}^2
\leq
C\norm{x}^2.
\]
Moreover,
Lemma~\ref{L:factor_second_moment_limits} and
Assumption~\ref{ass:Eigenvalue_Sigma_u_tilde} imply that the
eigenvalues of
\[
\begin{pmatrix}
	\widetilde{\Sigma}_w & 0\\
	0 & \widetilde{\Sigma}_u
\end{pmatrix}
\]
are bounded away from zero and infinity in probability.
Applying these bounds to the quadratic form associated with
\eqref{eq:scaled_lower_block_factorization} shows that the smallest
eigenvalue of the lower-right block of $W_N\tilde A W_N$ is bounded
away from zero in probability, while its largest eigenvalue is
bounded in probability.

Together with \eqref{eq:tilde_Azz_conditioning}, this gives
\begin{align}
	\kappa_N(W_N\tilde A W_N)
	&=
	O_e(1),
	&
	\kappa_1(W_N\tilde A W_N)
	&=
	O_p(1).
	\label{eq:tilde_A_scaled_conditioning}
\end{align}
Thus, $W_N\tilde A W_N$ is positive definite with probability
approaching one. Since congruence by the nonsingular matrix $W_N$
preserves positive definiteness, $\tilde A$ is also positive
definite with probability approaching one. Moreover,
\begin{align}
	\norm{
		\tilde A^{-1/2}W_N^{-1}
	}_2^2
	&=
	\norm{
		W_N^{-1}
		\tilde A^{-1}
		W_N^{-1}
	}_2
=
	\norm{
		\left(
		W_N\tilde A W_N
		\right)^{-1}
	}_2
	=
	O_p(1).
	\label{eq:tilde_AW_convergence}
\end{align}

We next show that the perturbation vanishes under the same
scaling. By \eqref{eq:Delta_A_zperp_rate},
\begin{align}
	\sqrt N
	\norm{\Delta A_{z\perp}}_F
	&\leq
	O_p\left(
	\sqrt{\frac{N}{T}}
	\right)
	+
	O_p\left(
	\sqrt N\,T^{-d_{0,s_0^*}}
	\right)
+
	O_p\left(
	N^{-1/2}
	T^{-d_{0,s_0^*}+1/2}
	\right)
	=
	o_p(1),
	\label{eq:scaled_Delta_A_zperp_rate}
\end{align}
because $N/T\to0$ and $d_{0,s_0^*}>1/2$.

Similarly, \eqref{eq:Delta_A_wperp_rate} gives
\begin{align}
	\sqrt N
	\norm{\Delta A_{w\perp}}_F
	&=
	O_p\left(
	\sqrt{\frac{N}{T}}
	\right)
	=
	o_p(1).
	\label{eq:scaled_Delta_A_wperp_rate}
\end{align}

Using the block structure of $W_N\Delta A W_N$ together with
\eqref{eq:Delta_A_zz_rate},
\eqref{eq:Delta_A_zw_rate},
\eqref{eq:Delta_A_ww_rate},
\eqref{eq:scaled_Delta_A_zperp_rate}, and
\eqref{eq:scaled_Delta_A_wperp_rate}, we obtain
\begin{align*}
	\norm{
		W_N\Delta A W_N
	}_F^2
	&=
	\norm{\Delta A_{zz}}_F^2
	+
	2\norm{\Delta A_{zw}}_F^2
	+
	\norm{\Delta A_{ww}}_F^2
	\nonumber\\
	&\quad+
	2N\norm{\Delta A_{z\perp}}_F^2
	+
	2N\norm{\Delta A_{w\perp}}_F^2
	\nonumber\\
	&=
	o_p(1).
\end{align*}
Consequently,
\begin{align}
	\norm{
		W_N\Delta A W_N
	}_F
	&=
	o_p(1).
	\label{eq:weighted_Delta_A_rate}
\end{align}
	
	Therefore,
	\begin{align}
		\norm{
			\tilde A^{-1/2}
			\Delta A
			\tilde A^{-1/2}
		}_F
		&=
		\norm{
			\left(
			\tilde A^{-1/2}W_N^{-1}
			\right)
			\left(
			W_N\Delta A W_N
			\right)
			\left(
			W_N^{-1}\tilde A^{-1/2}
			\right)
		}_F
		\nonumber\\
		&\leq
		\norm{
			\tilde A^{-1/2}W_N^{-1}
		}_2^2
		\norm{
			W_N\Delta A W_N
		}_F
		\nonumber\\
		&=
		o_p(1).
		\label{eq:diff_A_tilde_A}
	\end{align}
	
	Moreover,
	\begin{align}
		\hat A
		&=
		\tilde A+\Delta A
=
		\tilde A^{1/2}
		\left(
		I+
		\tilde A^{-1/2}
		\Delta A
		\tilde A^{-1/2}
		\right)
		\tilde A^{1/2}.
		\label{eq:hat_A}
	\end{align}
	By \eqref{eq:diff_A_tilde_A},
	\[
	\norm{
		\tilde A^{-1/2}
		\Delta A
		\tilde A^{-1/2}
	}_2
	\leq
	\norm{
		\tilde A^{-1/2}
		\Delta A
		\tilde A^{-1/2}
	}_F
	=
	o_p(1).
	\]
	Therefore, the eigenvalues of
	\[
	I+
	\tilde A^{-1/2}
	\Delta A
	\tilde A^{-1/2}
	\]
	are bounded away from zero with probability approaching one. Hence,
	\begin{align}
		\norm{
			\left(
			I+
			\tilde A^{-1/2}
			\Delta A
			\tilde A^{-1/2}
			\right)^{-1}
		}_2
		&=
		O_p(1).
		\label{eq:Norm_I_diffA}
	\end{align}
	In particular, \eqref{eq:hat_A} shows that $\hat A$ is positive
	definite with probability approaching one, and
	\begin{align}
		\hat A^{-1}
		&=
		\tilde A^{-1/2}
		\left(
		I+
		\tilde A^{-1/2}
		\Delta A
		\tilde A^{-1/2}
		\right)^{-1}
		\tilde A^{-1/2}.
		\label{eq:hat_A_tilde_A}
	\end{align}
	
	We now return to the trace representation described at the
	beginning of the proof. The definitions of $\hat A$ and $\tilde A$
	give
	\[
	\widetilde{\Sigma}_{\tilde y,T}^{-1}
	=
	N^{-1}Q_{all,0}D_{T,0}^{-1}
	\tilde A^{-1}
	D_{T,0}^{-1}Q_{all,0}^T,
	\]
	and
	\[
	\widehat{\Sigma}_{y,T}^{-1}
	=
	N^{-1}Q_{all,0}D_{T,0}^{-1}
	\hat A^{-1}
	D_{T,0}^{-1}Q_{all,0}^T.
	\]
Using these identities and
\eqref{eq:Delta_Sigma_y_Delta_A}, we obtain
\begin{align}
	&
	\norm{
		\widetilde{\Sigma}_{\tilde y,T}^{-1/2}
		\Delta\Sigma_{y,T}
		\widehat{\Sigma}_{y,T}^{-1/2}
	}_F^2
	\nonumber\\
	&\quad=
	\tr\left(
	\widetilde{\Sigma}_{\tilde y,T}^{-1}
	\Delta\Sigma_{y,T}
	\widehat{\Sigma}_{y,T}^{-1}
	\Delta\Sigma_{y,T}
	\right)
	\nonumber\\
	&\quad=
	\tr\Big[
	N^{-1}Q_{all,0}D_{T,0}^{-1}
	\tilde A^{-1}
	D_{T,0}^{-1}Q_{all,0}^T
	NQ_{all,0}D_{T,0}
	\Delta A
	D_{T,0}Q_{all,0}^T
	\nonumber\\
	&\qquad\qquad\times
	N^{-1}Q_{all,0}D_{T,0}^{-1}
	\hat A^{-1}
	D_{T,0}^{-1}Q_{all,0}^T
	NQ_{all,0}D_{T,0}
	\Delta A
	D_{T,0}Q_{all,0}^T
	\Big]
	\nonumber\\
	&\quad=
	\tr\left(
	Q_{all,0}D_{T,0}^{-1}
	\tilde A^{-1}
	\Delta A
	\hat A^{-1}
	\Delta A
	D_{T,0}Q_{all,0}^T
	\right)
	\nonumber\\
	&\quad=
	\tr\left(
	D_{T,0}^{-1}
	\tilde A^{-1}
	\Delta A
	\hat A^{-1}
	\Delta A
	D_{T,0}
	Q_{all,0}^TQ_{all,0}
	\right)
	\nonumber\\
	&\quad=
	\tr\left(
	D_{T,0}^{-1}
	\tilde A^{-1}
	\Delta A
	\hat A^{-1}
	\Delta A
	D_{T,0}
	\right)
	\nonumber\\
	&\quad=
	\tr\left(
	\tilde A^{-1}
	\Delta A
	\hat A^{-1}
	\Delta A
	\right),
	\label{eq:trace_reduction_A}
\end{align}
where we used $Q_{all,0}^TQ_{all,0}=I$ and cyclicity of the trace.

	Substituting \eqref{eq:hat_A_tilde_A} yields
	\begin{align*}
		&
		\tr\left(
		\tilde A^{-1}
		\Delta A
		\hat A^{-1}
		\Delta A
		\right)
		\\
		&\quad=
		\tr\left[
		\tilde A^{-1/2}
		\Delta A
		\tilde A^{-1/2}
		\left(
		I+
		\tilde A^{-1/2}
		\Delta A
		\tilde A^{-1/2}
		\right)^{-1}
		\tilde A^{-1/2}
		\Delta A
		\tilde A^{-1/2}
		\right]
		\\
		&\quad\leq
		\norm{
			\left(
			I+
			\tilde A^{-1/2}
			\Delta A
			\tilde A^{-1/2}
			\right)^{-1}
		}_2
		\norm{
			\tilde A^{-1/2}
			\Delta A
			\tilde A^{-1/2}
		}_F^2
=
		o_p(1),
	\end{align*}
	by \eqref{eq:diff_A_tilde_A} and
	\eqref{eq:Norm_I_diffA}. Thus,
	\[
	\norm{
		\widetilde{\Sigma}_{\tilde y,T}^{-1/2}
		\Delta\Sigma_{y,T}
		\widehat{\Sigma}_{y,T}^{-1/2}
	}_F^2
	=
	o_p(1).
	\]
	Taking square roots proves the claim.
\end{proof}

\begin{lemma}[Group-refined relative covariance perturbation]
	\label{L:Sigma_y_hat_tildetilde_diff_N}
	Suppose Assumptions~\ref{ass:factors}-\ref{ass:memory} hold and
	that $N,T \to \infty$ with $N/T\to0$. Then
	\[
	\norm{
		\widetilde{\widetilde{\Sigma}}_{\tilde{\tilde y},T}^{-1/2}
		\left(
		\widehat{\Sigma}_{y,T}
		-
		\widetilde{\widetilde{\Sigma}}_{\tilde{\tilde y},T}
		\right)
		\widehat{\Sigma}_{y,T}^{-1/2}
	}_F
	=
	O_p(1).
	\]
\end{lemma}

\begin{proof}
	If $q_0^*\leq1$, then
	$\widetilde{\widetilde{\Sigma}}_{\tilde{\tilde y},T}
	=
	\widetilde{\Sigma}_{\tilde y,T}$, and the claim follows directly
	from Lemma~\ref{L:Sigma_y_hat_tilde_diff_N}. We therefore suppose
	that $q_0^*\geq2$. Empty stationary blocks are omitted throughout.
	
	Define
\begin{align}
	\hat A
	&=
	N^{-1}
	D_{T,0}^{-1}
	Q_{all,0}^T
	\widehat{\Sigma}_{y,T}
	Q_{all,0}
	D_{T,0}^{-1},
	\label{eq:hat_A_definition} \\
	\tilde A
	&=
	N^{-1}
	D_{T,0}^{-1}
	Q_{all,0}^T
	\widetilde{\Sigma}_{\tilde y,T}
	Q_{all,0}
	D_{T,0}^{-1},
	\label{eq:tilde_A_definition} 
\end{align}

	and
	\[
	\tilde{\tilde A}
	=
	N^{-1}
	D_{T,0}^{-1}
	Q_{all,0}^T
	\widetilde{\widetilde{\Sigma}}_{\tilde{\tilde y},T}
	Q_{all,0}
	D_{T,0}^{-1}.
	\]
	Furthermore, let
	\begin{align}
		\Delta A
		&=
		\hat A-\tilde A,
		\label{eq:Delta_A_definition}
		\\
		\widetilde{\Delta}A
		&=
		\tilde A-\tilde{\tilde A}.
		\label{eq:tildetilde_delta_definitions}
	\end{align}
	Then
	\begin{align}
		\hat A-\tilde{\tilde A}
		&=
		\Delta A+\widetilde{\Delta}A.
		\label{eq:tildetilde_delta_decomposition}
	\end{align}
	
	We use the scaling matrix from the proof of
	Lemma~\ref{L:Sigma_y_hat_tilde_diff_N},
	\[
	W_N
	=
	\begin{pmatrix}
		I_{s_0^*} & 0 & 0\\
		0 & I_{s_0-s_0^*} & 0\\
		0 & 0 & \sqrt N I_{N-s_0}
	\end{pmatrix}.
	\]
	We first establish the required inverse bounds for
	$\tilde{\tilde A}$ and $\hat A$.
	
	For each nonstationary group $j=1,\ldots,q_0^*$, the corresponding
	diagonal block of $\tilde{\tilde A}$ is
	\begin{align}
		\left(
		\tilde{\tilde A}_{zz}
		\right)_{jj}
		&=
		\frac{1}{N}
		T^{-2d_0^{(j)}+1}
		R_{zz,0}^{(j,j)}
		\left(
		\frac{1}{T}
		\sum_{t=1}^{T}
		z_t^{(j)}
		\left(z_t^{(j)}\right)^T
		\right)
		R_{zz,0}^{(j,j)T}
		\nonumber\\
		&=
		\left(
		N^{-1/2}R_{zz,0}^{(j,j)}
		\right)
		\left[
		T^{-2d_0^{(j)}+1}
		\left(
		\frac{1}{T}
		\sum_{t=1}^{T}
		z_t^{(j)}
		\left(z_t^{(j)}\right)^T
		\right)
		\right]
		\left(
		N^{-1/2}R_{zz,0}^{(j,j)}
		\right)^T.
		\label{eq:tildetilde_group_block_factorization}
	\end{align}
	By \eqref{eq:nonstationary_factor_covariance_limit},
	\[
	T^{-2d_0^{(j)}+1}
	\left(
	\frac{1}{T}
	\sum_{t=1}^{T}
	z_t^{(j)}
	\left(z_t^{(j)}\right)^T
	\right)
	\]
	converges weakly to the corresponding principal diagonal block of
	$S_z$. Since $S_z$ is positive definite almost surely, the
	eigenvalues of this block are bounded away from zero and infinity
	in probability.
	
	Moreover, \eqref{eq:scaled_Rzz_bounds} implies
	\[
	\norm{
		N^{-1/2}R_{zz,0}^{(j,j)}
	}_2
	+
	\norm{
		\left(
		N^{-1/2}R_{zz,0}^{(j,j)}
		\right)^{-1}
	}_2
	=
	O(1).
	\]
	It follows from
	\eqref{eq:tildetilde_group_block_factorization}
	that the eigenvalues of each nonstationary diagonal block of
	$\tilde{\tilde A}$ are bounded away from zero and infinity in
	probability.
	
	The joint stationary-idiosyncratic block of
	$W_N\tilde{\tilde A}W_N$ is identical to the corresponding block
	of $W_N\tilde A W_N$. Hence, the conditioning argument in the
	proof of Lemma~\ref{L:Sigma_y_hat_tilde_diff_N} applies unchanged
	to this block. Since $\tilde{\tilde A}$ is block diagonal between
	the individual nonstationary groups and the joint
	stationary-idiosyncratic block, we obtain
	\begin{align}
		\kappa_N\left(
		W_N\tilde{\tilde A}W_N
		\right)
		&=
		 O_e(1),
		&
		\kappa_1\left(
		W_N\tilde{\tilde A}W_N
		\right)
		&=
		O_p(1).
		\label{eq:tildetilde_A_scaled_conditioning}
	\end{align}
	Thus, $\tilde{\tilde A}$ is positive definite with probability
	approaching one, and
	\begin{align}
		\norm{
			\tilde{\tilde A}^{-1/2}W_N^{-1}
		}_2^2
		&=
		\norm{
			W_N^{-1}\tilde{\tilde A}^{-1}W_N^{-1}
		}_2 =
		\norm{
			\left(
			W_N\tilde{\tilde A}W_N
			\right)^{-1}
		}_2
		=
		O_p(1).
		\label{eq:tildetilde_AW_convergence}
	\end{align}
	
	Next,
	\[
	W_N\hat A W_N
	=
	W_N\tilde A W_N
	+
	W_N\Delta A W_N.
	\]
	Therefore,
	\eqref{eq:tilde_A_scaled_conditioning} and
	\eqref{eq:weighted_Delta_A_rate}, together with Weyl's inequality,
	imply
	\begin{align*}
		\kappa_N\left(
		W_N\hat A W_N
		\right)
		&=
	 O_e(1),
		&
		\kappa_1\left(
		W_N\hat A W_N
		\right)
		&=
		O_p(1).
	\end{align*}
	Consequently, $\hat A$ is positive definite with probability
	approaching one, and
	\begin{align}
		\norm{
			W_N^{-1}\hat A^{-1/2}
		}_2^2
		&=
		\norm{
			W_N^{-1}\hat A^{-1}W_N^{-1}
		}_2 =
		\norm{
			\left(
			W_N\hat A W_N
			\right)^{-1}
		}_2
		=
		O_p(1).
		\label{eq:tildetilde_hat_A_scaling}
	\end{align}
	
	We next bound the perturbation under the same scaling. By
	\eqref{eq:weighted_Delta_A_rate},
	\begin{align*}
		\norm{
			W_N\Delta A W_N
		}_F
		&=
		o_p(1).
	\end{align*}
	By \eqref{eq:tildetilde_delta_definitions} and the construction of
	the two artificial covariance matrices, $\tilde A$ and
	$\tilde{\tilde A}$ differ only in the nonstationary block.
	Therefore,
	\begin{align*}
		\widetilde{\Delta}A
		&=
		\begin{pmatrix}
			\tilde A_{zz}-\tilde{\tilde A}_{zz} & 0 & 0\\
			0 & 0 & 0\\
			0 & 0 & 0
		\end{pmatrix}.
	\end{align*}
	Since $W_N$ leaves the nonstationary block unchanged,
	\begin{align}
		W_N\widetilde{\Delta}A W_N
		&=
		\widetilde{\Delta}A.
		\label{eq:tildetilde_delta_A_W}
	\end{align}
	
	It remains to bound
	$\tilde A_{zz}-\tilde{\tilde A}_{zz}$. For
	$j,j'=1,\ldots,q_0^*$, block upper triangularity of $R_{zz,0}$
	gives
	\begin{align}
		\left(
		\tilde A_{zz}
		\right)_{jj'}
		&=
		\frac{1}{N}
		T^{-d_0^{(j)}-d_0^{(j')}+1}
		\sum_{k=j}^{q_0^*}
		\sum_{k'=j'}^{q_0^*}
		R_{zz,0}^{(j,k)}
		\left(
		\frac{1}{T}
		\sum_{t=1}^{T}
		z_t^{(k)}
		\left(z_t^{(k')}\right)^T
		\right)
		R_{zz,0}^{(j',k')T}.
		\label{eq:tilde_A_group_block}
	\end{align}
	For any $k,k'=1,\ldots,q_0^*$,
	\eqref{eq:nonstationary_factor_covariance_limit} implies
	\begin{align*}
		\norm{
			T^{-d_0^{(k)}-d_0^{(k')}+1}
			\left(
			\frac{1}{T}
			\sum_{t=1}^{T}
			z_t^{(k)}
			\left(z_t^{(k')}\right)^T
			\right)
		}_F
		&=
		O_p(1).
	\end{align*}
	Together with \eqref{eq:loading_block_bounds}, this gives, for
	each summand in \eqref{eq:tilde_A_group_block},
	\begin{align*}
		&\norm{
			\frac{1}{N}
			T^{-d_0^{(j)}-d_0^{(j')}+1}
			R_{zz,0}^{(j,k)}
			\left(
			\frac{1}{T}
			\sum_{t=1}^{T}
			z_t^{(k)}
			\left(z_t^{(k')}\right)^T
			\right)
			R_{zz,0}^{(j',k')T}
		}_F
		\nonumber\\
		&\quad\leq
		T^{d_0^{(k)}-d_0^{(j)}}
		T^{d_0^{(k')}-d_0^{(j')}}
		\norm{
			N^{-1/2}R_{zz,0}^{(j,k)}
		}_F
		\nonumber\\
		&\qquad\quad\times
		\norm{
			T^{-d_0^{(k)}-d_0^{(k')}+1}
			\left(
			\frac{1}{T}
			\sum_{t=1}^{T}
			z_t^{(k)}
			\left(z_t^{(k')}\right)^T
			\right)
		}_F
		\norm{
			N^{-1/2}R_{zz,0}^{(j',k')}
		}_F
		\nonumber\\
		&\quad=
		T^{d_0^{(k)}-d_0^{(j)}}
		T^{d_0^{(k')}-d_0^{(j')}}
		O_p(1).
	\end{align*}
	Since $k\geq j$ and $k'\geq j'$, the ordering
	\[
	d_0^{(1)}
	>
	\cdots
	>
	d_0^{(q_0^*)}
	\]
	implies
	\[
	T^{d_0^{(k)}-d_0^{(j)}}
	T^{d_0^{(k')}-d_0^{(j')}}
	=
	\begin{cases}
		1,
		& k=j\ \text{and}\ k'=j',
		\\
		o(1),
		& \text{otherwise}.
	\end{cases}
	\]
	Thus, only the summand with $k=j$ and $k'=j'$ need not vanish.
	Because the number of nonstationary groups is fixed,
	\eqref{eq:tilde_A_group_block} gives
	\begin{align*}
		\left(
		\tilde A_{zz}
		\right)_{jj'}
		&=
		\frac{1}{N}
		T^{-d_0^{(j)}-d_0^{(j')}+1}
		R_{zz,0}^{(j,j)}
		\left(
		\frac{1}{T}
		\sum_{t=1}^{T}
		z_t^{(j)}
		\left(z_t^{(j')}\right)^T
		\right)
		R_{zz,0}^{(j',j')T}
		+
		o_p(1).
	\end{align*}
	By construction, $\tilde{\tilde A}_{zz}$ retains this leading term
	when $j=j'$ and sets the block equal to zero when $j\neq j'$.
	Consequently,
	\[
	\left(
	\tilde A_{zz}
	-
	\tilde{\tilde A}_{zz}
	\right)_{jj'}
	=
	\begin{cases}
		o_p(1),
		& j=j',
		\\
		O_p(1),
		& j\neq j'.
	\end{cases}
	\]
	Since the number of nonstationary groups and their dimensions are
	fixed, it follows that
	\begin{align}
		\norm{
			\tilde A_{zz}
			-
			\tilde{\tilde A}_{zz}
		}_F
		&=
		O_p(1).
		\label{eq:tildetilde_zblock_difference}
	\end{align}
	Together with \eqref{eq:tildetilde_delta_A_W} and
\eqref{eq:tildetilde_zblock_difference}, this yields
	\begin{align*}
		\norm{
			W_N\widetilde{\Delta}A W_N
		}_F
		&=
		O_p(1).
	\end{align*}
	Consequently,
	\begin{align}
		\norm{
			W_N
			\left(
			\Delta A+\widetilde{\Delta}A
			\right)
			W_N
		}_F
		&\leq
		\norm{
			W_N\Delta A W_N
		}_F
		+
		\norm{
			W_N\widetilde{\Delta}A W_N
		}_F
		=
		O_p(1).
		\label{eq:tildetilde_scaled_full_delta}
	\end{align}
	
	Using
	\eqref{eq:tildetilde_AW_convergence},
	\eqref{eq:tildetilde_hat_A_scaling}, and
	\eqref{eq:tildetilde_scaled_full_delta}, we obtain
	\begin{align*}
		&\norm{
			\tilde{\tilde A}^{-1/2}
			\left(
			\Delta A+\widetilde{\Delta}A
			\right)
			\hat A^{-1/2}
		}_F
		\\
		&\quad=
		\norm{
			\left(
			\tilde{\tilde A}^{-1/2}W_N^{-1}
			\right)
			W_N
			\left(
			\Delta A+\widetilde{\Delta}A
			\right)
			W_N
			\left(
			W_N^{-1}\hat A^{-1/2}
			\right)
		}_F
		\\
		&\quad\leq
		\norm{
			\tilde{\tilde A}^{-1/2}W_N^{-1}
		}_2
		\norm{
			W_N
			\left(
			\Delta A+\widetilde{\Delta}A
			\right)
			W_N
		}_F
		\norm{
			W_N^{-1}\hat A^{-1/2}
		}_2
		\\
		&\quad=
		O_p(1).
	\end{align*}
	
	Finally, by \eqref{eq:tildetilde_delta_decomposition}, the same
	trace reduction as in \eqref{eq:trace_reduction_A} gives
	\begin{align*}
		\norm{
			\widetilde{\widetilde{\Sigma}}_{\tilde{\tilde y},T}^{-1/2}
			\left(
			\widehat{\Sigma}_{y,T}
			-
			\widetilde{\widetilde{\Sigma}}_{\tilde{\tilde y},T}
			\right)
			\widehat{\Sigma}_{y,T}^{-1/2}
		}_F^2
		&=
		\tr\left[
		\tilde{\tilde A}^{-1}
		\left(
		\Delta A+\widetilde{\Delta}A
		\right)
		\hat A^{-1}
		\left(
		\Delta A+\widetilde{\Delta}A
		\right)
		\right]
		\\
		&=
		\norm{
			\tilde{\tilde A}^{-1/2}
			\left(
			\Delta A+\widetilde{\Delta}A
			\right)
			\hat A^{-1/2}
		}_F^2
		=
		O_p(1),
	\end{align*}
	where we used orthogonality of $Q_{all,0}$ and cyclicity of the
	trace. Taking square roots proves the claim.
\end{proof}

\begin{lemma}[Fully block-diagonal relative covariance perturbation]
	\label{L:Sigma_y_hat_tilde_bd_diff_N}
	Suppose Assumptions~\ref{ass:factors}-\ref{ass:memory} hold and
	that $N,T\to\infty$ with $N/T\to0$. Then
	\[
	\norm{
		\left(
		\widetilde{\Sigma}_{\tilde y,T}^{\mathrm{bd}}
		\right)^{-1/2}
		\Delta\Sigma_{y,T}^{\mathrm{bd}}
		\widehat{\Sigma}_{y,T}^{-1/2}
	}_F
	=
	o_p(1),
	\]
	where
	\[
	\Delta\Sigma_{y,T}^{\mathrm{bd}}
	=
	\widehat{\Sigma}_{y,T}
	-
	\widetilde{\Sigma}_{\tilde y,T}^{\mathrm{bd}}.
	\]
\end{lemma}

\begin{proof}
	If no stationary factors are present, then
	\[
	\widetilde{\Sigma}_{\tilde y,T}^{\mathrm{bd}}
	=
	\widetilde{\Sigma}_{\tilde y,T},
	\]
	and the claim follows directly from
	Lemma~\ref{L:Sigma_y_hat_tilde_diff_N}. We therefore suppose that
	$s_0^*<s_0$.
	
	The proof proceeds in three steps. We first show that the
	stationary factor-idiosyncratic blocks removed from
	$\widetilde{\Sigma}_{\tilde y,T}$ are negligible under the scaling
	$W_N$. We then establish the required inverse bounds for the
	block-diagonal reference matrix and $\widehat{\Sigma}_{y,T}$.
	Finally, we apply the trace reduction used in the proof of
	Lemma~\ref{L:Sigma_y_hat_tilde_diff_N}.
	
	Retain the matrices $\hat A$, $\tilde A$, and $\Delta A$ defined in
	\eqref{eq:hat_A_definition}, \eqref{eq:tilde_A_definition}, \eqref{eq:Delta_A_definition}, and the
	scaling matrix $W_N$ defined in \eqref{eq:W_N_def}. Define
	\begin{align*}
		\widetilde A^{\mathrm{bd}}
		&=
		\frac{1}{N}
		D_{T,0}^{-1}
		Q_{all,0}^T
		\widetilde{\Sigma}_{\tilde y,T}^{\mathrm{bd}}
		Q_{all,0}
		D_{T,0}^{-1}.
	\end{align*}
	Since $\Delta A=\hat A-\tilde A$,
	\begin{align}
		\hat A-\widetilde A^{\mathrm{bd}}
		&=
		\Delta A
		+
		\left(
		\tilde A-\widetilde A^{\mathrm{bd}}
		\right).
		\label{eq:hat_A_tilde_A_bd_decomposition}
	\end{align}
	
By \eqref{eq:Sigma_tilde_Qy_bd_difference}, and because
$D_{T,0}^{-1}$ acts as the identity on the $w$- and
$\perp$-coordinates,
	\[
	\tilde A-\widetilde A^{\mathrm{bd}}
	=
	\frac{1}{N}
	\begin{pmatrix}
		0
		&
		0
		&
		0
		\\
		0
		&
		0
		&
		Q_{w,0}^T
		\widetilde{\Sigma}_{u}
		Q_{\perp,0}
		\\
		0
		&
		Q_{\perp,0}^T
		\widetilde{\Sigma}_{u}
		Q_{w,0}
		&
		0
	\end{pmatrix}.
	\]
	Hence,
	\begin{align}
		W_N
		\left(
		\tilde A-\widetilde A^{\mathrm{bd}}
		\right)
		W_N
		&=
		\begin{pmatrix}
			0
			&
			0
			&
			0
			\\
			0
			&
			0
			&
			N^{-1/2}
			Q_{w,0}^T
			\widetilde{\Sigma}_{u}
			Q_{\perp,0}
			\\
			0
			&
			N^{-1/2}
			Q_{\perp,0}^T
			\widetilde{\Sigma}_{u}
			Q_{w,0}
			&
			0
		\end{pmatrix}.
		\label{eq:scaled_tilde_A_bd_difference}
	\end{align}
	Since $Q_{w,0}$ has $s_0-s_0^*$ orthonormal columns,
	\begin{align*}
		\norm{
			Q_{w,0}^T
			\widetilde{\Sigma}_{u}
			Q_{\perp,0}
		}_F
		&\leq
		\sqrt{s_0-s_0^*}
		\norm{
			\widetilde{\Sigma}_{u}
		}_2
		=
		O_p(1),
	\end{align*}
	where the final equality follows from
	Assumption~\ref{ass:Eigenvalue_Sigma_u_tilde}. It follows from
	\eqref{eq:scaled_tilde_A_bd_difference} that
	\begin{align*}
		\norm{
			W_N
			\left(
			\tilde A-\widetilde A^{\mathrm{bd}}
			\right)
			W_N
		}_F^2
		&=
		\frac{2}{N}
		\norm{
			Q_{w,0}^T
			\widetilde{\Sigma}_{u}
			Q_{\perp,0}
		}_F^2
		=
		O_p\left(
		N^{-1}
		\right).
	\end{align*}
	Therefore, since $N\to\infty$,
	\begin{align}
		\norm{
			W_N
			\left(
			\tilde A-\widetilde A^{\mathrm{bd}}
			\right)
			W_N
		}_F
		&=
		o_p(1).
		\label{eq:scaled_tilde_A_bd_difference_rate}
	\end{align}
	
	Combining
	\eqref{eq:hat_A_tilde_A_bd_decomposition},
	\eqref{eq:weighted_Delta_A_rate}, and
	\eqref{eq:scaled_tilde_A_bd_difference_rate} gives
	\begin{align}
		\norm{
			W_N
			\left(
			\hat A-\widetilde A^{\mathrm{bd}}
			\right)
			W_N
		}_F
		&\leq
		\norm{
			W_N\Delta A W_N
		}_F
		+
		\norm{
			W_N
			\left(
			\tilde A-\widetilde A^{\mathrm{bd}}
			\right)
			W_N
		}_F
		=
		o_p(1).
		\label{eq:weighted_hat_A_tilde_A_bd_rate}
	\end{align}
	
We next establish the inverse bound. By construction,
$W_N\widetilde A^{\mathrm{bd}}W_N$ is block diagonal.Its nonstationary block is well-conditioned by \eqref{eq:tilde_Azz_conditioning}, its stationary factor block is well-conditioned by \eqref{eq:tilde_Aww_conditioning}, and its scaled idiosyncratic block is well-conditioned by \eqref{eq:tilde_Sigma_u_EW}. Since the eigenvalues of a block-diagonal matrix are the eigenvalues of its diagonal blocks,
it follows that
\begin{align}
	\EW{N}{
		W_N\widetilde A^{\mathrm{bd}}W_N
	}
	&=
	O_e(1),
	&
	\EW{1}{
		W_N\widetilde A^{\mathrm{bd}}W_N
	}
	&=
	O_e(1).
	\label{eq:tilde_A_bd_scaled_conditioning}
\end{align}
Consequently,
\begin{align}
	\norm{
		\left(
		W_N\widetilde A^{\mathrm{bd}}W_N
		\right)^{-1}
	}_2
	&=
	\frac{1}{
		\EW{N}{
			W_N\widetilde A^{\mathrm{bd}}W_N
		}
	}
	=
	O_p(1).
	\label{eq:tilde_A_bd_scaled_inverse}
\end{align}
	
	Moreover, \eqref{eq:hat_A_tilde_A},
	\eqref{eq:Norm_I_diffA}, and
	\eqref{eq:tilde_AW_convergence} give
	\begin{align}
		&\norm{
			\left(
			W_N\hat A W_N
			\right)^{-1}
		}_2 =
		\norm{
			W_N^{-1}\hat A^{-1}W_N^{-1}
		}_2 \leq
		\norm{
			W_N^{-1}\tilde A^{-1/2}
		}_2^2
		\norm{
			\left(
			I+
			\tilde A^{-1/2}
			\Delta A
			\tilde A^{-1/2}
			\right)^{-1}
		}_2
		=
		O_p(1).
		\label{eq:hat_A_scaled_inverse_bd}
	\end{align}
	
	Finally, the same trace reduction as in
	\eqref{eq:trace_reduction_A} yields
	\begin{align*}
		&\norm{
			\left(
			\widetilde{\Sigma}_{\tilde y,T}^{\mathrm{bd}}
			\right)^{-1/2}
			\Delta\Sigma_{y,T}^{\mathrm{bd}}
			\widehat{\Sigma}_{y,T}^{-1/2}
		}_F^2
		\\
		&\quad=
		\tr\left[
		\left(
		W_N\widetilde A^{\mathrm{bd}}W_N
		\right)^{-1}
		W_N
		\left(
		\hat A-\widetilde A^{\mathrm{bd}}
		\right)
		W_N
		\left(
		W_N\hat A W_N
		\right)^{-1}
		W_N
		\left(
		\hat A-\widetilde A^{\mathrm{bd}}
		\right)
		W_N
		\right]
		\\
		&\quad\leq
		\norm{
			\left(
			W_N\widetilde A^{\mathrm{bd}}W_N
			\right)^{-1}
		}_2
		\norm{
			\left(
			W_N\hat A W_N
			\right)^{-1}
		}_2
		\norm{
			W_N
			\left(
			\hat A-\widetilde A^{\mathrm{bd}}
			\right)
			W_N
		}_F^2
		\\
		&=
		o_p(1),
	\end{align*}
	where we used
	\eqref{eq:weighted_hat_A_tilde_A_bd_rate},
	\eqref{eq:tilde_A_bd_scaled_inverse}, and
	\eqref{eq:hat_A_scaled_inverse_bd}.
	Taking square roots proves the claim.
\end{proof}

\subsubsection{Eigenvalue orders and relative gaps}

The relative eigenspace perturbation bounds used in
Theorems~\ref{T:Factor_space} and \ref{T:Factor_space_2} require
two ingredients. First, the perturbations between the empirical and
artificial covariance matrices must be small relative to the scales of
their blocks. These bounds were established in
Lemmas~\ref{L:Sigma_y_hat_tilde_diff_N}-%
\ref{L:Sigma_y_hat_tilde_bd_diff_N}. Second, the eigenvalue orders of
the empirical and artificial covariance matrices must be determined.
The following lemma collects these orders.

\begin{lemma}[Eigenvalue orders of the level covariance]
	\label{L:level_covariance_eigenvalue_orders}
	Suppose Assumptions~\ref{ass:factors}-\ref{ass:memory} hold and
	that $N,T\to\infty$ with $N/T\to0$. Then, for every
	nonstationary group $j=1,\ldots,q_0^*$,
	\begin{align}
		\EW{i}{
			\widehat{\Sigma}_{y,T}
		}
		&=
		O_e\left(
		NT^{2d_0^{(j)}-1}
		\right),
		\qquad
		s^{(1:j-1)}<i\leq s^{(1:j)}.
		\label{eq:level_covariance_nonstationary_eigenvalue_orders}
	\end{align}
	If stationary factors are present, then
	\begin{align}
		\EW{i}{
			\widehat{\Sigma}_{y,T}
		}
		&=
		O_e(N),
		\qquad
		s_0^*<i\leq s_0.
		\label{eq:level_covariance_stationary_eigenvalue_orders}
	\end{align}
	Moreover,
	\begin{align}
		\EW{i}{
			\widehat{\Sigma}_{y,T}
		}
		&=
		O_e(1),
		\qquad
		s_0<i\leq N.
		\label{eq:level_covariance_idiosyncratic_eigenvalue_orders}
	\end{align}
	The same eigenvalue orders hold with
	$\widehat{\Sigma}_{y,T}$ replaced by any of the artificial
	covariance matrices
	\[
	\widetilde{\Sigma}_{\tilde y,T},
	\qquad
	\widetilde{\widetilde{\Sigma}}_{\tilde{\tilde y},T},
	\qquad
	\widetilde{\Sigma}_{\tilde y,T}^{\mathrm{bd}}.
	\]
\end{lemma}

\begin{proof}
	We first establish the eigenvalue orders of
	$\widetilde{\Sigma}_{\tilde y,T}$. By the definition of $\tilde A$ in
	\eqref{eq:tilde_A_definition},
	\begin{align*}
		Q_{all,0}^T
		\widetilde{\Sigma}_{\tilde y,T}
		Q_{all,0}
		&=
		ND_{T,0}\tilde A D_{T,0}
		\nonumber\\
		&=
		\left(
		\sqrt N D_{T,0}W_N^{-1}
		\right)
		\left(
		W_N\tilde A W_N
		\right)
		\left(
		W_N^{-1}D_{T,0}\sqrt N
		\right).
	\end{align*}
	
	By \eqref{eq:tilde_A_scaled_conditioning}, the eigenvalues of
	$W_N\tilde A W_N$ are bounded away from zero and infinity in
	probability. Hence, for every $\varepsilon>0$, there exist constants
	$0<c<C<\infty$ such that
	\begin{align}
		\Pr\left(
		cI_N
		\leq
		W_N\tilde A W_N
		\leq
		CI_N
		\right)
		&\geq
		1-\varepsilon
		\label{eq:tilde_A_scaled_Loewner_bound}
	\end{align}
	for all sufficiently large $N$ and $T$. Premultiplying the matrix
	inequality in \eqref{eq:tilde_A_scaled_Loewner_bound} by
	$\sqrt N D_{T,0}W_N^{-1}$ and postmultiplying it by its transpose
	gives, on this event,
	\begin{align}
		cND_{T,0}W_N^{-2}D_{T,0}
		&\leq
		Q_{all,0}^T
		\widetilde{\Sigma}_{\tilde y,T}
		Q_{all,0}
		\leq
		CND_{T,0}W_N^{-2}D_{T,0}.
		\label{eq:tilde_covariance_Loewner_bound}
	\end{align}
	
	By the definitions of $D_{T,0}$ and $W_N$,
	\begin{align}
		ND_{T,0}W_N^{-2}D_{T,0}
		&=
		\operatorname{diag}
		\left(
		ND_{z,T,0}^2,
		NI_{s_0-s_0^*},
		I_{N-s_0}
		\right).
		\label{eq:tilde_covariance_order_matrix}
	\end{align}
	The diagonal entries corresponding to nonstationary group $j$ are
	of order
	\[
	NT^{2d_0^{(j)}-1},
	\]
	the entries corresponding to the stationary factor block are of
	order $N$, and the entries corresponding to the idiosyncratic block
	are of order one.
	
	Because $Q_{all,0}$ is orthogonal, it does not change the
	eigenvalues. Moreover, the factors are ordered by decreasing memory,
	so the diagonal entries in
	\eqref{eq:tilde_covariance_order_matrix} are ordered by their
	asymptotic magnitude. By the Courant-Fischer minimax
	characterization, the matrix inequalities in
	\eqref{eq:tilde_covariance_Loewner_bound} imply the corresponding
	inequalities for each ordered eigenvalue; see
	\citet[Theorem~1.3.2]{Tao2012}. Therefore, for every
	$j=1,\ldots,q_0^*$,
	\begin{align}
		\EW{i}{
			\widetilde{\Sigma}_{\tilde y,T}
		}
		&=
		O_e\left(
		NT^{2d_0^{(j)}-1}
		\right),
		&&
		\sum_{\ell=1}^{j-1}s_\ell
		<
		i
		\leq
		\sum_{\ell=1}^{j}s_\ell,
		\label{eq:tilde_covariance_nonstationary_eigenvalue_orders}
	\end{align}
	while, if stationary factors are present,
	\begin{align*}
		\EW{i}{
			\widetilde{\Sigma}_{\tilde y,T}
		}
		&=
		O_e(N),
		&&
		s_0^*<i\leq s_0,
	\end{align*}
	and
	\begin{align}
		\EW{i}{
			\widetilde{\Sigma}_{\tilde y,T}
		}
		&=
		O_e(1),
		&&
		s_0<i\leq N.
		\label{eq:tilde_covariance_idiosyncratic_eigenvalue_orders}
	\end{align}
	
	The same argument applies to
	$\widetilde{\widetilde{\Sigma}}_{\tilde{\tilde y},T}$. Indeed,
	\eqref{eq:tildetilde_A_scaled_conditioning} shows that the
	eigenvalues of
	\[
	W_N\tilde{\tilde A}W_N
	\]
	are bounded away from zero and infinity in probability. Moreover,
	by the definition of $\tilde{\tilde A}$,
	\begin{align*}
		Q_{all,0}^T
		\widetilde{\widetilde{\Sigma}}_{\tilde{\tilde y},T}
		Q_{all,0}
		&=
		\left(
		\sqrt N D_{T,0}W_N^{-1}
		\right)
		\left(
		W_N\tilde{\tilde A}W_N
		\right)
		\left(
		W_N^{-1}D_{T,0}\sqrt N
		\right).
	\end{align*}
	The diagonal order matrix generated by this congruence is again
	\[
	ND_{T,0}W_N^{-2}D_{T,0}.
	\]
	Hence,
	$\widetilde{\widetilde{\Sigma}}_{\tilde{\tilde y},T}$ has the same
	eigenvalue orders as those in
	\eqref{eq:tilde_covariance_nonstationary_eigenvalue_orders}-%
	\eqref{eq:tilde_covariance_idiosyncratic_eigenvalue_orders}.
	
	Similarly, \eqref{eq:tilde_A_bd_scaled_conditioning} shows that the
	eigenvalues of
	\[
	W_N\tilde A^{\mathrm{bd}}W_N
	\]
	are bounded away from zero and infinity in probability. Since
	\begin{align*}
		Q_{all,0}^T
		\widetilde{\Sigma}_{\tilde y,T}^{\mathrm{bd}}
		Q_{all,0}
		&=
		\left(
		\sqrt N D_{T,0}W_N^{-1}
		\right)
		\left(
		W_N\tilde A^{\mathrm{bd}}W_N
		\right)
		\left(
		W_N^{-1}D_{T,0}\sqrt N
		\right),
	\end{align*}
	the eigenvalues of
	$\widetilde{\Sigma}_{\tilde y,T}^{\mathrm{bd}}$ have the same
	orders as well.
	
	It remains to transfer these orders to
	$\widehat{\Sigma}_{y,T}$. By \eqref{eq:diff_A_tilde_A},
	\begin{align*}
		\norm{
			\tilde A^{-1/2}
			\left(
			\hat A-\tilde A
			\right)
			\tilde A^{-1/2}
		}_2
		&\leq
		\norm{
			\tilde A^{-1/2}
			\left(
			\hat A-\tilde A
			\right)
			\tilde A^{-1/2}
		}_F
		=
		o_p(1).
	\end{align*}
	Hence, for every fixed $\varepsilon\in(0,1)$, with probability
	approaching one,
	\begin{align}
		-\varepsilon I_N
		&\leq
		\tilde A^{-1/2}
		\left(
		\hat A-\tilde A
		\right)
		\tilde A^{-1/2}
		\leq
		\varepsilon I_N.
		\label{eq:hat_tilde_relative_Loewner_bound}
	\end{align}
	Premultiplying the matrix inequality in
	\eqref{eq:hat_tilde_relative_Loewner_bound} by $\tilde A^{1/2}$ and
	postmultiplying it by its transpose gives
	\begin{align}
		(1-\varepsilon)\tilde A
		&\leq
		\hat A
		\leq
		(1+\varepsilon)\tilde A.
		\label{eq:hat_tilde_A_Loewner_equivalence}
	\end{align}
	Using
	\[
	\widehat{\Sigma}_{y,T}
	=
	NQ_{all,0}D_{T,0}
	\hat A
	D_{T,0}Q_{all,0}^T
	\]
	and
	\[
	\widetilde{\Sigma}_{\tilde y,T}
	=
	NQ_{all,0}D_{T,0}
	\tilde A
	D_{T,0}Q_{all,0}^T,
	\]
	premultiplying the matrix inequality in
	\eqref{eq:hat_tilde_A_Loewner_equivalence} by
	$\sqrt N Q_{all,0}D_{T,0}$ and postmultiplying it by its transpose
	yields
	\begin{align*}
		(1-\varepsilon)
		\widetilde{\Sigma}_{\tilde y,T}
		&\leq
		\widehat{\Sigma}_{y,T}
		\leq
		(1+\varepsilon)
		\widetilde{\Sigma}_{\tilde y,T}.
	\end{align*}
	Another application of the Courant-Fischer characterization gives
	\begin{align}
		(1-\varepsilon)
		\EW{i}{
			\widetilde{\Sigma}_{\tilde y,T}
		}
		&\leq
		\EW{i}{
			\widehat{\Sigma}_{y,T}
		}
		\leq
		(1+\varepsilon)
		\EW{i}{
			\widetilde{\Sigma}_{\tilde y,T}
		}
		\label{eq:hat_tilde_eigenvalue_equivalence}
	\end{align}
	for every $i=1,\ldots,N$, with probability approaching one.
	Combining
	\eqref{eq:hat_tilde_eigenvalue_equivalence} with
	\eqref{eq:tilde_covariance_nonstationary_eigenvalue_orders}-%
	\eqref{eq:tilde_covariance_idiosyncratic_eigenvalue_orders}
	proves the stated eigenvalue orders for
	$\widehat{\Sigma}_{y,T}$.
\end{proof}

We next record the relative eigenvalue gap rates used in
Theorems~\ref{T:Factor_space} and \ref{T:Factor_space_2}.

For the separation of the nonstationary factor space from its
complement, define
\begin{align}
	\rho_{z,T}
	&=
	\min_{\substack{
			\hat\kappa
			\in
			\EW{1:s_0^*}{
				\widehat{\Sigma}_{y,T}
			}
			\\
			\tilde\kappa
			\in
			\EW{(s_0^*+1):N}{
				\widetilde{\Sigma}_{\tilde y,T}
			}
	}}
	\frac{
		\left|
		\tilde\kappa-\hat\kappa
		\right|
	}{
		\sqrt{
			\tilde\kappa\hat\kappa
		}
	}.
	\label{eq:relative_gap_z_definition}
\end{align}
If stationary factors are present, Lemma
\ref{L:level_covariance_eigenvalue_orders} shows that the smallest
eigenvalue in the nonstationary factor block is of order
\[
NT^{2d_0^{(q_0^*)}-1},
\]
whereas the largest eigenvalue in its complement is of order $N$.
Therefore,
\begin{align*}
	\rho_{z,T}
	&=
	O_e\left(
	T^{d_0^{(q_0^*)}-1/2}
	\right),
	&
	\rho_{z,T}^{-1}
	&=
	O_p\left(
	T^{-d_0^{(q_0^*)}+1/2}
	\right).
\end{align*}
If all factors are nonstationary, the complementary eigenvalues are
of order one. In that case, the relative gap is larger and
\begin{align*}
	\rho_{z,T}
	&=
	O_e\left(
	N^{1/2}T^{d_0^{(q_0^*)}-1/2}
	\right),
	&
	\rho_{z,T}^{-1}
	&=
	O_p\left(
	N^{-1/2}T^{-d_0^{(q_0^*)}+1/2}
	\right).
\end{align*}
Thus, in both cases,
\begin{align}
	\rho_{z,T}^{-1}
	&=
	O_p\left(
	T^{-d_0^{(q_0^*)}+1/2}
	\right).
	\label{eq:eigenvalue_order_gap_z_uniform}
\end{align}

Suppose next that stationary factors are present. For the separation
of the full factor space from the idiosyncratic space, define
\begin{align}
	\rho_{0,N}
	&=
	\min_{\substack{
			\hat\kappa
			\in
			\EW{1:s_0}{
				\widehat{\Sigma}_{y,T}
			}
			\\
			\tilde\kappa
			\in
			\EW{(s_0+1):N}{
				\widetilde{\Sigma}_{\tilde y,T}^{\mathrm{bd}}
			}
	}}
	\frac{
		\left|
		\tilde\kappa-\hat\kappa
		\right|
	}{
		\sqrt{
			\tilde\kappa\hat\kappa
		}
	}.
	\label{eq:relative_gap_factor_idiosyncratic_definition}
\end{align}
By Lemma~\ref{L:level_covariance_eigenvalue_orders}, the smallest
eigenvalue in the full factor block is of order $N$, whereas the
largest eigenvalue in the idiosyncratic block is of order one.
Consequently,
\begin{align}
	\rho_{0,N}
	&=
	O_e\left(
	N^{1/2}
	\right),
	&
	\rho_{0,N}^{-1}
	&=
	O_p\left(
	N^{-1/2}
	\right).
	\label{eq:eigenvalue_order_gap_factor_idiosyncratic}
\end{align}

Finally, fix $j\in\{1,\ldots,q_0^*-1\}$. For the separation of the
first $j$ nonstationary factor groups from their complement, define
\begin{align}
	\label{eq:relative_gap_adjacent_groups_definition}
\rho_{j,T}
=
\min_{\substack{
		\hat\kappa
		\in
		\EW{1:s^{(1:j)}}{
			\widehat{\Sigma}_{y,T}
		}
		\\
		\tilde\kappa
		\in
		\EW{(s^{(1:j)}+1):N}{
			\widetilde{\widetilde{\Sigma}}_{\tilde{\tilde y},T}
		}
}}
\frac{
	\left|
	\tilde\kappa-\hat\kappa
	\right|
}{
	\sqrt{
		\tilde\kappa\hat\kappa
	}
}.
\end{align}
Lemma~\ref{L:level_covariance_eigenvalue_orders} shows that the
smallest eigenvalue associated with the first $j$ nonstationary
groups is of order
\[
NT^{2d_0^{(j)}-1},
\]
whereas the largest eigenvalue in their complement is of order
\[
NT^{2d_0^{(j+1)}-1}.
\]
Since $d_0^{(j)}>d_0^{(j+1)}$, it follows that
\begin{align}
	\rho_{j,T}
	&=
	O_e\left(
	T^{d_0^{(j)}-d_0^{(j+1)}}
	\right),
	&
	\rho_{j,T}^{-1}
	&=
	O_p\left(
	T^{-d_0^{(j)}+d_0^{(j+1)}}
	\right).
	\label{eq:eigenvalue_order_gap_adjacent_groups}
\end{align}

The preceding relative gaps are used for cumulative eigenspace
comparisons. For Corollary~\ref{cor:factorspaces}, we additionally
require direct bounds for the overlap between individual reference and
estimated eigenvalue blocks. The following lemma gives the
corresponding blockwise perturbation bound. Its proof follows the
Sylvester-equation argument in
\citet[Theorem~2.4]{Dopico2000}.

\begin{lemma}[Blockwise Dopico bound]
	\label{L:Dopico_Extension}
	Let $j,k\in\{1,\ldots,q_0+1\}$ with $j\neq k$, and define
	\[
	\rho^{(j,k)}
	=
	\min_{\substack{
			\tilde\kappa
			\in
			\EW{(j)}{
				\widetilde{\widetilde{\Sigma}}_{\tilde{\tilde y},T}
			}
			\\
			\hat\kappa
			\in
			\EW{(k)}{
				\widehat{\Sigma}_{y,T}
			}
	}}
	\frac{
		\left|
		\tilde\kappa-\hat\kappa
		\right|
	}{
		\sqrt{
			\tilde\kappa\hat\kappa
		}
	}.
	\]
	If $\rho^{(j,k)}>0$, then
	\begin{align}
		\norm{
			V\left(
			\widetilde{\widetilde{\Sigma}}_{\tilde{\tilde y},T}
			\right)^{(j)T}
			V\left(
			\widehat{\Sigma}_{y,T}
			\right)^{(k)}
		}_F
		&\leq
		\frac{1}{\rho^{(j,k)}}
		\norm{
			\widetilde{\widetilde{\Sigma}}_{\tilde{\tilde y},T}^{-1/2}
			\left(
			\widehat{\Sigma}_{y,T}
			-
			\widetilde{\widetilde{\Sigma}}_{\tilde{\tilde y},T}
			\right)
			\widehat{\Sigma}_{y,T}^{-1/2}
		}_F.
		\label{eq:Dopico_Extension}
	\end{align}
	The same bound holds with
	$\widetilde{\widetilde{\Sigma}}_{\tilde{\tilde y},T}$
	replaced throughout by
	$\widetilde{\Sigma}^{\mathrm{bd}}_{\tilde y,T}$.
\end{lemma}

\begin{proof}
	We prove the result for
	$\widetilde{\widetilde{\Sigma}}_{\tilde{\tilde y},T}$; the proof
	for $\widetilde{\Sigma}_{\tilde y,T}^{\mathrm{bd}}$ is identical.
	Fix distinct $j,k\in\{1,\ldots,q_0+1\}$. To simplify notation
	within the proof, write
	\[
	\tilde V^{(j)}
	=
	V\left(
	\widetilde{\widetilde{\Sigma}}_{\tilde{\tilde y},T}
	\right)^{(j)},
	\qquad
	\hat V^{(k)}
	=
	V\left(
	\widehat{\Sigma}_{y,T}
	\right)^{(k)},
	\]
	and suppress the matrix arguments of the corresponding eigenvalue
	vectors by setting
	\[
	\EWg{j}
	=
	\EWg{j}\left(
	\widetilde{\widetilde{\Sigma}}_{\tilde{\tilde y},T}
	\right),
	\qquad
	\EWg{k}
	=
	\EWg{k}\left(
	\widehat{\Sigma}_{y,T}
	\right).
	\]
	Thus,
	\begin{align}
		\widetilde{\widetilde{\Sigma}}_{\tilde{\tilde y},T}
		\tilde V^{(j)}
		&=
		\tilde V^{(j)}
		\EWgtD{j},
		&
		\widehat{\Sigma}_{y,T}
		\hat V^{(k)}
		&=
		\hat V^{(k)}
		\EWghD{k}.
		\label{eq:blockwise_Dopico_eigenvalue_equations}
	\end{align}
	
	The second identity in
	\eqref{eq:blockwise_Dopico_eigenvalue_equations} implies
	\begin{align*}
		\widetilde{\widetilde{\Sigma}}_{\tilde{\tilde y},T}
		\hat V^{(k)}
		-
		\hat V^{(k)}
		\EWghD{k}
		&=
		-
		\left(
		\widehat{\Sigma}_{y,T}
		-
		\widetilde{\widetilde{\Sigma}}_{\tilde{\tilde y},T}
		\right)
		\hat V^{(k)}.
	\end{align*}
	Premultiplying by $\tilde V^{(j)T}$ and using the first identity in
	\eqref{eq:blockwise_Dopico_eigenvalue_equations} gives
	\begin{align}
		\EWgtD{j}
		\tilde V^{(j)T}\hat V^{(k)}
		-
		\tilde V^{(j)T}\hat V^{(k)}
		\EWghD{k}
		&=
		-
		\tilde V^{(j)T}
		\left(
		\widehat{\Sigma}_{y,T}
		-
		\widetilde{\widetilde{\Sigma}}_{\tilde{\tilde y},T}
		\right)
		\hat V^{(k)}.
		\label{eq:blockwise_Dopico_residual_first}
	\end{align}
	
	Moreover,
	\begin{align}
		\tilde V^{(j)T}
		\left(
		\widehat{\Sigma}_{y,T}
		-
		\widetilde{\widetilde{\Sigma}}_{\tilde{\tilde y},T}
		\right)
		\hat V^{(k)}
		&=
		\tilde V^{(j)T}
		\widetilde{\widetilde{\Sigma}}_{\tilde{\tilde y},T}^{1/2}
		\left[
		\widetilde{\widetilde{\Sigma}}_{\tilde{\tilde y},T}^{-1/2}
		\left(
		\widehat{\Sigma}_{y,T}
		-
		\widetilde{\widetilde{\Sigma}}_{\tilde{\tilde y},T}
		\right)
		\widehat{\Sigma}_{y,T}^{-1/2}
		\right]
		\widehat{\Sigma}_{y,T}^{1/2}
		\hat V^{(k)}
		\nonumber\\
		&=
		\EWgtD{j}^{1/2}
		G^{(j,k)}
		\EWghD{k}^{1/2},
		\label{eq:blockwise_Dopico_weighted_perturbation}
	\end{align}
	where
	\[
	G^{(j,k)}
	=
	\tilde V^{(j)T}
	\widetilde{\widetilde{\Sigma}}_{\tilde{\tilde y},T}^{-1/2}
	\left(
	\widehat{\Sigma}_{y,T}
	-
	\widetilde{\widetilde{\Sigma}}_{\tilde{\tilde y},T}
	\right)
	\widehat{\Sigma}_{y,T}^{-1/2}
	\hat V^{(k)}.
	\]
	Combining
	\eqref{eq:blockwise_Dopico_residual_first} and
	\eqref{eq:blockwise_Dopico_weighted_perturbation} yields the
	Sylvester equation
	\begin{align}
		\EWgtD{j}
		\tilde V^{(j)T}\hat V^{(k)}
		-
		\tilde V^{(j)T}\hat V^{(k)}
		\EWghD{k}
		&=
		-
		\EWgtD{j}^{1/2}
		G^{(j,k)}
		\EWghD{k}^{1/2}.
		\label{eq:blockwise_Dopico_Sylvester}
	\end{align}
	
	Since $\EWgtD{j}$ and $\EWghD{k}$ are diagonal,
	\eqref{eq:blockwise_Dopico_Sylvester} holds entrywise. For the
	entry corresponding to the $i$-th eigenvalue in reference block
	$j$ and the $\ell$-th eigenvalue in estimated block $k$,
	\begin{align*}
		\left(
		\EWgti{j}{i}
		-
		\EWghi{k}{\ell}
		\right)
		\left(
		\tilde V^{(j)T}\hat V^{(k)}
		\right)_{i\ell}
		&=
		-
		\sqrt{
			\EWgti{j}{i}
			\EWghi{k}{\ell}
		}
		G_{i\ell}^{(j,k)}.
	\end{align*}
	Therefore,
	\begin{align*}
		\left|
		\left(
		\tilde V^{(j)T}\hat V^{(k)}
		\right)_{i\ell}
		\right|
		&=
		\frac{
			\sqrt{
				\EWgti{j}{i}
				\EWghi{k}{\ell}
			}
		}{
			\left|
			\EWgti{j}{i}
			-
			\EWghi{k}{\ell}
			\right|
		}
		\left|
		G_{i\ell}^{(j,k)}
		\right|
		\\
		&\leq
		\frac{1}{\rho^{(j,k)}}
		\left|
		G_{i\ell}^{(j,k)}
		\right|.
	\end{align*}
	Summing over all entries gives
	\begin{align}
		\norm{
			\tilde V^{(j)T}\hat V^{(k)}
		}_F
		&\leq
		\frac{1}{\rho^{(j,k)}}
		\norm{
			G^{(j,k)}
		}_F.
		\label{eq:blockwise_Dopico_G_bound}
	\end{align}
	Finally, since $\tilde V^{(j)}$ and $\hat V^{(k)}$ have
	orthonormal columns,
	\begin{align}
		\norm{
			G^{(j,k)}
		}_F
		&\leq
		\norm{
			\widetilde{\widetilde{\Sigma}}_{\tilde{\tilde y},T}^{-1/2}
			\left(
			\widehat{\Sigma}_{y,T}
			-
			\widetilde{\widetilde{\Sigma}}_{\tilde{\tilde y},T}
			\right)
			\widehat{\Sigma}_{y,T}^{-1/2}
		}_F.
		\label{eq:blockwise_Dopico_middle_bound}
	\end{align}
	Combining
	\eqref{eq:blockwise_Dopico_G_bound} and
	\eqref{eq:blockwise_Dopico_middle_bound} proves
	\eqref{eq:Dopico_Extension}.
\end{proof}
\clearpage
\section{Diminishing signals}
\label{sec:diminishing_signals_appendix}

This section establishes the factor-space and model-selection results
under diminishing signals. In contrast to the standard-signal regime,
the innovations of the nonstationary factors are scaled with $T$ so
that their sample second moments no longer diverge. Consequently, the
nonstationary and stationary factor spaces are no longer separated by
different orders in $T$. The full factor space remains separated from
the idiosyncratic component, however, because the factors are pervasive
in the cross section.

We first derive the covariance orders that replace the corresponding
standard-signal results. We then use these orders to establish the block
structure of the sample covariance matrix and the resulting PCA and
model-selection properties.

For readability, we suppress the additional $T$-subscript used in the
diminishing-signal regime and write $x_{i,t}$ instead of $x_{i,t,T}$.

\begin{lemma}[Sample covariance matrix of the factors under diminishing signals]
	\label{L:diminishing_factor_second_moments}
	Suppose Assumptions \ref{ass:factors}, \ref{ass:one_indep_all},
	\ref{ass:memory}, and \ref{ass:diminishing_signals} hold. Then
	\[
	\widetilde{\Sigma}_{x}
	=
	\frac{1}{T}
	\sum_{t=1}^T
	x_tx_t^T
	\ \Rightarrow\
	\begin{pmatrix}
		S_z & 0
		\\
		0 & S_w
	\end{pmatrix},
	\]
where $S_z$ and $S_w$ are the limiting matrices in \eqref{eq:nonstationary_factor_covariance_limit} and \eqref{eq:stationary_factor_covariance_limit}, respectively. The limiting matrix is positive definite almost
	surely. Consequently,
	\[
	\EW{j}{
		\widetilde{\Sigma}_{x}
	}
	=
	O_e(1),
	\qquad
	j=1,\ldots,s_0.
	\]
\end{lemma}

\begin{proof}
	Let $z_t^{\circ}$ denote the corresponding nonstationary factor vector
	under the standard-signal regime, that is, with unscaled innovations,
	and let $x_{i,t}^{\circ}$ denote its $i$th coordinate.
	Under Assumption~\ref{ass:diminishing_signals},
	\[
	z_t
	=
	D_{z,T,0}^{-1}z_t^{\circ}.
	\]
	Therefore, the nonstationary block of the factor sample covariance matrix
	satisfies
	\[
	\frac{1}{T}
	\sum_{t=1}^T
	z_tz_t^T
	=
	D_{z,T,0}^{-1}
	\left(
	\frac{1}{T}
	\sum_{t=1}^T
	z_t^{\circ}z_t^{\circ T}
	\right)
	D_{z,T,0}^{-1}
	\Rightarrow
	S_z
	\]
by \eqref{eq:nonstationary_factor_covariance_limit}. The stationary factors are unaffected by the diminishing-signal scaling, so \eqref{eq:stationary_factor_covariance_limit} gives
	\[
	\frac{1}{T}
	\sum_{t=1}^T
	w_tw_t^T
	\ \stackrel{p}{\longrightarrow}\
	S_w.
	\]
	
	It remains to consider the covariance between the nonstationary and
	stationary factors. For $i=1,\ldots,s_0^*$ and
	$i'=s_0^*+1,\ldots,s_0$, Assumption~\ref{ass:diminishing_signals} and
 Lemma~\ref{L:zw},
	specifically \eqref{eq:zw_scaled}, imply
	\begin{align*}
		\frac{1}{T}
		\sum_{t=1}^T
		x_{i,t}x_{i',t}
		&=
		T^{-d_{0,i}+1/2}
		\frac{1}{T}
		\sum_{t=1}^T
		x_{i,t}^{\circ}x_{i',t}
		\\
		&=
		O_p\left(
		T^{\max\{d_{0,i'},0\}-1/2}
		\right)
		=
		o_p(1),
	\end{align*}
	because $d_{0,i'}<1/2$. Since $s_0$ is fixed, this also gives
	\[
	\norm{
		\frac{1}{T}
		\sum_{t=1}^T
		z_tw_t^T
	}_F
	=
	o_p(1).
	\]
	Combining the three blockwise results yields
	\[
	\widetilde{\Sigma}_{x}
	\ \Rightarrow\
	\begin{pmatrix}
		S_z & 0
		\\
		0 & S_w
	\end{pmatrix}.
	\]
	
	The almost-sure positive definiteness of $S_z$ and the positive definiteness of $S_w$ were established together with \eqref{eq:nonstationary_factor_covariance_limit} and \eqref{eq:stationary_factor_covariance_limit}. Hence, the limiting block-diagonal matrix is positive definite almost surely. By the continuous mapping theorem, the eigenvalues of $\widetilde{\Sigma}_{x}$ converge to the eigenvalues of the limiting block-diagonal matrix. Since the latter are finite and strictly positive almost surely,
	\[
	\EW{j}{
		\widetilde{\Sigma}_{x}
	}
	=
	O_e(1),
	\qquad
	j=1,\ldots,s_0.
	\]
\end{proof}

\begin{lemma}[Factor-idiosyncratic covariance under diminishing signals]
	\label{L:diminishing_factor_idiosyncratic_covariance}
	Suppose Assumptions~\ref{ass:factors},
	\ref{ass:idio_errors}, \ref{ass:one_indep_all},
	\ref{ass:memory}, and \ref{ass:diminishing_signals} hold.
	Then, uniformly over $i=1,\ldots,s_0$ and
	$l=1,\ldots,N$,
	\begin{align}
		\frac{1}{T}
		\sum_{t=1}^T
		x_{i,t}u_{l,t}
		&=
		O_p\left(
		T^{-1/2}
		\right).
		\label{eq:diminishing_factor_idiosyncratic_covariance}
	\end{align}
	Consequently,
	\begin{align}
		\norm{
			\widetilde{\Sigma}_{xu}
		}_F
		&=
		O_p\left(
		\sqrt{\frac{N}{T}}
		\right).
		\label{eq:diminishing_factor_idiosyncratic_Frobenius}
	\end{align}
\end{lemma}

\begin{proof}
For a nonstationary coordinate, let $x_{i,t}^{\circ}$ denote the
corresponding factor under the standard-signal regime. By
Assumption~\ref{ass:diminishing_signals},
\[
x_{i,t}
=
T^{-d_{0,i}+1/2}x_{i,t}^{\circ}.
\]
Hence, Lemma~\ref{L:sum_z_u} gives
\[
\frac{1}{T}
\sum_{t=1}^T
x_{i,t}u_{l,t}
=
T^{-d_{0,i}+1/2}
\frac{1}{T}
\sum_{t=1}^T
x_{i,t}^{\circ}u_{l,t}
=
O_p(T^{-1/2}).
\]
	Moreover, the moment bounds underlying Lemmas~\ref{L:sum_z_u} and \ref{L:wu} imply
	\begin{align*}
		\sup_{\substack{i=1,\ldots,s_0\\l=1,\ldots,N}}
		\EE{
			\left|
			\frac{1}{T}
			\sum_{t=1}^T
			x_{i,t}u_{l,t}
			\right|^2
		}
		&=
		O\left(
		T^{-1}
		\right).
	\end{align*}
	Since $s_0$ is fixed,
	\[
	\begin{aligned}
		\EE{
			\norm{
				\widetilde{\Sigma}_{xu}
			}_F^2
		}
		&=
		\sum_{i=1}^{s_0}
		\sum_{l=1}^N
		\EE{
			\left|
			\frac{1}{T}
			\sum_{t=1}^T
			x_{i,t}u_{l,t}
			\right|^2
		}
		=
		O\left(
		\frac{N}{T}
		\right).
	\end{aligned}
	\]
	Markov's inequality therefore yields \eqref{eq:diminishing_factor_idiosyncratic_Frobenius}.
\end{proof}
The preceding lemmas provide the covariance bounds needed for the
perturbation argument. Since the nonstationary and stationary factors
are no longer separated by different covariance orders, we treat the
full factor space as a single block.

With the orthogonal basis $Q_{all,0}$ defined in Appendix~\ref{sec:notation},
\[
Q_{all,0}^Ty_t
=
\begin{pmatrix}
	R_0x_t+Q_0^Tu_t
	\\
	Q_{\perp,0}^Tu_t
\end{pmatrix}.
\]
Hence, in terms of the empirical covariance matrices,
\begin{align}
	\begin{split}
		Q_{all,0}^T
		\widehat{\Sigma}_{y,T}
		Q_{all,0}
		&=
		\begin{pmatrix}
			R_0\widetilde{\Sigma}_{x}R_0^T
			+
			R_0\widetilde{\Sigma}_{xu}Q_0
			+
			Q_0^T\widetilde{\Sigma}_{ux}R_0^T
			+
			Q_0^T\widetilde{\Sigma}_{u}Q_0
			&
			R_0\widetilde{\Sigma}_{xu}Q_{\perp,0}
			+
			Q_0^T\widetilde{\Sigma}_{u}Q_{\perp,0}
			\\
			Q_{\perp,0}^T\widetilde{\Sigma}_{ux}R_0^T
			+
			Q_{\perp,0}^T\widetilde{\Sigma}_{u}Q_0
			&
			Q_{\perp,0}^T\widetilde{\Sigma}_{u}Q_{\perp,0}
		\end{pmatrix}.
		\label{eq:scaling_hat_sigma_y}
	\end{split}
\end{align}
For the subsequent perturbation argument, define
\begin{align}
	\widehat A
	=
	\frac{1}{N}
	Q_{all,0}^T
	\widehat{\Sigma}_{y,T}
	Q_{all,0}
	&=
	\begin{pmatrix}
		\widehat{A}_{xx}(N,T)
		&
		\widehat{A}_{xu}(N,T)^T
		\\
		\widehat{A}_{xu}(N,T)
		&
		\widehat{A}_{uu}(N,T)
	\end{pmatrix}.
	\label{eq:scaling_hat_sigma_y_normalized}
\end{align}
\begin{lemma}[Block convergence under diminishing signals]
	\label{L:diminishing_block_convergence}
	Suppose Assumptions~\ref{ass:factors}-\ref{ass:memory} and
	\ref{ass:diminishing_signals} hold, and let $N,T\to\infty$ with
	$N/T\to0$. Then
	\begin{align}
		\norm{
			\widehat{A}_{xx}(N,T)
			-
			\frac{1}{N}
			R_0\widetilde{\Sigma}_{x}R_0^T
		}_F
		&=
		o_p(1),
		\label{eq:diminishing_xx_block_convergence}
		\\
		\norm{
			\widehat{A}_{xu}(N,T)
		}_F
		&=
		o_p(1),
		\label{eq:diminishing_xu_block_convergence}
		\\
		\norm{
			\widehat{A}_{uu}(N,T)
		}_F
		&=
		o_p(1).
		\label{eq:diminishing_uu_block_convergence}
	\end{align}
	Moreover,
	\begin{align}
		\EW{j}{
			\widehat{A}_{xx}(N,T)
		}
		&=
		O_e(1),
		\qquad
		j=1,\ldots,s_0,
		\label{eq:diminishing_sample_factor_block_eigenvalues}
		\\
		\EW{j}{
			\widehat{A}_{uu}(N,T)
		}
		&=
		O_e\left(
		N^{-1}
		\right),
		\qquad
		j=1,\ldots,N-s_0.
		\label{eq:diminishing_idiosyncratic_block_eigenvalues}
	\end{align}
\end{lemma}

\begin{proof}
	We first establish the three block convergence statements. By
	\eqref{eq:strong_factors_RR},
	\[
	\frac{1}{N}R_0^TR_0
	\longrightarrow
	\Sigma_{\Lambda,0},
	\]
	where $\Sigma_{\Lambda,0}$ is positive definite. Hence,
	\[
	\frac{1}{N}\norm{R_0}_F^2
	=
	\tr\left(
	\frac{1}{N}R_0^TR_0
	\right)
	=
	O(1),
	\]
	and therefore
	\[
	\norm{R_0}_F
	=
	O\left(
	N^{1/2}
	\right).
	\]
	
	By Lemma~\ref{L:diminishing_factor_idiosyncratic_covariance},
	\[
	\norm{
		\widetilde{\Sigma}_{xu}
	}_F
	=
	O_p\left(
	\sqrt{\frac{N}{T}}
	\right).
	\]
	Thus, for either $Q=Q_0$ or $Q=Q_{\perp,0}$, using
	$\norm{Q}=1$ gives
	\[
	\norm{
		\frac{1}{N}
		R_0\widetilde{\Sigma}_{xu}Q
	}_F
	\leq
	\frac{1}{N}
	\norm{R_0}_F
	\norm{
		\widetilde{\Sigma}_{xu}
	}_F
	=
	O_p\left(
	T^{-1/2}
	\right).
	\]
	The corresponding transpose terms satisfy the same bound.
	
	Assumption~\ref{ass:Eigenvalue_Sigma_u_tilde} and $N/T\to0$
	imply
	\[
	\norm{
		\widetilde{\Sigma}_{u}
	}
	=
	O_p(1).
	\]
	Since $Q_0$ and $Q_{\perp,0}$ have orthonormal columns, the
	projected idiosyncratic blocks satisfy
	\begin{align*}
		\norm{
			\frac{1}{N}
			Q_0^T\widetilde{\Sigma}_{u}Q_0
		}_F
		&\leq
		\frac{\sqrt{s_0}}{N}
		\norm{
			\widetilde{\Sigma}_{u}
		}
		=
		O_p\left(
		N^{-1}
		\right),
		\\
		\norm{
			\frac{1}{N}
			Q_{\perp,0}^T\widetilde{\Sigma}_{u}Q_0
		}_F
		&\leq
		\frac{\sqrt{s_0}}{N}
		\norm{
			\widetilde{\Sigma}_{u}
		}
		=
		O_p\left(
		N^{-1}
		\right),
		\\
		\norm{
			\frac{1}{N}
			Q_{\perp,0}^T
			\widetilde{\Sigma}_{u}
			Q_{\perp,0}
		}_F
		&\leq
		\frac{\sqrt{N-s_0}}{N}
		\norm{
			\widetilde{\Sigma}_{u}
		}
		=
		O_p\left(
		N^{-1/2}
		\right).
	\end{align*}
	
	Applying these bounds to the block representation in
	\eqref{eq:scaling_hat_sigma_y}-%
	\eqref{eq:scaling_hat_sigma_y_normalized} yields
	\begin{align*}
		\norm{
			\widehat{A}_{xx}(N,T)
			-
			\frac{1}{N}
			R_0\widetilde{\Sigma}_{x}R_0^T
		}_F
		&=
		O_p\left(
		T^{-1/2}
		\right)
		+
		O_p\left(
		N^{-1}
		\right)
		=
		o_p(1),
		\\
		\norm{
			\widehat{A}_{xu}(N,T)
		}_F
		&=
		O_p\left(
		T^{-1/2}
		\right)
		+
		O_p\left(
		N^{-1}
		\right)
		=
		o_p(1),
		\\
		\norm{
			\widehat{A}_{uu}(N,T)
		}_F
		&=
		O_p\left(
		N^{-1/2}
		\right)
		=
		o_p(1).
	\end{align*}
	
	It remains to establish the eigenvalue statements. By
	Lemma~\ref{L:diminishing_factor_second_moments},
	\[
	\EW{j}{
		\widetilde{\Sigma}_{x}
	}
	=
	O_e(1),
	\qquad
	j=1,\ldots,s_0.
	\]
	Together with \eqref{eq:strong_factors_RR}, the eigenvalue bounds
	for a congruence transformation give, for every
	$j=1,\ldots,s_0$,
	\begin{align*}
		\EW{s_0}{
			\frac{1}{N}R_0^TR_0
		}
		\EW{s_0}{
			\widetilde{\Sigma}_{x}
		}
		&\leq
		\EW{j}{
			\frac{1}{N}
			R_0\widetilde{\Sigma}_{x}R_0^T
		}
		\leq
		\EW{1}{
			\frac{1}{N}R_0^TR_0
		}
		\EW{1}{
			\widetilde{\Sigma}_{x}
		}.
	\end{align*}
	Hence,
	\[
	\EW{j}{
		\frac{1}{N}
		R_0\widetilde{\Sigma}_{x}R_0^T
	}
	=
	O_e(1),
	\qquad
	j=1,\ldots,s_0.
	\]
	Weyl's inequality and
	\eqref{eq:diminishing_xx_block_convergence} then imply
	\[
	\max_{1\leq j\leq s_0}
	\left|
	\EW{j}{
		\widehat{A}_{xx}(N,T)
	}
	-
	\EW{j}{
		\frac{1}{N}
		R_0\widetilde{\Sigma}_{x}R_0^T
	}
	\right|
	=
	o_p(1).
	\]
	Therefore,
	\[
	\EW{j}{
		\widehat{A}_{xx}(N,T)
	}
	=
	O_e(1),
	\qquad
	j=1,\ldots,s_0.
	\]
	
	Finally, by the definition of the lower-right block,
	\[
	N\widehat{A}_{uu}(N,T)
	=
	Q_{\perp,0}^T
	\widetilde{\Sigma}_{u}
	Q_{\perp,0}.
	\]
	The Poincar\'e separation theorem gives
	\[
	\EW{N}{
		\widetilde{\Sigma}_{u}
	}
	\leq
	\EW{j}{
		N\widehat{A}_{uu}(N,T)
	}
	\leq
	\EW{1}{
		\widetilde{\Sigma}_{u}
	},
	\qquad
	j=1,\ldots,N-s_0.
	\]
	Assumption~\ref{ass:Eigenvalue_Sigma_u_tilde} therefore yields
	\[
	\EW{j}{
		\widehat{A}_{uu}(N,T)
	}
	=
	O_e\left(
	N^{-1}
	\right),
	\qquad
	j=1,\ldots,N-s_0.
	\]
	This completes the proof.
\end{proof}

The preceding lemma describes the asymptotic block structure of the
normalized covariance matrix in the basis $Q_{all,0}$. In particular,
\eqref{eq:diminishing_sample_factor_block_eigenvalues} shows that the
factor block is well-conditioned,
\eqref{eq:diminishing_xu_block_convergence} shows that the
off-diagonal block vanishes, and
\eqref{eq:diminishing_idiosyncratic_block_eigenvalues} shows that the
idiosyncratic block has eigenvalues of order $N^{-1}$. Hence, after
rescaling the orthogonal-complement block by $N^{1/2}$, both diagonal
blocks have eigenvalues of order one.

For the relative perturbation argument, we now construct an exactly
block-diagonal reference matrix in the basis $Q_{all,0}$. In the
factor block, we retain the factor covariance and the projected
idiosyncratic covariance but omit the factor-idiosyncratic cross
terms. We set the two off-diagonal blocks equal to zero and leave the
idiosyncratic block unchanged. The resulting matrix has $\operatorname{span}(Q_0)$ and
$\operatorname{span}(Q_{\perp,0})$ as exact invariant subspaces. The
next lemma shows that the removed terms form a vanishing relative
perturbation. This allows the inverse-weighted perturbation argument
used later to compare the empirical factor space with
$\operatorname{span}(Q_0)$.

Using the rotated representation of $\widehat{\Sigma}_{y,T}$ in
\eqref{eq:scaling_hat_sigma_y}, define
\begin{align}
	\widetilde{\Sigma}_{\widetilde{Qy},T}^{DS}
	&=
	\begin{pmatrix}
		R_0\widetilde{\Sigma}_{x}R_0^T
		+
		Q_0^T\widetilde{\Sigma}_{u}Q_0
		&
		0
		\\
		0
		&
		Q_{\perp,0}^T
		\widetilde{\Sigma}_{u}
		Q_{\perp,0}
	\end{pmatrix}.
	\label{eq:scaling_tilde_Sigma_Qy}
\end{align}
Moreover, let
\[
\widetilde{\Sigma}_{\tilde y,T}^{DS}
=
Q_{all,0}
\widetilde{\Sigma}_{\widetilde{Qy},T}^{DS}
Q_{all,0}^T,
\qquad
\Delta\Sigma_{y,T}^{DS}
=
\widehat{\Sigma}_{y,T}
-
\widetilde{\Sigma}_{\tilde y,T}^{DS}.
\]
Using \eqref{eq:scaling_hat_sigma_y}, the perturbation in rotated
coordinates is
\begin{align}
	Q_{all,0}^T
	\Delta\Sigma_{y,T}^{DS}
	Q_{all,0}
	&=
	\begin{pmatrix}
		R_0\widetilde{\Sigma}_{xu}Q_0
		+
		Q_0^T\widetilde{\Sigma}_{ux}R_0^T
		&
		R_0\widetilde{\Sigma}_{xu}Q_{\perp,0}
		+
		Q_0^T\widetilde{\Sigma}_{u}Q_{\perp,0}
		\\
		Q_{\perp,0}^T\widetilde{\Sigma}_{ux}R_0^T
		+
		Q_{\perp,0}^T\widetilde{\Sigma}_{u}Q_0
		&
		0
	\end{pmatrix}.
	\label{eq:DS_rotated_perturbation}
\end{align}
The lower-right block is zero because the reference matrix retains
the corresponding block of
$Q_{all,0}^T\widehat{\Sigma}_{y,T}Q_{all,0}$ unchanged.

\begin{lemma}[Relative perturbation bound under diminishing signals]
	\label{L:Sigma_y_hat_tilde_diff_DS}
	Suppose Assumptions~\ref{ass:factors}-\ref{ass:memory} and
	\ref{ass:diminishing_signals} hold, and let $N,T\to\infty$ with
	$N/T\to0$. Then
	\[
	\norm{
		\left(
		\widetilde{\Sigma}_{\tilde y,T}^{DS}
		\right)^{-1/2}
		\Delta\Sigma_{y,T}^{DS}
		\widehat{\Sigma}_{y,T}^{-1/2}
	}_F
	=
	o_p(1).
	\]
\end{lemma}

\begin{proof}
	The proof follows the relative perturbation argument of
	Lemma~\ref{L:Sigma_y_hat_tilde_bd_diff_N}. By
	\eqref{eq:diminishing_sample_factor_block_eigenvalues}, the factor
	block of $\widehat A$ in
	\eqref{eq:scaling_hat_sigma_y_normalized} has eigenvalues of order
	one, so no additional $T$-dependent normalization is required.
	
	Define
	\[
	\widetilde A
	=
	\frac{1}{N}
	\widetilde{\Sigma}_{\widetilde{Qy},T}^{DS},
	\qquad
	\Delta A
	=
	\widehat A-\widetilde A
	=
	\frac{1}{N}
	Q_{all,0}^T
	\Delta\Sigma_{y,T}^{DS}
	Q_{all,0}.
	\]
	As in Lemma~\ref{L:Sigma_y_hat_tilde_bd_diff_N}, let
	\[
	W_N
	=
	\begin{pmatrix}
		I_{s_0} & 0
		\\
		0 & N^{1/2}I_{N-s_0}
	\end{pmatrix}.
	\]
	
	By \eqref{eq:strong_factors_RR} and
	\eqref{eq:diminishing_factor_idiosyncratic_Frobenius},
	\[
	\norm{R_0}_F
	=
	O\left(
	N^{1/2}
	\right),
	\qquad
	\norm{
		\widetilde{\Sigma}_{xu}
	}_F
	=
	O_p\left(
	\sqrt{\frac{N}{T}}
	\right).
	\]
	Hence, the factor block of $\Delta A$ satisfies
	\[
	\norm{
		\frac{1}{N}
		\left(
		R_0\widetilde{\Sigma}_{xu}Q_0
		+
		Q_0^T\widetilde{\Sigma}_{ux}R_0^T
		\right)
	}_F
	\leq
	\frac{2}{N}
	\norm{R_0}_F
	\norm{
		\widetilde{\Sigma}_{xu}
	}_F
	=
	O_p\left(
	T^{-1/2}
	\right).
	\]
	
	For the factor-idiosyncratic contribution to the off-diagonal
	block,
	\[
	N^{1/2}
	\norm{
		\frac{1}{N}
		Q_{\perp,0}^T
		\widetilde{\Sigma}_{ux}
		R_0^T
	}_F
	\leq
	\frac{1}{N^{1/2}}
	\norm{
		\widetilde{\Sigma}_{ux}
	}_F
	\norm{R_0}_F
	=
	O_p\left(
	\sqrt{\frac{N}{T}}
	\right).
	\]
	Moreover, Assumption~\ref{ass:Eigenvalue_Sigma_u_tilde} implies
	$\norm{\widetilde{\Sigma}_{u}}=O_p(1)$ and therefore
	\[
	N^{1/2}
	\norm{
		\frac{1}{N}
		Q_{\perp,0}^T
		\widetilde{\Sigma}_{u}
		Q_0
	}_F
	\leq
	\sqrt{\frac{s_0}{N}}
	\norm{
		\widetilde{\Sigma}_{u}
	}
	=
	O_p\left(
	N^{-1/2}
	\right).
	\]
	
	Using \eqref{eq:DS_rotated_perturbation}, symmetry, and the
	blockwise Frobenius norm identity, it follows that
	\[
	\norm{
		W_N\Delta A W_N
	}_F^2
	=
	O_p\left(
	T^{-1}
	\right)
	+
	O_p\left(
	\frac{N}{T}
	\right)
	+
	O_p\left(
	N^{-1}
	\right)
	=
	o_p(1),
	\]
	as $N/T \to 0$.
	
	Consequently,
	\begin{align}
		\norm{
			W_N\Delta A W_N
		}_F
		&=
		o_p(1).
		\label{eq:DS_scaled_perturbation}
	\end{align}
	
	Next,
	\[
	W_N\widetilde A W_N
	=
	\begin{pmatrix}
		\displaystyle
		\frac{1}{N}
		\left(
		R_0\widetilde{\Sigma}_{x}R_0^T
		+
		Q_0^T\widetilde{\Sigma}_{u}Q_0
		\right)
		&
		0
		\\
		0
		&
		Q_{\perp,0}^T
		\widetilde{\Sigma}_{u}
		Q_{\perp,0}
	\end{pmatrix}.
	\]
	The factor block of this matrix differs from
	$\widehat A_{xx}(N,T)$ by the factor block of $\Delta A$, whose
	Frobenius norm is $O_p(T^{-1/2})$. Hence,
	\eqref{eq:diminishing_sample_factor_block_eigenvalues} and Weyl's
	inequality imply that its eigenvalues are bounded away from zero
	and infinity in probability.
	
	The idiosyncratic block satisfies
	\[
	Q_{\perp,0}^T
	\widetilde{\Sigma}_{u}
	Q_{\perp,0}
	=
	N\widehat A_{uu}(N,T).
	\]
	Thus, \eqref{eq:diminishing_idiosyncratic_block_eigenvalues}
	implies that its eigenvalues are also bounded away from zero and
	infinity in probability. Therefore,
	\begin{align}
		\EW{1}{
			\left(
			W_N\widetilde A W_N
			\right)^{-1}
		}
		&=
		O_p(1).
		\label{eq:DS_reference_inverse}
	\end{align}
	
	Since
	\[
	W_N\widehat A W_N
	=
	W_N\widetilde A W_N
	+
	W_N\Delta A W_N,
	\]
	Weyl's inequality gives
	\[
	\EW{N}{
		W_N\widehat A W_N
	}
	\geq
	\EW{N}{
		W_N\widetilde A W_N
	}
	-
	\norm{
		W_N\Delta A W_N
	}_F.
	\]
	Combining this with
	\eqref{eq:DS_scaled_perturbation} and
	\eqref{eq:DS_reference_inverse} yields
	\begin{align*}
		\EW{1}{
			\left(
			W_N\widehat A W_N
			\right)^{-1}
		}
		&=
		O_p(1).
	\end{align*}
	
	By orthogonality of $Q_{all,0}$, cyclicity of the trace, and the
	same trace reduction as in \eqref{eq:trace_reduction_A},
	\[
	\begin{aligned}
		&\norm{
			\left(
			\widetilde{\Sigma}_{\tilde y,T}^{DS}
			\right)^{-1/2}
			\Delta\Sigma_{y,T}^{DS}
			\widehat{\Sigma}_{y,T}^{-1/2}
		}_F^2
		\\
		&\qquad=
		\tr\left[
		\left(
		W_N\widetilde A W_N
		\right)^{-1}
		W_N\Delta A W_N
		\left(
		W_N\widehat A W_N
		\right)^{-1}
		W_N\Delta A W_N
		\right]
		\\
		&\qquad\leq
		\EW{1}{
			\left(
			W_N\widetilde A W_N
			\right)^{-1}
		}
		\EW{1}{
			\left(
			W_N\widehat A W_N
			\right)^{-1}
		}
		\norm{
			W_N\Delta A W_N
		}_F^2
		\\
		&\qquad=
		O_p(1)\,
		O_p(1)\,
		o_p(1)
		=
		o_p(1).
	\end{aligned}
	\]
	Taking square roots proves the claim.
\end{proof}
We now use the preceding relative perturbation bound to establish
consistency of the full factor space under diminishing signals. Recall
from \eqref{eq:scaling_tilde_Sigma_Qy} that
$\widetilde{\Sigma}_{\tilde y,T}^{DS}$ is block diagonal in the basis
$Q_{all,0}$, with factor block
\[
R_0\widetilde{\Sigma}_{x}R_0^T
+
Q_0^T\widetilde{\Sigma}_{u}Q_0
\]
and idiosyncratic block
\[
Q_{\perp,0}^T
\widetilde{\Sigma}_{u}
Q_{\perp,0}.
\]

We first determine the eigenvalue orders of these two blocks. From
\eqref{eq:strong_factors_RR} and
\eqref{eq:diminishing_factor_idiosyncratic_Frobenius},
\[
\norm{R_0}_F
=
O\left(N^{1/2}\right),
\qquad
\norm{\widetilde{\Sigma}_{xu}}_F
=
O_p\left(\sqrt{\frac{N}{T}}\right).
\]
Hence, by the block representation of $\widehat A_{xx}(N,T)$,
\[
\begin{aligned}
	&\norm{
		\widehat A_{xx}(N,T)
		-
		\frac{1}{N}
		\left(
		R_0\widetilde{\Sigma}_{x}R_0^T
		+
		Q_0^T\widetilde{\Sigma}_{u}Q_0
		\right)
	}_F
	\\
	&\qquad\leq
	\frac{2}{N}
	\norm{R_0}_F
	\norm{\widetilde{\Sigma}_{xu}}_F
	=
	O_p\left(T^{-1/2}\right)
	=
	o_p(1).
\end{aligned}
\]
Together with
\eqref{eq:diminishing_sample_factor_block_eigenvalues} and Weyl's
inequality, this gives
\begin{align*}
	\EW{j}{
		R_0\widetilde{\Sigma}_{x}R_0^T
		+
		Q_0^T\widetilde{\Sigma}_{u}Q_0
	}
	&=
	O_e(N),
	\qquad
	j=1,\ldots,s_0.
\end{align*}

For the idiosyncratic block, by definition,
\[
Q_{\perp,0}^T
\widetilde{\Sigma}_{u}
Q_{\perp,0}
=
N\widehat A_{uu}(N,T).
\]
Therefore,
\eqref{eq:diminishing_idiosyncratic_block_eigenvalues} directly yields
\begin{align}
	\EW{j}{
		Q_{\perp,0}^T
		\widetilde{\Sigma}_{u}
		Q_{\perp,0}
	}
	&=
	O_e(1),
	\qquad
	j=1,\ldots,N-s_0.
	\label{eq:DS_reference_idiosyncratic_eigenvalues}
\end{align}

Thus, the factor block contains $s_0$ eigenvalues of order $N$,
whereas all eigenvalues of the idiosyncratic block are of order one.
Since $\widetilde{\Sigma}_{\tilde y,T}^{DS}$ is block diagonal in the
basis $Q_{all,0}$, its $s_0$ leading eigenvectors therefore belong to
the factor block with probability approaching one. Rotating this
eigenspace back to the original coordinates gives
\begin{align}
	\operatorname{span}\left[
	V\left(
	\widetilde{\Sigma}_{\tilde y,T}^{DS}
	\right)_{1:s_0}
	\right]
	&=
	\operatorname{span}(Q_0)
	\label{eq:DS_reference_factor_space}
\end{align}
with probability approaching one.

We also require the order of the leading eigenvalues of the sample
covariance matrix. By
\eqref{eq:diminishing_xu_block_convergence} and
\eqref{eq:diminishing_uu_block_convergence},
\[
\norm{
	\widehat A
	-
	\begin{pmatrix}
		\widehat A_{xx}(N,T) & 0
		\\
		0 & 0
	\end{pmatrix}
}_F
=
o_p(1).
\]
Hence, Weyl's inequality and
\eqref{eq:diminishing_sample_factor_block_eigenvalues} give
\[
\EW{j}{\widehat A}
=
O_e(1),
\qquad
j=1,\ldots,s_0.
\]
Since
\[
\widehat{\Sigma}_{y,T}
=
NQ_{all,0}\widehat A Q_{all,0}^T,
\]
and $Q_{all,0}$ is orthogonal, it follows that
\begin{align}
	\EW{j}{
		\widehat{\Sigma}_{y,T}
	}
	&=
	O_e(N),
	\qquad
	j=1,\ldots,s_0.
	\label{eq:DS_sample_factor_eigenvalues}
\end{align}

Accordingly, define the relative gap between the leading sample
eigenvalues and the complementary eigenvalues of the reference matrix
by
\begin{align*}
	\rho_{0,N}^{DS}
	=
	\min_{\substack{
			\hat\kappa\in\EW{1:s_0}{\widehat{\Sigma}_{y,T}}
			\\
			\tilde\kappa\in
			\EW{(s_0+1):N}{\widetilde{\Sigma}_{\tilde y,T}^{DS}}
	}}
	\frac{
		\left|
		\tilde\kappa-\hat\kappa
		\right|
	}{
		\sqrt{\tilde\kappa\hat\kappa}
	}.
\end{align*}
By \eqref{eq:DS_reference_idiosyncratic_eigenvalues} and
\eqref{eq:DS_sample_factor_eigenvalues},
$\hat\kappa=O_e(N)$ and $\tilde\kappa=O_e(1)$. Hence,
\[
\left|
\tilde\kappa-\hat\kappa
\right|
=
O_e(N),
\qquad
\sqrt{\tilde\kappa\hat\kappa}
=
O_e\left(N^{1/2}\right),
\]
and therefore
\begin{align}
	\rho_{0,N}^{DS}
	&=
	O_e\left(N^{1/2}\right),
	\qquad
	\left(\rho_{0,N}^{DS}\right)^{-1}
	=
	O_p\left(N^{-1/2}\right).
	\label{eq:DS_relative_gap}
\end{align}

We can now establish consistency of the full factor space under
diminishing signals.

\begin{theorem}[Factor-space consistency under diminishing signals]
	\label{T:Factor_space_DS}
	Suppose Assumptions~\ref{ass:factors}-\ref{ass:memory} and
	\ref{ass:diminishing_signals} hold, and let $N,T\to\infty$ with
	$N/T\to0$. Then
	\[
	\norm{
		\sin\Theta\left(
		V\left(
		\widehat{\Sigma}_{y,T}
		\right)_{1:s_0},
		Q_0
		\right)
	}_F
	=
	o_p\left(N^{-1/2}\right).
	\]
\end{theorem}

\begin{proof}
	By \eqref{eq:DS_reference_factor_space} and
	\citet[Theorem~2.4]{Dopico2000},
	\[
	\norm{
		\sin\Theta\left(
		V\left(
		\widehat{\Sigma}_{y,T}
		\right)_{1:s_0},
		Q_0
		\right)
	}_F
	\leq
	\left(\rho_{0,N}^{DS}\right)^{-1}
	\norm{
		\left(
		\widetilde{\Sigma}_{\tilde y,T}^{DS}
		\right)^{-1/2}
		\Delta\Sigma_{y,T}^{DS}
		\widehat{\Sigma}_{y,T}^{-1/2}
	}_F.
	\]
	By \eqref{eq:DS_relative_gap}, the first term on the right-hand
	side is $O_p(N^{-1/2})$, while
	Lemma~\ref{L:Sigma_y_hat_tilde_diff_DS} gives $o_p(1)$ for the
	second term. Therefore,
	\[
	\norm{
		\sin\Theta\left(
		V\left(
		\widehat{\Sigma}_{y,T}
		\right)_{1:s_0},
		Q_0
		\right)
	}_F
	=
	o_p\left(N^{-1/2}\right).
	\]
	This proves the claim.
\end{proof}
\begin{proof}[Proof of Theorem~\ref{T:factor_convergence_DS}]
	Define
	\[
	H_{0,T}
	=
	N^{-1/2}
	V\left(
	\widehat{\Sigma}_{y,T}
	\right)_{1:s_0}^T
	\Lambda_0.
	\]
	Since $y_t=\Lambda_0x_t+u_t$,
	\[
	\widehat{x}_{s_0,t}
	=
	H_{0,T}x_t+e_{0,t},
	\qquad
	e_{0,t}
	=
	N^{-1/2}
	V\left(
	\widehat{\Sigma}_{y,T}
	\right)_{1:s_0}^T
	u_t.
	\]
	
	To obtain the pointwise bound, use
	$I_N=Q_0Q_0^T+Q_{\perp,0}Q_{\perp,0}^T$ to write
	\begin{align*}
		e_{0,t}
		&=
		N^{-1/2}
		V\left(
		\widehat{\Sigma}_{y,T}
		\right)_{1:s_0}^T
		Q_0Q_0^Tu_t
		+
		N^{-1/2}
		V\left(
		\widehat{\Sigma}_{y,T}
		\right)_{1:s_0}^T
		Q_{\perp,0}Q_{\perp,0}^Tu_t.
	\end{align*}
	The bounds in
	\eqref{eq:factor_convergence_projected_u_pointwise} remain valid
	under diminishing signals because the idiosyncratic component is
	not rescaled. Hence,
	\[
	\norm{Q_0^Tu_t}_F=O_p(1),
	\qquad
	\norm{Q_{\perp,0}^Tu_t}_F=O_p\left(N^{1/2}\right).
	\]
	Moreover, Theorem~\ref{T:Factor_space_DS} gives
	\[
	\norm{
		V\left(
		\widehat{\Sigma}_{y,T}
		\right)_{1:s_0}^T
		Q_{\perp,0}
	}_F
	=
	o_p\left(N^{-1/2}\right).
	\]
	Since
	\[
	\norm{
		V\left(
		\widehat{\Sigma}_{y,T}
		\right)_{1:s_0}^T
		Q_0
	}
	\leq
	1,
	\]
	the two terms in the decomposition of $e_{0,t}$ are respectively
	$O_p(N^{-1/2})$ and $o_p(N^{-1/2})$. Therefore,
	\[
	\norm{e_{0,t}}_F
	=
	O_p\left(N^{-1/2}\right).
	\]
	
	For the average error,
	\begin{align*}
		\frac{1}{T}
		\sum_{t=1}^T
		\norm{e_{0,t}}_F^2
		&=
		\frac{1}{N}
		\tr\left[
		V\left(
		\widehat{\Sigma}_{y,T}
		\right)_{1:s_0}^T
		\widetilde{\Sigma}_{u}
		V\left(
		\widehat{\Sigma}_{y,T}
		\right)_{1:s_0}
		\right]
		\\
		&\leq
		\frac{s_0}{N}
		\EW{1}{
			\widetilde{\Sigma}_{u}
		}
		=
		O_p\left(N^{-1}\right),
	\end{align*}
	by Assumption~\ref{ass:Eigenvalue_Sigma_u_tilde}.
	
	It remains to verify the well-conditioning of $H_{0,T}$. Writing
	$\Lambda_0=Q_0R_0$ gives
	\[
	H_{0,T}
	=
	V\left(
	\widehat{\Sigma}_{y,T}
	\right)_{1:s_0}^T
	Q_0
	\left(
	N^{-1/2}R_0
	\right).
	\]
	By Theorem~\ref{T:Factor_space_DS}, the canonical angles between
	$V(\widehat{\Sigma}_{y,T})_{1:s_0}$ and $Q_0$ converge to zero.
	Hence, the singular values of
	\[
	V\left(
	\widehat{\Sigma}_{y,T}
	\right)_{1:s_0}^T
	Q_0
	\]
	converge to one in probability. Moreover,
	\eqref{eq:factor_convergence_loading_limit} gives
	\[
	\frac{1}{N}R_0^TR_0
	\longrightarrow
	\Sigma_{\Lambda,0}.
	\]
	Since $\Sigma_{\Lambda,0}$ is positive definite, the singular
	values of $N^{-1/2}R_0$ are bounded away from zero and infinity.
	It follows that the singular values of $H_{0,T}$ are bounded away
	from zero and infinity in probability.
\end{proof}

\begin{proof}[Proof of Corollary~\ref{C:Bai_Ng_DS}]
	Since $N/T\to0$, we have $\min\{N,T\}=N$ for all sufficiently
	large $N,T$. Hence, the penalty conditions imply
	\[
	g(N,T)\to0,
	\qquad
	Ng(N,T)\to\infty.
	\]
	
	We first rule out underestimation. For any $s<s_0$,
	\begin{align*}
		PC(s)-PC(s+1)
		&=
		\frac{1}{N}
		\EW{s+1}{
			\widehat{\Sigma}_{y,T}
		}
		-
		g(N,T)
		\\
		&\geq
		\frac{1}{N}
		\EW{s_0}{
			\widehat{\Sigma}_{y,T}
		}
		-
		g(N,T).
	\end{align*}
	By \eqref{eq:DS_sample_factor_eigenvalues},
	\[
	\frac{1}{N}
	\EW{s_0}{
		\widehat{\Sigma}_{y,T}
	}
	=
	O_e(1).
	\]
	Since $g(N,T)\to0$,
	\[
	PC(s)-PC(s+1)>0
	\]
	with probability approaching one. Since $s_0$ is fixed, this
	holds simultaneously for all $s=0,\ldots,s_0-1$. Therefore,
	\[
	\Pr\left(
	\widehat{s}<s_0
	\right)
	\longrightarrow
	0.
	\]
	
	We next rule out overestimation. For any candidate dimension $s$,
	define
	\[
	\widehat P_s
	=
	V\left(
	\widehat{\Sigma}_{y,T}
	\right)_{1:s}
	V\left(
	\widehat{\Sigma}_{y,T}
	\right)_{1:s}^T.
	\]
	The matrix $\widehat P_s$ is the orthogonal projector onto the
	space spanned by the first $s$ empirical principal components. In
	particular,
	\[
	\widehat P_s^T=\widehat P_s,
	\qquad
	\widehat P_s^2=\widehat P_s,
	\qquad
	\operatorname{rank}\left(
	\widehat P_s
	\right)
	=
	s.
	\]
	The fitted common component associated with $s$ principal
	components is
	\[
	\widehat{\chi}_{s,t}
	=
	\widehat P_s y_t.
	\]
	At the true factor dimension $s_0$, using
	$y_t=\Lambda_0x_t+u_t$ gives
	\begin{align}
		\Lambda_0x_t-\widehat{\chi}_{s_0,t}
		&=
		\Lambda_0x_t
		-
		\widehat P_{s_0}
		\left(
		\Lambda_0x_t+u_t
		\right)
		\nonumber\\
		&=
		\left(
		I_N-\widehat P_{s_0}
		\right)
		\Lambda_0x_t
		-
		\widehat P_{s_0}u_t.
		\label{eq:DS_common_component_decomposition}
	\end{align}
	Thus, the common-component estimation error consists of the part of
	the true common component that lies outside the estimated factor
	space and the idiosyncratic component that is projected into that
	space.
	
	We bound these two terms separately. Since
	$\Lambda_0=Q_0R_0$,
	\begin{align*}
		&\frac{1}{NT}
		\sum_{t=1}^T
		\norm{
			\left(
			I_N-\widehat P_{s_0}
			\right)
			\Lambda_0x_t
		}_F^2
		\\
		&\qquad=
		\frac{1}{N}
		\tr\left[
		\left(
		I_N-\widehat P_{s_0}
		\right)
		Q_0
		R_0\widetilde{\Sigma}_{x}R_0^T
		Q_0^T
		\left(
		I_N-\widehat P_{s_0}
		\right)
		\right]
		\\
		&\qquad\leq
		\frac{1}{N}
		\norm{
			\left(
			I_N-\widehat P_{s_0}
			\right)
			Q_0
		}_F^2
		\norm{
			R_0\widetilde{\Sigma}_{x}R_0^T
		}_2.
	\end{align*}
	By Theorem~\ref{T:Factor_space_DS},
	\[
	\norm{
		\left(
		I_N-\widehat P_{s_0}
		\right)
		Q_0
	}_F
	=
	o_p\left(
	N^{-1/2}
	\right).
	\]
	Moreover, Lemma~\ref{L:diminishing_factor_second_moments} and
	Assumption~\ref{ass:strong_factors} imply
	\[
	\norm{
		R_0\widetilde{\Sigma}_{x}R_0^T
	}_2
	\leq
	\norm{R_0}_2^2
	\norm{
		\widetilde{\Sigma}_{x}
	}_2
	=
	O_p(N).
	\]
	Hence,
	\[
	\frac{1}{NT}
	\sum_{t=1}^T
	\norm{
		\left(
		I_N-\widehat P_{s_0}
		\right)
		\Lambda_0x_t
	}_F^2
	=
	o_p\left(
	N^{-1}
	\right).
	\]
	
	For the projected idiosyncratic component,
	\begin{align*}
		\frac{1}{NT}
		\sum_{t=1}^T
		\norm{
			\widehat P_{s_0}u_t
		}_F^2
		&=
		\frac{1}{N}
		\tr\left(
		\widehat P_{s_0}
		\widetilde{\Sigma}_{u}
		\right)
		\leq
		\frac{s_0}{N}
		\EW{1}{
			\widetilde{\Sigma}_{u}
		}
		=
		O_p\left(
		N^{-1}
		\right),
	\end{align*}
	by Assumption~\ref{ass:Eigenvalue_Sigma_u_tilde}. Combining these
	bounds with \eqref{eq:DS_common_component_decomposition} yields
	\begin{align}
		\frac{1}{NT}
		\sum_{t=1}^T
		\norm{
			\Lambda_0x_t-\widehat{\chi}_{s_0,t}
		}_F^2
		&=
		O_p\left(
		N^{-1}
		\right).
		\label{eq:DS_common_component_rate}
	\end{align}
	
	Now fix $s>s_0$. Since the empirical principal-component spaces
	are nested,
	\[
	\widehat P_s\widehat P_{s_0}
	=
	\widehat P_{s_0}\widehat P_s
	=
	\widehat P_{s_0}.
	\]
	Therefore,
	$\widehat P_s-\widehat P_{s_0}$ is the orthogonal projector onto
	the additional $s-s_0$ empirical principal components and has rank
	$s-s_0$. Moreover,
	\[
	\widehat{\chi}_{s,t}
	-
	\widehat{\chi}_{s_0,t}
	=
	\left(
	\widehat P_s-\widehat P_{s_0}
	\right)
	y_t.
	\]
	The residual
	\[
	y_t-\widehat{\chi}_{s,t}
	=
	\left(
	I_N-\widehat P_s
	\right)
	y_t
	\]
	is orthogonal to
	$\widehat{\chi}_{s,t}-\widehat{\chi}_{s_0,t}$. Hence, the
	Pythagorean identity gives
	\begin{align*}
		\mathcal V(s_0)-\mathcal V(s)
		&=
		\frac{1}{NT}
		\sum_{t=1}^T
		\norm{
			\widehat{\chi}_{s,t}
			-
			\widehat{\chi}_{s_0,t}
		}_F^2.
	\end{align*}
	
	To bound this difference, use $y_t=\Lambda_0x_t+u_t$ and add and
	subtract $\widehat{\chi}_{s_0,t}$. Since
	\[
	\left(
	\widehat P_s-\widehat P_{s_0}
	\right)
	\widehat{\chi}_{s_0,t}
	=
	\left(
	\widehat P_s-\widehat P_{s_0}
	\right)
	\widehat P_{s_0}y_t
	=
	0,
	\]
	we obtain
	\begin{align*}
		\widehat{\chi}_{s,t}
		-
		\widehat{\chi}_{s_0,t}
		&=
		\left(
		\widehat P_s-\widehat P_{s_0}
		\right)
		\left(
		\Lambda_0x_t-\widehat{\chi}_{s_0,t}
		\right)
		\\
		&\quad+
		\left(
		\widehat P_s-\widehat P_{s_0}
		\right)
		u_t.
	\end{align*}
	Because an orthogonal projector is a contraction,
	\begin{align*}
		\mathcal V(s_0)- \mathcal V(s)
		&\leq
		\frac{2}{NT}
		\sum_{t=1}^T
		\norm{
			\Lambda_0x_t-\widehat{\chi}_{s_0,t}
		}_F^2
		\\
		&\quad+
		\frac{2}{NT}
		\sum_{t=1}^T
		\norm{
			\left(
			\widehat P_s-\widehat P_{s_0}
			\right)
			u_t
		}_F^2.
	\end{align*}
	The first term is $O_p(N^{-1})$ by
	\eqref{eq:DS_common_component_rate}. For the second term,
	\begin{align*}
		&\frac{1}{NT}
		\sum_{t=1}^T
		\norm{
			\left(
			\widehat P_s-\widehat P_{s_0}
			\right)
			u_t
		}_F^2
		=
		\frac{1}{N}
		\tr\left[
		\left(
		\widehat P_s-\widehat P_{s_0}
		\right)
		\widetilde{\Sigma}_{u}
		\right]
		\leq
		\frac{s-s_0}{N}
		\EW{1}{
			\widetilde{\Sigma}_{u}
		}
		=
		O_p\left(
		N^{-1}
		\right),
	\end{align*}
	by Assumption~\ref{ass:Eigenvalue_Sigma_u_tilde}. Thus,
	\[
	\mathcal V(s_0)-\mathcal V(s)
	=
	O_p\left(
	N^{-1}
	\right).
	\]
	Since $s_{\max}$ is fixed,
	\[
	\max_{s_0<s\leq s_{\max}}
	\left[
	\mathcal V(s_0)- \mathcal V(s)
	\right]
	=
	O_p\left(
	N^{-1}
	\right).
	\]
	Consequently, uniformly over $s=s_0+1,\ldots,s_{\max}$,
	\[
	PC(s)-PC(s_0)
	=
	(s-s_0)g(N,T)
	-
	O_p\left(
	N^{-1}
	\right).
	\]
	Since $Ng(N,T)\to\infty$, the penalty dominates the reduction in
	the value function. Therefore,
	\[
	\Pr\left(
	\widehat{s}>s_0
	\right)
	\longrightarrow
	0.
	\]
	
	Combining the underestimation and overestimation results gives
	\[
	\Pr\left(
	\widehat{s}=s_0
	\right)
	\longrightarrow
	1.
	\]
\end{proof}

\begin{proof}[Proof of Corollary~\ref{C:IC_criteria_DS}]
The proof of Corollary~\ref{C:IC_criteria} applies with two replacements: the lower bound on $N^{-1}\EW{s_0}{\widehat{\Sigma}_{y,T}}$ follows from \eqref{eq:DS_sample_factor_eigenvalues}, and $\max_{s_0<s\leq s_{\max}}\left[\mathcal V(s_0)-\mathcal V(s)\right]=O_p\left(N^{-1}\right)$ is established in the proof of Corollary~\ref{C:Bai_Ng_DS}.
The two bounds on the value function use only Assumptions~\ref{ass:Lambda_full_rank} and \ref{ass:Eigenvalue_Sigma_u_tilde} and the fact that the common component $\chi_t=\Lambda_0x_{t,T}$ lies in the column space of $\Lambda_0$; with $x_{t,T}$ in place of $x_t$, they hold unchanged.
\end{proof}
 
\subsection{Differencing under diminishing signals}

We finally consider the effect of fractionally differencing the panel
by the largest memory parameter $d_0^{(1)}$. For a nonstationary factor
$i=1,\ldots,s_0^*$, \eqref{eq:diminishing_factor_scaling} gives
\[
\Delta_+^{d_0^{(1)}}x_{i,t}
=
T^{-d_{0,i}+1/2}
\Delta_+^{d_0^{(1)}}
\mathcal B_{i,+}(L;d_{0,i})\eta_{i,t}.
\]
Since
\[
\mathcal B_i(z;d_{0,i})
=
\psi_i(z)(1-z)^{-d_{0,i}},
\]
the corresponding infinite filter has transfer function
\[
\psi_i(z)(1-z)^{d_0^{(1)}-d_{0,i}}.
\]
By Assumption~\ref{ass:memory},
$d_0^{(1)}-d_{0,i}\geq0$. Hence, the coefficients of
$\Delta^{d_0^{(1)}-d_{0,i}}$ are absolutely summable
\citep[Lemma~5.2]{Hassler2018}. Together with
Assumption~\ref{ass:short_run_filters}, this implies that the
convolution
$\Delta^{d_0^{(1)}-d_{0,i}}\psi_i(L)$ also has absolutely summable
coefficients. Moreover, its coefficient at lag zero equals one because
$\psi_{i,0}=1$.

To make the role of the type-II initialization explicit, consider the
stationary type-I process obtained by applying the same infinite-past
filter $\Delta^{d_0^{(1)}-d_{0,i}}\psi_i(L)$ to $\eta_{i,t}$. The
coefficients of this filter are square summable, so temporal
uncorrelatedness of the innovations implies that this process has
finite variance. Its variance is strictly positive because the
coefficient at lag zero equals one and $\EE{\eta_{i,t}^2}>0$. At time
$t$, the observed type-II process retains only the first $t$ filter
coefficients, whereas the type-I process also contains the pre-sample
terms. The mean-square contribution of these omitted terms vanishes
on average because the filter coefficients are square summable, by the
same truncation argument as in
Lemma~\ref{L:stationary_filter_coefficients}. Stationarity and
ergodicity therefore give
\begin{align*}
	T^{2d_{0,i}-1}
	\frac{1}{T}
	\sum_{t=1}^T
	\left(
	\Delta_+^{d_0^{(1)}}x_{i,t}
	\right)^2
	&\xrightarrow{p}
	\EE{
		\left(
		\Delta^{d_0^{(1)}-d_{0,i}}
		\psi_i(L)\eta_{i,t}
		\right)^2
	}
	\in(0,\infty).
\end{align*}
Consequently,
\begin{align}
	\frac{1}{T}
	\sum_{t=1}^T
	\left(
	\Delta_+^{d_0^{(1)}}x_{i,t}
	\right)^2
	&=
	O_e\left(
	T^{1-2d_{0,i}}
	\right).
	\label{eq:DS_differenced_factor_second_moment}
\end{align}

By Assumption~\ref{ass:strong_factors}, the squared norm of the loading
vector corresponding to factor $i$ is of order $N$. Hence, the unique
nonzero eigenvalue of the rank-one common covariance contribution
generated by this factor is of order
\begin{align*}
	O_e\left(
	NT^{1-2d_{0,i}}
	\right)
	&=
	O_e\left(
	\frac{N}{T^{2d_{0,i}-1}}
	\right).
\end{align*}
Thus, differencing removes the variance accumulation that keeps the
nonstationary factors of order one under diminishing signals in
levels. In particular, if
\[
\frac{N}{T^{2d_{0,i}-1}}
\longrightarrow
0,
\]
the common covariance contribution generated by factor $i$ vanishes
in spectral norm.

It remains to compare this contribution with the idiosyncratic
spectrum after differencing. No additional upper eigenvalue condition
is required. Since $d_0^{(1)}>0$, the coefficients of
$\Delta^{d_0^{(1)}}$ are absolutely summable
\citep[Lemma~5.2]{Hassler2018}. Therefore, after extending
$a^Tu_t$ by zero outside the sample, the convolution inequality gives,
for any $a\in\mathbb R^N$ with $\norm{a}_2=1$,
\begin{align*}
	\frac{1}{T}
	\sum_{t=1}^T
	\left(
	a^T\Delta_+^{d_0^{(1)}}u_t
	\right)^2
	&\leq
	\left(
	\sum_{\ell=0}^{\infty}
	\left|
	\pi_\ell\left(d_0^{(1)}\right)
	\right|
	\right)^2
	\frac{1}{T}
	\sum_{t=1}^T
	\left(
	a^Tu_t
	\right)^2.
\end{align*}
Taking the supremum over $\norm{a}_2=1$ and using
Assumption~\ref{ass:Eigenvalue_Sigma_u_tilde} yields
\begin{align}
	\EW{1}{
		\frac{1}{T}
		\sum_{t=1}^T
		\left(
		\Delta_+^{d_0^{(1)}}u_t
		\right)
		\left(
		\Delta_+^{d_0^{(1)}}u_t
		\right)^T
	}
	&=
	O_p(1).
	\label{eq:DS_differenced_idiosyncratic_upper_bound}
\end{align}

\begin{proof}[Proof of Corollary~\ref{C:differencing_DS_non_detectability}]
	Fix a nonstationary factor $i\leq s_0^*$ such that
	\[
	\frac{N}{T^{2d_{0,i}-1}}\longrightarrow0.
	\]
	By \eqref{eq:DS_differenced_factor_second_moment},
	\[
	\frac{1}{T}\sum_{t=1}^T
	\left(\Delta_+^{d_0^{(1)}}x_{i,t}\right)^2
	=
	O_e\left(T^{1-2d_{0,i}}\right).
	\]
	Moreover, Assumption~\ref{ass:strong_factors} implies
	$\norm{\Lambda_0}_2^2=O(N)$. Hence,
	\begin{align}
		\norm{\Lambda_0}_2^2
		\frac{1}{T}\sum_{t=1}^T
		\left(\Delta_+^{d_0^{(1)}}x_{i,t}\right)^2
		&=
		O_p\left(\frac{N}{T^{2d_{0,i}-1}}\right)
		=
		o_p(1).
		\label{eq:DS_differenced_factor_vanishing}
	\end{align}
	
	We now relate this vanishing factor contribution to the eigenvalues
	of the sample covariance matrix of the differenced data. The
	Courant-Fischer characterization implies that, to obtain an upper
	bound for its $s_0$th eigenvalue, it is sufficient to exhibit an
	$(N-s_0+1)$-dimensional subspace on which the corresponding
	quadratic form is bounded.
	
	Choose the subspace orthogonal to the loading vectors of all
	factors except factor $i$. By
	Assumption~\ref{ass:Lambda_full_rank}, these $s_0-1$ loading
	vectors are linearly independent, so this subspace has dimension
	$N-s_0+1$. The important point is that, along any direction in
	this subspace, the common contributions of all factors other than
	$i$ vanish. Hence, only the contribution of factor $i$ and the
	idiosyncratic component remain.
	
	Formally, the Courant-Fischer characterization gives
	\begin{align*}
		&\EW{s_0}{
			\frac{1}{T}\sum_{t=1}^T
			\left(\Delta_+^{d_0^{(1)}}y_t\right)
			\left(\Delta_+^{d_0^{(1)}}y_t\right)^T
		}
		=
		\min_{\substack{\mathcal S\subset\mathbb R^N\\
				\dim(\mathcal S)=N-s_0+1}}
		\max_{\substack{a\in\mathcal S\\ \norm{a}_2=1}}
		\frac{1}{T}\sum_{t=1}^T
		\left(a^T\Delta_+^{d_0^{(1)}}y_t\right)^2.
	\end{align*}
	Since the minimum cannot exceed the value obtained for the
	particular subspace chosen above,
	\begin{align*}
		&\EW{s_0}{
			\frac{1}{T}\sum_{t=1}^T
			\left(\Delta_+^{d_0^{(1)}}y_t\right)
			\left(\Delta_+^{d_0^{(1)}}y_t\right)^T
		}
		\leq
		\max_{\substack{a\in\mathcal S\\ \norm{a}_2=1}}
		\frac{1}{T}\sum_{t=1}^T
		\left(a^T\Delta_+^{d_0^{(1)}}y_t\right)^2.
	\end{align*}
	For every such $a$, orthogonality to the other loading vectors
	implies
	\[
	a^T\Delta_+^{d_0^{(1)}}y_t
	=
	(a^T\Lambda_0)_i
	\Delta_+^{d_0^{(1)}}x_{i,t}
	+
	a^T\Delta_+^{d_0^{(1)}}u_t.
	\]
	Therefore,
	\begin{align*}
		&\max_{\substack{a\in\mathcal S\\ \norm{a}_2=1}}
		\frac{1}{T}\sum_{t=1}^T
		\left(a^T\Delta_+^{d_0^{(1)}}y_t\right)^2
		\\
		&\qquad\leq
		2\norm{\Lambda_0}_2^2
		\frac{1}{T}\sum_{t=1}^T
		\left(\Delta_+^{d_0^{(1)}}x_{i,t}\right)^2
		+
		2\EW{1}{
			\frac{1}{T}\sum_{t=1}^T
			\left(\Delta_+^{d_0^{(1)}}u_t\right)
			\left(\Delta_+^{d_0^{(1)}}u_t\right)^T
		}.
	\end{align*}
	The first term is $o_p(1)$ by
	\eqref{eq:DS_differenced_factor_vanishing}, whereas the second is
	$O_p(1)$ by
	\eqref{eq:DS_differenced_idiosyncratic_upper_bound}. Consequently,
	\begin{align*}
		\EW{s_0}{
			\frac{1}{T}\sum_{t=1}^T
			\left(\Delta_+^{d_0^{(1)}}y_t\right)
			\left(\Delta_+^{d_0^{(1)}}y_t\right)^T
		}
		&=
		O_p(1).
	\end{align*}
	
	The implication for factor-number selection now follows directly.
	By \eqref{eq:value_function_drop}, adding the $s_0$th principal
	component reduces the PCA value function by
	\[
	\frac{1}{N}
	\EW{s_0}{
		\frac{1}{T}\sum_{t=1}^T
		\left(\Delta_+^{d_0^{(1)}}y_t\right)
		\left(\Delta_+^{d_0^{(1)}}y_t\right)^T
	}
	=
	O_p(N^{-1}).
	\]
	Since $N/T\to0$, the penalty condition
	\eqref{eq:BN_penalty_conditions} implies
	$Ng(N,T)\to\infty$. Hence,
	\begin{align*}
		PC(s_0)-PC(s_0-1)
		&=
		g(N,T)
		-
		\frac{1}{N}
		\EW{s_0}{
			\frac{1}{T}\sum_{t=1}^T
			\left(\Delta_+^{d_0^{(1)}}y_t\right)
			\left(\Delta_+^{d_0^{(1)}}y_t\right)^T
		}
		\\
		&>
		0
	\end{align*}
	with probability approaching one. Thus, $s_0$ cannot minimize the
	Bai-Ng criterion with probability approaching one, and therefore
	\[
	\Pr\left(\widehat{s}=s_0\right)\longrightarrow0.
	\]
	This proves that the Bai-Ng criterion applied to
	$\Delta_+^{d_0^{(1)}}y_t$ cannot consistently recover the full
	factor dimension.
\end{proof}

\clearpage
\begin{spacing}{1}
\bibliographystyle{dcu}
\bibliography{literatur.bib}
\end{spacing}
\end{document}